\pdfoutput=1 
\documentclass[ twoside,openright,titlepage,numbers=noenddot,headinclude,
                footinclude=true,cleardoublepage=plain,abstractoff, 
                BCOR=5mm,paper=letter,pagesize,fontsize=11pt,
                american,%
                ]{scrreprt}

\PassOptionsToPackage{eulerchapternumbers,
				 pdfspacing,floatperchapter,
				 subfig,
                 eulermath,
                 parts}{classicthesis}										

\usepackage{ifthen}
\newboolean{enable-backrefs} 
\setboolean{enable-backrefs}{false} 

\newcommand{\myTitle}{Flips and Spanners\xspace}

\newcommand{\myName}{Alexander Jozef Hubertus Verdonschot\xspace}

\newcommand{\myUni}{Carleton University\xspace}

\newcommand{\myVersion}{version 4.1\xspace} 

\newenvironment{shortitemize}
    {\begin{itemize}}
    {\end{itemize}}
\newenvironment{shortenumerate}
    {\begin{enumerate}}
    {\end{enumerate}}

\newcounter{dummy} 
\providecommand{\mLyX}{L\kern-.1667em\lower.25em\hbox{Y}\kern-.125emX\@}

\PassOptionsToPackage{latin9}{inputenc}	
 \usepackage{inputenc}				

\usepackage{chapterbib}
 
\PassOptionsToPackage{canadian}{babel}   
 \usepackage{babel}					

\PassOptionsToPackage{square,numbers}{natbib}
 \usepackage{natbib}				

\PassOptionsToPackage{fleqn}{amsmath}		
 \usepackage{amsmath}

\usepackage{amssymb}
\usepackage{amsthm}

\newtheoremstyle{caps}
  {\topsep}   
  {\topsep}   
  {\itshape}  
  {0pt}       
  {}          
  {.}         
  {5pt plus 1pt minus 1pt} 
  {\spacedlowsmallcaps{\thmname{#1}\thmnumber{ #2}}\thmnote{ (#3)}} 

\theoremstyle{caps}
\newtheorem{definition}{Definition}
\newtheorem{theorem}[definition]{Theorem}
\newtheorem{lemma}[definition]{Lemma}
\newtheorem{observation}[definition]{Observation}
\newtheorem{corollary}[definition]{Corollary}
\newtheorem{conjecture}{Conjecture}

\newtheoremstyle{algo}
  {\topsep}   
  {\topsep}   
  {\upshape}  
  {0pt}       
  {}          
  {}          
  {\newline}  
  {\spacedlowsmallcaps{\thmname{#1}\thmnumber{ #2}}\thmnote{ (#3)}} 

\theoremstyle{algo}
\newtheorem{algorithm}{Algorithm}

\newenvironment{proofof}[1]{\textit{Proof of #1.}}{\hfill$\Box$}

\numberwithin{definition}{chapter} 
\numberwithin{figure}{chapter} 

\PassOptionsToPackage{T1}{fontenc} 
	\usepackage{fontenc}     
\usepackage[full]{textcomp} 
\usepackage{scrhack} 
\usepackage{xspace} 
\usepackage{mparhack} 
\usepackage{fixltx2e} 
\PassOptionsToPackage{printonlyused,smaller}{acronym}
	\usepackage{acronym} 

\usepackage{tabularx} 
\usepackage{longtable} 
\usepackage{caption}
\usepackage{subcaption}  

\usepackage{wrapfig}

\usepackage{listings} 
\PassOptionsToPackage{pdftex,hyperfootnotes=false,pdfpagelabels}{hyperref}
	\usepackage{hyperref}  
\PassOptionsToPackage{pdftex}{graphicx}
	\usepackage{graphicx} 

\graphicspath{{figures/}}

\newcommand{\backrefnotcitedstring}{\relax}
\newcommand{\backrefcitedsinglestring}[1]{(Cited on page~#1.)}
\newcommand{\backrefcitedmultistring}[1]{(Cited on pages~#1.)}
\ifthenelse{\boolean{enable-backrefs}}%
{%
		\PassOptionsToPackage{hyperpageref}{backref}
		\usepackage{backref} 
		   \renewcommand*{\backref}[1]{}  
		   \renewcommand*{\backrefalt}[4]{
		      \ifcase #1 %
		         \backrefnotcitedstring%
		      \or%
		         \backrefcitedsinglestring{#2}%
		      \else%
		         \backrefcitedmultistring{#2}%
		      \fi}%
}{\relax}    

\hypersetup{%
    colorlinks=true, linktocpage=true, pdfstartpage=3, pdfstartview=FitV,%
    breaklinks=true, pdfpagemode=UseNone, pageanchor=true, pdfpagemode=UseOutlines,%
    plainpages=false, bookmarksnumbered, bookmarksopen=true, bookmarksopenlevel=1,%
    hypertexnames=true, pdfhighlight=/O,
    urlcolor=webbrown, linkcolor=RoyalBlue, citecolor=webgreen, 
    pdftitle={\myTitle},%
    pdfauthor={\textcopyright\ 2015 \myName},%
    pdfsubject={},%
    pdfkeywords={},%
    pdfcreator={pdfLaTeX},%
    pdfproducer={LaTeX with hyperref and classicthesis}%
}   

\makeatletter
\@ifpackageloaded{babel}%
    {%
       \addto\extrasamerican{%
				}%
       \addto\extrasngerman{%
				}%
    }{\relax}
\makeatother

\usepackage{classicthesis} 

\makeatletter
\def\ttl@tocpart{%
  \def\ttl@a{\protect\numberline{\thepart}\@gobble{}}}
\makeatother

\KOMAoption{bibliography}{leveldown}

\newboolean{carleton-version}
\setboolean{carleton-version}{false}
\ifthenelse{\boolean{carleton-version}}{

\renewcommand{\href}[2]{#2}
\renewcommand{\url}[1]{#1}

}{
}

\newcommand{\etal}{et~al.\xspace}

\newcommand{\case}[1]{\paragraph{Case {\upshape #1}.}}
\newenvironment{extraStretch}[1]{
  \par
  \setlength{\emergencystretch}{#1}
}{
  \par
}

\newcommand{\T}[2]{\ensuremath{\triangle_{#1 #2}}}

\newcommand{\graph}{\ensuremath{\Theta_5}-graph\xspace}
\newcommand{\valc}{\ensuremath{2 \left( 2 + \sqrt{5} \right) \approx 8.473}\xspace}
\newcommand{\valsr}{\ensuremath{\sqrt{50 + 22 \sqrt{5}} \approx 9.960}\xspace}
\newcommand{\vallb}{\ensuremath{\frac{1}{2}(11\sqrt{5} - 17) \approx 3.799}\xspace}

\newcommand{\hts}{half-\ensuremath{\Theta_6}-graph\xspace}
\newcommand{\length}[1]{\ensuremath{|#1|}}

\newcommand{\dtw}{\ensuremath{G_{12}}\xspace}
\newcommand{\dn}{\ensuremath{G_{9}}\xspace}

\renewcommand{\c}[1]{\ensuremath{C_#1}}
\newcommand{\nc}[1]{\ensuremath{\overline{C}_#1}}

\newcommand{\sA}{\ensuremath{\mathcal{A}}\xspace}
\newcommand{\sB}{\ensuremath{\mathcal{B}}\xspace}
\newcommand{\sC}{\ensuremath{\mathcal{C}}\xspace}

\newcommand{\type}[1]{\smallskip\spacedlowsmallcaps{Type #1 (}\includegraphics{f4c/Type#1}\spacedlowsmallcaps{):}}

\newcommand{\gS}{\ensuremath{\mathcal{S}}\xspace}
\newcommand{\gI}{\ensuremath{\mathcal{C}_{\text{in}}}\xspace}
\newcommand{\gO}{\ensuremath{\mathcal{C}_{\text{out}}}\xspace}
\newcommand{\vI}{\ensuremath{v_{\text{in}}}\xspace}
\newcommand{\vO}{\ensuremath{v_{\text{out}}}\xspace}

\newcommand{\vl}{\ensuremath{v_{\textsc{l}}}}
\newcommand{\vr}{\ensuremath{v_{\textsc{r}}}}
\newcommand{\free}{\ensuremath{\mathcal{F}}}
\newcommand{\nf}{\ensuremath{\overline{F}}}

\begin{document}
\frenchspacing
\raggedbottom
\selectlanguage{canadian} 
\pagenumbering{roman}
\pagestyle{plain}
\begin{titlepage}
  \begin{addmargin}[-1cm]{-3cm}
    \begin{center}
        \large  

        \hfill

        \vfill

        \begingroup
            \color{Maroon}\spacedallcaps{\myTitle} \\ \bigskip
        \endgroup

        \spacedlowsmallcaps{\myName}

        \vfill


        {\footnotesize A thesis submitted to the Faculty of Graduate and Post Doctoral Affairs\\
         in partial fulfillment of the requirements for the degree of} \\ \medskip
        Doctor of Philosophy {\footnotesize in} Computer Science \\ \bigskip
        \myUni \\ 
        {\footnotesize Ottawa, Ontario, Canada}

        \vfill                      

        {\footnotesize \textcopyright\ 2015 \myName}
    \end{center}  
  \end{addmargin}       
\end{titlepage}   

\cleardoublepage
\phantomsection
\addcontentsline{toc}{chapter}{\tocEntry{Abstract}}
\begingroup
\let\clearpage\relax
\let\cleardoublepage\relax
\let\cleardoublepage\relax

\chapter*{Abstract}
In this thesis, we study two different graph problems.

The first problem revolves around \emph{geometric spanners}. Here, we have a set of points in the plane and we want to connect them with straight line segments, such that there is a path between each pair of points and these paths do not require large detours. If we achieve this, the resulting graph is called a spanner. We focus our attention on two graphs (the $\Theta$-graph and Yao-graph) that are constructed by connecting each point with its nearest neighbour in a number of cones. Although this construction is very straight-forward, it has proven challenging to fully determine the properties of the resulting graphs. We show that if the construction uses 5 cones, the resulting graphs are still spanners. This was the only number of cones for which this question remained unanswered. We also present a routing strategy (a way to decide where to go next, based only on our current location, its direct neighbourhood, and our destination) on the \hts, a variant of the graph with 6 cones. We show that our routing strategy avoids large detours: it finds a path whose length is at most a constant factor from the straight-line distance between the endpoints. Moreover, we show that this routing strategy is optimal.

In the second part, we turn our attention to flips in triangulations. A flip is a simple operation that transforms one triangulation into another. It turns out that with enough flips, we can transform any triangulation into any other. But how many flips is enough? We present an improved upper bound of $5.2 n - 33.6$ on the maximum flip distance between any pair of triangulations with $n$ vertices. Along the way, we prove matching lower bounds on each step in the current algorithm, including a tight bound of $\lfloor(3n - 9)/5\rfloor$ flips needed to make a triangulation 4-connected. In addition, we prove tight $\Theta(n \log n)$ bounds on the number of flips required in several settings where the edges have unique labels.

\endgroup			

\vfill

\cleardoublepage
\phantomsection
\addcontentsline{toc}{chapter}{\tocEntry{Acknowledgments}}


\bigskip

\begingroup
\let\clearpage\relax
\let\cleardoublepage\relax
\let\cleardoublepage\relax
\chapter*{Acknowledgments}

I would not have been able to write this thesis without the help and support of many people.

First of all, I would like to thank my supervisors -- Prosenjit Bose, Pat Morin, and Vida Dujmovi\'c. They are all I could have wished for in my supervisors and more. Combining a wealth of knowledge with a burning curiosity and a penchant for finding fascinating, yet approachable, open problems, they made these past five years into a journey of exploration and excitement.

I also want to thank the other members and students of the Computational Geometry lab for making it such a nice place to work (and occasionally not work). I am especially grateful to fellow PhD students Andr\'e, Carsten, Dana, and Luis, for being great friends and collaborators. In fact, I am very grateful to all the researchers and students who I got to work with during my PhD studies. Working together was always a pleasure, and they taught me more than classes ever could.

Finally, I would like to thank my family and friends for their support during these long, and at times stressful, years. I am especially grateful to my mother for encouraging me to take the leap of faith that is an international PhD, and to my partner, Gehana, for her unfailing love and support.

Thank you!

\endgroup

\pagestyle{scrheadings}
\cleardoublepage
\refstepcounter{dummy}
\addcontentsline{toc}{chapter}{\tocEntry{\contentsname}}
\setcounter{tocdepth}{2} 
\setcounter{secnumdepth}{3} 
\manualmark
\markboth{\spacedlowsmallcaps{\contentsname}}{\spacedlowsmallcaps{\contentsname}}
\tableofcontents 
\automark[section]{chapter}
\renewcommand{\chaptermark}[1]{\markboth{\spacedlowsmallcaps{#1}}{\spacedlowsmallcaps{#1}}}
\renewcommand{\sectionmark}[1]{\markright{\thesection\enspace\spacedlowsmallcaps{#1}}}

\cleardoublepage
\pagenumbering{arabic}
\cleardoublepage

\chapter{Summary of the thesis}
\label{ch:summary}

This thesis is comprised of two main parts. The first part, found in Chapters~\ref{ch:csi} through~\ref{ch:cr}, deals with geometric spanners. Chapters~\ref{ch:fh} through~\ref{ch:el} contain the second part, which focuses on flips in triangulations. A brief introduction and summary of each part is given below. The first chapter of each part provides a more detailed introduction.

The common theme in the two parts is that both deal with \emph{graphs}. A graph consists of a set of \emph{vertices}, some of which are connected by \emph{edges}. In this thesis, all graphs will be simple, which means that there is at most one edge connecting each pair of vertices, and edges cannot connect a vertex to itself.

\section{Geometric spanners}

Spanners can be informally described as graphs in which one never needs to make a large detour. That is, the shortest path between two vertices is proportional to their actual distance. Road networks are a good example; nearby cities are typically connected by a direct road, so that the total distance travelled is not much more than the distance `as the crow flies'. Spanners have been studied in many different contexts, but we will focus on \emph{geometric spanners}, where the vertices are points in the plane, and the length of an edge is the Euclidean distance between its endpoints. The \emph{spanning ratio} is the maximum ratio between the shortest path in the graph and the straight-line distance between any pair of vertices.

Chapter~\ref{ch:csi} gives an in-depth introduction to geometric spanners in general, and simple cone-based spanners in particular. The $\Theta$-graph is one such cone-based spanner. To construct it, we partition the plane around each vertex into a number of equiangular cones and add an edge to the `closest' vertex in each cone, where the closest vertex is defined as the vertex whose projection on the bisector of the cone is closest. It has been shown that for any desired spanning ratio $t$, there is a number of cones $k$ such that the $\Theta$-graph with $k$ cones (typically written as $\Theta_k$) is guaranteed to have spanning ratio $t$.

However, it was not known exactly for which values of $k$ the spanning ratio of $\Theta_k$ is bounded by a constant. It was known that $\Theta_3$ and below are not constant spanners, while $\Theta_6$ and up are. Recently, $\Theta_4$ was shown to be a constant spanner as well, leaving the question unanswered only for $\Theta_5$. In Chapter~\ref{ch:t5}, we prove that $\Theta_5$ is, indeed, a constant spanner. With the earlier results, this implies that $\Theta_k$ is a spanner for all $k \geq 4$. This result was first published in the proceedings of the 39th International Workshop on Graph-Theoretic Concepts in Computer Science (WG 2013)~\cite{bose2013theta5} and later appeared in Computational Geometry: Theory and Applications~\cite{bose2013theta5journal}.

Of course, knowing that there exists a short path to where you want to go is not the end of the story: you also have to know how to find it. This is called \emph{routing}, or \emph{competitive routing} if the spanning ratio of the resulting path is bounded by a constant. If you know the entire graph, routing is nothing more than computing a path, but most settings consider the more restricted scenario where you know your destination, but you can only see your current location and its neighbours. This is referred to as \emph{local} routing. In Chapter~\ref{ch:cr}, we present a local, competitive routing strategy for the \hts, which is closely related to $\Theta_6$. Our strategy achieves a routing ratio of $5/\sqrt{3} = 2.886 \dots$, which seems slightly disappointing compared to the spanning ratio of 2. This makes it all the more surprising that we managed to show that our algorithm is, in fact, optimal: no other routing strategy can achieve a better routing ratio, under the same restrictions. This is the first such separation between the spanning and routing ratios on a graph. These results were first published in the proceedings of the 23rd ACM-SIAM Symposium on Discrete Algorithms (SODA 2012)~\cite{bose2012competitive}, and the proceedings of the 24th Canadian Conference on Computational Geometry (CCCG 2012)~\cite{bose2012competitive2}, and have recently been accepted for publication in the SIAM Journal on Computing~\cite{bose2015optimal}.

\section{Flips in triangulations}

A triangulation is a planar graph where each face is a triangle (a cycle of three edges). A \emph{flip} is a simple, local operation that transforms one triangulation into another. Specifically, we can flip an edge $e$ by removing it, leaving an empty quadrilateral, and inserting the other diagonal of this quadrilateral. Flips were introduced by Wagner in 1936 in an attempt to make progress on the famous four-colour-theorem, and have been actively studied ever since. Applications of flips range from enumeration~\cite{avis1996reverse} and optimization of triangulations~\cite{bern1992mesh} to correcting errors in 3-dimensional terrains generated from height measurements~\cite{dekok2007generating}. Similar local operations that transform one graph into another in the same class have been used to build robust peer-to-peer network toplogies~\cite{cooper2009flip} and to find heuristic solutions to the Traveling Salesman Problem~\cite{lin1965computer}.

Wagner showed that, using flips, it is possible to transform any triangulation into any other. One question that has received a great deal of attention since then is: how many flips does this take, in the worst case? Chapter~\ref{ch:fh} presents a detailed history of various attempts to answer this question. This survey was published as an invited chapter in the proceedings of the XIV Spanish Meeting on Computational Geometry (EGC 2011)~\cite{bose2012history}.

Chapter~\ref{ch:f4c} details our own contribution to answering this question. In particular, we prove a tight bound of $\lfloor(3n - 9)/5\rfloor$ on the number of flips required to make an $n$-vertex triangulation 4-connected. And since the best known algorithm to transform any triangulation into any other first makes the triangulations in question 4-connected, this improves the upper bound on the total number of flips required from $6 n - 30$ to $5.2 n - 33.6$. These results were first published in the proceedings of the 23rd Canadian Conference on Computational Geometry (CCCG 2011)~\cite{bose2011making}, and subsequently appeared in a special issue of Computational Geometry: Theory and Applications~\cite{bose2012making}.

All of the research on flips thus far has assumed that edges are indistinguishable. But what happens when we give each edge a unique label, that is carried over to the new edge when an edge is flipped? This is the question studied in Chapter~\ref{ch:el}. We prove the first upper and lower bounds on the number of flips required in this setting. In particular, we show that $\Theta(n \log n)$ flips are required for edge-labelled triangulations of a convex polygon, edge-labelled combinatorial triangulations, and edge-labelled pseudo-triangulations. The results on pseudo-triangulations have been accepted to the 27th Canadian Conference on Computational Geometry (CCCG 2015)~\cite{bose2015flips}.

\bibliographystyle{plain}
\bibliography{../thesis}

\part{Geometric spanners}
\chapter{An introduction to Yao- and \texorpdfstring{$\Theta$}{Theta}-graphs}
\label{ch:csi}

In the past thirty years, geometric spanners have become an important field of study in computational geometry. This chapter serves as an introduction to the field, with a focus on two closely related families of geometric spanners: Yao-graphs and $\Theta$-graphs.

Most of the material in this chapter was already known, but the improvement for Yao-graphs with an odd number of cones (Theorem~\ref{thm:csi-yaoodd}) is new, although it was discovered independently by Keng and Xia~\cite{keng2013yao}. The proof of the spanning ratio of $\Theta$-graphs (Theorem~\ref{thm:csi-theta}) is also new.

\section{Geometric spanners}
\label{sec:csi-spanners}

 
Many practical geometric problems can be modelled as connecting a set of points in the plane. Examples include building roads to connect cities, or creating a communications network among wireless sensors. For these problems, we typically want to achieve good connectivity between the points, while using only a small number of connections. In the case of road networks in particular, we would like to avoid large detours: if cities $A$ and $B$ are fairly close, people should not have to drive to a distant city $C$ to travel from $A$ to $B$. This is what geometric spanners try to achieve: the shortest path between any two points in the network should be proportional to the distance between the points.
 
More formally, given a set $P$ of points in the plane, a geometric \emph{$t$-spanner} of $P$ is a graph $G$ with vertex set $P$, such that for each pair of points, the length of the shortest path between the corresponding vertices in $G$ is at most $t$ times the Euclidean distance between them. The \emph{spanning ratio} of $G$ is the smallest $t$ for which it is a $t$-spanner (in other texts, the spanning ratio is also called the \emph{dilation} or \emph{stretch factor}).

This definition is often applied to families of graphs. A family of graphs is called a $t$-spanner if every graph in the family is a $t$-spanner, and the spanning ratio of the family is the smallest $t$ such that every graph in the family is a $t$-spanner. A family of graphs is called a \emph{spanner} if there exists some finite $t$ for which it is a $t$-spanner.

As a first example, consider the complete graph on $P$. As it contains an edge between every pair of points in $P$, this family of graphs is a 1-spanner. And if $P$ does not contain three co-linear points, it is also the only 1-spanner, since the removal of any edge would increase the distance between its endpoints. Of course, the large drawback of the complete graph is that the number of edges is quadratic in the number of vertices. We would like to find sparser graphs (typically with a linear number of edges) that still have a small spanning ratio.

\begin{figure}[ht]
 \centering
 \includegraphics{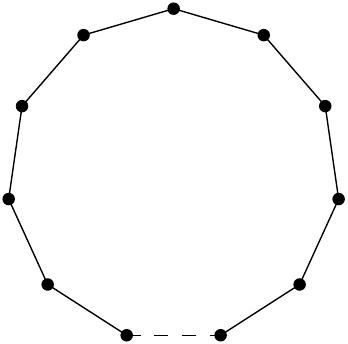}
 \caption{The minimum spanning tree of the vertices of a regular $n$-gon has spanning ratio $n - 1$.}
 \label{fig:csi-mst}
\end{figure}

The minimum spanning tree is at the other end of the spectrum. In order to be a spanner, a graph clearly needs to be connected (otherwise the spanning ratio is infinite). The minimum spanning tree is the connected graph on $P$ with lowest total edge length. Unfortunately, this family of graphs is not a spanner. To see this, imagine $n$ points spread equally on a circle. The minimum spanning tree of these points will include every edge between two consecutive points, except for one (see Figure~\ref{fig:csi-mst}). The endpoints of this non-edge are at distance $x$, but the only path between them in the graph follows the entire path around the circle, which has length $(n - 1) \cdot x$. Thus for every constant $t$, we can construct a point set with $\lceil t \rceil + 2$ vertices whose minimum spanning tree has spanning ratio $\lceil t \rceil + 1 > t$, meaning that there does not exist a constant $t$ such that every minimum spanning tree is a $t$-spanner. In fact, for this particular point set, \emph{every} tree has spanning ratio $\Omega(n)$.

\begin{theorem}[Eppstein~\cite{eppstein1999spanning}, Lemma 15]
 Any spanning tree $T$ on $n \geq 6$ points spread evenly on a circle has spanning ratio at least $\frac{n}{2\pi}$.
\end{theorem}
\begin{proof}
 Every tree has a vertex separator: a vertex $v$ such that removing $v$ splits $T$ into connected components with at most $n / 2$ vertices each. Consider the $n / 2 + 1$ vertices that lie opposite $v$ on the circle. Since each connected component has size at most $n / 2$, there must be a pair of vertices $x$ and $y$ from different components that are adjacent on the circle. Since they are in different connected components, the shortest path in $T$ from $x$ to $y$ passes through $v$. Thus, the spanning ratio of $T$ is at least:
 \begin{align*}
   \frac{|xv| + |yv|}{|xy|}~~&\geq~~\frac{2 \cdot 2\sin\left(\left(\frac{n}{4} - 1\right) \frac{\pi}{n}\right)}{2\sin\left(\frac{\pi}{n}\right)}\\
   ~~&\geq~~\frac{2 \cdot \frac{1}{2}}{2\frac{\pi}{n}} \hspace{8em} \text{(for $n \geq 6$)}\\
   ~~&=~~\frac{n}{2\pi} \qedhere
 \end{align*}
\end{proof}

Note that spanners have also been studied for general weighted graphs (where the shortest path in the spanner is compared to the shortest path in the original graph), or for point sets in higher dimensions. In this thesis, we deal almost exclusively with spanners of two-dimensional point sets; any exceptions will be mentioned explicitly. For a broader overview of geometric spanners, we recommend the book by Narasimhan and Smid~\cite{narasimhan2007geometric}.

\section{Preliminaries}
\label{sec:csi-prelim}

The proofs in this part of the thesis make extensive use of trigonometry. This section contains a short review of the basic properties used throughout the next chapters. Here, and in the rest of this thesis, we use $|ab|$ to denote the Euclidean distance between two points (or vertices) $a$ and $b$.

\begin{figure}[htb]
 \centering
 \begin{subfigure}[b]{0.48\textwidth}
  \centering
  \includegraphics{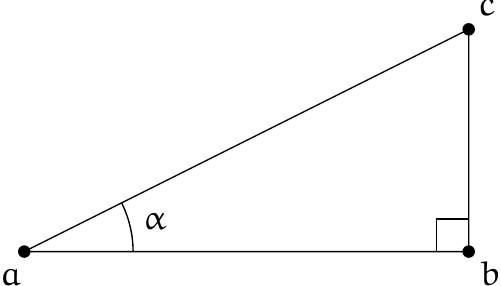}
  \caption{}
  \label{fig:csi-prelim-triangle-1}
 \end{subfigure}
 \begin{subfigure}[b]{0.48\textwidth}
  \centering
  \includegraphics{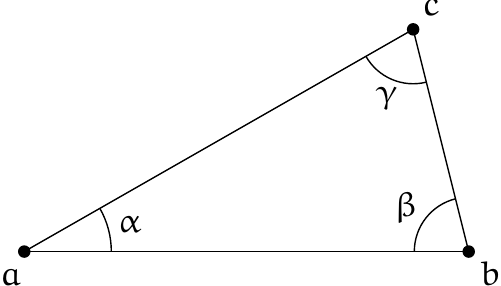}
  \caption{}
  \label{fig:csi-prelim-triangle-2}
 \end{subfigure}
 \caption{(a) A right triangle. (b) A general triangle.}
\end{figure}

\paragraph{Trigonometric functions.} The basic trigonometric functions are the \emph{sine}, \emph{cosine}, and \emph{tangent}. They are defined as the ratio of the sides in a right triangle. Consider a triangle $abc$ such that $\angle abc$ is a right angle (see Figure~\ref{fig:csi-prelim-triangle-1}). If we let $\alpha = \angle cab$, then
\begin{align*}
    \sin \alpha~~&=~~\frac{|bc|}{|ac|}, & \cos \alpha~~&=~~\frac{|ab|}{|ac|},\text{ and} & \tan \alpha~~&=~~\frac{|bc|}{|ab|}.
\end{align*}

\paragraph{Trigonometric identities.} There are several more complex equalities that can be derived from these basic functions. The two we use most often are called the \emph{law of sines} and the \emph{law of cosines}. They have the advantage that they apply to all triangles, not only right triangles. In a triangle $abc$ with $\alpha = \angle cab$, $\beta = \angle abc$, and $\gamma = \angle bca$ (see Figure~\ref{fig:csi-prelim-triangle-2}), these identities are expressed as follows.

\begin{equation}
    \tag{law of sines}
    \frac{|ab|}{\sin \gamma}~~=~~\frac{|ac|}{\sin \beta}~~=~~\frac{|bc|}{\sin \alpha}
\end{equation}
\begin{align*}
    |ab|^2~~&=~~|ac|^2 + |bc|^2 - 2|ac||bc|\cos \gamma\\ \tag{law of cosines}
    |ac|^2~~&=~~|ab|^2 + |bc|^2 - 2|ab||bc|\cos \beta\\
    |bc|^2~~&=~~|ab|^2 + |ac|^2 - 2|ab||ac|\cos \alpha
\end{align*}

\paragraph{Triangle inequality.} The last equation we cover here is not an equality, but rather an inequality known as the \emph{triangle inequality}. It states that one side of a triangle is never longer than the other two sides combined. Using the notation of the triangle depicted in Figure~\ref{fig:csi-prelim-triangle-2}, it can be expressed as follows.
\begin{align*}
    |ab|~~&\leq~~|ac| + |bc|\\ \tag{triangle inequality}
    |ac|~~&\leq~~|ab| + |bc|\\
    |bc|~~&\leq~~|ab| + |ac|
\end{align*}

\section{Yao-graphs}
\label{sec:csi-yao}

 
\begin{figure}[ht]
 \centering
 \includegraphics{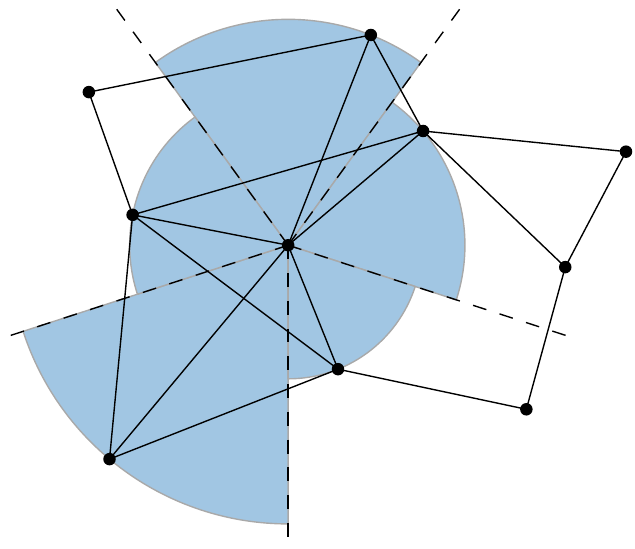}
 \caption{An example $Y_5$-graph. Each vertex adds an edge to the closest vertex in each of five equiangular cones.}
 \label{fig:csi-yaograph}
\end{figure}

One simple way to build a geometric spanner is to take each vertex, partition the plane around it into a fixed number of cones with equal angles, and add an edge between the vertex and the closest vertex in each cone (see Figure~\ref{fig:csi-yaograph}). The resulting graph is called a Yao-graph, and is typically denoted by $Y_k$, where $k$ is the number of cones around each vertex. This construction guarantees that a Yao-graph with $k$ cones has at most $kn$ edges, where $n$ is the number of vertices. Furthermore, if the cones are narrow enough, we can find a path between any two vertices by starting at one and walking to the closest vertex in the cone that contains the other, repeating this until we end up at our destination. Intuitively, this results in a short path because we are always walking approximately in the right direction, and, since our neighbour is the closest vertex in that direction, never too far.

\begin{figure}[ht]
 \centering
 \includegraphics{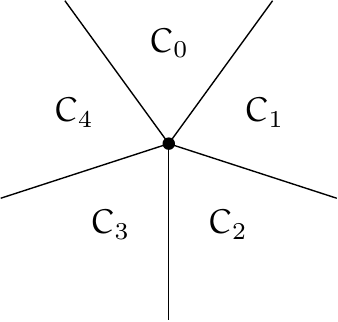}
 \caption{The cones used to construct the $Y_5$-graph. The orientation is the same for all vertices.}
 \label{fig:csi-cones}
\end{figure}

A Yao-graph whose cones are narrower will typically give a better approximation of the shortest path. If we want to prove this formally, we need to iron out a few details in the definition of a Yao-graph. First, we assume that points are in general position; in particular, we assume that for every vertex $a$, there are no two points at the exact same distance from $a$. This means that each vertex has a unique closest vertex in each cone. (This is not strictly necessary, but it makes our proofs simpler. If we don't assume general position, the same properties hold by breaking ties arbitrarily.) Second, we label the cones $C_0$ through $C_{k-1}$ in clockwise order, and we orient them such that the bisector of $C_0$ aligns with the positive $y$-axis (see Figure~\ref{fig:csi-cones}). This orientation is the same for each vertex. If the apex is not clear from the context, we use $C^a_i$ to denote cone $C_i$ with apex $a$. The boundary between two cones belongs to the counter-clockwise one (so the boundary between $C_0$ and $C_1$ is part of $C_0$). The crux of the proof lies in the following small geometric lemma that captures our earlier intuition that taking a small step in approximately the right direction makes meaningful progress towards our destination.

\begin{figure}[ht]
 \centering
 \includegraphics{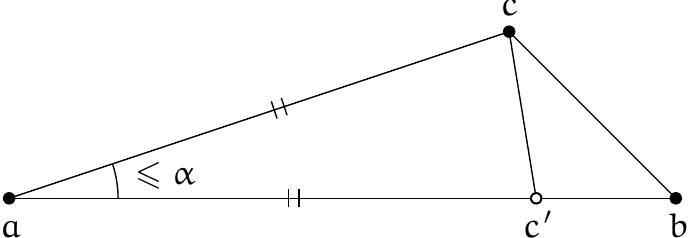}
 \caption{We are at a vertex $a$ and want to go to $b$, but $c$ is the closest vertex in the cone that contains $b$.}
 \label{fig:csi-basicYaoLemma}
\end{figure}

\begin{lemma}
 \label{lem:csi-basicyao}
 Given three points $a$, $b$, and $c$, such that $|ac| \leq |ab|$ and $\angle bac \leq \alpha < \pi/3$, then
 \[ 
  |bc|~~\leq~~|ab| - \left( 1 - 2 \sin (\alpha/2) \right) \cdot |ac|.
 \]
\end{lemma}
\begin{proof}
 Let $c'$ be the point on $ab$ such that $|ac| = |ac'|$ (see Figure~\ref{fig:csi-basicYaoLemma}). Since $acc'$ forms an isosceles triangle, we can express $|cc'|$ in terms of $|ac|$:
 \[
  |cc'|
  ~~=~~2 \sin (\angle bac / 2) \cdot |ac|
  ~~\leq~~2 \sin (\alpha / 2) \cdot |ac|. 
 \]
 The inequality holds since $\sin x$ is increasing in this range. Now we just need the triangle inequality:
 \begin{align*}
  |bc|~~&\leq~~|bc'| + |c'c|\\
      ~~&\leq~~|ab| - |ac'| + 2 \sin (\alpha / 2) \cdot |ac|\\
      ~~&=~~|ab| - (1 - 2 \sin (\alpha / 2)) \cdot |ac|. \qedhere
 \end{align*}
\end{proof}

Imagine that we are at a vertex $a$ and we want to go to $b$, but $c$ is the closest vertex in the cone of $a$ that contains $b$. Then this lemma essentially tells us that if the angle between our destination and the edge we follow ($\alpha$) is small, the amount of progress we make ($|ab| - |bc|$) is directly proportional to the distance we travel ($|ac|$):
\begin{align*}
 |ab| - |bc|~~&\geq~~|ab| - (|ab| - \left( 1 - 2 \sin (\alpha/2) \right) \cdot |ac|)\\
 ~~&\geq~~\left( 1 - 2 \sin (\alpha/2) \right) \cdot |ac|.
\end{align*}
Now that we have this lemma, we can use an inductive argument to show that Yao-graphs are spanners.

\begin{theorem}
 \label{thm:csi-yao}
 For any integer $k \geq 7$, the graph $Y_k$ has spanning ratio at most $1 / (1 - 2 \sin(\theta/2))$, where $\theta = 2\pi/k$.
\end{theorem}
\begin{proof}
 Let $a$ and $b$ be two arbitrary vertices in our point set. We show that the obvious way to get from $a$ to $b$ -- keep following the edge in the cone that contains $b$ -- not only works, it even gives us a short path. To start off, consider all pairs of vertices $(u, v)$ and sort them by their distance $|uv|$. Our proof proceeds by induction on the index of $(a, b)$ in this sorted order.

 In the base case, $(a, b)$ is the closest pair. This means that $b$ must be the closest vertex in the cone of $a$ that contains $b$, so the edge $(a,b)$ is in the graph. Thus, the shortest path between $a$ and $b$ has length exactly $|ab|$, giving a spanning ratio of $1$. Since $1 / (1 - 2 \sin(\theta/2)) > 1$ for $0 < \theta < 2\pi$, this proves the base case.

 For the inductive step, assume that for any pair of vertices $(u, v)$ such that $|uv| < |ab|$, there exists a path from $u$ to $v$ with length at most $1 / (1 - 2 \sin(\theta/2)) \cdot |uv|$. Now consider the cone of $a$ that contains $b$. If $b$ is the closest vertex, the edge $(a,b)$ is in the graph and we can use the same argument as in the base case. Otherwise, let $c$ be the closest vertex to $a$. Note that, because $\angle bac \leq \theta \leq 2\pi/7 < \pi/3$, we know that $\angle bac$ is not the largest angle in triangle $abc$. Since the largest angle lies opposite the longest edge, $bc$ is not the longest edge, so $|bc| < |ab|$. Using our inductive hypothesis, this means that there is a path between $b$ and $c$ with length at most $1 / (1 - 2 \sin(\theta/2)) \cdot |bc|$. So to go from $a$ to $b$, we can first take the direct edge to $c$ and then follow the path to $b$. Since $a$, $b$, and $c$ satisfy all the conditions for Lemma~\ref{lem:csi-basicyao}, we can use it to bound the length of the resulting path:
\begin{align*}
 &~~~~~~|ac| + \frac{1}{1 - 2 \sin(\theta/2)} \cdot |bc| \\
 &\leq~~|ac| + \frac{1}{1 - 2 \sin(\theta/2)} \cdot (|ab| - (1 - 2 \sin (\theta/2)) \cdot |ac|) \\
 &=~~|ac| + \frac{1}{1 - 2 \sin(\theta/2)} \cdot |ab| - |ac| \\
 &=~~\frac{1}{1 - 2 \sin(\theta/2)} \cdot |ab|.
\end{align*}
Which is exactly what we needed to show.
\end{proof}

\begin{figure}[ht]
 \centering
 \includegraphics{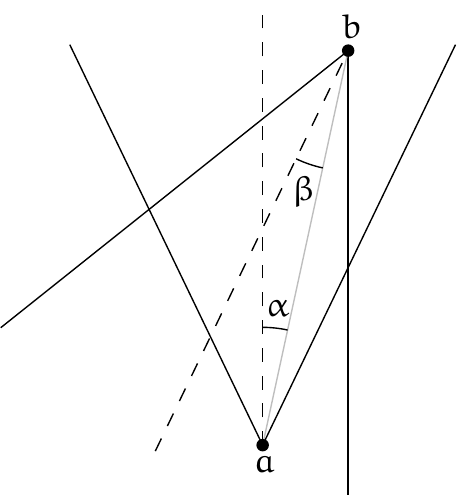}
 \caption{If the number of cones is odd, opposite cones are not symmetric and either $\alpha$ or $\beta$ is small.}
 \label{fig:csi-asymmetric}
\end{figure}

Interestingly, we can do a little better when the number of cones is odd. This is caused by the asymmetry in the cones. To see why, consider the situation where we have two vertices, $a$ and $b$, and the number of cones is odd. Let $C^a$ be the cone of $a$ that contains $b$ and let $C^b$ be the analogous cone for $b$. Let $\alpha$ and $\beta$ be the angles between $ab$ and the bisectors of $C^a$ and $C^b$, respectively (see Figure~\ref{fig:csi-asymmetric}). Since the bisector of $C^a$ is parallel to one of the sides of $C^b$, the transversal $ab$ creates equal angles at $a$ and $b$, showing that $\alpha + \beta = \theta / 2$. Therefore, the smaller of $\alpha$ and $\beta$ can be at most $\theta / 4$. If we assume that $\alpha$ is the smaller of the two, and let $c$ be the closest vertex in $C^a$, then $\angle bac \leq \alpha + \theta / 2 \leq 3\theta/4$. Plugging this into the proof of Theorem~\ref{thm:csi-yao} gives the following result.

\begin{theorem}
 \label{thm:csi-yaoodd}
 For any odd integer $k \geq 5$, the graph $Y_k$ has spanning ratio at most $1 / (1 - 2 \sin(3/8 \cdot \theta))$, where $\theta = 2\pi/k$.
\end{theorem}

Note that this theorem extends to $Y_5$, as $3\theta/4 = 3/4 \cdot 2\pi/5 = 3\pi/10 < \pi/3$, whereas Theorem~\ref{thm:csi-yao} does not.

\section{\texorpdfstring{$\Theta$}{Theta}-graphs}
\label{sec:csi-theta}

 
\begin{figure}[ht]
 \centering
 \includegraphics{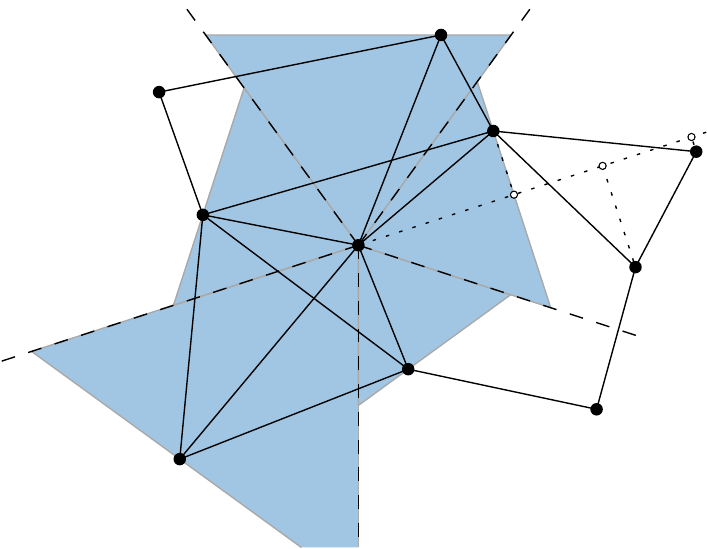}
 \caption{An example $\Theta_5$-graph.}
 \label{fig:csi-theta}
\end{figure}

If we modify the definition of Yao-graphs slightly we obtain another type of geometric spanner, called a $\Theta$-graph. The only difference lies in the way the closest vertex is determined: for each vertex $u$, the closest vertex in a cone $C$ is the vertex $v$ whose orthogonal projection on the bisector of $C$ is closest to $u$ (see Figure~\ref{fig:csi-theta}). We again assume general position to simplify our proofs; in particular, we assume that no two vertices lie on a line parallel or perpendicular to a cone boundary, guaranteeing that each vertex connects to at most one vertex in each cone, and thus that the graph has at most $kn$ edges. Another way to look at the construction is that we sweep $C$ with a line perpendicular to the bisector, and add an edge to the first vertex we hit. Note that this creates an empty triangle (the shaded regions in Figure~\ref{fig:csi-theta}). Given two vertices $a$ and $b$, we can define their \emph{canonical triangle} $\T{a}{b}$ as the triangle formed by the boundaries of the cone $C$ of $a$ that contains $b$ and the line through $b$ perpendicular to the bisector of $C$. Note that the canonical triangle $\T{b}{a}$ also exists: it is the same size as $\T{a}{b}$, but is has apex $b$ and is oriented towards $a$ instead. This gives a third way to describe the construction of the $\Theta$-graph, by adding an edge between two vertices if one of their canonical triangles is empty. These canonical triangles play an important role in Chapters~\ref{ch:t5} and~\ref{ch:cr}. As one might expect from the similarity in construction, $\Theta$-graphs share many of the properties that make Yao-graphs interesting. Their key advantage, however, is that they can be constructed by an easy sweep-line algorithm, whereas all known algorithms for constructing Yao-graphs are more complex. Here we prove that $\Theta$-graphs are spanners as well.

\begin{figure}[htb]
 \centering
 \includegraphics{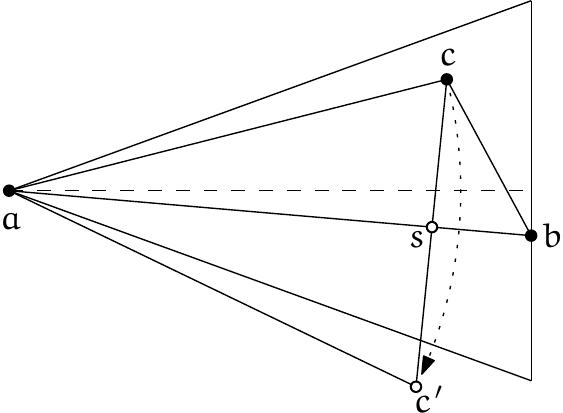}
 \caption{The canonical triangle $\T{a}{b}$, with the closest vertex to $a$ rotated over an angle $\theta$.}
 \label{fig:csi-theta-rotation}
\end{figure}

\begin{theorem}
 \label{thm:csi-theta}
 For any integer $k \geq 7$, the graph $\Theta_k$ has spanning ratio at most $t = 1 / (1 - 2 \sin(\theta/2))$, where $\theta = 2\pi/k$.
\end{theorem}
\begin{proof}
 This proof is similar to the proof of Theorem~\ref{thm:csi-yao}; we have two vertices $a$ and $b$ and we show that there is a path between them of length at most $t |ab|$. The proof is again by induction on the relative position of $(a, b)$ among all pairs of points when ordered by distance. For convenience, we translate and rotate the point set such that $a$ is in the origin, and the bisector of the cone of $a$ that contains $b$ coincides with the positive $x$-axis. We start by considering the inductive step, and prove the base case at the end.

 We assume that there is a path from $u$ to $v$ of length at most $t |uv|$ for all pairs $(u, v)$ with $|uv| < |ab|$. If the edge $(a,b)$ is in the graph, we have a path from $a$ to $b$ of length $|ab| \leq t |ab|$ and are done, since $t > 1$, so assume that this is not the case. Then there is another vertex $c$, whose projection on the bisector is closest to $a$ in the cone containing $b$. Without loss of generality, we assume that $c$ lies above $ab$ (if it does not, we can mirror everything in the $x$-axis).

 Now imagine rotating $c$ clockwise around $a$ by an angle of $\theta$, and let $c'$ be the resulting position (see Figure~\ref{fig:csi-theta-rotation}). Note that $c'$ lies below $ab$, as the angle between $ac$ and $ab$ is at most $\theta$. Furthermore, rotating $c$ by the angle between $ac$ and the positive $x$-axis would move it to a point with the same $x$-coordinate, but since we rotated it further, $c'$ lies to the left of $c$ and therefore to the left of $b$. Since two line segments intersect if and only if for both segments, the endpoints lie on opposite sides of the other segment, $ab$ and $cc'$ intersect, and we call their intersection point $s$. Now we can use the triangle inequality to obtain the following inequalities:
\begin{align*}
 |ac|~~&=~~|ac'|~~\leq~~|as| + |sc'|, \text{ and}\\
 |bc|~~&\leq~~|bs| + |sc|.
\end{align*}
 Thus, we get that
\begin{align*}
 |ac| + |bc|~~&\leq~~|as| + |sc'| + |bs| + |sc|\\
  &=~~|ab| + |cc'|\\
  &=~~|ab| + 2 \sin \frac{\theta}{2} |ac|\\
 |bc|~~&\leq~~|ab| + 2 \sin \frac{\theta}{2} |ac| - |ac|\\
  &=~~|ab| - \frac{1}{t} |ac|
\end{align*}
 For $k \geq 7$, this implies that $|bc| < |ab|$, which means that we can apply our inductive hypothesis to $bc$. By first following $ac$, this gives us a path from $a$ to $b$ of length at most:
\begin{align*}
 |ac| + t |bc|~~&\leq~~|ac| + t \left(|ab| -  \frac{1}{t} |ac|\right)\\
  &=~~|ac| + t |ab| - |ac|\\
  &=~~t |ab|
\end{align*}
 This settles the inductive step. For the base case, if the edge $(a,b)$ is in the graph we are again done. The only way this edge could be absent is if another vertex $c$ had a projection on the bisector closer to $a$. But we just derived that in that case $|bc| < |ab|$, which contradicts the fact that $|ab|$ is minimal. Therefore this cannot happen, and $b$ must be the closest vertex to $a$, proving the theorem.
\end{proof}

\section{History}
\label{sec:csi-hist}


Research on geometric spanners was sparked by a paper by Paul Chew in 1986~\cite{chew1986there}, titled ``There is a planar graph almost as good as the complete graph''. In that paper and the subsequent journal version~\cite{chew1989there}, he showed that certain Delaunay triangulations are geometric spanners with few edges. In particular, he showed this for Delaunay triangulations whose empty regions are the square and the equilateral triangle. The traditional Delaunay triangulation, which uses a circle, was quickly shown to be a spanner as well~\cite{dobkin1987delaunay}. Although its true spanning ratio remains a mystery, the upper bound has been improved multiple times; from the initial bound of 5.08 in 1987, to 2.42 in 1989~\cite{keil1989delaunay,keil1992classes}, and recently to just below 2~\cite{xia2011improved}.

Yao-graphs were introduced independently by Flinchbaugh and Jones~\cite{flinchbaugh1981strong} and Yao~\cite{yao1982constructing} around 1981, before the concept of spanners was even introduced by Chew. Yao showed that $Y_8$ is a supergraph of the minimum spanning tree and that this still holds in higher dimensions. This gave an efficient algorithm to compute the minimum spanning tree in higher dimensions.

To the best of our knowledge, the first proof that Yao-graphs are geometric spanners was published in 1993, by Alth{\"o}fer~\etal~\cite{althofer1993sparse}. In particular, they showed that for every spanning ratio $t > 1$, there exists a number of cones $k$ such that $Y_k$ is a $t$-spanner. It appears that some form of this result was known earlier, as Clarkson~\cite{clarkson1987approximation} already remarked in 1987 that $Y_{12}$ is a $1 + \sqrt{3}$-spanner, albeit without providing a proof or reference. In 2004, Bose~\etal~\cite{bose2004approximating} provided a more specific bound on the spanning ratio, by showing that for $k > 8$, $Y_k$ is a geometric spanner with spanning ratio at most $1 / (\cos \theta - \sin \theta)$, where $\theta = 2\pi/k$. This bound was later improved to $1 / (1 - 2 \sin(\theta / 2))$, for $k > 6$~\cite{bose2012piArxiv}. We presented a simplified version of this proof for Theorem~\ref{thm:csi-yao}. The improvement for odd $k$ given in Theorem~\ref{thm:csi-yaoodd} is a recent development by Barba~\etal~\cite{barba2013new}.

The $\Theta$-graph was introduced independently by Clarkson~\cite{clarkson1987approximation} and Keil~\cite{keil1988approximating,keil1992classes}, as an alternative to Yao-graphs that was easier to compute. Both papers prove a spanning ratio of $1 / (\cos \theta - \sin \theta)$, which was later improved to $1 / (1 - 2 \sin(\theta / 2))$ by Ruppert and Seidel~\cite{ruppert1991approximating}. The proof of Theorem~\ref{thm:csi-theta} is significantly simpler than their proof, and is based on another proof by Lukovski~\cite[p.~11]{lukovski1999new}.

This bound of $1 / (1 - 2 \sin(\theta/2))$ was the best known upper bound on the spanning ratio for over twenty years. Only very recently have researchers been able to prove that the true bound is lower. In 2012, Bose~\etal~\cite{bose2012optimal} showed that $\Theta$-graphs with $4m + 2$ cones ($m \geq 1$) have a spanning ratio of $1 + 2 \sin(\theta/2)$. Surprisingly, they were also able to give a matching lower bound, making this the first family of $\Theta$-graphs for which a tight bound on the spanning ratio is known. Later, they used similar techniques to improve the upper bound on the spanning ratio of all other $\Theta$-graphs~\cite{bose2014towards}, although these do not yet match the best known lower bounds. A good overview of these results, and more, can be found in the thesis of Andr\'e van~Renssen~\cite{R2014ConstrainedSpanners}.

\bibliographystyle{plain}
\bibliography{../thesis}
\chapter{The \texorpdfstring{\graph}{Theta-5-graph} is a spanner}
\label{ch:t5}

Given any set of points in the plane, we show that the $\Theta$-graph with 5 cones is a geometric spanner with spanning ratio at most \valsr. This is the first constant upper bound on the spanning ratio of this graph. The upper bound uses a constructive argument that gives a (possibly self-intersecting) path between any two vertices, of length at most $\sqrt{50 + 22 \sqrt{5}}$ times the Euclidean distance between the vertices. We also prove that \vallb is a lower bound on the spanning ratio.

The results in this chapter were first published in the proceedings of the 39th International Workshop on Graph-Theoretic Concepts in Computer Science (WG 2013)~\cite{bose2013theta5}, and have subsequently been published in Computational Geometry: Theory and Applications~\cite{bose2013theta5journal}. This chapter contains joint work with Prosenjit Bose, Pat Morin, and Andr\'e van Renssen.

\section{Introduction}
\label{sec:t5-introduction}

As described in Section~\ref{sec:csi-hist}, most early research focused on Yao- and $\Theta$-graphs with a large number of cones. However, using the smallest possible number of cones is important for many practical applications, where the cost of a network is mostly determined by the number of edges. One such example is point-to-point wireless networks. These networks use narrow directional wireless transceivers that can transmit over long distances (up to 50km \cite{dist1,dist2}). The cost of an edge in such a network is therefore equal to the cost of the two transceivers that are used at each endpoint of that edge. In such networks, the cost of building $\Theta_6$ is approximately 29\% higher than the cost of building $\Theta_5$ if the transceivers are randomly distributed~\cite{morin2014average}. Assuming that we still want our network to be a spanner, this leads to the natural question: for which values of $k$ are $Y_k$ and $\Theta_k$ spanners? Kanj~\cite{kanj2013geometric} presented this question as one of the main open problems in the area of geometric spanners.

Surprisingly, this question was not studied until quite recently. In 2009, El~Molla~\cite{el2009yao} showed that both $Y_2$ and $Y_3$ are not spanners, and these proofs translate to $\Theta_2$ and $\Theta_3$ as well. Since the general proofs (presented in Theorems~\ref{thm:csi-yao} and~\ref{thm:csi-theta}) work for $k \geq 7$, this left the question open for graphs with 4, 5 and 6 cones. A surprising connection between $\Theta$-graphs and Delaunay triangulations led to the first positive result on this question, when Bonichon~\etal~\cite{bonichon2010connections} showed that $\Theta_6$ is the union of two rotated copies of the empty equilateral triangle Delaunay triangulation. This graph had been shown to be a 2-spanner by Chew~\cite{chew1989there} over 20 years earlier. This result was then used by Damian and Raudonis~\cite{damian2012yao} to show that $Y_6$ is a spanner as well. The next graphs to fall were $Y_4$~\cite{bose2012pi} and $\Theta_4$~\cite{barba2013stretch}, both of which were shown to be spanners, albeit with very loose upper bounds on the spanning ratio of 663 ($Y_4$) and 237 ($\Theta_4$). The improvement on the spanning ratio of Yao-graphs with an odd number of cones presented in Theorem~\ref{thm:csi-yaoodd}, discovered by Barba~\etal~\cite{barba2013new}, settled the matter for $Y_5$, leaving only $\Theta_5$.

Note that this problem was already claimed to be solved in 1991, by Ruppert and Seidel~\cite{ruppert1991approximating}. Specifically, they wrote:

\begin{quote}
    In the planar case, some improvement can be made on the constants. In particular, when $k$ is odd, there is an asymmetry between the cones [\dots] that we can take advantage of by growing paths from both ends. Interestingly, this asymmetry allows us to prove a bound near 10 on the path lengths even for the case $k=5$. [\dots] The details are omitted here due to lack of space.
\end{quote}

However, to the best of our knowledge they never published a proof of this claim.

\begin{figure}[ht]
  \centering
  \includegraphics{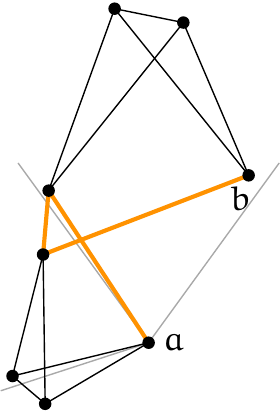}
  \caption{An example where the shortest path between two vertices (in bold) in $\Theta_5$ crosses itself.}
  \label{fig:t5-cross}
 \end{figure}

In this chapter we present the final piece of this puzzle, by giving the first constant upper bound on the spanning ratio of $\Theta_5$, thereby proving that it is a geometric spanner. We show that the spanning ratio is at most $\valsr$. Note that this bound is slightly better than the bound for $Y_5$ given by Theorem~\ref{thm:csi-yaoodd}, although Barba~\etal~\cite{barba2013new} improved the bound for $Y_5$ to $2 + \sqrt{3} \approx 3.74$ using a different technique. Since the proof for $\Theta_5$ is constructive, it gives us a path between any two vertices, $u$ and $w$, of length at most $9.960 \cdot |uw|$. Surprisingly, this path can cross itself, a property we observed for the shortest path as well (see Figure~\ref{fig:t5-cross}). We also prove that \vallb is a lower bound on the spanning ratio.

\section{Connectivity}
\label{sec:t5-connectivity}

Recall that the \emph{canonical triangle} $\T{u}{v}$ of two vertices $u$ and $v$ is the triangle bounded by the cone of $u$ that contains $v$ and the line through $v$ perpendicular to the bisector of that cone. We define the size $|\T{u}{v}|$ of a canonical triangle as the length of one of the sides incident to the apex $u$. This gives us the useful property that any line segment between $u$ and a point inside the triangle has length at most $|\T{u}{v}|$.

To introduce the structure of the proof that the spanning ratio of $\Theta_5$ is bounded, we first show that the \graph is connected.

\begin{theorem}
 \label{thm:t5-connected}
 The \graph is connected.
\end{theorem}
\begin{proof}
 We prove that there is a path between any (ordered) pair of vertices in $\Theta_5$, using induction on the size of their canonical triangle. Formally, given two vertices $u$ and $w$, we perform induction on the rank (relative position) of $\T{u}{w}$ among the canonical triangles of all pairs of vertices, when ordered by size. For ease of description, we assume that $w$ lies in the right half of $C_0^u$. The other cases are analogous.

 If $\T{u}{w}$ has rank 1, it is the smallest canonical triangle. Therefore there can be no point closer to $u$ in $C_0^u$, so the edge $(u, w)$ must be in the graph. This proves the base case.

 If $\T{u}{w}$ has a larger rank, our inductive hypothesis is that there exists a path between any pair of vertices with a smaller canonical triangle. Let $a$ and $b$ be the left and right corners of $\T{u}{w}$. Let $m$ be the midpoint of $ab$ and let $x$ be the intersection of $ab$ and the bisector of $\angle mub$ (see~Figure~\ref{fig:t5-canon-a}).

\begin{figure}[htb]
 \centering
 \begin{subfigure}[b]{0.48\textwidth}
  \centering
  \includegraphics{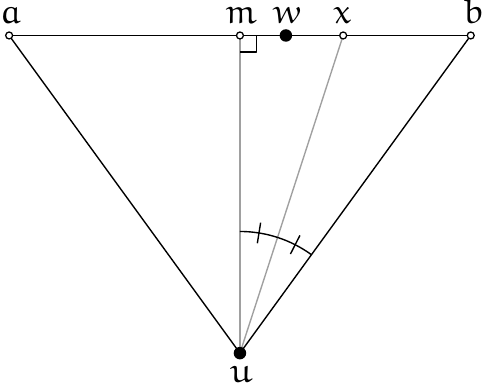}
  \caption{}
  \label{fig:t5-canon-a}
 \end{subfigure}
 \begin{subfigure}[b]{0.48\textwidth}
  \centering
  \includegraphics{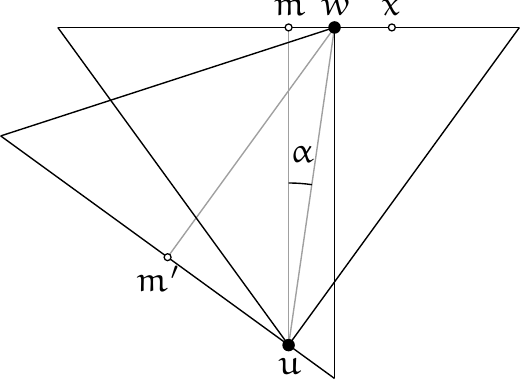}
  \caption{}
  \label{fig:t5-canon-b}
 \end{subfigure}
 \caption{(a) The canonical triangle $\T{u}{w}$. (b) If $w$ lies between $m$ and $x$, then $\T{w}{u}$ is smaller than $\T{u}{w}$.}
\end{figure}

 If $w$ lies to the left of $x$, consider the canonical triangle $\T{w}{u}$. Let $m'$ be the midpoint of the side of $\T{w}{u}$ opposite $w$ and let $\alpha = \angle muw$ (see~Figure~\ref{fig:t5-canon-b}). Note that $\angle uwm' = \frac{\pi}{5} - \alpha$, since $um$ and the vertical border of $\T{w}{u}$ are parallel and both are intersected by $uw$. Using basic trigonometry, we can express the size of $\T{w}{u}$ as follows.
\begin{align*}
 |\T{w}{u}|~~&=~~\frac{|wm'|}{\cos \frac{\pi}{5}}~~=~~\frac{\cos \angle uwm' \cdot |uw|}{\cos \frac{\pi}{5}}\\
             &=~~\frac{\cos \left( \frac{\pi}{5} - \alpha \right) \cdot \frac{|um|}{\cos \alpha}}{\cos \frac{\pi}{5}}~~=~~\frac{\cos \left( \frac{\pi}{5} - \alpha \right)}{\cos \alpha} \cdot |\T{u}{w}|
\end{align*}
 Since $w$ lies to the left of $x$, the angle $\alpha$ is less than $\pi/10$, which means that $\cos ( \frac{\pi}{5} - \alpha ) / \cos \alpha$ is less than 1. Hence $\T{w}{u}$ is smaller than $\T{u}{w}$ and by induction, there is a path between $w$ and $u$. Since the graph is undirected, we are done in this case. The rest of the proof deals with the case where $w$ lies on or to the right of $x$.

 If $\T{w}{u}$ is empty, there is an edge between $u$ and $w$ and we are done, so assume that this is not the case. Then there is a vertex $v_w$ that is closest to $w$ in $C_3^w$ (the cone of $w$ that contains $u$). This gives rise to four cases, depending on the location of $v_w$ (see Figure~\ref{fig:t5-cases-a}). In each case, we will show that $\T{u}{v_w}$ is smaller than $\T{u}{w}$ and hence we can apply induction to obtain a path between $u$ and $v_w$. Since $v_w$ is the closest vertex to $w$ in $C_3$, there is an edge between $v_w$ and $w$, completing the path between $u$ and $w$.

\begin{figure}[htb]
 \centering
 \begin{subfigure}[b]{0.46\textwidth}
  \centering
  \includegraphics{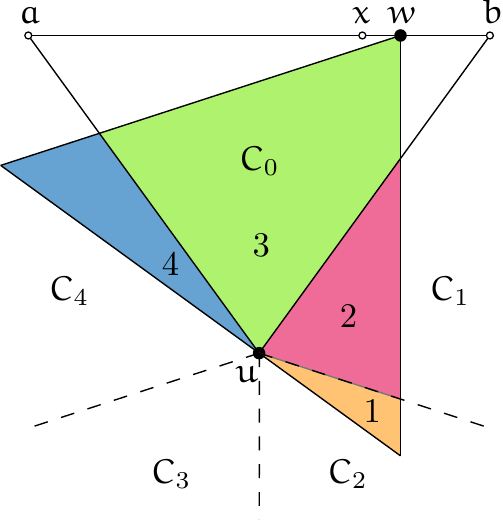}
  \caption{}
  \label{fig:t5-cases-a}
 \end{subfigure}
 \begin{subfigure}[b]{0.50\textwidth}
  \centering
  \includegraphics{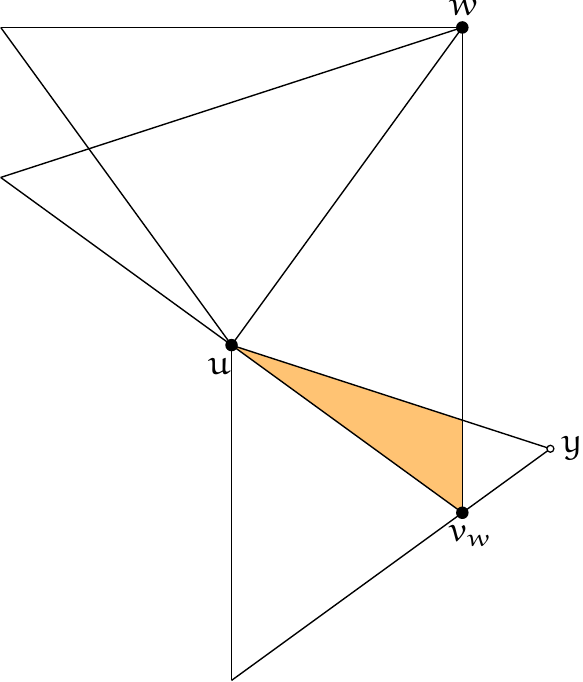}
  \caption{}
  \label{fig:t5-cases-b}
 \end{subfigure}
 \caption{(a) The four cases for $v_w$. (b) Case 1: The situation that maximizes $|\T{u}{v_w}|$ when $v_w$ lies in $C_2^u$.}
\end{figure}

 \case{1} $v_w$ lies in $C_2^u$. In this case, the size of $\T{u}{v_w}$ is maximized when $v_w$ lies in the bottom right corner of $\T{w}{u}$ and $w$ lies on $b$. Let $y$ be the rightmost corner of $\T{u}{v_w}$ (see~Figure~\ref{fig:t5-cases-b}). Using the law of sines, we can express the size of $\T{u}{v_w}$ as follows.
 \begin{align*}
  |\T{u}{v_w}|
~~&=~~|uy|\\
~~&=~~\frac{\sin \angle uv_wy}{\sin \angle uyv_w} \cdot |uv_w|\\
~~&=~~\frac{\sin \frac{3\pi}{5}}{\sin \frac{3\pi}{10}} \cdot \tan \frac{\pi}{5} \cdot |\T{u}{w}|\\
~~&<~~|\T{u}{w}|
 \end{align*}
 \case{2} $v_w$ lies in $C_1^u$. In this case, the size of $\T{u}{v_w}$ is maximized when $w$ lies on $b$ and $v_w$ lies almost on $w$. By symmetry, this gives $|\T{u}{v_w}| = |\T{u}{w}|$. However, $v_w$ cannot lie precisely on $w$ and must therefore lie a little closer to $u$, giving us that $|\T{u}{v_w}| < |\T{u}{w}|$.
 \case{3} $v_w$ lies in $C_0^u$. As in the previous case, the size of $\T{u}{v_w}$ is maximized when $v_w$ lies almost on $w$, but since $v_w$ must lie closer to $u$, we have that $|\T{u}{v_w}| < |\T{u}{w}|$.

 \begin{figure}[htb]
  \centering
  \includegraphics{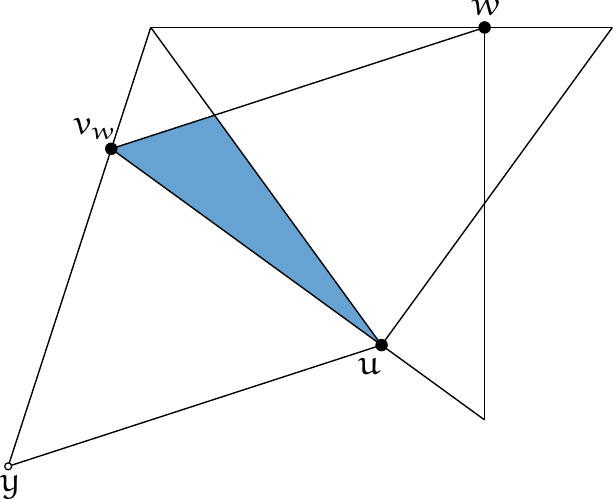}
  \caption{Case 4: The situation that maximizes $|\T{u}{v_w}|$ when $v_w$ lies in $C_4^u$.}
  \label{fig:t5-cases-c}
 \end{figure}

 \case{4} $v_w$ lies in $C_4^u$. In this case, the size of $\T{u}{v_w}$ is maximized when $v_w$ lies in the left corner of $\T{w}{u}$ and $w$ lies on $x$. Let $y$ be the bottom corner of $\T{u}{v_w}$ (see~Figure~\ref{fig:t5-cases-c}). Since $x$ is the point where $|\T{u}{w}| = |\T{w}{u}|$, and $v_wyuw$ forms a parallelogram, $|\T{u}{v_w}| = |\T{u}{w}|$. However, by general position, $v_w$ cannot lie on the boundary of $\T{w}{u}$, so it must lie a little closer to $u$, giving us that $|\T{u}{v_w}| < |\T{u}{w}|$.
 
 \bigskip
 Since any vertex in $C_3^u$ would be further from $w$ than $u$ itself, these four cases are exhaustive.
\end{proof}

\section{Spanning ratio}
\label{sec:t5-SpanningRatio}

In this section, we prove an upper bound on the spanning ratio of $\Theta_5$.

\begin{lemma}
 \label{lem:t5-spanningPath}
 Between any pair of vertices $u$ and $w$ of $\Theta_5$, there is a path of length at most $c \cdot |\T{u}{w}|$, where $c = \valc$.
\end{lemma}
\begin{proof}
 We begin in a way similar to the proof of Theorem~\ref{thm:t5-connected}. Given an ordered pair of vertices $u$ and $w$, we perform induction on the size of their canonical triangle. If $|\T{u}{w}|$ is minimal, there must be a direct edge between them. Since $c > 1$ and any edge inside $\T{u}{w}$ with endpoint $u$ has length at most $|\T{u}{w}|$, this proves the base case. The rest of the proof deals with the inductive step, where we assume that there exists a path of length at most $c \cdot |\triangle|$ between every pair of vertices whose canonical triangle $\triangle$ is smaller than $\T{u}{w}$. As in the proof of Theorem~\ref{thm:t5-connected}, we assume that $w$ lies in the right half of $C_0^u$. If $w$ lies to the left of $x$, we have seen that $\T{w}{u}$ is smaller than $\T{u}{w}$. Therefore we can apply induction to obtain a path of length at most $c \cdot |\T{w}{u}| < c \cdot |\T{u}{w}|$ between $u$ and $w$. Hence we need to concern ourselves only with the case where $w$ lies on or to the right of $x$.

 If $u$ is the vertex closest to $w$ in $C_3^w$ or $w$ is the closest vertex to $u$ in $C_0^u$, there is a direct edge between them and we are done by the same reasoning as in the base case. Therefore assume that this is not the case and let $v_w$ be the vertex closest to $w$ in $C_3^w$. We distinguish the same four cases for the location of $v_w$ (see~Figure~\ref{fig:t5-cases-a}). We already showed that we can apply induction on $\T{u}{v_w}$ in each case. This is a crucial part of the proof for the first three cases.

 The basic strategy for the rest of the proof is as follows. If we can find a path of length $g \cdot |\T{u}{w}|$ that leaves us with a strictly smaller canonical triangle of size $h \cdot |\T{u}{w}|$, where $h < 1$, we can then apply induction to obtain a path of length $g \cdot |\T{u}{w}| + c \cdot h \cdot |\T{u}{w}|$. Since we aim to show that there is a path of length at most $c \cdot |\T{u}{w}|$, we can derive:
 \begin{align*}
    g \cdot |\T{u}{w}| + c \cdot h \cdot |\T{u}{w}|~~&\leq~~c \cdot |\T{u}{w}|\\
                                      g + c \cdot h~~&\leq~~c\\
                                                  g~~&\leq~~(1 - h) \cdot c\\
                                      \frac{g}{1-h}~~&\leq~~c.
 \end{align*}
 Therefore we are done if $g/(1-h) \leq \valc$.

 \case{1} $v_w$ lies in $C_2^u$. By induction, there exists a path between $u$ and $v_w$ of length at most $c \cdot |\T{u}{v_w}|$. Since $v_w$ is the closest vertex to $w$ in $C_3^w$, there is a direct edge between them, giving a path between $u$ and $w$ of length at most $|wv_w| + c \cdot |\T{u}{v_w}|$.
 
 Given any initial position of $v_w$ in $C_2^u$, we can increase $|wv_w|$ by moving $w$ to the right. Since this does not change $|\T{u}{v_w}|$, the worst case occurs when $w$ lies on $b$. Then we can increase both $|wv_w|$ and $|\T{u}{v_w}|$ by moving $v_w$ into the bottom corner of $\T{w}{u}$. This gives rise to the same worst-case configuration as in the proof of Theorem~\ref{thm:t5-connected}, depicted in Figure~\ref{fig:t5-cases-b}. Building on the analysis there, we can bound the worst-case length of the path as follows.
 \[
     |wv_w| + c \cdot |\T{u}{v_w}|
~~=~~\frac{|\T{u}{w}|}{\cos \frac{\pi}{5}} + c \cdot \frac{\sin \frac{3\pi}{5}}{\sin \frac{3\pi}{10}} \cdot \tan \frac{\pi}{5} \cdot |\T{u}{w}|
 \]
 This is at most $c \cdot |\T{u}{w}|$ for $c \geq 2 ( 2 + \sqrt{5} )$. Since we picked $c = 2 ( 2 + \sqrt{5} )$,  the theorem holds in this case. Note that this is one of the cases that determines the value of $c$.
 
\begin{figure}[htb]
 \centering
 \begin{subfigure}[b]{0.48\textwidth}
  \centering
  \includegraphics{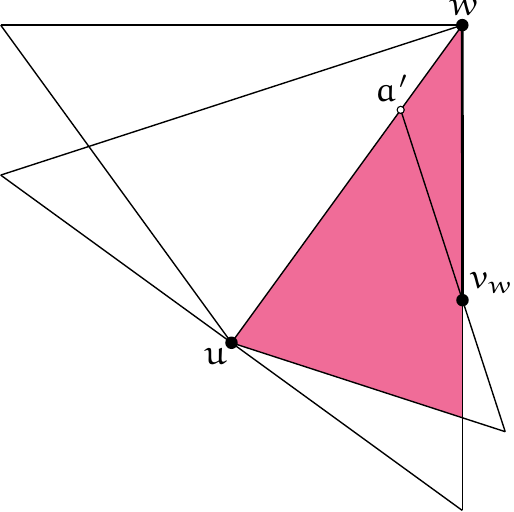}
  \caption{}
  \label{fig:t5-worst-a}
 \end{subfigure}
 \begin{subfigure}[b]{0.48\textwidth}
  \centering
  \includegraphics{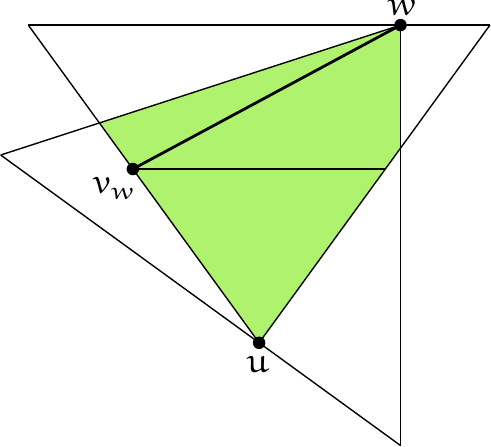}
  \caption{}
  \label{fig:t5-worst-b}
 \end{subfigure}
 \caption{(a) Case 2: Vertex $v_w$ lies on the boundary of $C_3^w$ after moving it down along the side of $\T{u}{v_w}$. (b) Case 3: Vertex $v_w$ lies on the boundary of $C_0^u$ after moving it left along the side of $\T{u}{v_w}$.}
\end{figure}

 \case{2} $v_w$ lies in $C_1^u$. By the same reasoning as in the previous case, we have a path of length at most $|wv_w| + c \cdot |\T{u}{v_w}|$ between $u$ and $w$ and we need to bound this length by $c \cdot |\T{u}{w}|$.
 
 Given any initial position of $v_w$ in $C_1^u$, we can increase $|wv_w|$ by moving $w$ to the right. Since this does not change $|\T{u}{v_w}|$, the worst case occurs when $w$ lies on $b$. We can further increase $|wv_w|$ by moving $v_w$ down along the side of $\T{u}{v_w}$ opposite $u$ until it hits the boundary of $C_1^u$ or $C_3^w$, whichever comes first (see~Figure~\ref{fig:t5-worst-a}).
 
 Now consider what happens when we move $v_w$ along these boundaries. If $v_w$ lies on the boundary of $C_1^u$ and we move it away from $u$ by $\varepsilon$, $|\T{u}{v_w}|$ increases by $\varepsilon$. At the same time, $|wv_w|$ might decrease, but not by more than $\varepsilon$. Since $c > 1$, the total path length is maximized by moving $v_w$ as far from $u$ as possible, until it hits the boundary of $C_3^w$. Once $v_w$ lies on the boundary of $C_3^w$, we can express the size of $\T{u}{v_w}$ as follows, where $a'$ is the top corner of $\T{u}{v_w}$.
\begin{align*}
  |\T{u}{v_w}|
  ~~&=~~ |\T{u}{w}| - |wa'|\\
  &=~~ |\T{u}{w}| - |wv_w| \cdot \frac{\sin \angle wv_wa'}{\sin \angle wa'v_w}\\
  &=~~ |\T{u}{w}| - |wv_w| \cdot \frac{\sin \frac{\pi}{10}}{\sin \frac{7\pi}{10}}
\end{align*}
Now we can express the length of the complete path as follows.
\begin{align*}
  |wv_w| + c \cdot |\T{u}{v_w}|
  ~~&=~~|wv_w| + c \cdot \left( |\T{u}{w}| - |wv_w| \cdot \frac{\sin \frac{\pi}{10}}{\sin \frac{7\pi}{10}} \right)\\
  &=~~c \cdot |\T{u}{w}| - \left( c \cdot \frac{\sin \frac{\pi}{10}}{\sin \frac{7\pi}{10}} - 1 \right) \cdot |wv_w|
\end{align*}
 Since $c > \sin \frac{7\pi}{10} / \sin \frac{\pi}{10} \approx 2.618$, we have that $c \cdot (\sin \frac{\pi}{10} / \sin \frac{7\pi}{10}) - 1 > 0$. Therefore $|wv_w| + c \cdot |\T{u}{v_w}| < c \cdot |\T{u}{w}|$.
 
 \case{3} $v_w$ lies in $C_0^u$. Again, we have a path of length at most $|wv_w| + c \cdot |\T{u}{v_w}|$ between $u$ and $w$ and we need to bound this length by $c \cdot |\T{u}{w}|$.
 
 Given any initial position of $v_w$ in $C_0^u$, moving $v_w$ to the left increases $|wv_w|$ while leaving $|\T{u}{v_w}|$ unchanged. Therefore the path length is maximized when $v_w$ lies on the boundary of either $C_0^u$ or $C_3^w$, whichever it hits first (see~Figure~\ref{fig:t5-worst-b}).
 
 \extraStretch{1em}{Again, consider what happens when we move $v_w$ along these boundaries. Similar to the previous case, if $v_w$ lies on the boundary of $C_0^u$ and we move it away from $u$ by $\varepsilon$, $|\T{u}{v_w}|$ increases by $\varepsilon$, while $|wv_w|$ might decrease by at most $\varepsilon$. Since $c > 1$, the total path length is maximized by moving $v_w$ as far from $u$ as possible, until it hits the boundary of $C_3^w$. Once there, the situation is symmetric to the previous case, with $|\T{u}{v_w}| = |\T{u}{w}| - |wv_w| \cdot (\sin \frac{\pi}{10} / \sin \frac{7\pi}{10})$. Therefore the theorem holds in this case as well.}
 
 \case{4} $v_w$ lies in $C_4^u$. This is the hardest case. Similar to the previous two cases, the size of $\T{u}{v_w}$ can be arbitrarily close to that of $\T{u}{w}$, but in this case $|wv_w|$ does not approach $0$. This means that simply invoking the inductive hypothesis on $\T{u}{v_w}$ does not work, so another strategy is required. We first look at a sub-case where we \emph{can} apply induction directly, before considering the position of $v_u$, the closest vertex to $u$ in $C_0$.
 
 \begin{figure}[htb]
  \centering
  \includegraphics{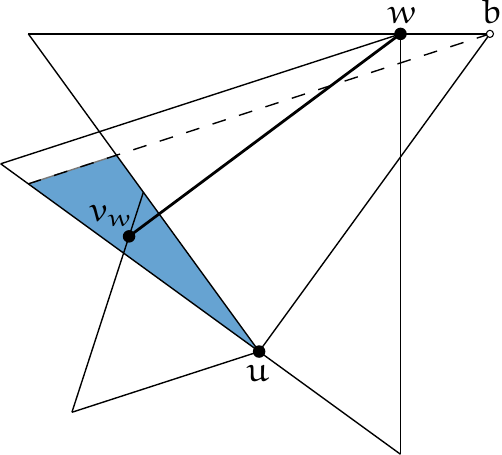}
  \caption{Case 4a: Vertex $v_w$ lies in $C_4^u \cap C_3^b$.}
  \label{fig:t5-worst-c}
 \end{figure}

 \case{4a} $v_w$ lies in $C_4^u \cap C_3^b$. This situation is illustrated in Figure~\ref{fig:t5-worst-c}. Given any initial position of $v_w$, moving $w$ to the right onto $b$ increases the total path length by increasing $|wv_w|$ while not affecting $|\T{u}{v_w}|$. Here we use the fact that $v_w$ already lies in $C_3^b$, otherwise we would not be able to move $w$ onto $b$ while keeping $v_w$ in $C_3^w$. Now the total path length is maximized by placing $v_w$ on the left corner of $\T{w}{u}$. Since this situation is symmetrical to the worst-case situation in Case 1, the theorem holds by the same analysis.

 \begin{figure}[htb]
  \centering
  \includegraphics{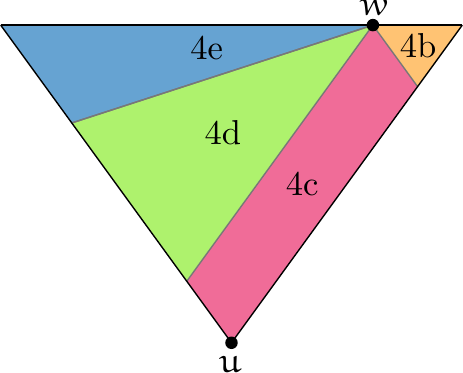}
  \caption{The four different cases for the position of $v_u$.}
  \label{fig:t5-closestcases-a}
 \end{figure}

 \bigskip
 Next, we distinguish four cases for the position of $v_u$ (the closest vertex to $u$ in $C_0$), illustrated in Figure~\ref{fig:t5-closestcases-a}. The cases are: (4b) $w$ lies in $C_4^{v_u}$, (4c) $w$ lies in $C_0^{v_u}$, (4d) $w$ lies in $C_1^{v_u}$ and $v_u$ lies in $C_3^w$, and (4e) $w$ lies in $C_1^{v_u}$ and $v_u$ lies in $C_4^w$. These are exhaustive, since the cones $C_4$, $C_0$ and $C_1$ are the only ones that can contain a vertex above the current vertex, and $w$ must lie above $v_u$, as $v_u$ is closer to $u$. Further, if $w$ lies in $C_1^{v_u}$, $v_u$ must lie in one of the two opposite cones of $w$. We can solve the first two cases by applying our inductive hypothesis to $\T{v_u}{w}$.
 
\begin{figure}[htb]
 \centering
 \begin{subfigure}[b]{0.48\textwidth}
  \centering
  \includegraphics{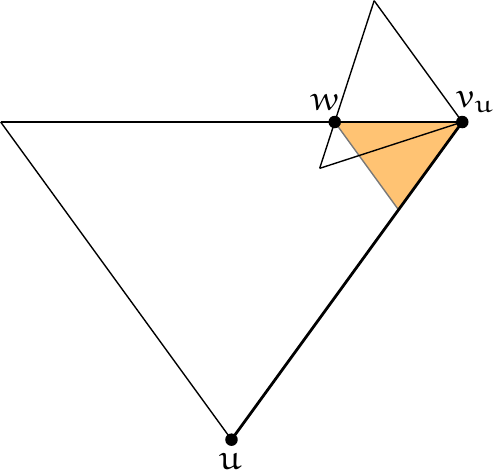}
  \caption{}
  \label{fig:t5-closestcases-b}
 \end{subfigure}
 \begin{subfigure}[b]{0.48\textwidth}
  \centering
  \includegraphics{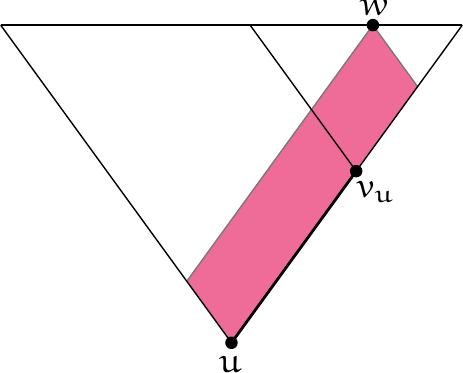}
  \caption{}
  \label{fig:t5-closestcases-c}
 \end{subfigure}
 \caption{(a) The worst-case configuration with $w$ in $C_4^{v_u}$. (b) A configuration with $w$ in $C_0^{v_u}$, after moving $v_u$ onto the right side of $C_0^u$.}
\end{figure}

 \case{4b} $w$ lies in $C_4^{v_u}$. To apply our inductive hypothesis, we need to show that $|\T{v_u}{w}| < |\T{u}{w}|$. If that is the case, we obtain a path between $v_u$ and $w$ of length at most $c \cdot |\T{v_u}{w}|$. Since $v_u$ is the closest vertex to $u$, there is a direct edge from $u$ to $v_u$, resulting in a path between $u$ and $w$ of length at most $|uv_u| + c \cdot |\T{v_u}{w}|$.
 
 Given any initial positions for $v_u$ and $w$, moving $w$ to the left increases $|\T{v_u}{w}|$ while leaving $|uv_u|$ unchanged. Moving $v_u$ closer to $b$ increases both. Therefore the path length is maximal when $w$ lies on $x$ and $v_u$ lies on $b$ (see~Figure~\ref{fig:t5-closestcases-b}). Using the law of sines, we can express $|\T{v_u}{w}|$ as follows.
 \begin{align*}
   |\T{v_u}{w}|~~&=~~\frac{\sin \frac{3\pi}{5}}{\sin \frac{3\pi}{10}} \cdot |wv_u|~~=~~\frac{\sin \frac{3\pi}{5}}{\sin \frac{3\pi}{10}} \cdot \frac{\sin \frac{\pi}{10}}{\sin \frac{3\pi}{5}} \cdot |\T{u}{w}|\\
   &=~~ \frac{\sin \frac{\pi}{10}}{\sin \frac{3\pi}{10}} \cdot |\T{u}{w}|~~=~~\frac{1}{2} \left( 3 - \sqrt{5} \right) \cdot |\T{u}{w}|
 \end{align*}
 Since $\frac{1}{2} \left( 3 - \sqrt{5} \right) < 1$, we have that $|\T{v_u}{w}| < |\T{u}{w}|$ and we can apply our inductive hypothesis to $\T{v_u}{w}$. Since $|uv_u| = |\T{u}{w}|$, the complete path has length at most $c \cdot |\T{u}{w}|$ for
 \[
  c
  ~~\geq~~ \frac{1}{1 - \frac{1}{2} \left( 3 - \sqrt{5} \right)}
  ~~=~~ \frac{1}{2} \left( 1 + \sqrt{5} \right)
  ~~\approx~~ 1.618.
 \]
 
 \case{4c} $w$ lies in $C_0^{v_u}$. Since $v_u$ lies in $C_0^u$, it is clear that $|\T{v_u}{w}| < |\T{u}{w}|$, which allows us to apply our inductive hypothesis. This gives us a path between $u$ and $w$ of length at most $|uv_u| + c \cdot |\T{v_u}{w}|$. For any initial location of $v_u$, we can increase the total path length by moving $v_u$ to the right until it hits the side of $C_0^u$ (see~Figure~\ref{fig:t5-closestcases-c}), since $|\T{v_u}{w}|$ stays the same and $|uv_u|$ increases. Once there, we have that $|uv_u| + |\T{v_u}{w}| = |\T{u}{w}|$. Since $c > 1$, this immediately implies that $|uv_u| + c \cdot |\T{v_u}{w}| \leq c \cdot |\T{u}{w}|$.

 \bigskip
 To solve the last two cases, we need to consider the positions of both $v_u$ and $v_w$. Recall that for $v_w$, there is only a small region left where we have not yet proved the existence of a short path between $u$ and $w$. In particular, this is the case when $v_w$ lies in cone $C_4^u$, but not in $C_3^b$.

 \begin{figure}[htb]
 \centering
 \begin{subfigure}[b]{0.48\textwidth}
  \centering
  \includegraphics{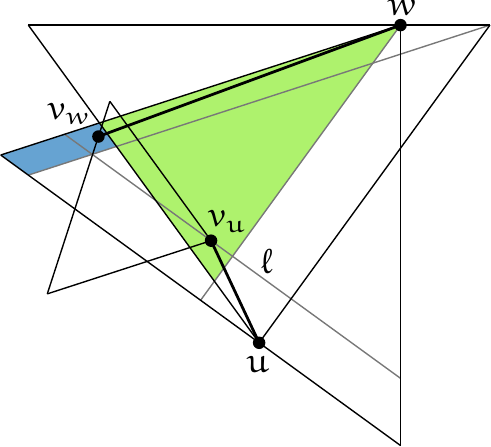}
  \caption{}
  \label{fig:t5-complex-a}
 \end{subfigure}
 \begin{subfigure}[b]{0.48\textwidth}
  \centering
  \includegraphics{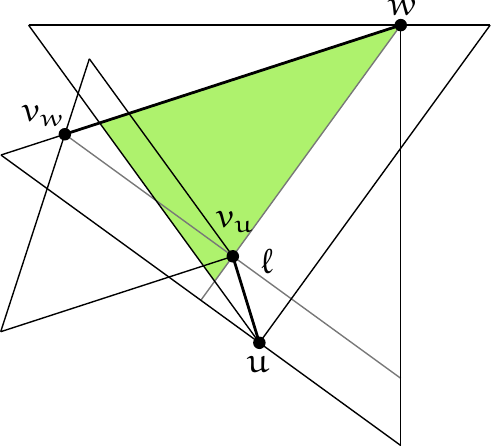}
  \caption{}
  \label{fig:t5-complex-b}
 \end{subfigure}
 \caption{(a) The regions where $v_u$ (light) and $v_w$ (dark) can lie. (b) The worst case when $v_u$ lies on a given line $\ell$.}
\end{figure}

 \case{4d} $w$ lies in $C_1^{v_u}$ and $v_u$ lies in $C_3^w$. We would like to apply our inductive hypothesis to $\T{v_u}{v_w}$, resulting in a path between $v_u$ and $v_w$ of length at most $c \cdot |\T{v_u}{v_w}|$. The edges $(w, v_w)$ and $(u, v_u)$ complete this to a path between $u$ and $w$, giving a total length of at most $|uv_u| + c \cdot |\T{v_u}{v_w}| + |v_ww|$.

 First, note that $v_u$ cannot lie in $\T{w}{v_w}$, as this region is empty by definition. Since $v_w$ lies in $C_4^u$, this means that $v_w$ must lie in $C_4^{v_u}$. We first show that $\T{v_u}{v_w}$ is always smaller than $\T{u}{w}$, which means that we are allowed to use induction. Given any initial position for $v_u$, consider the line $\ell$ through $v_u$, perpendicular to the bisector of $C_3$ (see~Figure~\ref{fig:t5-complex-a}). Since $v_w$ cannot be further from $w$ than $v_u$, the size of $\T{v_u}{v_w}$ is maximized when $v_w$ lies on the intersection of $\ell$ and the top boundary of $\T{w}{u}$. We can increase $|\T{v_u}{v_w}|$ further by moving $v_u$ along $\ell$ until it reaches the bisector of $C_3^w$ (see~Figure~\ref{fig:t5-complex-b}). Since the top boundary of $\T{w}{u}$ and the bisector of $C_3^w$ approach each other as they get closer to $w$, the size of $\T{v_u}{v_w}$ is maximized when $v_u$ lies on the bottom boundary of $\T{w}{u}$ (ignoring for now that this would move $v_u$ out of $\T{u}{w}$). Now it is clear that $|\T{v_u}{v_w}| < |\T{u}{v_w}|$. Since we already established that $\T{u}{v_w}$ is smaller than $\T{u}{w}$ in the proof of Theorem~\ref{thm:t5-connected}, this holds for $\T{v_u}{v_w}$ as well and we can use induction.

 \begin{figure}[htb]
  \centering
  \includegraphics{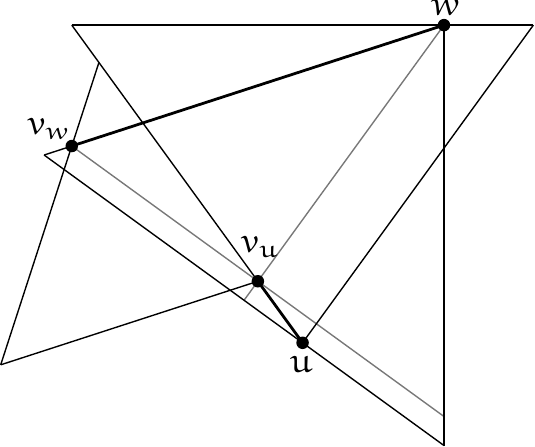}
  \caption{The worst case for a fixed position of $w$.}
  \label{fig:t5-complex-c}
 \end{figure}

 All that is left is to bound the total length of the path. Given any initial position of $v_u$, the path length is maximized when we place $v_w$ at the intersection of $\ell$ and the top boundary of $\T{w}{u}$, as this maximizes both $|\T{v_u}{v_w}|$ and $|wv_w|$. When we move $v_u$ away from $v_w$ along $\ell$ by $\varepsilon$, $|uv_u|$ decreases by at most $\varepsilon$, while $|\T{v_u}{v_w}|$ increases by $\sin \frac{3\pi}{5} / \sin \frac{3\pi}{10} \cdot \varepsilon > \varepsilon$. Since $c > 1$, this increases the total path length. Therefore the worst case again occurs when $v_u$ lies on the bisector of $C_3^w$, as depicted in Figure~\ref{fig:t5-complex-b}. Moving $v_u$ down along the bisector of $\T{w}{u}$ by $\varepsilon$ decreases $|uv_u|$ by at most $\varepsilon$, while increasing $|wv_w|$ by $1 / \sin \frac{3\pi}{10} \cdot \varepsilon > \varepsilon$ and increasing $|\T{v_u}{v_w}|$. Therefore this increases the total path length and the worst case occurs when $v_u$ lies on the left boundary of $\T{u}{w}$ (see~Figure~\ref{fig:t5-complex-c}).

 Finally, consider what happens when we move $v_u$ $\varepsilon$ towards $u$, while moving $w$ and $v_w$ such that the construction stays intact. This causes $w$ to move to the right. Since $v_u$, $w$ and the left corner of $\T{u}{w}$ form an isosceles triangle with apex $v_u$, this also moves $v_u$ $\varepsilon$ further from $w$. We saw before that moving $v_u$ away from $w$ increases the size of $\T{v_u}{v_w}$. Finally, it also increases $|wv_w|$ by $1 / \sin \frac{3\pi}{10} \cdot \varepsilon > \varepsilon$. Thus, the increase in $|wv_w|$ cancels the decrease in $|uv_u|$ and the total path length increases. Therefore the worst case occurs when $v_u$ lies almost on $u$ and $v_w$ lies in the corner of $\T{w}{u}$, which is symmetric to the worst case of Case~1. Thus the theorem holds by the same analysis.

 \case{4e} $w$ lies in $C_1^{v_u}$ and $v_u$ lies in $C_4^w$. We split this case into three final sub-cases, based on the position of $v_u$. These cases are illustrated in Figure~\ref{fig:t5-complex2-a}. Note that $v_u$ cannot lie in $C_2$ or $C_3$ of $v_w$, as it lies above $v_w$. It also cannot lie in $C_4^{v_w}$, as $C_4^{v_w}$ is completely contained in $C_4^u$, whereas $v_u$ lies in $C_0^u$. Thus the cases presented below are exhaustive.

 \begin{figure}[ht]
  \centering
  \includegraphics{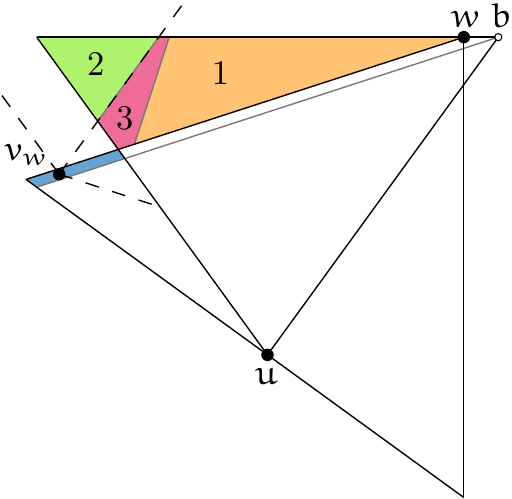}
  \caption{The three sub-cases for the position of $v_u$.}
  \label{fig:t5-complex2-a}
 \end{figure}

 \case{4e-1} $|\T{w}{v_u}| \leq \frac{c - 1}{c} \cdot |\T{u}{w}|$. If $\T{w}{v_u}$ is small enough, we can apply our inductive hypothesis to obtain a path between $v_u$ and $w$ of length at most $c \cdot |\T{w}{v_u}|$. Since there is a direct edge between $u$ and $v_u$, we obtain a path between $u$ and $w$ of length at most $|uv_u| + c \cdot |\T{w}{v_u}|$. Any edge from $u$ to a point inside $\T{u}{w}$ has length at most $|\T{u}{w}|$, so we can bound the length of the path as follows.
 \begin{align*}
  |uv_u| + c \cdot |\T{w}{v_u}|
  ~~&\leq~~ |\T{u}{w}| + c \cdot \frac{c - 1}{c} \cdot |\T{u}{w}|\\
  ~~&=~~ |\T{u}{w}| + (c - 1) \cdot |\T{u}{w}|\\
  ~~&=~~ c \cdot |\T{u}{w}|
 \end{align*}

 \bigskip
 In the other two cases, we use induction on $\T{v_w}{v_u}$ to obtain a path between $v_w$ and $v_u$ of length at most $c \cdot |\T{v_w}{v_u}|$. The edges $(u, v_u)$ and $(w, v_w)$ complete this to a (self-intersecting) path between $u$ and $w$. We can bound the length of these edges by the size of the canonical triangles that contain them, as follows.
 \begin{align*}
  |uv_u| + |wv_w|
  ~~&\leq~~ |\T{u}{w}| + |\T{w}{u}|\\
  ~~&\leq~~ |\T{u}{w}| + \frac{1}{\cos \frac{\pi}{5}} \cdot |\T{u}{w}|\\
  ~~&=~~ \sqrt{5} \cdot |\T{u}{w}|
 \end{align*}
 All that is left now is to bound the size of $\T{v_w}{v_u}$ and express it in terms of $\T{u}{w}$.

 \begin{figure}[htb]
 \centering
 \begin{subfigure}[b]{0.51\textwidth}
  \centering
  \includegraphics{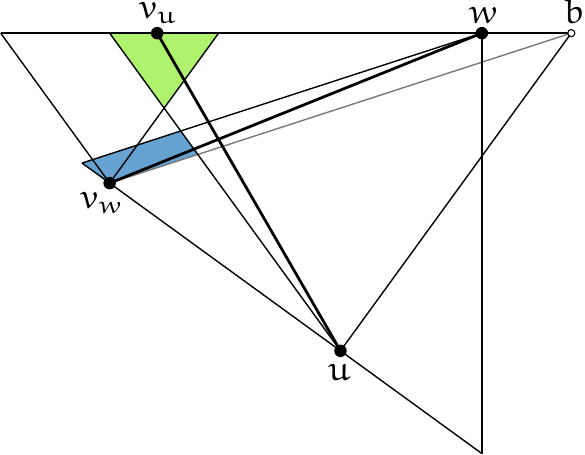}
  \caption{}
  \label{fig:t5-complex2-b}
 \end{subfigure}
 \begin{subfigure}[b]{0.45\textwidth}
  \centering
  \includegraphics{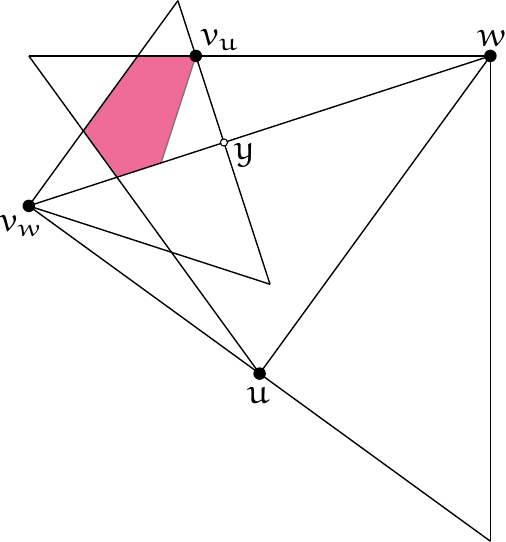}
  \caption{}
  \label{fig:t5-complex2-c}
 \end{subfigure}
 \caption{(a) The situation that maximizes $\T{v_w}{v_u}$ when $v_u$ lies in $C_0^{v_w}$. (b) The worst case when $v_u$ lies in $C_1^{v_w}$.}
\end{figure}

 \case{4e-2} $v_u$ lies in $C_0^{v_w}$. In this case, the size of $\T{v_w}{v_u}$ is maximal when $v_u$ lies on the top boundary of $\T{u}{w}$ and $v_w$ lies at the lowest point in its possible region: the left corner of $\T{b}{u}$ (see~Figure~\ref{fig:t5-complex2-b}). Now we can express $|\T{v_w}{v_u}|$ as follows.
 \begin{align*}
  |\T{v_w}{v_u}|
  ~~&=~~ \frac{\sin \frac{\pi}{10}}{\sin \frac{7\pi}{10}} \cdot |bv_w|\\
  ~~&=~~ \frac{\sin \frac{\pi}{10}}{\sin \frac{7\pi}{10}} \cdot \frac{1}{\cos \frac{\pi}{5}} \cdot |\T{u}{w}|\\
  ~~&=~~ 2 \left( \sqrt{5} - 2 \right) \cdot |\T{u}{w}|
 \end{align*}
 Since $2 \left( \sqrt{5} - 2 \right) < 1$, we can use induction. The total path length is bounded by $c \cdot |\T{u}{w}|$ for
 \[
  c
  ~~\geq~~ \frac{\sqrt{5}}{1 - 2 \left( \sqrt{5} - 2 \right)}
  ~~=~~ 2 + \sqrt{5}
  ~~\approx~~ 4.236.
 \]

 \case{4e-3} $v_u$ lies in $C_1^{v_w}$. Since $|\T{w}{v_u}| > \frac{c - 1}{c} \cdot |\T{u}{w}|$, $\T{v_w}{v_u}$ is maximal when $v_w$ lies on the left corner of $\T{w}{u}$ and $v_u$ lies on the top boundary of $\T{u}{w}$, such that $|\T{w}{v_u}| = \frac{c - 1}{c} \cdot |\T{u}{w}|$ (see~Figure~\ref{fig:t5-complex2-c}). Let $y$ be the intersection of $\T{v_w}{v_u}$ and $\T{w}{u}$. Note that since $v_w$ lies on the corner of $\T{w}{u}$, $y$ is also the midpoint of the side of $\T{v_w}{v_u}$ opposite $v_w$. We can express the size of $\T{v_w}{v_u}$ as follows.
 \begin{align*}
  |\T{v_w}{v_u}|
  ~~&=~~\frac{|v_wy|}{\cos \frac{\pi}{5}}\\
  ~~&=~~\frac{|wv_w| - |wy|}{\cos \frac{\pi}{5}}\\
  ~~&=~~\frac{\displaystyle\frac{|\T{u}{w}|}{\cos \frac{\pi}{5}} - \cos {\textstyle \frac{\pi}{10}} \cdot |wv_u|}{\cos \frac{\pi}{5}}\\
  &=~~\frac{\displaystyle\frac{|\T{u}{w}|}{\cos \frac{\pi}{5}} - \cos {\textstyle \frac{\pi}{10}} \cdot \frac{\sin \frac{3\pi}{10}}{\sin \frac{3\pi}{5}} \cdot |\T{w}{v_u}|}{\cos \frac{\pi}{5}}\\
  ~~&=~~\frac{\displaystyle\frac{|\T{u}{w}|}{\cos \frac{\pi}{5}} - \cos {\textstyle \frac{\pi}{10}} \cdot \frac{\sin \frac{3\pi}{10}}{\sin \frac{3\pi}{5}} \cdot \frac{c - 1}{c} \cdot |\T{u}{w}|}{\cos \frac{\pi}{5}}\\
  &=~~\left( \frac{1}{c} + 5 - 2 \sqrt{5} \right) \cdot |\T{u}{w}|
 \end{align*}
 Thus we can use induction for $c > 1 / \left(2 \sqrt{5} - 4\right) \approx 2.118$ and the total path length can be bounded by $c \cdot |\T{u}{w}|$ for
 \[
  c
  ~~\geq~~\frac{\sqrt{5} + 1}{2 \sqrt{5} - 4}
  ~~=~~ \frac{1}{2}\left( 7 + 3 \sqrt{5} \right)
  ~~\approx~~ 6.854. \qedhere
 \]
\end{proof}

Using this result, we can compute the exact spanning ratio.

\begin{theorem}
 \label{thm:t5-spanner}
 The \graph has spanning ratio at most \valsr.
\end{theorem}
\begin{proof}
 Given two vertices $u$ and $w$, we know from Lemma~\ref{lem:t5-spanningPath} that there is a path between them of length at most $c \cdot \min\left(|\T{u}{w}|, |\T{w}{u}|\right)$, where $c = \valc$. This gives an upper bound on the spanning ratio of $c \cdot \min\left(|\T{u}{w}|, |\T{w}{u}|\right) / |uw|$. We assume without loss of generality that $w$ lies in the right half of $C_0^u$. Let $\alpha$ be the angle between the bisector of $C_0^u$ and the line $uw$ (see~Figure~\ref{fig:t5-canon-b}). In the proof of Theorem~\ref{thm:t5-connected}, we saw that we can express $|\T{w}{u}|$ and $|uw|$ in terms of $\alpha$ and $|\T{u}{w}|$, as $|\T{w}{u}| = (\cos (\frac{\pi}{5} - \alpha) / \cos \alpha) \cdot |\T{u}{w}|$ and $|uw| = (\cos \frac{\pi}{5} / \cos \alpha) \cdot |\T{u}{w}|$, respectively. Using these expressions, we can write the spanning ratio in terms of $\alpha$.
 \begin{align*}
   \frac{c \cdot \min\left(|\T{u}{w}|, |\T{w}{u}|\right)}{|uw|}~~&=~~\frac{c \cdot \min\left(|\T{u}{w}|, \frac{\cos \left( \frac{\pi}{5} - \alpha \right)}{\cos \alpha} \cdot |\T{u}{w}|\right)}{\frac{\cos \frac{\pi}{5}}{\cos \alpha} \cdot |\T{u}{w}|}\\
   &=~~\frac{c}{\cos \frac{\pi}{5}} \cdot \min\left(\cos \alpha, \cos \left( {\textstyle \frac{\pi}{5}} - \alpha \right)\right)
 \end{align*}
 To get an upper bound on the spanning ratio, we need to maximize the minimum of $\cos \alpha$ and $\cos \left(\frac{\pi}{5} - \alpha\right)$. Since for $\alpha \in [0, \pi / 5]$, one is increasing and the other is decreasing, this maximum occurs at $\alpha = \pi/10$, where they are equal. Thus, our upper bound becomes
 \[
  \frac{c}{\cos \frac{\pi}{5}} \cdot \cos {\textstyle \frac{\pi}{10}}
  ~~=~~\sqrt{50 + 22 \sqrt{5}}. \qedhere
 \]
\end{proof}

\section{Lower bound}
\label{sec:t5-lowerbound}

In this section, we derive a lower bound on the spanning ratio of the \graph.

\begin{figure}[ht]
  \begin{center}
    \includegraphics{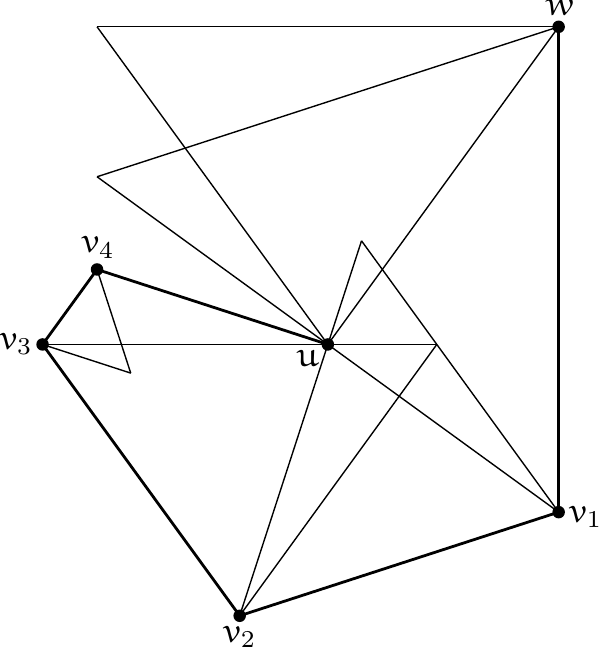}
  \end{center}
  \caption{A path with a large spanning ratio.}
  \label{fig:t5-lbpath}
\end{figure}

\begin{figure}[ht]
  \centering
  \includegraphics{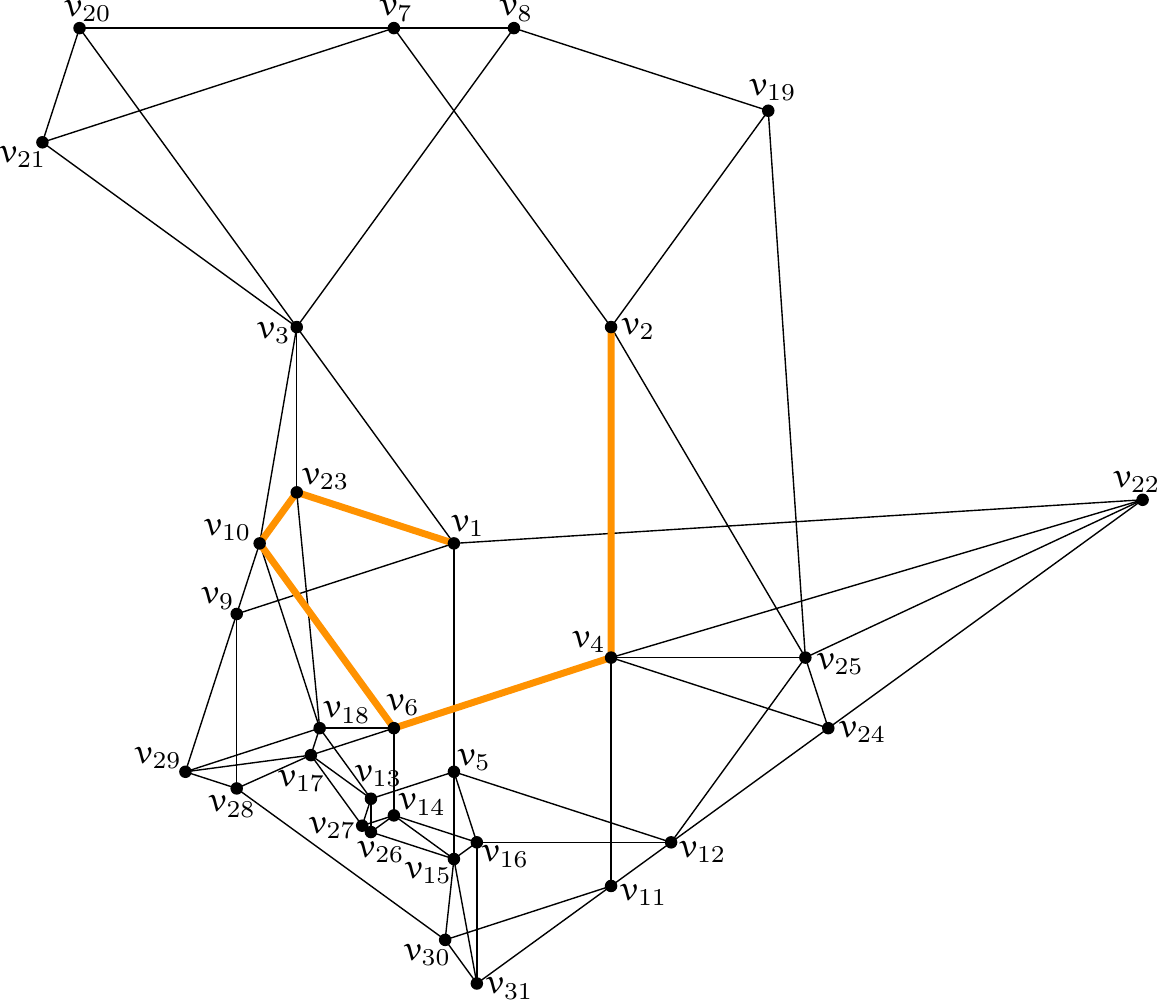}
  \caption{A \graph with a spanning ratio that matches the lower bound. The shortest path between $v_1$ and $v_2$ is indicated in bold.}
  \label{fig:t5-lbgraph}
 \end{figure}

\begin{theorem}
 \label{thm:t5-lowerbound}
 The \graph has spanning ratio at least $\vallb$.
\end{theorem}
\begin{proof}
 For the lower bound, we present and analyze a path between two vertices that has a large spanning ratio. The path has the following structure (illustrated in Figure~\ref{fig:t5-lbpath}).

 The path can be thought of as being directed from $w$ to $u$. First, we place $w$ in the right corner of $\T{u}{w}$. Then we add a vertex $v_1$ in the bottom corner of $\T{w}{u}$. We repeat this two more times, each time adding a new vertex in the corner of $\T{v_i}{u}$ furthest from $u$. The final vertex $v_4$ is placed on the top boundary of $C_1^{v_3}$, such that $u$ lies in $C_1^{v_4}$. Since we know all the angles involved, we can compute the length of each edge, taking $|uw| = 1$ as baseline.
 \begin{align*}
   |wv_1|~~&=~~\frac{1}{\cos \frac{\pi}{5}} & |v_1v_2|~~=~~|v_2v_3|~~&=~~2 \sin {\textstyle \frac{\pi}{5}} \tan {\textstyle \frac{\pi}{5}}\\
   |v_3v_4|~~&=~~\frac{\sin \frac{\pi}{10}}{\sin \frac{3\pi}{5}} \tan {\textstyle \frac{\pi}{5}} & |v_4u|~~&=~~\frac{\sin \frac{3\pi}{10}}{\sin \frac{3\pi}{5}} \tan {\textstyle \frac{\pi}{5}}
 \end{align*}
 Since we set $|uw| = 1$, the spanning ratio is simply $|wv_1| + |v_1v_2| + |v_2v_3| + |v_3v_4| + |v_4u| = \textstyle\frac{1}{2}(11\sqrt{5} -17) \approx 3.798$. Note that the \graph with just these 5 vertices would have a far smaller spanning ratio, as there would be a lot of shortcut edges. However, a graph where this path is the shortest path between two vertices can be found in Figure~\ref{fig:t5-lbgraph}, and its construction is described in Table~\ref{tab:t5-lbconstruction}.
\end{proof}

\begin{longtable}{cp{0.598\textwidth}c}
    \caption{Stepwise construction of a \graph with a spanning ratio that matches the lower bound (see Figure~\ref{fig:t5-lbgraph}).\label{tab:t5-lbconstruction}}\\
    \hline
    \spacedlowsmallcaps{\#} & \spacedlowsmallcaps{Action} & \spacedlowsmallcaps{Shortest path} \\
    \hline
    \endfirsthead
    \hline
    \spacedlowsmallcaps{\#} & \spacedlowsmallcaps{Action} & \spacedlowsmallcaps{Shortest path} \\
    \hline
    \endhead
    \hline
    \endfoot
    1 & Start with a vertex $v_{1}$. & - \\
    2 & Add $v_{2}$ in $C_0^u$, such that $v_{2}$ is arbitrarily close to the top right corner of $\T{v_{1}}{v_{2}}$. & $v_{1}v_{2}$\\
    3 & Remove edge $(v_{1}, v_{2})$ by adding two vertices, $v_{3}$ and $v_{4}$, arbitrarily close to the counter-clockwise corners of $\T{v_{1}}{v_{2}}$ and $\T{v_{2}}{v_{1}}$. & $v_{1}v_{4}v_{2}$\\
    4 & Remove edge $(v_{1}, v_{4})$ by adding two vertices, $v_{5}$ and $v_{6}$, arbitrarily close to the clockwise corner of $\T{v_{1}}{v_{4}}$ and the counter-clockwise corner of $\T{v_{4}}{v_{1}}$. & $v_{1}v_{3}v_{2}$\\
    5 & Remove edge $(v_{2}, v_{3})$ by adding two vertices, $v_{7}$ and $v_{8}$, arbitrarily close to the clockwise corner of $\T{v_{2}}{v_{3}}$ and the counter-clockwise corner of $\T{v_{3}}{v_{2}}$. & $v_{1}v_{6}v_{4}v_{2}$\\
    6 & Remove edge $(v_{1}, v_{6})$ by adding two vertices, $v_{9}$ and $v_{10}$, arbitrarily close to the clockwise corner of $\T{v_{1}}{v_{6}}$ and the counter-clockwise corner of $\T{v_{6}}{v_{1}}$. & $v_{1}v_{5}v_{4}v_{2}$\\
    7 & Remove edge $(v_{4}, v_{5})$ by adding two vertices, $v_{11}$ and $v_{12}$, arbitrarily close to the counter-clockwise corner of $\T{v_{4}}{v_{5}}$ and the clockwise corner of $\T{v_{5}}{v_{4}}$. & $v_{1}v_{5}v_{6}v_{4}v_{2}$\\
    8 & Remove edge $(v_{5}, v_{6})$ by adding two vertices, $v_{13}$ and $v_{14}$, arbitrarily close to the counter-clockwise corner of $\T{v_{5}}{v_{6}}$ and the clockwise corner of $\T{v_{6}}{v_{5}}$. & $v_{1}v_{5}v_{14}v_{6}v_{4}v_{2}$\\
    9 & Remove edge $(v_{5}, v_{14})$ by adding two vertices, $v_{15}$ and $v_{16}$, arbitrarily close to the counter-clockwise corner of $\T{v_{5}}{v_{14}}$ and the clockwise corner of $\T{v_{14}}{v_{5}}$. & $v_{1}v_{5}v_{13}v_{6}v_{4}v_{2}$\\
    10 & Remove edge $(v_{6}, v_{13})$ by adding two vertices, $v_{17}$ and $v_{18}$, arbitrarily close to the clockwise corner of $\T{v_{6}}{v_{13}}$ and the counter-clockwise corner of $\T{v_{13}}{v_{6}}$. & $v_{1}v_{3}v_{8}v_{2}$\\
    11 & Remove edge $(v_{2}, v_{8})$ by adding a vertex $v_{19}$ in the union of, and arbitrarily close to the intersection point of $\T{v_{2}}{v_{8}}$ and $\T{v_{8}}{v_{2}}$. & $v_{1}v_{3}v_{7}v_{2}$\\
    12 & Remove edge $(v_{3}, v_{7})$ by adding two vertices, $v_{20}$ and $v_{21}$, arbitrarily close to the counter-clockwise corner of $\T{v_{3}}{v_{7}}$ and the clockwise corner of $\T{v_{7}}{v_{3}}$. & $v_{1}v_{5}v_{12}v_{2}$\\
    13 & Remove edge $(v_{2}, v_{12})$ by adding a vertex $v_{22}$ arbitrarily close to the counter-clockwise corner of $\T{v_{2}}{v_{12}}$. & $v_{1}v_{10}v_{6}v_{4}v_{2}$\\
    14 & Remove edge $(v_{1}, v_{10})$ by adding a vertex $v_{23}$ in the union of $\T{v_{1}}{v_{10}}$ and $\T{v_{10}}{v_{1}}$, arbitrarily close to the top boundary of $C_1^{v_{10}}$, and such that $v_{1}$ lies in $C_1^{v_{23}}$, arbitrarily close to the bottom boundary. & $v_{1}v_{5}v_{12}v_{4}v_{2}$\\
    15 & Remove edge $(v_{4}, v_{12})$ by adding two vertices, $v_{24}$ and $v_{25}$, arbitrarily close to the counter-clockwise corner of $\T{v_{4}}{v_{12}}$ and the clockwise corner of $\T{v_{12}}{v_{4}}$. & $v_{1}v_{5}v_{13}v_{14}v_{6}v_{4}v_{2}$\\
    16 & Remove edge $(v_{13}, v_{14})$ by adding two vertices, $v_{26}$ and $v_{27}$, arbitrarily close to the clockwise corner of $\T{v_{13}}{v_{14}}$ and the counter-clockwise corner of $\T{v_{14}}{v_{13}}$. & $v_{1}v_{9}v_{18}v_{6}v_{4}v_{2}$\\
    17 & Remove edge $(v_{9}, v_{18})$ by adding two vertices, $v_{28}$ and $v_{29}$, arbitrarily close to the clockwise corner of $\T{v_{9}}{v_{18}}$ and the counter-clockwise corner of $\T{v_{18}}{v_{9}}$. & $v_{1}v_{5}v_{16}v_{11}v_{4}v_{2}$\\
    18 & Remove edge $(v_{11}, v_{16})$ by adding two vertices, $v_{30}$ and $v_{31}$, arbitrarily close to the counter-clockwise corner of $\T{v_{11}}{v_{16}}$ and the clockwise corner of $\T{v_{16}}{v_{11}}$. & $v_{1}v_{23}v_{10}v_{6}v_{4}v_{2}$\\
\end{longtable}

\section{Conclusions}
\label{sec:t5-conclusions}

We showed that there is a path between every pair of vertices in $\Theta_5$, and this path has length at most \valsr times the straight-line distance between the vertices. This is the first constant upper bound on the spanning ratio of the \graph, proving that it is a geometric spanner. We also presented a \graph with spanning ratio arbitrarily close to \vallb, thereby giving a lower bound on the spanning ratio. There is still a significant gap between these bounds, which is caused by the upper bound proof mostly ignoring the main obstacle to improving the lower bound: that every edge requires at least one of its canonical triangles to be empty. Hence we believe that the true spanning ratio is closer to the lower bound.

While our proof for the upper bound on the spanning ratio returns a spanning path between the two vertices, it requires knowledge of the neighbours of both the current vertex and the destination vertex. This means that the proof does not lead to a local routing strategy that can be applied in, say, a wireless setting. This raises the question whether it is possible to route \emph{competitively} on this graph, i.e. to discover a spanning path from one vertex to another by using only information local to the current vertices visited so far.

\bibliographystyle{plain}
\bibliography{../thesis}
\chapter{Competitive routing in the half-\texorpdfstring{$\Theta_6$}{Theta-6}-graph}
\label{ch:cr}

In this chapter, we present a deterministic local routing algorithm that is guaranteed to find a path between any pair of vertices in a half-$\Theta_6$-graph (the half-$\Theta_6$-graph is equivalent to the Delaunay triangulation where the empty region is an equilateral triangle). The length of the path is at most $5/\sqrt{3} \approx 2.887$ times the Euclidean distance between the pair of vertices. Moreover, we show that no local routing algorithm can achieve a better routing ratio, thereby proving that our routing algorithm is optimal. This is somewhat surprising because the spanning ratio of the half-$\Theta_6$-graph is 2, meaning that even though there always exists a path whose lengths is at most twice the Euclidean distance, we cannot always find such a path when routing locally.

Since every triangulation can be embedded in the plane as a half-$\Theta_6$-graph using $O(\log n)$ bits per vertex coordinate via Schnyder's embedding scheme~\cite{schnyder1990embedding}, our result provides a competitive local routing algorithm for every such embedded triangulation. Finally, we show how our routing algorithm can be adapted to provide a routing ratio of $15/\sqrt{3} \approx 8.661$ on two bounded degree subgraphs of the half-$\Theta_6$-graph.

The results in this chapter were first published in the proceedings of the 23rd ACM-SIAM Symposium on Discrete Algorithms (SODA 2012)~\cite{bose2012competitive}, and the proceedings of the 24th Canadian Conference on Computational Geometry (CCCG 2012)~\cite{bose2012competitive2}. A paper based on this chapter has been accepted for publication in the SIAM Journal on Computing~\cite{bose2015optimal}. This chapter is the result of joint work with Prosenjit Bose, Rolf Fagerberg, and Andr\'e van Renssen.

\section{Introduction}

A fundamental problem in networking is the routing of a message from one vertex to another in a graph. What makes routing more challenging is that often in a network the routing strategy must be \emph{local}. Informally, a routing strategy is \emph{local} when the routing algorithm must choose the next vertex to forward a message to based solely on knowledge of the current and destination vertex, and all vertices directly connected to the current vertex. Routing algorithms are considered \emph{geometric} when the underlying graph is embedded in the plane, with edges being straight line segments connecting pairs of points and weighted by the Euclidean distance between their endpoints. Geometric routing algorithms are important in wireless sensor networks \mbox{(see \cite{misra2009guide} and \cite{racke2009survey}} for surveys of the area), since they offer routing strategies that use the coordinates of the vertices to help guide the search as opposed to using the more traditional routing tables.  

Papadimitriou and Ratajczak~\cite{papadimitriou2005conjecture} posed a tantalizing question in this area that led to a flurry of activity: Does every 3-connected planar graph have a straight-line embedding in the plane that admits a local routing strategy? They were particularly interested in embeddings that admit a \emph{greedy} strategy, where a message is always forwarded to the vertex whose distance to the destination is the smallest among all vertices in the neighbourhood of the current vertex, including the current vertex. They provided a partial answer by showing that 3-connected planar graphs can always be embedded in $\mathbb{R}^3$ such that they admit a greedy routing strategy. They also showed that the class of complete bipartite graphs, $K_{k,6k+1}$ for all $k\geq 1$ cannot be embedded such that greedy routing always succeeds since every embedding has at least one vertex that is not connected to its nearest neighbour. Bose and Morin~\cite{bose2004online} showed that greedy routing always succeeds on Delaunay triangulations. In fact, a slightly restricted greedy routing strategy known as {\em greedy-compass} is the first local routing strategy shown to succeed on all triangulations~\cite{bose2002online}. Dhandapani~\cite{dhandapani2010greedy} proved the existence of an embedding that admits greedy routing for every triangulation and Angelini~\etal~\cite{angelini2010algorithm} provided a constructive proof. Leighton and Moitra~\cite{leighton2010some} settled Papadimitriou and Ratajczak's question by showing that every 3-connected planar graph can be embedded in the plane such that greedy routing succeeds. One drawback of these embedding algorithms is that the coordinates require $\Omega(n \log n)$ bits per vertex. To address this, He and Zhang~\cite{he2010schnyder} and Goodrich and Strash~\cite{goodrich2009succinct} gave succinct embeddings using only $O(\log n)$ bits per vertex. Recently, He and Zhang~\cite{he2011succinct} showed that every 3-connected plane graph admits a succinct embedding with convex faces on which a slightly modified greedy routing strategy always succeeds.

In light of these recent successes, it is surprising to note that the above routing strategies have solely concentrated on finding an embedding that guarantees that a local routing strategy will succeed, but pay little attention to the quality of the resulting path. For example, none of the above routing strategies have been shown to be \emph{competitive}. A geometric routing strategy is said to be competitive if the length of the path found by the routing strategy is not more than a constant times the Euclidean distance between its endpoints. This constant is called the \emph{routing ratio}. Bose and Morin~\cite{bose2004online} show that many local routing strategies are not competitive, but show how to route competitively on the Delaunay triangulation. However, Dillencourt~\cite{dillencourt1990realizability} showed that not all triangulations can be embedded in the plane as Delaunay triangulations. This raises the following question: can \emph{every} triangulation be embedded in the plane such that it admits a competitive local routing strategy? We answer this question in the affirmative.

\extraStretch{1em}{The \hts was introduced by Bonichon~\etal~\cite{bonichon2010connections}, who showed that it is identical to the Delaunay triangulation where the empty region is an equilateral triangle. Although both graphs are identical, the local definition of the \hts makes it more useful in the context of routing. We formally define the \hts in the next section. Our main result is a deterministic local routing algorithm that is guaranteed to find a path between any pair of vertices in a \hts whose length is at most $5/\sqrt{3} \approx 2.887$ times the Euclidean distance between the pair of vertices. On the way to proving our main result, we uncover some local properties of spanning paths in the \hts. Since Schnyder~\cite{schnyder1990embedding} showed that every triangulation can be embedded in the plane as a \hts using $O(\log n)$ bits per vertex coordinate, our main result implies that every triangulation has an embedding that admits a competitive local routing algorithm. Moreover, we show that no local routing algorithm can achieve a better routing ratio on a \hts, implying that our routing algorithm is optimal. This is somewhat surprising because Chew~\cite{chew1989there} showed that the spanning ratio of the \hts is 2. Thus, our lower bound provides a separation between the spanning ratio of the \hts and the best achievable routing ratio on the \hts. We believe that this is the first separation between the spanning ratio and routing ratio of any graph. It also makes the \hts one of the few graphs for which tight spanning and routing ratios are known. Finally, we show how our routing algorithm can be adapted to provide a routing ratio of $15/\sqrt{3} \approx 8.661$ on two bounded degree subgraphs of the \hts introduced by Bonichon~\etal~\cite{bonichon2010plane}. To the best of our knowledge, this is the first competitive routing algorithm on a bounded-degree plane graph.}

\section{Preliminaries} \label{sec:cr-prelims}

In this section we describe the construction of the \hts and introduce a few related concepts. Readers who are not familiar with general $\Theta$-graphs may want to read Chapter~\ref{ch:csi} first. Note that some of the notation in this chapter differs from the notation introduced in Chapter~\ref{ch:csi}. All such differences will be explained in this section.

\begin{figure}[htb]
 \centering
 \begin{subfigure}[b]{0.48\textwidth}
  \centering
  \includegraphics{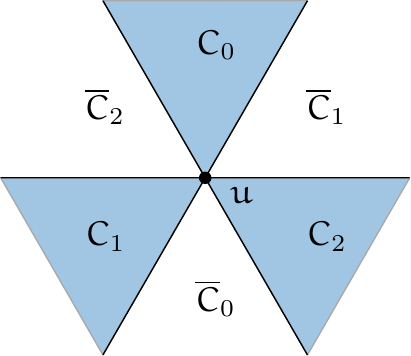}
  \caption{}
  \label{fig:cr-cones-a}
 \end{subfigure}
 \begin{subfigure}[b]{0.48\textwidth}
  \centering
  \includegraphics{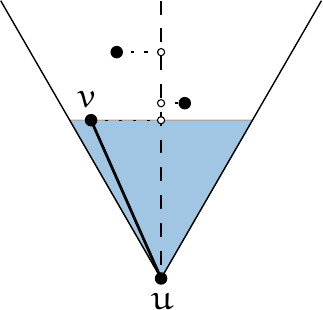}
  \caption{}
  \label{fig:cr-cones-b}
 \end{subfigure}
 \caption{(a) The positive (highlighted) and negative cones around a vertex $u$. (b) In each positive cone, $u$ connects to the vertex with the closest projection on the bisector of that cone.}
\end{figure}

As the name implies, the \hts is closely related to the $\Theta_6$-graph. The difference is that every other cone is ignored. To reflect this, the cones are relabelled from $C_0, \dots, C_5$ to $\c{0}, \nc{1}, \c{2}, \nc{0}, \c{1}, \nc{2}$ (see Figure~\ref{fig:cr-cones-a}). The cones $\c{0}$, $\c{1}$ and $\c{2}$ are called \emph{positive}, while the others are called \emph{negative}. Note that corresponding positive and negative cones are opposite each other. Combined with their symmetry, this implies that if $u$ lies in $\nc{0}^v$ (shorthand for cone $\nc{0}$ with apex $v$), then $v$ must lie in $\c{0}^u$.

\begin{figure}[htb]
 \centering
 \includegraphics{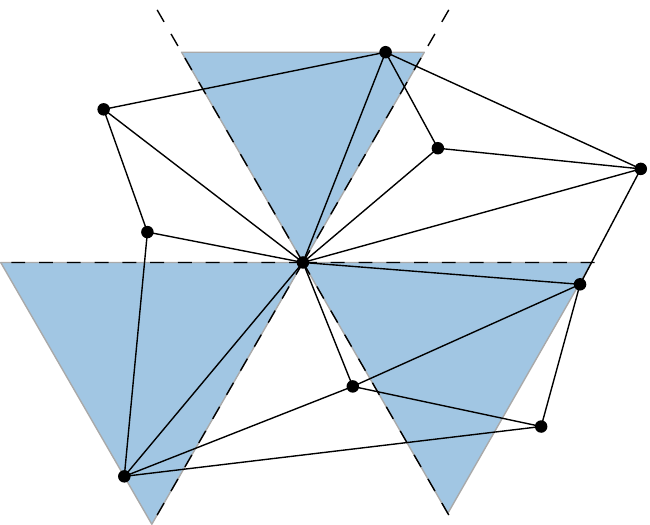}
 \caption{An example \hts.}
 \label{fig:cr-cones-c}
\end{figure}

To build the \hts, we consider each positive cone of every vertex, and add an edge to the closest vertex in that cone (according to the projection onto the bisector, see Figure~\ref{fig:cr-cones-b}). That is, edges are added to the closest vertex in $\c{0}$, $\c{1}$, and $\c{2}$, but not in the other cones. See Figure~\ref{fig:cr-cones-c} for an example \hts. For simplicity, we assume that no two points lie on a line parallel to a cone boundary, guaranteeing that each vertex connects to exactly one vertex in each positive cone. Hence the graph has at most $3n$ edges in total.

We slightly modify the concept of \emph{canonical triangle} to take the distinction between positive and negative cones into account. Given two vertices $u$ and $v$, we now define their canonical triangle as $\T{u}{v}$ if $v$ lies in a positive cone of $u$, and $\T{v}{u}$ if $u$ lies in a positive cone of $v$. Note that either $v$ lies in a positive cone of $u$, or $u$ lies in a positive cone of $v$, so there is exactly one canonical triangle (either \T{u}{v} or \T{v}{u}) for the pair. With this definition, the construction of the \hts can alternatively be described as adding an edge between two vertices if and only if their canonical triangle is empty. This property will play an important role in our proofs.

\section{Spanning ratio of the \texorpdfstring{\hts}{half-theta-6-graph}} \label{sec:cr-spanning}

Bonichon~\etal~\cite{bonichon2010connections} showed that the \hts is a geometric spanner with spanning ratio~2 by showing it is equivalent to the Delaunay triangulation based on empty equilateral triangles, which is known to have spanning ratio~2~\cite{chew1989there}. This correspondence also shows that the \hts is internally triangulated: every face except for the outer face is a triangle (this follows from the duality with the Voronoi diagram, along with the fact that all vertices in the Voronoi diagram have degree 3, provided that no 4 points lie on the same equilateral triangle). In this section, we provide an alternative proof of the spanning ratio of the \hts. Our proof shows that between any pair of points, there always exists a path with spanning ratio 2 that lies in the canonical triangle. This property plays an important role in our routing algorithm, which we describe in Section~\ref{sec:cr-routing}.

\begin{figure}[ht]
    \centering
    \includegraphics{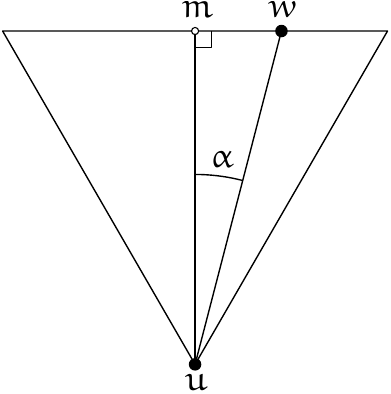}
    \caption{Two vertices $u$ and $w$ with their canonical triangle $\T{u}{w}$. The angle $\alpha$ is the unsigned angle between the line $uw$ and the bisector of the cone containing $w$.}
    \label{fig:cr-angleFigure}
  \end{figure}

For a pair of vertices~$u$ and $w$, our bound is expressed in terms of the angle~$\alpha$ between the line from $u$ to $w$ and the bisector of their canonical triangle (see~Figure~\ref{fig:cr-angleFigure}).

\begin{theorem}
\label{theo:cr-UnconstrainedSpanningRatio}
Let $u$ and $w$ be vertices with $w$ in a positive cone of $u$. Let $m$ be the midpoint of the side of \T{u}{w} opposing $u$, and let $\alpha \leq \pi / 6$ be the smaller of the two unsigned angles between the segments $uw$ and $um$. Then the \hts contains a path between $u$ and $w$ of length at most
\[
 (\sqrt{3} \cdot \cos \alpha + \sin \alpha) \cdot |u w|,
\]
where all vertices on this path lie in $\T{u}{w}$.
\end{theorem}
The expression $\sqrt{3} \cdot \cos \alpha + \sin \alpha$ is increasing for $\alpha \in [0,\pi/6]$. By inserting the extreme value $\pi/6$ for $\alpha$, we arrive at the following.
\begin{corollary}
\label{cor:cr-spanning}
The spanning ratio of the \hts is 2.
\end{corollary}
We note that the bounds of Theorem~\ref{theo:cr-UnconstrainedSpanningRatio} and Corollary~\ref{cor:cr-spanning} are tight: for all values of $\alpha \in [0,\pi/6]$ there exists a point set for which the shortest path in the \hts for some pair of vertices $u$ and $w$ has length arbitrarily close to $(\sqrt{3} \cdot \cos \alpha + \sin \alpha) \cdot |u w|$. A simple example appears later in the proof of Theorem~\ref{thm:cr-routing}.

\vspace{\topsep}
\begin{proofof}{Theorem~\ref{theo:cr-UnconstrainedSpanningRatio}}
  Given two vertices $u$ and $w$, we assume without loss of generality that $w$ lies in $C^u_0$. We prove the theorem by induction on the rank, when ordered by area, of the triangles $\T{x}{y}$ for all pairs of points $x$ and $y$ where $y$ lies in a positive cone of $x$. Let $a$ and $b$ be the upper left and right corner of $\T{u}{w}$, and let $A = \T{u}{w} \cap C^w_1$ and $B = \T{u}{w} \cap C^w_2$, as illustrated in Figure~\ref{fig:cr-Triangle}.

  \begin{figure}[ht]
    \begin{center}
      \includegraphics{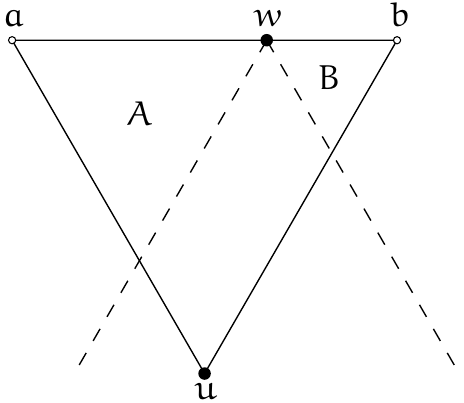}
    \end{center}
    \caption{The corners $a$ and $b$, and the regions $A$ and $B$.}
    \label{fig:cr-Triangle}
  \end{figure}

  Our inductive hypothesis is the following, where $\delta(u,w)$ denotes the length of the shortest path from~$u$ to $w$ in the part of the \hts induced by the vertices in \T{u}{w}.
  \begin{shortenumerate}
    \item If $A$ is empty, then $\delta(u, w) \leq |u b| + |b w|$.
    \item If $B$ is empty, then $\delta(u, w) \leq |u a| + |a w|$.
    \item If neither $A$ nor $B$ is empty, then $\delta(u, w) \leq \max\{|u a| + |a w|,$ \newline $|u b| + |b w|\}$.
  \end{shortenumerate}

  We first note that this induction hypothesis implies Theorem~\ref{theo:cr-UnconstrainedSpanningRatio}: using the side of  \T{u}{w} as the unit of length, we have from Figure~\ref{fig:cr-angleFigure} that $\length{wm} = \length{uw}\cdot\sin\alpha$ and $\sqrt{3}/2 = \length{um} = \length{uw}\cdot\cos\alpha$. Hence the induction hypothesis gives us that $\delta(u, w)$ is at most $1+1/2+\length{wm} = \sqrt{3} \cdot (\sqrt{3}/2) +\length{wm} = (\sqrt{3} \cdot \cos\alpha + \sin\alpha) \cdot \length{uw}$, as required.

  \paragraph{Base case.} $\T{u}{w}$ has rank 1. Since there are no smaller canonical triangles, $w$ must be the closest vertex to $u$. Hence the edge $(u,w)$ is in the \hts, and $\delta(u, w) = |u w|$. Using the triangle inequality, we have $|u w| \leq \min\{|u a| + |a w|, |u b| + |b w|\}$, so the induction hypothesis holds.

  \paragraph{Induction step.} We assume that the induction hypothesis holds for all pairs of points with canonical triangles of rank up to $i$. Let $\T{u}{w}$ be a canonical triangle of rank $i+1$.

  If $(u,w)$ is an edge in the \hts, the induction hypothesis follows by the same argument as in the base case. If there is no edge between $u$ and $w$, let $v$ be the vertex closest to $u$ in the positive cone $C^u_0$, and let $a'$ and $b'$ be the upper left and right corner of $\T{u}{v}$. By definition, $\delta(u, w) \leq |u v| + \delta(v, w)$, and by the triangle inequality, $|u v| \leq \min\{|u a'| + |a' v|, |u b'| + |b' v|\}$.

  We perform a case distinction on the location of $v$: (a) $v$ lies  neither in $A$ nor in $B$, (b) $v$ lies inside $A$, and (c) $v$ lies inside $B$. The case where $v$ lies inside $B$ is analogous to the case where $v$ lies inside $A$, so we only discuss the first two cases, which are illustrated in Figure~\ref{fig:cr-TriangleCases2}.
  
\begin{figure}[htb]
 \centering
 \begin{subfigure}[b]{0.48\textwidth}
  \centering
  \includegraphics{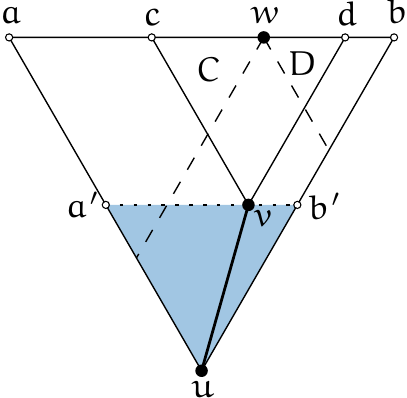}
  \caption{}
  \label{fig:cr-TriangleCases2-a}
 \end{subfigure}
 \begin{subfigure}[b]{0.48\textwidth}
  \centering
  \includegraphics{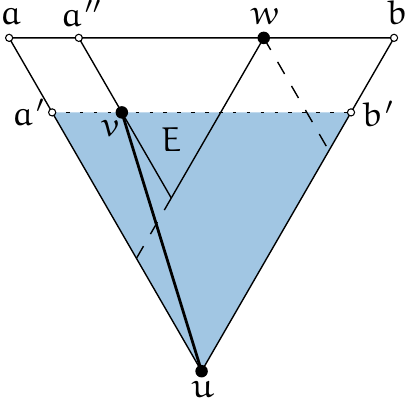}
  \caption{}
  \label{fig:cr-TriangleCases2-b}
 \end{subfigure}
 \caption{The two cases: (a) $v$ lies in neither $A$ nor $B$, (b) $v$ lies in $A$.}
 \label{fig:cr-TriangleCases2}
\end{figure}

 \case{(a)} Let $c$ and $d$ be the upper left and right corner of $\T{v}{w}$, and let $C = \T{v}{w} \cap C^w_1$ and $D = \T{v}{w} \cap C^w_2$ (see~Figure~\ref{fig:cr-TriangleCases2-a}). Since $\T{v}{w}$ has smaller area than $\T{u}{w}$, we apply the inductive hypothesis on $\T{v}{w}$. Our task is to prove all three statements of the inductive hypothesis for $\T{u}{w}$.

  \begin{enumerate}
  \item If $A$ is empty, then $C$ is also empty, so by induction $\delta(v, w) \leq |v d| + |d w|$. Since $v$, $d$, $b$, and $b'$ form a parallelogram, we have:
   \begin{align*}
    \delta(u, w)~~&\leq~~|u v| + \delta(v, w) \\
    ~~&\leq~~|u b'| + |b' v| + |v d| + |d w| \\
    ~~&=~~|u b| + |b w|,
   \end{align*}
  which proves the first statement of the induction hypothesis. This argument is illustrated in Figure~\ref{fig:cr-SpanningProofVisualization-a}.

\begin{figure}[htb]
 \centering
 \begin{subfigure}[b]{0.48\textwidth}
  \centering
  \includegraphics{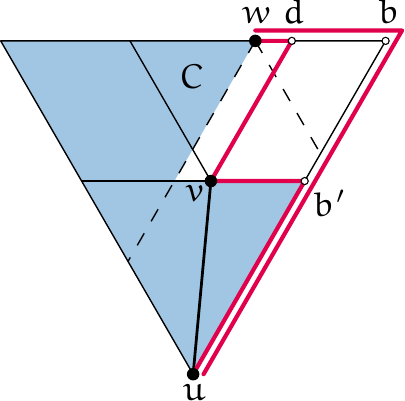}
  \caption{}
  \label{fig:cr-SpanningProofVisualization-a}
 \end{subfigure}
 \begin{subfigure}[b]{0.48\textwidth}
  \centering
  \includegraphics{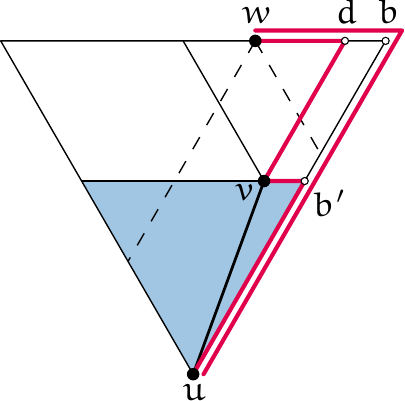}
  \caption{}
  \label{fig:cr-SpanningProofVisualization-b}
 \end{subfigure}
 \caption{Visualization of the path inequalities in two cases: (a) $v$ lies in neither $A$ nor $B$ and one of $A$ or $B$ is empty (cases a.1 and a.2 in our proof), (b) $v$ lies in neither $A$ nor $B$ and neither is empty (case a.3). The paths occurring in the equations are drawn with thick red lines, and light blue areas indicate empty regions.}
\end{figure}

      \item If $B$ is empty, an analogous argument proves the second statement of the induction hypothesis.

      \item If neither $A$ nor $B$ is empty, by induction we have $\delta(v, w) \leq \max\{|v c| + |c w|, |v d| + |d w|\}$. Assume, without loss of generality, that the maximum of the right hand side is attained by its second argument $|v d| + |d w|$ (the other case is analogous).

      Since vertices $v$, $d$, $b$, and $b'$ form a parallelogram, we have that:
      \begin{align*}
	\delta(u, w)~~&\leq~~|uv| + \delta(v, w) \\
	~~&\leq~~|u b'| + |b' v| +  |v d| + |d w| \\
	~~&\leq~~|u b| + |b w| \\
	~~&\leq~~\max\{|u a| + |a w|, |u b| + |b w|\},
      \end{align*}
      which proves the third statement of the induction hypothesis. This argument is illustrated in Figure~\ref{fig:cr-SpanningProofVisualization-b}.
  \end{enumerate}

  \case{(b)} Let $E = \T{u}{v} \,\cap\, \T{w}{v}$, and let $a''$ be the upper left corner of $\T{w}{v}$ (see~Figure~\ref{fig:cr-TriangleCases2-b}). Since $v$ is the closest vertex to $u$ in one of its positive cones, \T{u}{v} is empty and hence $E$ is also empty. Since $\T{w}{v}$ is smaller than $\T{u}{w}$, we can apply induction on it. As $E$ is empty, the first statement of the induction hypothesis for $\T{w}{v}$ applies, giving us that $\delta(v, w) \leq |v a''| + |a'' w|$. Since $|u v| \leq |u a'| + |a' v|$ and $v$, $a''$, $a$, and $a'$ form a parallelogram, we have that $\delta(u, w) \leq |u a| + |a w|$, proving the second and third statement in the induction hypothesis for \T{u}{w}. This argument is illustrated in Figure~\ref{fig:cr-SpanningProofVisualization-c}. Since $v$ lies in $A$, the first statement in the induction hypothesis for \T{u}{w} is vacuously true.
\end{proofof}
  
  \begin{figure}[htb]
    \centering
    \includegraphics{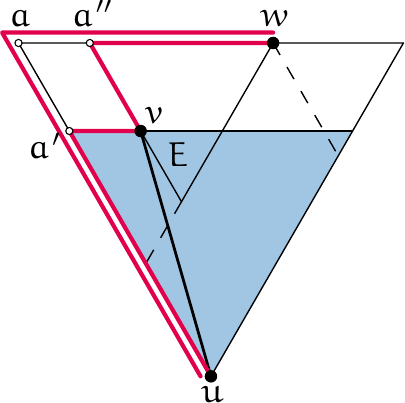}
    \caption{Visualization of the path inequalities when $v$ lies in $A$ or $B$ (case b).}
    \label{fig:cr-SpanningProofVisualization-c}
  \end{figure}

\section{Remarks on the spanning ratio} \label{sec:cr-remarks}

The $\Theta_6$-graph, introduced by Keil and Gutwin~\cite{keil1992classes}, is similar to the \hts except that all 6 cones are positive cones. Thus, $\Theta_6$ is the union of two copies of the \hts, where one \hts is rotated by $\pi/3$ radians. The \hts and $\Theta_6$ both have a spanning ratio of 2, with lower bound examples showing that it is tight for both graphs. This is surprising since $\Theta_6$ can have twice the number of edges of the \hts.

Note that since $\Theta_6$ consists of two rotated copies of the \hts, one question that comes to mind is what is the best spanning ratio if one is to construct a graph consisting of two rotated copies of the \hts ? Can one do better than a spanning ratio of 2? Consider the following construction. Build two \hts{}s as described in Section~\ref{sec:cr-prelims}, but rotate each cone of the second graph by $\pi/6$ radians. For each pair of vertices, there is a path of length at most $\sqrt{3}\cos{\alpha} + \sin{\alpha}$ times the Euclidean distance between them, where $\alpha$ is the angle between the line connecting the vertices in question, and the closest bisector. Since this function is increasing, the spanning ratio is defined by the maximum possible angle to the closest bisector, which is $\pi/12$ radians, giving a spanning ratio of roughly 1.932.

By using $k$ copies, we improve the spanning ratio even further: if each is rotated by $\pi/(3k)$ radians, we get a spanning ratio of $\sqrt{3}\cos{\frac{\pi}{6k}} + \sin{\frac{\pi}{6k}}$. This is better than the known upper bounds for $\Theta_{3k}$~\cite{bose2013spanning} and $Y_{3k}$~\cite{barba2013new} for $k \leq 4$.

\section{Routing in the \texorpdfstring{\hts}{half-theta-6-graph}} \label{sec:cr-routing}

In this section, we give matching upper and lower bounds for the routing ratio on the \hts. We begin by defining our model. Formally, a routing algorithm $A$ is a deterministic $k$-local, $m$-memory routing algorithm, if the vertex to which a message is forwarded from the current vertex $s$ is a function of $s$, $t$, $N_k(s)$, and $M$, where $t$ is the destination vertex, $N_k(s)$ is the $k$-neighbourhood of $s$ and $M$ is a memory of size $m$, stored with the message. The $k$-neighbourhood of a vertex $s$ is the set of vertices in the graph that can be reached from $s$ by following at most $k$ edges. For our purposes, we consider a unit of memory to consist of a $\log_2 n$ bit integer or a point in $\mathbb{R}^2$. Our model also assumes that the only information stored at each vertex of the graph is $N_k(s)$. Since our graphs are geometric, we identify each vertex by its coordinates in the plane. Note that while many local routing models allow the algorithm to use the location of the source vertex (where the routing algorithm started) in addition to the current vertex and destination vertex, our model does not.

A routing algorithm is {\em $d$-competitive} provided that the total distance travelled by the message is never more than $d$ times the Euclidean distance between source and destination. Analogous to the spanning ratio, the \emph{routing ratio} of an algorithm is the smallest $d$ for which it is $d$-competitive.

We present a deterministic $1$-local $0$-memory routing algorithm that achieves the upper bounds, but our lower bounds hold for any deterministic $k$-local $0$-memory algorithm, provided $k$ is a constant. Our bounds are expressed in terms of the angle~$\alpha$ between the line from the source to the destination and the bisector of their canonical triangle (see~Figure~\ref{fig:cr-angleFigure}).

\begin{theorem}\label{thm:cr-routing}
Let $u$ and $w$ be two vertices, with $w$ in a positive cone of $u$. Let $m$ be the midpoint of the side of \T{u}{w} opposing $u$, and let $\alpha$ be the unsigned angle between the lines $uw$ and $um$. There is a deterministic $1$-local $0$-memory routing algorithm on the \hts for which every path followed has length at most
\renewcommand{\labelenumi}{{\upshape\roman{enumi})}}
\begin{enumerate}
\label{thm:cr-routing:item:posbound}
\item $(\sqrt{3} \cdot \cos \alpha + \sin \alpha) \cdot |u w|$ when routing from $u$ to $w$,
\label{thm:cr-routing:item:negbound}
\item $(5/\sqrt{3} \cdot \cos \alpha - \sin \alpha) \cdot |u w|$ when routing from $w$ to $u$,
\end{enumerate}
and this is best possible for deterministic $k$-local, $0$-memory routing algorithms, where $k$ is constant.
\end{theorem}
The first expression is increasing for $\alpha \in [0,\pi/6]$, while the second expression is decreasing. Inserting the extreme values $\pi/6$ and $0$ for $\alpha$, we get the following worst case version of Theorem~\ref{thm:cr-routing}.
\begin{corollary}
Let $u$ and $w$ be two vertices, with $w$ in a positive cone of $u$. There is a deterministic $1$-local $0$-memory routing algorithm on the \hts with routing ratio
\renewcommand{\labelenumi}{{\upshape\roman{enumi})}}
\begin{enumerate}
\item $2$ when routing from $u$ to $w$,
\item $5/\sqrt{3} \approx 2.887$ when routing from $w$ to $u$,
\end{enumerate}
and this is best possible for deterministic $k$-local, $0$-memory routing algorithms, where $k$ is constant.
\end{corollary}
Since the spanning ratio of the \hts is~2, the second lower bound shows a separation between the spanning ratio and the best possible routing ratio in the \hts.

Since every triangulation can be embedded in the plane as a half-$\Theta_6$-graph using $O(\log n)$ bits per vertex via Schnyder's embedding scheme~\cite{schnyder1990embedding}, an important implication of Theorem~\ref{thm:cr-routing} is the following.

\begin{corollary}
Every $n$-vertex triangulation can be embedded in the plane using $O(\log n)$ bits per coordinate such that the embedded triangulation admits a deterministic $1$-local routing algorithm with routing ratio at most $5/\sqrt{3} \approx 2.887$.
\end{corollary}

\subsection{Positive routing}

In the remainder of this section we prove Theorem~\ref{thm:cr-routing}. We first consider the case where the destination lies in a positive cone of the source. We start with a proof of the lower bound, followed by a description of the routing algorithm and a proof of the upper bound.

\begin{figure}[htb]
  \centering
   \includegraphics{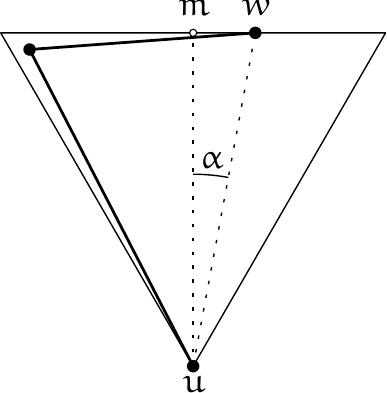}
  \caption{The lower bound example when routing to a vertex in a positive cone.}
  \label{fig:cr-lowerBoundInstancesPos}
\end{figure}

\begin{lemma}[Lower bound for positive routing] \label{lem:cr-poslb}
 Let $u$ and $w$ be two vertices, with $w$ in a positive cone of $u$. Let $m$ be the midpoint of the side of \T{u}{w} opposing $u$, and let $\alpha$ be the unsigned angle between the lines $uw$ and $um$. For any deterministic $k$-local, $0$-memory routing algorithm, there are instances for which the path followed has length at least $(\sqrt{3} \cdot \cos \alpha + \sin \alpha) \cdot |u w|$ when routing from $u$ to $w$.
\end{lemma}
\begin{proof}
 Let the side of \T{u}{w} be the unit of length. From Figure~\ref{fig:cr-angleFigure}, we have $\length{wm} = \length{uw}\cdot\sin\alpha$ and $\sqrt{3}/2 = \length{um} = \length{uw}\cdot\cos\alpha$. From Figure~\ref{fig:cr-lowerBoundInstancesPos}, the spanning ratio of the \hts is at least $1+1/2+\length{wm} = \sqrt{3} \cdot (\sqrt{3}/2) +\length{wm} = (\sqrt{3} \cdot \cos\alpha + \sin\alpha) \cdot \length{uw}$, since the point in the upper left corner of~\T{u}{w} can be moved arbitrarily close to the corner. As there is no shorter path between $u$ and $w$, this is a lower bound for \emph{any} routing algorithm.
\end{proof}

\begin{figure}[htb]
  \centering
  \includegraphics{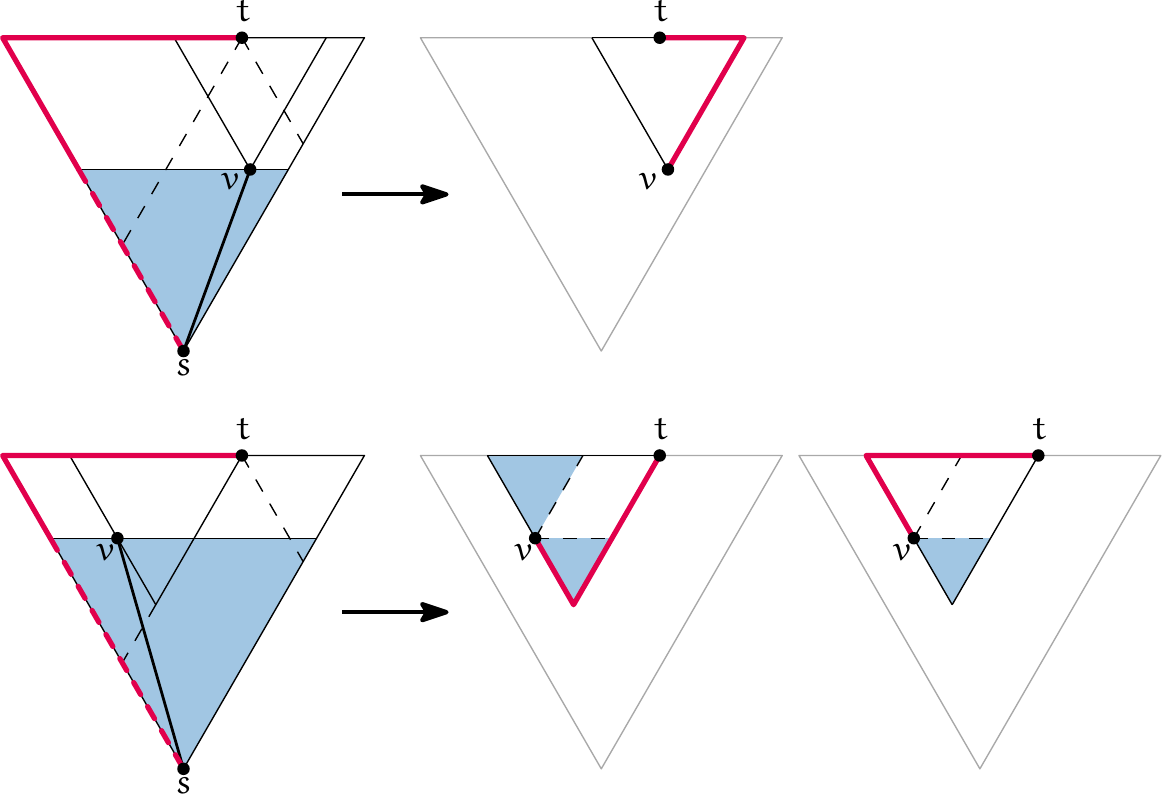}
  \caption{Routing in case A. (Top) $v$ lies in a negative cone of $t$, (Bottom) $v$ lies in a positive cone of $t$. Dashed red lines indicate which parts of the potential are used to pay for the edge.}
  \label{fig:cr-routingStateA}
\end{figure}

\paragraph{Routing algorithm.} While routing, let $s$ denote the current vertex and let $t$ denote the fixed destination (i.e. $t$ corresponds to $w$ in Theorem~\ref{thm:cr-routing}). To be deterministic, $1$-local, and $0$-memory, the routing algorithm needs to determine which edge $(s,v)$ to follow next based only on $s$, $t$, and the neighbours of $s$. We say we are \emph{routing positively} when $t$ is in a positive cone of $s$, and \emph{routing negatively} when $t$ is in a negative cone. (Note the distinction between ``positive routing'' and ``routing positively'': the first describes the conditions \emph{at the start} of the routing process, while the second does so \emph{during} the routing process. In other words, positive routing describes a routing process that starts by routing positively. It is very common for positive routing to include situations where we are routing negatively, see e.g. the bottom part of Figure~\ref{fig:cr-routingStateA}.)

\begin{figure}[htb]
 \centering
 \begin{subfigure}[b]{0.48\textwidth}
  \centering
  \includegraphics{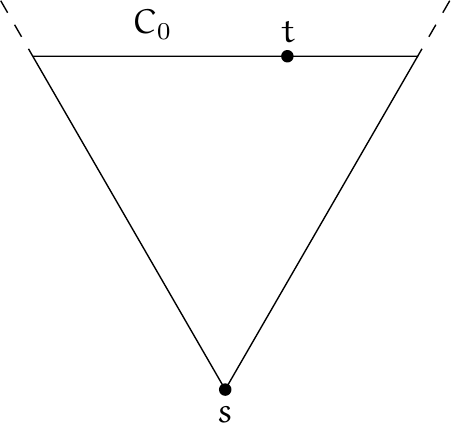}
  \caption{}
  \label{fig:cr-routingTerminology-a}
 \end{subfigure}
 \begin{subfigure}[b]{0.48\textwidth}
  \centering
  \includegraphics{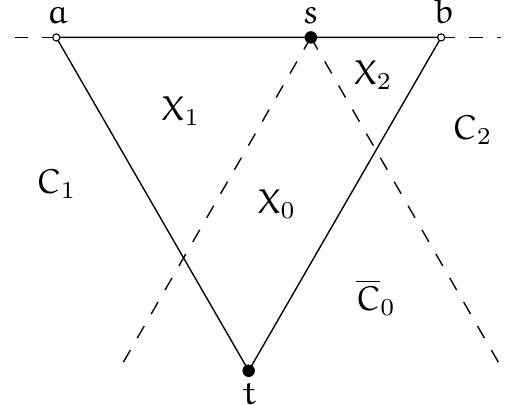}
  \caption{}
  \label{fig:cr-routingTerminology-b}
 \end{subfigure}
 \caption{Routing terminology when (a) routing positively and (b) routing negatively.}
 \label{fig:cr-routingTerminology}
\end{figure}

For ease of description, we assume without loss of generality that $t$ is in cone~$C_0^s$ when routing positively, and in cone~$\overline{C}_0^s$ when routing negatively. When routing positively, \T{s}{t} intersects only $C_0^s$ among the cones of~$s$. When routing negatively, \T{t}{s} intersects $\overline{C}_0^s$, as well as the two positive cones $C_1^s$ and $C_2^s$. Let $X_0 = \overline{C}_0^s \cap \T{t}{s}$, $X_1 = C_1^s \cap \T{t}{s}$, and $X_2 = C_2^s \cap \T{t}{s}$. Let $a$ be the corner of~\T{t}{s} contained in $X_1$ and $b$ the corner of~\T{t}{s} contained in $X_2$. These definitions are illustrated in Figure~\ref{fig:cr-routingTerminology}.

The routing algorithm will only follow edges~$(s,v)$ where $v$ lies in the canonical triangle of $s$ and $t$.  Routing positively is straightforward since there is exactly one edge $(s,v)$ with $v \in \T{s}{t}$, by the construction of the \hts. The challenge is to route negatively. When routing negatively, at least one edge $(s,v)$ with $v \in \T{t}{s}$ exists, since by Theorem~\ref{theo:cr-UnconstrainedSpanningRatio}, $s$ and $t$ are connected by a path in~\T{t}{s}. The core of our routing algorithm is how to choose which edge to follow when there is more than one. Intuitively, when routing negatively, our algorithm tries to select an edge that makes measurable progress towards the destination. When no such edge exists, we are forced to take an edge that does not make measurable progress, however we are able to then deduce that certain regions within the canonical triangle are empty. This allows us to bound the total distance travelled while not making measurable progress. We provide a formal description of our routing algorithm below.

Our routing algorithm can be in one of four cases. We call the situation when routing positively case~A, and divide the situation when routing negatively into three further cases: both $X_1$ and $X_2$ are empty (case~B), either $X_1$ or $X_2$ is empty (case~C), or neither is empty (case~D). Since $X_1$ and $X_2$ correspond to positive cones of $s$, each contains the endpoint of at most one edge $(s,v)$. These edges contain a lot of information about the regions $X_1$ and $X_2$. In particular, if there is no edge in the corresponding cone, then the entire cone must be empty. And if there is an edge, but its endpoint lies outside of the region, the region is guaranteed to be empty. This allows our algorithm to \emph{locally} determine if $X_1$ and $X_2$ are empty, and therefore which case we are in.
      
Since we are routing to a destination in a positive cone of the source, our routing algorithm starts in case~A. Routing in this case is straightforward, as there is only one edge $(s,v)$ with $v$ in \T{t}{s} that we can follow. We now turn our attention to routing in cases~B and~C (it turns out case~D never occurs when routing to a destination in a positive cone of the source; we come back to it when describing negative routing in Section~\ref{sec:cr-negativeRouting}).

In case~B, both $X_1$ and $X_2$ are empty, so there must be edges $(s,v)$ with $v \in X_0$, as $s$ and $t$ are connected by a path in~\T{t}{s} by Theorem~\ref{theo:cr-UnconstrainedSpanningRatio}. If $\length{as} \ge \length{sb}$, the routing algorithm follows the last edge in clockwise order around $s$; if $\length{as} < \length{sb}$, it follows the first edge. In short, when both sides of~\T{t}{s} are empty, the routing algorithm favours staying close to the largest empty side of~\T{t}{s}. Note that $\length{as}$ and $\length{sb}$ can be computed locally from the coordinates of $s$ and $t$.

In case~C, exactly one of $X_1$ or $X_2$ is empty. If there exist edges $(s,v)$ with $v \in X_0$, the routing algorithm will follow one of these, choosing among them in the following way: If $X_1$ is empty, it chooses the last edge in clockwise order around $s$. Else $X_2$ is empty, and it chooses the first edge in clockwise order around $s$. In short, the routing algorithm favours staying close to the empty side of~\T{t}{s}. If no edges $(s,v)$ with $v \in X_0$ exist, the routing algorithm follows the single edge $(s,v)$ with $v$ in $X_1$ or $X_2$.

\paragraph{Upper bound.} The proof of the upper bound uses a potential function~$\phi$, defined as follows for each of the cases~A, B, and~C. For the potential in case~C, $x \in \{a,b\}$ is the corner contained in the non-empty one of the two areas $X_1$ and $X_2$.

\begin{center}
\begin{tabular}{cl}
Case A:&$\phi~~=~~\length{sa} + \max (\length{at},\length{tb})$\\
Case B:&$\phi~~=~~\length{ta} + \min (\length{as},\length{sb})$\\
Case C:&$\phi~~=~~\length{ta} + \length{sx}$
\end{tabular}
\end{center}

\begin{figure}[htb]
 \centering
 \begin{subfigure}[b]{0.32\textwidth}
  \centering
  \includegraphics{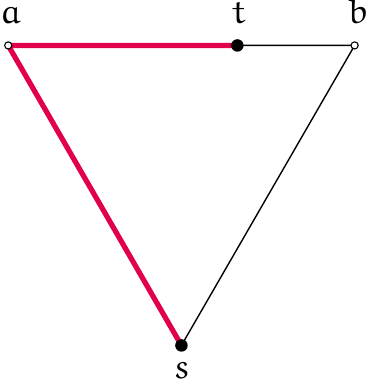}
  \caption*{Case A}
 \end{subfigure}
 \begin{subfigure}[b]{0.32\textwidth}
  \centering
  \includegraphics{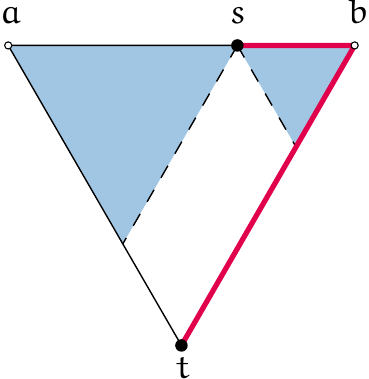}
  \caption*{Case B}
 \end{subfigure}
 \begin{subfigure}[b]{0.32\textwidth}
  \centering
  \includegraphics{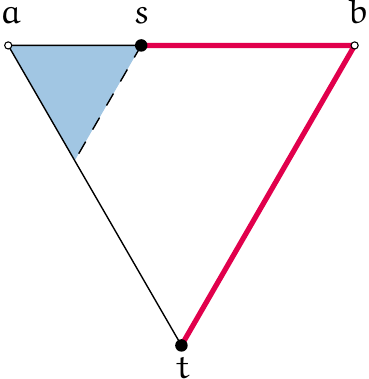}
  \caption*{Case C}
 \end{subfigure}
 \caption{The potential $\phi$ in each case. The thick lines designate potential and shaded areas are empty.}
 \label{fig:cr-potentialFunction}
\end{figure}

This definition is illustrated in Figure~\ref{fig:cr-potentialFunction}. We will refer to the first term of~$\phi$ (i.e., $\length{sa}$ in case~A, $\length{ta}$ in cases~B, and C) as the \emph{vertical part} of~$\phi$ and to the rest as the \emph{horizontal part}. Note that since all sides of the canonical triangle have equal length, $a$ and $b$ are interchangeable in the vertical part. The proof makes extensive use of the following observation about equilateral triangles:

\begin{observation} \label{obs:cr-simpleFact}
In an equilateral triangle, the diameter (the longest distance defined by any two points in the triangle) is equal to the side length.
\end{observation}

Our aim is to prove the following claim: for any routing step, the reduction in $\phi$ is at least as large as the length of the edge followed. This allows us to `pay' for each edge with the difference in potential, thereby bounding the total length of the path by the initial potential. We do this by case analysis of the possible routing steps.

\case{\MakeUppercase{a}} For a routing step starting in case~A, $v$ can be in a negative or a positive cone of~$t$. The first situation leads to case~A again. The second leads to case~B or C, since the area of~\T{s}{t} between $s$ and $v$ must be empty by construction of the \hts. These situations are illustrated in Figure~\ref{fig:cr-routingStateA}.

If we remain in case~A after following edge $(s, v)$, the reduction of the vertical part of~$\phi$ (dashed in Figure~\ref{fig:cr-routingStateA}a) is at least as large as $\length{sv}$ by Observation~\ref{obs:cr-simpleFact}. Therefore we can use it to pay for this step. Since \T{v}{t} is contained in \T{s}{t}, both $\length{at}$ and $\length{bt}$ decrease. Thus the horizontal part of~$\phi$ decreases too, as it is the maximum of the two. Hence the claim holds for this situation.

For the situation ending in case~C (the second illustration after the arrow in Figure~\ref{fig:cr-routingStateA}b), we again use the reduction of the vertical part of~$\phi$ to pay for the step. The rest of the vertical part precisely covers the new horizontal part. Since \T{t}{v} is contained in \T{s}{t}, the new vertical part is a portion of either $ta$ or $tb$. This can be covered by the current horizontal part, as it is the maximum of $\length{ta}$ and $\length{tb}$. Thus the claim holds for this situation as well. Finally, for the situation ending in case~B, the final value of $\phi$ is at most that of the situation ending in case~C, so again the claim holds.

\begin{figure}[htb]
  \centering
  \includegraphics{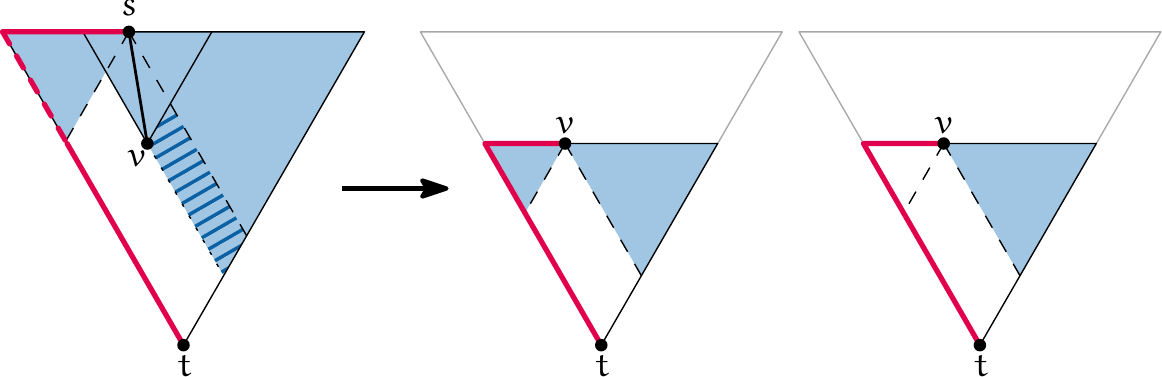}
  \caption{Routing in case B.}
  \label{fig:cr-routingStateB}
\end{figure}

\case{\MakeUppercase{b}} A routing step starting in case~B (illustrated in Figure~\ref{fig:cr-routingStateB}) cannot lead to case~A, as the step stays within~\T{t}{s}. We first show that it always results in Case~B or C, meaning that at least one of $X_1$ or $X_2$ is empty again. The algorithm follows an edge $(s,v)$ with $v \in X_0$. If $s$ is to the left of $t$, it follows the first edge in clockwise order around $s$, otherwise it follows the last one. We consider only the case where $s$ is to the left of $t$, the other case is symmetric. By the construction of the \hts, the existence of the edge $(s,v)$ implies that $\T{v}{s}$ is empty. It follows that the hatched area in Figure~\ref{fig:cr-routingStateB} is also empty: if not, the topmost point in it would have an edge to~$s$, while coming before $v$ in the clockwise order around~$s$, contradicting the choice of~$v$ by the routing algorithm. Therefore $X_2$ will again be empty, resulting in case~B or C.

By Observation~\ref{obs:cr-simpleFact}, the reduction in the vertical part of~$\phi$ is at least as large as $\length{sv}$. In addition, the horizontal part of $\phi$ can only decrease. If it remains on the same side of the triangle, this follows from the fact that $v$ lies in $X_0$ and \T{t}{v} is contained in \T{t}{s}. And the only case where the potential switches sides, is when we end up in case~B again but the other side is shorter than the current one, reducing the potential even further. Hence the claim holds.

\begin{figure}[htb]
  \centering
  \includegraphics{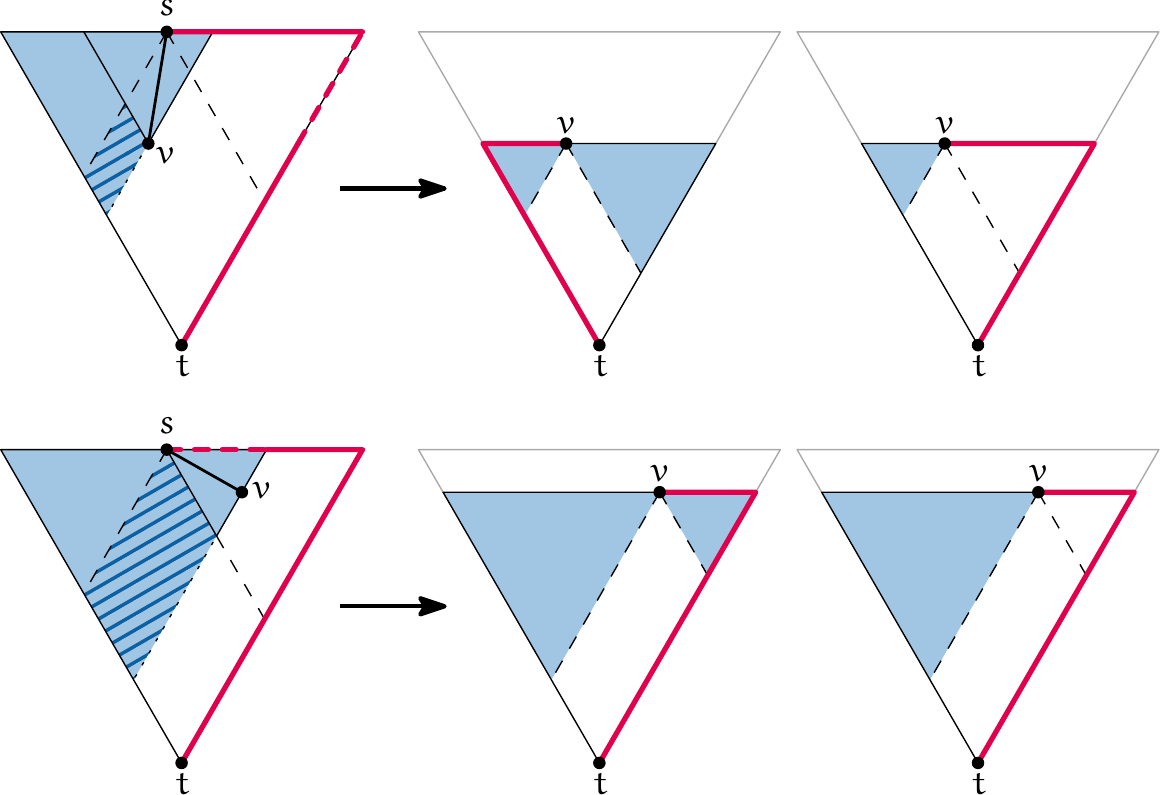}
  \caption{Routing in case C.}
  \label{fig:cr-routingStateC}
\end{figure}

\case{\MakeUppercase{c}} As in the previous case, a routing step starting in case~C cannot lead to case~A and we show that it cannot lead to case~D, either. There are two situations, depending on whether edges $(s,v)$ with \mbox{$v \in X_0$} exist. For the situation where such edges do exist (illustrated in the top part of Figure~\ref{fig:cr-routingStateC}), the analysis is exactly the same as for a routing step starting in case~B.

For the situation where edges $(s,v)$ with $v \in X_0$ do not exist, the start of the step is illustrated on the left of the arrow in the bottom part of Figure~\ref{fig:cr-routingStateC}. Again, $\T{s}{v}$ must be empty by the construction of the \hts, which implies that the hatched area must also be empty: if not, the topmost point in it would have an edge to~$s$, contradicting that edges $(s,v)$ with $v \in X_0$ do not exist. Thus, the routing step can only lead to case~B or C. Looking at the potential, the vertical part can only decrease, and by Observation~\ref{obs:cr-simpleFact}, the reduction of the horizontal part of~$\phi$ is at least as large as $\length{sv}$. Thus we can pay for this step as well and the claim holds in both situations.

\begin{lemma}[Upper bound for positive routing] \label{lem:cr-posub}
 Let $u$ and $w$ be two vertices, with $w$ in a positive cone of $u$. Let $m$ be the midpoint of the side of \T{u}{w} opposing $u$, and let $\alpha$ be the unsigned angle between the lines $uw$ and $um$. There is a deterministic $1$-local $0$-memory routing algorithm on the \hts for which every path followed has length at most $(\sqrt{3} \cdot \cos \alpha + \sin \alpha) \cdot |u w|$ when routing from $u$ to $w$.
\end{lemma}
\begin{proof}
 That the algorithm is deterministic, $1$-local, and $0$-memory follows from the description of the algorithm, so we only need to prove the bound on the distance. We showed that for any routing step, the reduction in $\phi$ is at least as large as the length of the edge followed. Since $\phi$ is always non-negative, this implies that no path followed can be longer than the initial value of $\phi$. As all edges have strictly positive length, the routing algorithm must terminate. Since we are routing to a vertex in a positive cone, we start in case~A, with an initial potential of $\length{ua} + \max (\length{aw},\length{wb})$. Taking the side of $\T{u}{w}$ as the unit of length reduces this to $1 + 1/2 + |wm|$, and using the same analysis as in Lemma~\ref{lem:cr-poslb}, we obtain the desired bound of $(\sqrt{3} \cdot \cos \alpha + \sin \alpha) \cdot |u w|$.
\end{proof}

\subsection{Negative routing}
\label{sec:cr-negativeRouting}

Next we turn our attention to the case when we are routing to a destination in a negative cone of the source. We start by deriving a lower bound, then present the required extensions to our routing algorithm and finish with the matching upper bound.

\begin{figure}[htb]
 \centering
 \begin{subfigure}[b]{0.48\textwidth}
  \centering
  \includegraphics{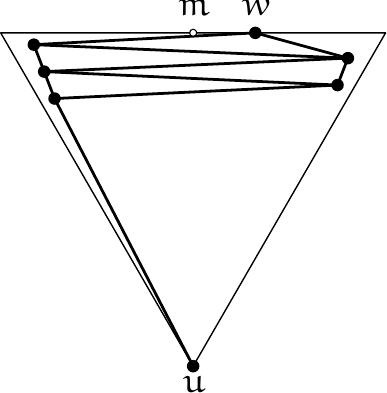}
  \caption{}
  \label{fig:cr-lowerBoundInstances-a}
 \end{subfigure}
 \begin{subfigure}[b]{0.48\textwidth}
  \centering
  \includegraphics{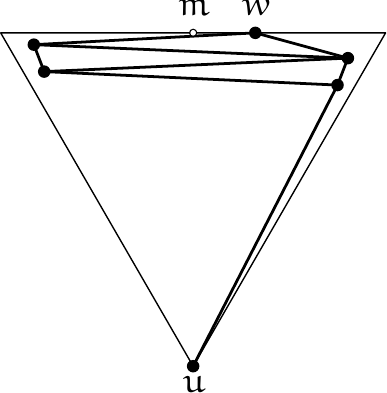}
  \caption{}
  \label{fig:cr-lowerBoundInstances-b}
 \end{subfigure}
 \caption{The lower bound instances for routing to a vertex in a negative cone.}
 \label{fig:cr-lowerBoundInstances}
\end{figure}

\begin{lemma}[Lower bound for negative routing] \label{lem:cr-neglb}
 Let $u$ and $w$ be two vertices, with $w$ in a positive cone of $u$. Let $m$ be the midpoint of the side of \T{u}{w} opposing $u$, and let $\alpha$ be the unsigned angle between the lines $uw$ and $um$, and let $k$ be a constant. For any deterministic $k$-local $0$-memory routing algorithm, there are instances for which the path followed has length at least $(5/\sqrt{3} \cdot \cos \alpha - \sin \alpha) \cdot |u w|$ when routing from $w$ to $u$.
\end{lemma}
\begin{proof}
 Consider the two instances in~Figure~\ref{fig:cr-lowerBoundInstances}. Any deterministic $1$-local $0$-memory routing algorithm has information about direct neighbours only. Hence, it cannot distinguish between the two instances when routing out of~$w$. This means that it routes to the same neighbour of $w$ in both instances, and either choice of neighbour leads to a non-optimal route in one of the two instances. The smallest loss occurs when the choice is towards the closest corner of~\T{u}{w}, for which Figure~\ref{fig:cr-lowerBoundInstances-a} is the bad instance. If we let the side of \T{u}{w} be the unit of length, this gives a lower bound of $(1/2-\length{wm})+1+1 = 5/2-\length{wm}$, since the points in the corners of~\T{u}{w} can be moved arbitrarily close to the corners while keeping their relative positions. Using that $\length{wm} = \length{uw}\cdot\sin\alpha$ and $\sqrt{3}/2 = \length{um} = \length{uw}\cdot\cos\alpha$, the lower bound reduces to $(5/\sqrt{3} \cdot \cos \alpha - \sin \alpha)\cdot\length{uw}$. By appropriately adding $\Omega(k)$ points close to the corners such that $u$ is not in the $k$-neighbourhood of $w$, the lower bound holds for any deterministic $k$-local $0$-memory routing algorithm.
\end{proof}

\paragraph{Routing algorithm.} The only difference with the routing algorithm we used for positive routing lies in the initial case. Since our destination is in a negative cone, we start in one of the negative cases. This time, besides cases~B and C, where both or one of $X_1$ and $X_2$ are empty, we also need case~D, where neither is empty. Recall that in the previous section, we showed that a routing step starting in case~A, B, or C can never result in case~D. Thus, if the routing process starts in case~D, it never returns there once it enters case~A, B, or C.

In case~D, the routing algorithm first tries to follow an edge $(s,v)$ with $v \in X_0$. If several such edges exist, an arbitrary one of these is followed. If no such edge exists, the routing algorithm follows the single edge $(s,v)$ with $v$ in the smaller of $X_1$ and $X_2$. In short, the routing algorithm favours moving towards the closest corner of~\T{t}{s} when it is not able to move towards~$t$. Note that, in the instances of Figure~\ref{fig:cr-lowerBoundInstances}, this choice ensures that the first routing step incurs the smallest loss in the worst case, making it possible to meet the lower bound of Lemma~\ref{lem:cr-neglb}. We now show that our algorithm achieves this lower bound in all cases.

\paragraph{Upper bound.} The potential in case~D is given below. It mirrors the lower bound path, in that it allows walking towards the closest corner, crossing the triangle, then walking down to $t$. This is the highest potential among the four cases.

\begin{center}
\begin{tabular}{cl}
Case D:&$\phi~~=~~\length{ta} + \length{ab} + \min (\length{as},\length{sb})$
\end{tabular}
\end{center}

\begin{figure}[htb]
 \centering
 \begin{subfigure}[b]{\textwidth}
  \centering
  \includegraphics{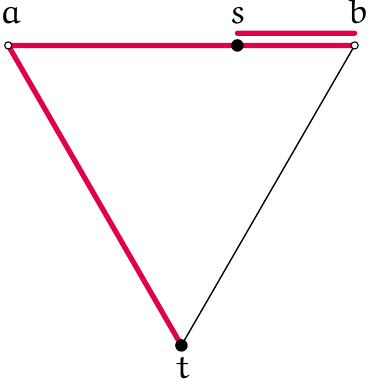}
  \caption*{Case~D}
  \label{fig:cr-FIRST}
 \end{subfigure}
 \caption{The potential $\phi$ in case~D.}
 \label{fig:cr-potentialFunctionNeg}
\end{figure}

As before, we want to show that for any routing step, the reduction in $\phi$ is at least as large as the length of the edge followed. Since we already did this for cases~A, B, and C, and none of them can lead to case~D, all that is left is to prove it for case~D.

\begin{figure}[htb]
  \centering
  \includegraphics{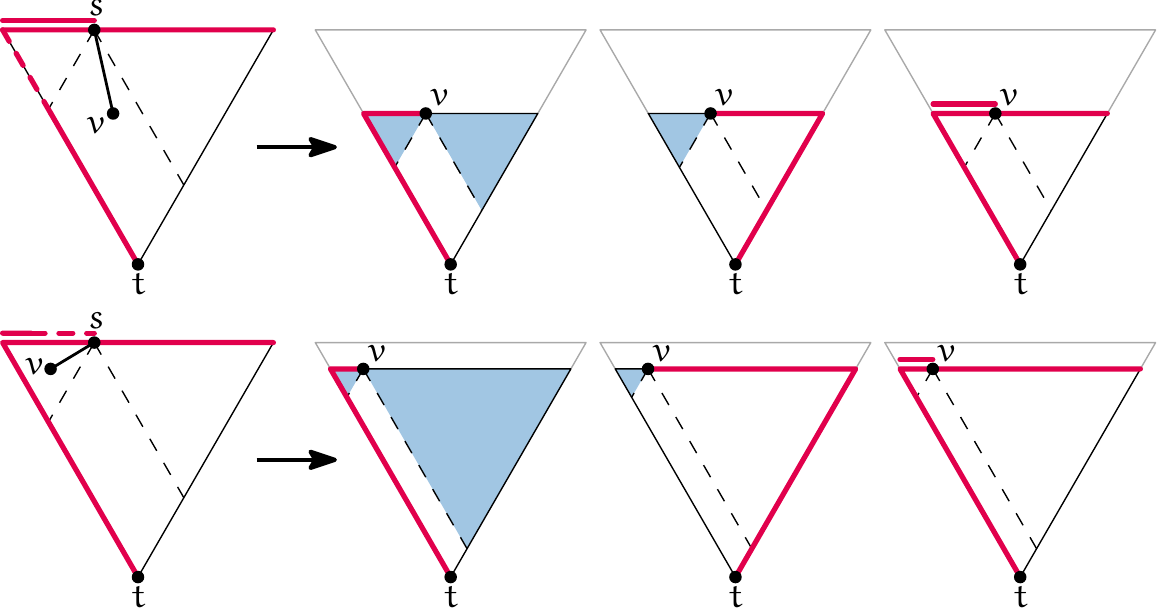}
  \caption{Routing in case D. The endpoint $v$ of the edge followed lies in $X_0$ (Top), or the smaller of $X_1$ and $X_2$ (Bottom).}
  \label{fig:cr-routingStateD}
\end{figure}

\case{\MakeUppercase{d}} A routing step starting in case~D cannot lead to case~A, as the step stays within~\T{t}{s}, but it may lead to case~B, C, or D. There are two situations, depending on whether edges $(s,v)$ with $v \in X_0$ exist or not. These are illustrated in Figure~\ref{fig:cr-routingStateD}.

In the first situation, where we follow an edge $(s, v)$ with $v \in X_0$, the reduction of the vertical part of~$\phi$ is at least as large as $\length{sv}$ by Observation~\ref{obs:cr-simpleFact}. The horizontal part of~$\phi$ can only decrease, as \T{t}{v} is fully contained in \T{t}{s} and $v$ lies in $X_0$. In the second situation, where the endpoint of our edge lies in the smaller of $X_1$ and $X_2$, these roles switch, with the reduction of the horizontal part of~$\phi$ being at least as large as $\length{sv}$ and the vertical part of~$\phi$ only decreasing. In both situations, the statement is proven.

\begin{lemma}[Upper bound for negative routing] \label{lem:cr-negub}
 Let $u$ and $w$ be two vertices, with $w$ in a positive cone of $u$. Let $m$ be the midpoint of the side of \T{u}{w} opposing $u$, and let $\alpha$ be the unsigned angle between the lines $uw$ and $um$. There is a deterministic $1$-local $0$-memory routing algorithm on the \hts for which every path followed has length at most $(5/\sqrt{3} \cdot \cos \alpha - \sin \alpha) \cdot |u w|$ when routing from $w$ to $u$.
\end{lemma}
\begin{proof}
 Since the choices that the routing algorithm makes are completely determined by the neighbours of $s$ and the location of $s$ and $t$, the algorithm is indeed deterministic, $1$-local, and $0$-memory. To bound the length of the resulting path, we again showed that for any routing step, the reduction in $\phi$ is at least as large as the length of the edge followed. As in the proof of Lemma~\ref{lem:cr-posub}, this implies that the routing algorithm terminates and that the total length of the path followed is bounded by the initial value of $\phi$. Since our destination lies in a negative cone, we start in one of the cases~B, C, or D. Of these three cases, case~D has the largest initial potential of $\length{ta} + \length{ab} + \min (\length{as},\length{sb})$. Taking the side of $\T{u}{w}$ as the unit of length reduces this to $1 + 1 + 1/2 - \length{wm} = 5/2 - \length{wm}$, and using the same analysis as in Lemma~\ref{lem:cr-neglb}, we obtain the desired bound of $(5/\sqrt{3} \cdot \cos \alpha - \sin \alpha) \cdot |u w|$.
\end{proof}

As Theorem~\ref{thm:cr-routing} follows from Lemmas~\ref{lem:cr-poslb}, \ref{lem:cr-posub}, \ref{lem:cr-neglb}, and \ref{lem:cr-negub}, this concludes our proof.

\section{A stateful algorithm}
\label{sec:cr-stateful}

Next we present a slightly different routing algorithm from the one in the previous section. The main difference between the two algorithms is that this one maintains one piece of information as state, making it $O(1)$-memory instead of $0$-memory. The information that is stored is a \emph{preferred side}, and it is either nil, $X_1$, or $X_2$. Intuitively, the new algorithm follows the original algorithm until it is routing negatively and determines that either $X_1$ or $X_2$ is empty. At that point, the algorithm sets the empty side as the preferred side and picks the rest of the edges in such a way that the preferred side remains empty. Thus, the algorithm maintains as invariant that if the preferred side is set (not nil), that region is empty. Furthermore, once the preferred side is set, it stays fixed until the algorithm reaches the destination. This algorithm simplifies the cases a little, but more importantly, it allows the algorithm to check far fewer edges while routing. This is crucial, as the new algorithm forms the basis for routing algorithms on versions of the \hts with some edges removed to bound the maximum degree, described in the next section.

We now present the details of this stateful version of the routing algorithm. Recall that we are trying to find a path from a current vertex $s$ to a destination vertex $t$. For ease of description, we again assume without loss of generality that $t$ lies in $\c{0}$ or $\nc{0}$ of $s$. If $t$ lies in $\nc{0}$, the cones around $s$ split $\T{t}{s}$ into three regions $X_0$, $X_1$, and $X_2$, as in Figure~\ref{fig:cr-routingTerminology}. For brevity, we use ``an edge in $X_0$'' to denote an edge incident to $s$ with the other endpoint in $X_0$. The cases are as follows:

\begin{shortitemize}
 \item If $t$ lies in a positive cone of $s$, we are in case~\sA.
 \item If $t$ lies in a negative cone of $s$ and no preferred side has been set yet, we are in case~\sB.
 \item If $t$ lies in a negative cone of $s$ and a preferred side has been set, we are in case~\sC.
\end{shortitemize}

These cases are closely related to the cases in the stateless algorithm. Cases~\sA and \sB correspond to cases~A and D, respectively, while case~\sC merges cases~B and C from the original algorithm into a single case, where only one side's emptiness is tracked. This is reflected in the routing strategy for each case:

\begin{shortitemize}
 \item In case~\sA, follow the unique edge $(s, v)$ in the positive cone containing $t$. If $t$ lies in a negative cone of $v$, set the preferred side to the region ($X_1$ or $X_2$ of $v$) that is contained in $\T{s}{v}$, as this is now known to be empty (see Figure~\ref{fig:cr-routingStateA}b).
 \item In case~\sB, if there are edges in $X_0$, follow an arbitrary one. Otherwise, if there is an edge in the smaller of $X_1$ and $X_2$, follow that edge. Otherwise, follow the edge in the larger of $X_1$ and $X_2$ and set the other as the preferred side. By Theorem~\ref{theo:cr-UnconstrainedSpanningRatio}, at least one of these edges must exist.
 \item In case~\sC, if there are edges in $X_0$, follow the one closest to the preferred side in cyclic order around $s$. Otherwise, follow the edge in the positive cone that is not on the preferred side. Again, at least one of these edges must exist.
\end{shortitemize}

The proof in Section~\ref{sec:cr-routing} can be adapted to show that this routing algorithm achieves the same upper bounds. In short, the proof is simplified to only use a potential as defined for cases~A, C, and D, and only a subset of the illustrations in Figures~\ref{fig:cr-routingStateA}, \ref{fig:cr-routingStateC}, and \ref{fig:cr-routingStateD} are relevant. We omit the repetitive details.

\section{Bounding the maximum degree}
\label{sec:cr-boundeddeg}

Each vertex in the \hts has at most one incident edge in each positive cone, but it can have an unbounded number of incident edges in its negative cones. In this section, we describe two transformations that allow us to bound the total degree of each vertex. The transformations are adapted from Bonichon~\etal~\cite{bonichon2010plane}.

\begin{figure}[htb]
 \centering
 \begin{subfigure}[b]{0.48\textwidth}
  \centering
  \includegraphics{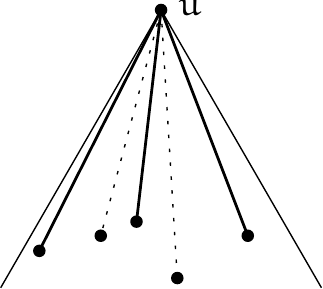}
  \caption{}
  \label{fig:cr-boundeddegree-a}
 \end{subfigure}
 \begin{subfigure}[b]{0.48\textwidth}
  \centering
  \includegraphics{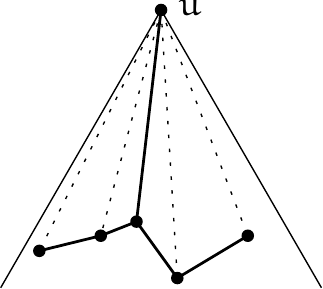}
  \caption{}
  \label{fig:cr-boundeddegree-b}
 \end{subfigure}
 \caption{The construction for \dtw (a) and \dn (b). Solid edges are kept, while dotted edges are discarded if no other vertex wants to keep them.}
\end{figure}

The first transformation discards all edges in each negative cone, except for three: the first and last edges in clockwise order around the vertex and the edge to the ``closest'' vertex, meaning the vertex whose projection on the bisector of the cone is closest (see Figure~\ref{fig:cr-boundeddegree-a}). This results in a subgraph with maximum degree 12, which we call \dtw.

To reduce the degree even further, we note that since the \hts is internally triangulated, consecutive neighbours of $u$ within a negative cone are connected by edges. We call the path formed by these edges the \emph{canonical path}. Instead of keeping three edges per negative cone, we now keep only the edge to the closest vertex, but force the edges of the canonical path to be kept as well (see Figure~\ref{fig:cr-boundeddegree-b}). We call the resulting graph \dn. Bonichon~\etal~\cite{bonichon2010plane} showed that all edges on the canonical path are either first or last in a negative cone, making \dn a subgraph of \dtw. Note that since the \hts is planar, both subgraphs are planar as well. They also proved that \dn is a 3-spanner of the \hts with maximum degree 9. Since the \hts is a 2-spanner and \dn is a subgraph of \dtw, this shows that both \dn and \dtw are 6-spanners of the complete Euclidean graph. We give an adapted version of the proof of the spanning ratio of \dn below.

\begin{theorem}
 \label{thm:cr-3spanner}
 \dn is a 3-spanner of the \hts.
\end{theorem}
\begin{proof}
 Consider an edge $(s, v)$ in the \hts and assume, without loss of generality, that $v$ lies in a negative cone of $s$ (if not, we can swap the roles of $s$ and $v$). Now consider the path between them in \dn{} consisting of the edge from $s$ to the vertex closest to $s$, followed by the edges on the canonical path between the closest vertex and $v$. We will refer to this path as the \emph{approximation path}, and we show that it has length at most $3 \cdot |sv|$.

 \begin{figure}[htb]
  \centering
  \includegraphics{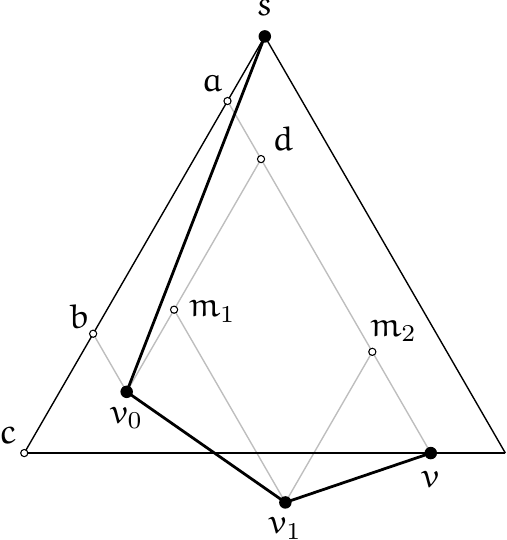}
  \caption{The approximation path.}
  \label{fig:cr-spanningratio}
 \end{figure}

 Let $v_0$ be the closest vertex and let $v_1, \dots, v_k = v$ be the other vertices on the approximation path. We assume without loss of generality that $s$ lies in $\c{0}$ of $v$ and that $v$ lies to the right of $v_0$. We shoot rays parallel to the boundaries of $\c{0}$ from each vertex on the approximation path. Let $m_i$ be the intersection of the right ray of $v_{i-1}$ and the left ray of $v_i$ (see Figure~\ref{fig:cr-spanningratio}). These intersections must exist, as $s$ is the closest vertex in $\c{0}^{v_i}$, for each $v_i$. Let $a$ and $b$ be the intersections of the left boundary of $\nc{0}^s$ with the left rays of $v$ and $v_0$, respectively, and let $c$ be the intersection of this left boundary with the horizontal line through $v$. Finally, let $d$ be the intersection of the right ray of $v_0$ and the left ray of $v$. We can bound the length of the approximation path as follows:
 \begin{align*}
    & |sv_0| + \sum_{i=1}^k |v_{i-1}v_i| \\
  ~~&\leq~~|sb| + |bv_0| + \sum_{i=1}^k |v_{i-1}m_i| + \sum_{i=1}^k |m_iv_i| \\
  ~~&=~~|sb| + |bv_0| + |ab| + |dv| \textmd{~~~~\{by projection\}}\\
  ~~&=~~|sb| + |ab| + |av| \\
  ~~&\leq~~|sc| + 2 \cdot |cv|.
 \end{align*}

 The last inequality follows from the fact that $v_0$ is the closest vertex to $s$. Let $\alpha$ be $\angle csv$. Some basic trigonometry gives us that $|sc| = \frac{2}{\sqrt{3}} \cdot \sin\left(\alpha + \frac{\pi}{3}\right) \cdot |sv|$ and $|cv| = \frac{2}{\sqrt{3}} \cdot \sin \alpha \cdot |sv|$. Thus the approximation path is at most $\frac{2}{\sqrt{3}} \cdot \left(\sin\left(\alpha + \frac{\pi}{3}\right) + 2 \cdot \sin \alpha \right)$ times as long as $(s, v)$. Since this function is increasing in $[0, \frac{\pi}{3}]$, the maximum is achieved for $\alpha = \pi/3$, where it is 3. Therefore every edge of the \hts can be approximated by a path that is at most 3 times as long and the theorem follows.
\end{proof}

Note that the part of the approximation path that lies on the canonical path has length at most $2 \cdot |cv| = \frac{4}{\sqrt{3}} \cdot \sin \alpha \cdot |sv|$. This function is also increasing in $[0, \frac{\pi}{3}]$ and its maximal value is 2, so the total length of this part is at most $2 \cdot |sv|$.

\subsection{Routing in \texorpdfstring{$\dtw$}{G-12}}
\label{sec:cr-dtw}

The stateful algorithm in Section~\ref{sec:cr-stateful} constructs a path between two vertices in the \hts. We cannot directly follow this path in \dtw, as some of the edges may have been removed. Hence, we need to find a new path in \dtw that approximates the path in the \hts, taking the missing edges into account. This often amounts to following the approximation path for edges that are in the path in the \hts, but were removed to create \dtw. In addition, some of the information the algorithm uses to decide which edge to follow relies on the presence or absence of edges in the \hts. Since the absence of these edges in \dtw does not tell us whether or not they were present in the \hts, we need to find a new way to make these decisions.

First, note that the only information we need to determine in which of the three cases we are, are the coordinates of $s$ and $t$ and whether the preferred side has been set or not. Therefore we can still make this distinction in \dtw. The following five headlines refer to steps of the stateful algorithm on the \hts, and the text after a headline describes how to simulate that step in \dtw. We discuss modifications for \dn in Section~\ref{sec:cr-dn}.

\paragraph{Follow an edge $(s, v)$ in a positive cone $C$.} If the edge of the \hts is still present in \dtw, we simply follow it. If it is not, the edge was removed because $s$ is on the canonical path of $v$ and it is not the closest, first or last vertex on the path. Since \dtw is a supergraph of \dn, we know that all of the edges of the canonical path are kept and every vertex on the path originally had an edge to $v$ in $C$. Therefore it suffices to traverse the canonical path in one direction until we reach a vertex with an edge in $C$, and follow this edge. Since the edges connecting $v$ to the first and last vertices on the path are always kept, the edge we find in this way must lead to~$v$. Note that the edges of the canonical path are easy to identify, as they are the closest edges to $C$ in cyclic order around $s$ (one on either side of $C$).

This method is guaranteed to reach $v$, but we want to find a \emph{competitive} path to $v$. Therefore we use exponential search along the canonical path: we start by following the shorter of the two edges of the canonical path incident to $s$. If the endpoint of this edge does not have an edge in $C$, we return to $s$ and travel twice the length of the first edge in the other direction. We keep returning to $s$ and doubling the maximum travel distance until we find a vertex $x$ that does have an edge in $C$. If $x$ is not the closest to $v$, by the triangle inequality, following its edge to $v$ is shorter than continuing our search until we reach the closest and following its edge. So for the purpose of bounding the distance travelled, we can assume that $x$ is closest to $v$. Let $d$ be the distance between $s$ and $x$ along the canonical path. By using exponential search to find $x$, we travel at most 9 times this distance~\cite{baezayates1993searching} and afterwards we follow $(x, v)$. From the proof of Theorem~\ref{thm:cr-3spanner}, we know that $d \leq 2 \cdot |sv|$ and $d + |xv| \leq 3 \cdot |sv|$. Thus the total length of our path is at most $9 \cdot d + |xv| = 8 \cdot d + (d + |xv|) \leq 16 \cdot |sv| + 3 \cdot |sv| = 19 \cdot |sv|$.

\begin{figure}[htb]
 \centering
 \includegraphics{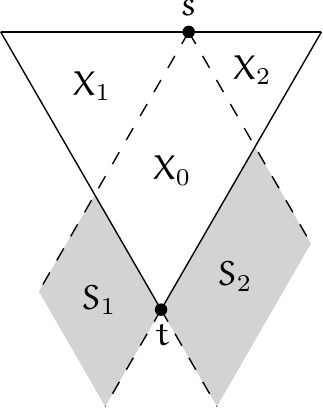}
 \caption{Possible regions for the first and last vertex.}
 \label{fig:cr-checkx0}
\end{figure}

\paragraph{Determine if there are edges in $X_0$.} In the regular \hts we can look at all our neighbours and see if any of them lie in $X_0$. However, in \dtw, these edges may have been removed. Fortunately, we can still determine if they existed in the original \hts. To do this, we look at the vertices of the canonical path in this cone that are first and last in clockwise order around $s$. If these vertices do not exist, $s$ did not have any incoming edges in this cone, so there can be no edges in $X_0$. If the first and last are the same vertex, this was the only incoming edge to $s$ from this cone, so we simply check if its endpoint lies in $X_0$. The interesting case is when the first and last exist and are distinct. If either of them lies in $X_0$, we have our answer, so assume that both lie outside of $X_0$. Since they were connected to $s$, they cannot have $t$ in their positive cone, so they must lie in one of two regions, which we call $S_1$ and $S_2$ (see Figure~\ref{fig:cr-checkx0}).

If both the first and last lie in $S_2$, there can be no edge in $X_0$, since any vertex of the canonical path in $X_0$ either lies in cone $\c{0}$ of the last vertex, or would come after the last vertex in clockwise order around $s$. Both yield a contradiction. If both lie in $S_1$, a similar argument using the first vertex applies.

On the other hand, if the first lies in $S_2$ and the last in $S_1$, both $X_1$ and $X_2$ have to be empty, since both vertices are connected to $s$. Now we are in one of two cases: either $X_0$ is also empty, or it is not. If there are no vertices in $X_0$ (different from~$t$ and $s$), $t$ must have had an edge to $s$. On the other hand, if there are other vertices in $X_0$, the topmost of these vertices must have had an edge to $s$. In either case, there must have been an edge in $X_0$. This shows that we can check whether there was an edge in $X_0$ in the \hts using only the coordinates of the first and last vertex.

\paragraph{Follow an arbitrary edge in $X_0$.} If the \hts has edges in $X_0$, we simulate following an arbitrary one of these by first following the edge to the closest vertex in the negative cone. If this vertex is in $X_0$, we are done. Otherwise, we follow the canonical path in the direction of $X_0$ and stop once we are inside. This traverses exactly the approximation path of the edge, and hence travels a distance of at most 3 times the length of the edge.

\begin{figure}[htb]
 \centering
 \includegraphics{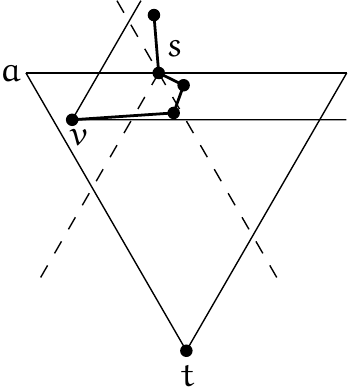}
 \caption{A vertex $v$ in $X_1$.}
 \label{fig:cr-checkx1}
\end{figure}

\paragraph{Determine if there is an edge in $X_1$ or $X_2$.} Since these regions are symmetric, we will consider only the case for $X_1$. Since $X_1$ is contained in a positive cone of $s$, it contains at most one edge incident to $s$. If the edge is present in \dtw, we can simply test whether the other endpoint lies in $X_1$. However, if $s$ does not have a neighbour in this cone (see Figure~\ref{fig:cr-checkx1}), we need to find out whether it used to have one in the original \hts and if so, whether it was in $X_1$. Since this step is only needed in case~\sB after we determine that there are no edges in $X_0$, we can use this information to guide our search. Specifically, we know that if we find an edge, we should follow it.

Therefore we simply attempt to follow the edge in this cone, using the exponential search method for following an edge in a positive cone described earlier. Let $x$ be the first vertex we encounter that still has an edge $(x,w)$ in $C_1$. If in the \hts, $s$ had an edge $(s, v)$ in $X_1$, then we know (from the arguments presented earlier for following an edge in a positive cone) that $w$ is $v$. As such, $w$ must lie in $X_1$. We also know (from the proof of Theorem~\ref{thm:cr-3spanner}) that the distance along the canonical path from $s$ to $x$ is at most $2 \cdot |sv|$, which is bounded by $2 \cdot |as|$ since $v$ lies in $X_1$. In this case, we follow the edge from $x$ to $v$. Conversely, if we do not find any vertex with an edge in $C_1$ within a distance of $2 \cdot |as|$ from $s$, or we do, but the endpoint $w$ of the edge does not lie in $X_1$, then we can return to $s$ and conclude that it did not have an edge in $X_1$ in the \hts and therefore $X_1$ must be empty.

If there was an edge in $X_1$, we travelled the same distance as if we were simply following the edge: at most $19 \cdot |sv|$. If we return to $s$ unsuccessfully, we travelled at most $20 \cdot |as|$: 9 times $2 \cdot |as|$ during the exponential search and $2 \cdot |as|$ to return to $s$.

\paragraph{Follow the edge in $X_0$ closest to the preferred side in clockwise order.} To follow this edge, we first follow the edge to the closest vertex. If this lands us in $X_0$, we then follow the canonical path towards the preferred side and stop at the last vertex on the canonical path that is in $X_0$. If the closest is not in $X_0$, we follow the canonical path towards $X_0$ and stop at the first or last vertex in $X_0$, depending on which side of $X_0$ we started on. This follows the approximation path of the edge, so the distance travelled is at most 3 times the length of the edge.

\paragraph{Routing ratio.} This shows that we can simulate the stateful routing algorithm on \dtw. As state in the message, we need to store not only the preferred side, but also information for the exponential search, including distance travelled. The exact routing ratios are as follows.

\begin{theorem}
 Let $u$ and $w$ be two vertices, with $w$ in a positive cone of $u$. There exists a deterministic $1$-local $O(1)$-memory routing algorithm on \dtw with routing ratio
 \renewcommand{\labelenumi}{{\upshape\roman{enumi})}}
 \begin{enumerate}
  \item $19 \cdot 2 = 38$ when routing from $u$ to $w$,
  \item $19 \cdot 5/\sqrt{3} \approx 54.849$ when routing from $w$ to $u$.
 \end{enumerate}
\end{theorem}
\begin{proof}
 As shown above, we can simulate every edge followed by the algorithm by travelling at most 19 times the length of the edge. The only additional cost is incurred in case~\sB, when we try to follow an edge in the smaller of $X_1$ and $X_2$, but this edge does not exist. In this case, we travel an additional $20 \cdot |as|$, where $a$ is the corner closest to $s$. Fortunately, this can happen at most once during the execution of the algorithm, as it prompts the transition to case~\sC, after which the algorithm never returns to case~\sB. Looking at the proof for the upper bound in Section~\ref{sec:cr-routing} (specifically, the second case in Figure~\ref{fig:cr-routingStateD}b), we observe that in the transition from case~$D$ to $C$, there is $2 \cdot |as|$ of unused potential. Since we are trying to show a routing ratio of 19 times the original, we can charge the additional $20 \cdot |as|$ to the $38 \cdot |as|$ of unused potential.
\end{proof}

\subsection{Routing in \texorpdfstring{$\dn$}{G-9}}
\label{sec:cr-dn}

In this subsection, we explain how to modify the previously described simulation strategies so that they work for \dn, where the first and last edges are not guaranteed to be present. We discuss only those steps that rely on the presence of these edges. To route successfully in this setting, we need to change our model slightly. We now let every vertex store a constant amount of information in addition to the information about its neighbours.

\paragraph{Follow an edge $(s, v)$ in a positive cone.} Because the first and last edges are not always kept, we cannot guarantee that the first vertex we reach with an edge in this positive cone is still part of the same canonical path. This means that the edge could connect to some arbitrary vertex, far away from $v$. Therefore our original exponential search solution does not work. Instead, we store one bit of information at $s$ (per positive cone), namely in which direction we have to follow the canonical path to reach the closest vertex to $v$. Knowing this, we just follow the canonical path in the indicated direction until we reach a vertex with an edge in this positive cone. This vertex must be the closest, so it gives us precisely the approximation path and therefore we travel at most $3 \cdot |sv|$.

\paragraph{Determine if there are edges in $X_0$.} In \dtw, this test was based on the coordinates of the endpoints of the first and last edge. Since these might be missing in \dn, we store the coordinates of these vertices at $s$. This allows us to perform the check without increasing the distance travelled.

\paragraph{Determine if there is an edge in $X_1$ or $X_2$.} As in the positive routing simulation, we now know where to go to find the closest. Therefore we simply follow the canonical path in this direction from $s$ and stop when we reach a vertex with an edge in the correct positive cone, or when we have travelled $2 \cdot |as|$. If there is an edge, we follow exactly the approximation path, giving us 3 times the length of the edge. If there is no edge, we travel $2 \cdot |as|$ back and forth, for a total of $4 \cdot |as|$.

\paragraph{Routing ratio.} Since the other simulation strategies do not rely on the presence of the first or last edges, we can now analyze the routing ratio obtained on \dn.

\begin{theorem}
 Let $u$ and $w$ be two vertices, with $w$ in a positive cone of $u$. By storing $O(1)$ additional information at each vertex, there exists a deterministic $1$-local $O(1)$-memory routing algorithm on \dn and \dtw with routing ratio
 \renewcommand{\labelenumi}{{\upshape\roman{enumi})}}
 \begin{enumerate}
  \item $3 \cdot 2 = 6$ when routing from $u$ to $w$,
  \item $3 \cdot 5/\sqrt{3} \approx 8.661$ when routing from $w$ to $u$.
 \end{enumerate}
\end{theorem}
\begin{proof}
 The simulation strategy for \dtw followed the approximation path for each edge, except when following an edge in a positive cone. Since our new strategy follows the approximation path there as well, our new routing ratio is only 3 times the one for the \hts. Note that this is still sufficient to charge the additional $4 \cdot |sa|$ travelled to the transition from case~\sB to~\sC, which has $3 \cdot 2 \cdot |as|$ of otherwise unused potential. Since \dn is a subgraph of \dtw, this strategy works on \dtw as well.
\end{proof}

\section{Conclusions}

We presented a competitive deterministic $1$-local $0$-memory routing algorithm on the \hts. We also presented matching lower bounds on the routing ratio for any deterministic $k$-local $0$-memory algorithm, showing that our algorithm is optimal. Since any triangulation can be embedded as a \hts using Schnyder's embedding~\cite{schnyder1990embedding}, this shows that any triangulation has an embedding that admits a competitive routing algorithm. An interesting open problem here is whether this approach can be extended to other theta-graphs. In particular, we recently extended the proof for the spanning ratio of the \hts to theta-graphs with $4k+2$ cones, for integer $k > 0$~\cite{bose2012optimal}. It would be interesting to see if it is possible to find optimal routing algorithms for these graphs as well.

We further extended our routing algorithm to work on versions of the \hts with bounded maximum degree. As far as we know, these are the first competitive routing algorithms on bounded-degree plane graphs. There are several problems here that are still open. For example, while we found a matching lower bound for negative routing in the regular \hts, we do not have one for the version with bounded degree. Can we find this, or is it possible to improve the routing algorithm further?

Bonichon~\etal~\cite{bonichon2010plane} also introduced a version of the \hts with maximum degree 6. This graph differs from \dtw and \dn in that it is not a subgraph of the \hts: to maintain the spanning ratio while removing even more edges, they add certain shortcut edges that were not part of the original \hts. It would be interesting to see if our routing algorithms could be extended to work on this graph. This would most likely require locally detecting shortcut edges, and finding a way to route `around' the newly removed edges.

\bibliographystyle{plain}
\bibliography{../thesis}

\part{Flips in triangulations}
\chapter{A history of flips in combinatorial triangulations}
\label{ch:fh}

Given two combinatorial triangulations, how many edge flips are necessary and sufficient to convert one into the other? This question has occupied researchers for over 75 years. We provide a comprehensive survey, including full proofs, of previous attempts to answer it, before presenting our own contribution in Chapter~\ref{ch:f4c}.

This chapter was first published as an invited chapter in the proceedings of the XIV Spanish Meeting on Computational Geometry (EGC 2011)~\cite{bose2012history}, and contains joint work with Prosenjit Bose.

\section{Introduction}
\label{sec:fh-introduction}

\begin{figure}[b]
 \centering
 \includegraphics{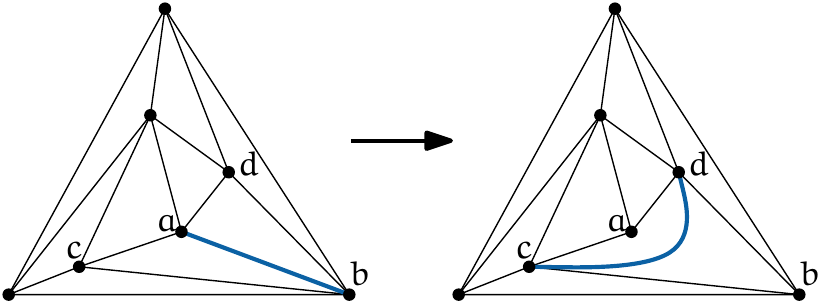}
 \caption{An example triangulation before and after flipping edge $(a, b)$.}
 \label{fig:fh-flip}
\end{figure}

A \emph{triangulation} is a simple planar graph that is \emph{maximal}, which means that adding any other edge would make the graph non-planar. This implies that every face is a triangle (a cycle of length 3). In any triangulation, an edge $e=(a,b)$ is adjacent to two faces: $abc$ and $abd$. An \emph{edge flip} consists of deleting the edge $e$ from the triangulation and adding the other diagonal of the resulting quadrilateral (in this case $(c,d)$) to the graph, so that it remains a triangulation. Figure~\ref{fig:fh-flip} shows an example of an edge flip. An edge $e$ is not flippable if $(c,d)$ is already an edge of the triangulation. If the vertices have fixed coordinates in the plane and edges are drawn as straight-line segments between their endpoints, the restriction that the new edge may not introduce any crossings is usually added. This is commonly referred to as the \emph{geometric} setting. However, we focus on the problem in the \emph{combinatorial} setting, where we are only given a combinatorial embedding of the graph (the clockwise order of edges around each vertex). Even in this setting, not all edges in a triangulation are flippable. Gao~\etal~\cite{gao2001diagonal} showed that in every $n$-vertex triangulation at least $n-2$ edges are always flippable and that there exist some triangulations where at most $n-2$ edges are flippable. If the triangulation has minimum degree at least 4, they showed that there are at least $2n+3$ flippable edges and the bound is tight in certain cases.

Note that by flipping an edge $e$, we transform one triangulation into another. This gives rise to the following question: Can any $n$-vertex triangulation be transformed into any other $n$-vertex triangulation through a finite sequence of flips? This question was first addressed by Wagner~\cite{wagner1936bemerkungen} in 1936, who answered it in the affirmative. Although it is well known that the number of $n$-vertex triangulations is exponential in $n$, Wagner's inductive proof gives rise to an algorithm that can achieve this transformation using at most $2n^2$ edge flips. The key element of Wagner's proof is that he circumvents the issue of graph isomorphism by showing how to convert any given triangulation into a fixed {\em canonical} triangulation that can be easily recognized. The downside of this approach is that one may use many more flips than necessary to convert one triangulation into another. In fact, it is possible that two triangulations are one edge flip away from each other, but Wagner's approach uses a quadratic number of flips to convert one into the other.

The notion of two triangulations being ``close" to each other in terms of number of flips can be expressed through a {\em flip graph}. The {\em flip graph} has a vertex for each distinct $n$-vertex triangulation and an edge between two vertices if their corresponding triangulations differ by a single flip. Two triangulations are considered distinct if they are not isomorphic. Questions about the flip operation can be viewed as questions on the flip graph. Asking whether any $n$-vertex triangulation can be converted into any other via flips is asking whether the flip graph is connected. Asking for the smallest number of flips required to convert one triangulation into another is asking for the shortest path in the flip graph between the two vertices representing the given triangulations. The maximum, minimum and average degree in the flip graph almost correspond to the maximum, minimum and average number of flippable edges, with the caveat that different edges might result in isomorphic triangulations when flipped. One can also ask what the chromatic number of the flip graph is, whether it is Hamiltonian, etc. Many of these questions have been addressed in the literature. The survey by Bose and Hurtado gives a good overview of the field~\cite{bose2009flips}. In this chapter, we focus mainly on attempts to determine the diameter of the flip graph. In other words, how many edge flips are sufficient and sometimes necessary to transform a given triangulation into any other? Sections~\ref{sec:fh-wagner}, \ref{sec:fh-komuro}, and \ref{sec:fh-mori} detail the techniques used to provide upper bounds for this question, while Section~\ref{sec:fh-lower} presents a lower bound. Our own contributions to this question are presented in the next chapter.

\section{Wagner's bound}
\label{sec:fh-wagner}

In 1936, Wagner~\cite{wagner1936bemerkungen} first addressed the problem of determining whether one can convert a given triangulation into another via edge flips. Although his paper is entitled ``Remarks on the four-colour problem'', it contains a proof that every planar graph has a straight-line embedding, defines the edge flip operation (or diagonal transformation, as Wagner calls it) and shows that any two triangulations can be transformed into each other by a finite series of edge flips before finishing with a result on the number of valid colourings of a graph.

\begin{figure}[htb]
 \centering
 \begin{subfigure}[b]{0.48\textwidth}
  \centering
  \includegraphics{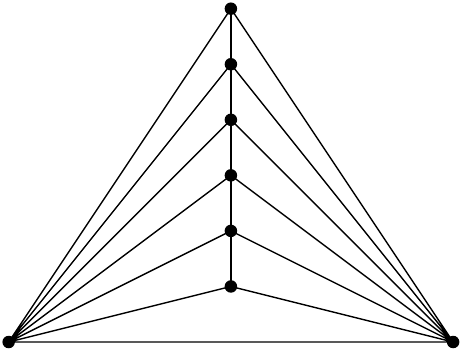}
  \caption{}
  \label{fig:fh-wagner_canonical}
 \end{subfigure}
 \begin{subfigure}[b]{0.48\textwidth}
  \centering
  \includegraphics{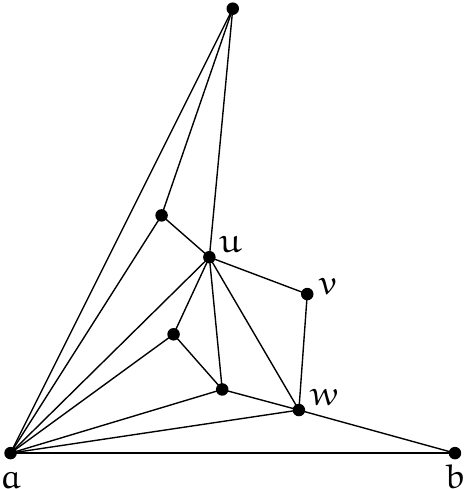}
  \caption{}
  \label{fig:fh-wagner_proof}
 \end{subfigure}
 \caption{(a) The canonical triangulation on 8 vertices. (b) A face $uwv$ such that $u$ and $w$ are neighbours of $a$, while $v$ is not. Flipping the edge $(u,w)$ brings us closer to the canonical triangulation.}
\end{figure}

To prove that any pair of triangulations can be transformed into each other via flips, Wagner first introduces the \emph{canonical triangulation}, which is the unique triangulation with two dominant vertices (see Figure~\ref{fig:fh-wagner_canonical}). We will denote the canonical triangulation on $n$ vertices by $\triangle_n$.

\begin{lemma}[Wagner \cite{wagner1936bemerkungen}, Theorem 4]
 \label{lem:fh-wagner}
 Any triangulation on $n$ vertices can be transformed into $\triangle_n$ by a sequence of at most $n^2 - 7n + 12$ flips.
\end{lemma}
\begin{proof}
 To transform a given triangulation into the canonical one, we fix an outer face and pick two of its vertices, say $a$ and $b$, to become the dominant vertices in the canonical triangulation. If $a$ is not adjacent to all other vertices, there exists a face $uwv$ such that $u$ and $w$ are neighbours of $a$, while $v$ is not. This situation is illustrated in Figure~\ref{fig:fh-wagner_proof}. We flip the edge $(u,w)$.

 In his original proof, Wagner argues that this gives a finite sequence of flips that increases the degree of $a$ by one. He simply states that this sequence is finite and does not argue why $(u,w)$ is flippable in the first place. We provide these additional arguments below.
\newpage 
 We consider two cases:
 \begin{itemize}
  \item $auw$ is a face. In this case, the flip will result in the edge $(a,v)$, increasing the degree of $a$ by one. This flip is valid, as $v$ was not adjacent to $a$ before the flip.

  \item $auw$ is not a face. In this case the flip is also valid, since $auw$ forms a triangle that separates $v$ from the vertices inside. The flip does not increase the degree of $a$, but it does increase the degree of $v$ and since the number of vertices is finite, the degree of $v$ cannot increase indefinitely. Therefore, we must eventually arrive to the first case, where we increase the degree of $a$ by one.
 \end{itemize}

 Since the same strategy can be used to increase the degree of $b$ as long as it is not dominant, this gives us a sequence of flips that transforms any triangulation into the canonical one. Every vertex of a triangulation has degree at least 3, so the degree of $a$ and $b$ needs to increase by at most $n - 4$. Since we might need to increase the degree of $v$ from 2 until it is adjacent to all but one of the neighbours of $a$ or $b$, the total flip sequence has length at most
\[
 2 \sum_{i=3}^{n-2} (i - 2)~~=~~n^2 - 7n + 12. \qedhere
\]
\end{proof}

By using the canonical triangulation as an intermediate form, the main result follows.

\begin{theorem}[Wagner \cite{wagner1936bemerkungen}, Theorem 4]
 Any pair of triangulations $T_1$ and $T_2$ on $n$ vertices can be transformed into each other by a sequence of at most $2n^2 - 14n + 24$ flips.
\end{theorem}
\begin{proof}
 By Lemma~\ref{lem:fh-wagner}, we have two sequences of flips, $S_1$ and $S_2$, that transform $T_1$ and $T_2$ into the canonical triangulation, respectively. Since a flip can be reversed, we can use $S_1$, followed by the reverse of $S_2$ to transform $T_1$ into $T_2$. Since both $S_1$ and $S_2$ have length at most $n^2 - 7n + 12$, the total sequence uses at most $2n^2 - 14n + 24$ flips.
\end{proof}

A simpler and more precise proof that also gives a quadratic upper bound was given by Negami and Nakamoto \cite{negami1993diagonal}.

\begin{figure}
 \centering
 \includegraphics{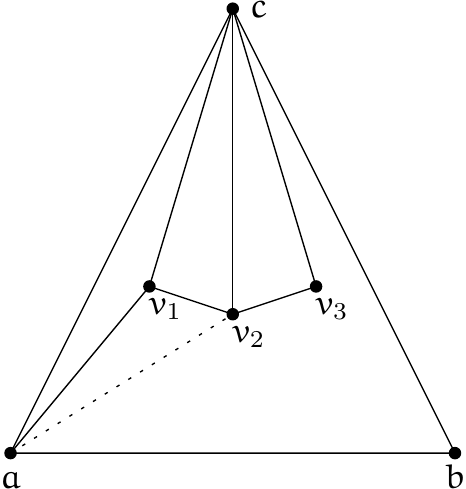}
 \caption{The exterior triangle $abc$ with the first three neighbours of $c$ in counter-clockwise order. Depending on the presence of edge $(a,v_2)$, either $(c,v_1)$ or $(c,v_2)$ is flipped.}
 \label{fig:fh-wagner_negami}
\end{figure}

\begin{lemma}[Negami and Nakamoto \cite{negami1993diagonal}, Theorem 1]
 Any triangulation on $n$ vertices can be transformed into $\triangle_n$ by a sequence of $O(n^2)$ flips.
\end{lemma}
\begin{proof}
 Let $abc$ be the outer face. Suppose we wish to make both $a$ and $b$ dominant. Instead of showing that a sequence of flips can always increase the degree of $a$ or $b$, we will show that it is always possible to find one flip that decreases the degree of $c$. Once $c$ has degree 3, the same argument can be used to find a flip that decreases the degree of $c$'s neighbour inside the triangle until it has degree 4, and so on.

 To determine which edge to flip, let $a, v_1, v_2, \dots, b$ be the neighbours of $c$ in counter-clockwise order. This situation is illustrated in Figure~\ref{fig:fh-wagner_negami}. If $a$ and $v_2$ are not adjacent, we can flip $(c,v_1)$ into $(a,v_2)$, reducing the degree of $c$. If $a$ and $v_2$ are adjacent, $av_2c$ forms a cycle that separates $v_1$ and $v_3$, so we can flip $(c,v_2)$ to reduce $c$'s degree. We continue this until $c$ has degree 3, at which point we apply the same argument to reduce the degree of $c$'s remaining neighbour inside the triangle until it has degree 4. Then we continue with the neighbour of $v_1$ inside the triangle $av_1b$, and so on, until all vertices except for $a$ and $b$ have degree 3 or 4, at which point we have obtained the canonical triangulation.
\end{proof}

\section{Komuro's bound}
\label{sec:fh-komuro}

Since Wagner's result, it remained an open problem whether the diameter of the flip graph was indeed quadratic in the number of vertices. Komuro~\cite{komuro1997diagonal} showed that in fact the diameter was linear by proving a linear upper and lower bound. We present the argument for the upper bound in this section and discuss the lower bound in Section~\ref{sec:fh-lower}.

\begin{figure}
 \centering
 \includegraphics{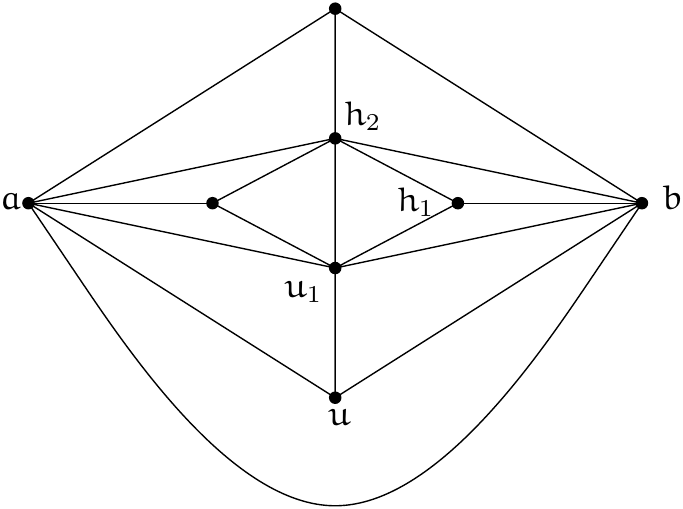}
 \caption{No single edge can be flipped to increase the degree of $a$ or $b$.}
 \label{fig:fh-komuro_proof}
\end{figure}

Komuro used Wagner's approach of converting a given triangulation into the canonical triangulation. Given an arbitrary triangulation, the key is to bound the number of flips needed to make two vertices, say $a$ and $b$, dominant. If there always exists one edge flip that increases the degree of $a$ or $b$ by 1, then at most $2n-8$ flips are sufficient since dominant vertices have degree $n-1$ and all vertices in a triangulation have degree at least 3. However, this is not always the case. Figure~\ref{fig:fh-komuro_proof} shows a triangulation where no single flip increases the degree of $a$ or $b$. Komuro used the following function to bound the number of flips: $d_G(a,b) = 3 \deg(a) + \deg(b)$. He showed that there always exists either one edge flip where $d_G(a,b)$ goes up by at least 1 or two edge flips where $d_G(a,b)$ goes up by at least 2. The cleverness of the function is that in some cases, two edge flips increase the degree of $a$ by 1 but decrease the degree of $b$ by 1. However, since the function increases by 2, it still increases by at least 1 per flip. Since $d_G(a,b) \leq 4n-4$, we have that $4n - 4 - d_G(a,b)$ is an upper bound on the number of flips required to make $a$ and $b$ dominant.

\begin{lemma}[Komuro~\cite{komuro1997diagonal}, Lemma 2]
 \label{lem:fh-komuro}
Let $G$ be a triangulation on $n$ vertices and let $a,b$ be any pair of adjacent vertices of $G$. Then $G$ can be transformed into the canonical triangulation $\triangle_n$ with $a$ and $b$ as dominant vertices with at most $4n-4-(3\deg(a) + \deg(b))$ edge flips. 
\end{lemma}
\begin{proof} 
In a triangulation, every vertex must have degree at least 3. Let $uab$ be a face adjacent to $ab$. We consider two cases: $\deg(u) = 3$ and $\deg(u) > 3$. We begin with the latter. Since $\deg(u) \geq 4$, let $a, b, w_1, w_2$ be four consecutive neighbours of $u$ in counter-clockwise order. If $b$ is not adjacent to $w_2$, then flipping edge $(u,w_1)$ increases $\deg(b)$ by 1 and thus $d_G(a,b)$ by 1. If $b$ is adjacent to $w_2$, then $ubw_2$ is a separating triangle (a cycle of length 3 whose removal disconnects the graph) that separates $a$ from $w_1$. Therefore, flipping edge $(u,b)$ decreases $\deg(b)$ by 1 and increases $\deg(a)$ by 1. Thus, with one flip $d_G(a,b)$ increases by 2.

Now consider the case when $\deg(u) = 3$. Let $u_1$ be the unique vertex adjacent to $u$, $a$, and $b$. We now have 3 cases to consider: $\deg(u_1) = 3$, $\deg(u_1) \geq 5$, or $\deg(u_1) = 4$. If $\deg(u_1) = 3$, then the graph is isomorphic to $K_4$, which is $\triangle_4$. If $\deg(u_1) \geq 5$, let $a$, $u$, $b$, $h_1$, and $h_2$ be five consecutive neighbours of $u_1$ in counter-clockwise order. If $b$ is not adjacent to $h_2$, then flipping the edge $(u_1,h_1)$ increases $\deg(b)$ by 1 and thus $d_G(a,b)$ by 1. If $b$ is adjacent to $h_2$, then $u_1bh_2$ is a separating triangle that separates $u$ and $a$ from $h_1$ (see Figure \ref{fig:fh-komuro_proof}). Therefore, flipping edges $(u_1,b)$ and $(u_1,u)$ decreases $\deg(b)$ by 1 and increases $\deg(a)$ by 1. Thus, with two flips $d_G(a,b)$ increases by 2.

Finally, if $\deg(u_1)=4$, then there is unique vertex $u_2$ adjacent to $a$, $u_1$, and $b$. If $\deg(u_2) = 3$, the graph is isomorphic to $\triangle_5$. If $\deg(u_2) \geq 5$ we apply the same argument as when $\deg(u_1) \geq 5$. If $\deg(u_2) = 4$, we obtain another unique vertex $u_3$. This process ends with $u_{n-3}$, at which point $a$ and $b$ are dominant.

Since $d_G(a,b)$ increases by at least 1 for one flip and at least 2 for two flips, we note that the total number of flips does not exceed $d_{\triangle_n}(a,b) - d_G(a,b) = 4n-4-(3\deg(a) + \deg(b))$ as required.
\end{proof}

Using this lemma, Komuro proved the following theorem.

\begin{theorem}[Komuro \cite{komuro1997diagonal}, Theorem 1]
\label{thm:fh-komuro}
Any two triangulations with $n$ vertices can be transformed into each other by at most $8n-54$ edge flips if $n\geq 13$ and at most $8n-48$ edge flips if $n\geq 7$.
\end{theorem}

\begin{proof}
Given a triangulation $G$ on $7 \leq n \leq 12$ vertices, one can prove by contradiction that either $G$ is one flip from $\triangle_n$ or there exists an edge $(a,b)$ where both vertices have degree at least 5, implying that $d_G(a,b) \geq 20$. This gives an upper bound of $4n-24$ to convert $G$ to $\triangle_n$, which gives an upper bound of $8n-48$ to convert any triangulation to any other via the canonical triangulation. Moreover, for $n\geq 13$, either $G$ is one flip from canonical or there exists an edge $(a,b)$ where $a$ has degree at least 6 and $b$ has degree at least 5. This means that $d_G(a,b) \geq 23$. The result follows.
\end{proof}

\section{Mori et al.'s bound}
\label{sec:fh-mori}

In 2001, Mori, Nakamoto and Ota \cite{mori2003diagonal} improved the bound by Komuro to $6n - 30$. They used a two-step approach by finding a short path to a strongly connected kernel, which consists of all Hamiltonian triangulations. An $n$-vertex triangulation is Hamiltonian if it contains a Hamiltonian cycle, i.e. a cycle of length $n$. The general idea of the proof is to find a fast way to make any triangulation Hamiltonian and then use the Hamiltonian cycle to decompose the graph into two outerplanar graphs. These have the following nice property.

\begin{lemma}[Mori \etal \cite{mori2003diagonal}, Lemma 8 and Proposition 9]
 \label{lem:fh-mori_outerplanar}
 Any vertex $v$ in a maximal outerplanar graph on $n$ vertices can be made dominant by $n - 1 - deg(v)$ flips.
\end{lemma}
\begin{proof}
 If $v$ is not dominant, there is a triangle $vxy$ where $(x,y)$ is not an edge of the outer face. Then we can flip $(x,y)$ into $(v,z)$, where $z$ is the other vertex of the quadrilateral formed by the two triangles that share $(x,y)$. This flip must be legal, since if $(v,z)$ was already an edge, the graph would have $K_4$ as a subgraph, which is impossible for outerplanar graphs. Since each such flip increases the degree of $v$ by one, $n - 1 - deg(v)$ flips are both necessary and sufficient.
\end{proof}

With this property, Mori~\etal~showed that it is possible to quickly transform any Hamiltonian triangulation into the canonical form by decomposing it along the Hamiltonian cycle into two outerplanar graphs. Interestingly, this exact approach was already used 10 years earlier by Sleator, Tarjan and Thurston~\cite{sleator1992short} to prove a $\Theta(n \log n)$ bound on the diameter of the flip graph when the vertices are labelled. Note that this was even before the linear bound by Komuro that was discussed in the previous section. However, they did not state their result in terms of unlabelled triangulations and it seems that both Komuro and Mori~\etal~were unaware of this earlier work.

\begin{figure}[ht]
 \centering
 \includegraphics{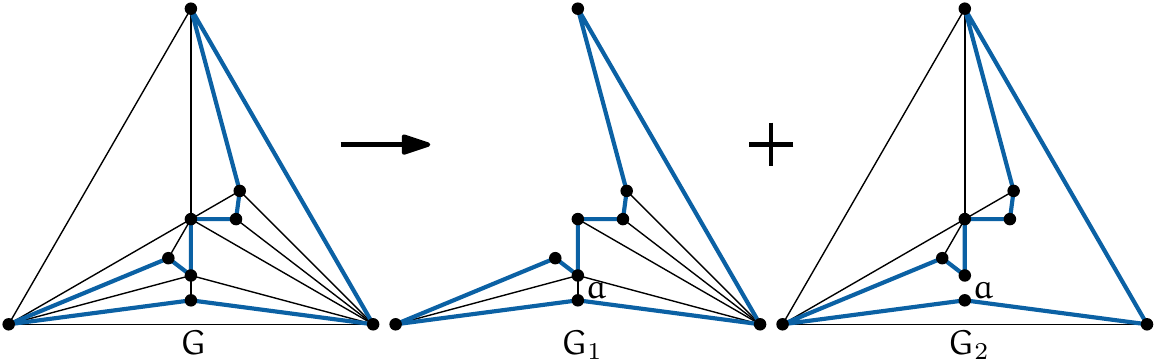}
 \caption{The decomposition of a Hamiltonian graph $G$ into two outerplanar graphs $G_1$ and $G_2$. The vertex $a$ has degree 2 in $G_2$.}
 \label{fig:fh-mori_decomposition}
\end{figure}

\begin{theorem}[Mori \etal \cite{mori2003diagonal}, Proposition 9]
 \label{thm:fh-mori_hamiltonian}
 Any Hamiltonian triangulation on $n$ vertices can be transformed into $\triangle_n$ by at most $2n - 10$ flips, preserving the existence of Hamiltonian cycles.
\end{theorem}
\begin{proof}
 Given a Hamiltonian triangulation $G$ with Hamiltonian cycle $C$, we can decompose it into two outerplanar graphs $G_1$ and $G_2$, such that each contains $C$ and all edges on one side of $C$. This is illustrated in Figure~\ref{fig:fh-mori_decomposition}. Let $a$ be a vertex of degree 2 in $G_2$. We are going to make $a$ dominant in $G_1$. Since $G$ is 3-connected and $a$ has no additional neighbours in $G_2$, the degree of $a$ in $G_1$ is at least 3. Thus by Lemma~\ref{lem:fh-mori_outerplanar}, we can make $a$ dominant by at most $n - 4$ flips. Each of these flips is valid, as $a$ is not connected to anyone in $G_2$, except for its neighbours on $C$.

 Now consider the subgraph $G_2' = G_2 \setminus \{a\}$. Since $a$ has degree 2 in $G_2$, $G_2'$ is still outerplanar, so by applying Lemma~\ref{lem:fh-mori_outerplanar} again we can make a vertex of $G_2'$ dominant as well, which gives us the canonical triangulation. Since $G_2'$ has $n - 1$ vertices and it always has a vertex of degree at least 4 (provided that $n \geq 6$), we need at most $n - 6$ flips for this. Since we did not flip any of the edges on $C$, the theorem follows.
\end{proof}

This shows that the Hamiltonian triangulations are closely connected, so all we need to figure out is how we can quickly make a triangulation Hamiltonian. Here, we turn to an old result by Whitney~\cite{whitney1931theorem} that shows that all 4-connected triangulations are Hamiltonian. Since a triangulation is 4-connected if and only if it does not have any separating triangles (cycles of length 3 whose removal disconnects the graph; see Lemma~\ref{lem:f4c-4connected} for a proof), by removing all separating triangles from a triangulation, we make it 4-connected and therefore Hamiltonian. Fortunately, separating triangles are easy to remove using flips, as the following lemmas show.

\begin{lemma}[Mori \etal \cite{mori2003diagonal}, Lemma 11]
 \label{lem:fh-mori_flip}
 In a triangulation with $n \geq 6$ vertices, flipping any edge of a separating triangle $D = abc$ will remove that separating triangle. This never introduces a new separating triangle, provided that the selected edge belongs to multiple separating triangles or none of the edges of $D$ belong to multiple separating triangles.
\end{lemma}
\begin{proof}
 Since $D$ is separating and the newly created edge connects a vertex on the inside to a vertex on the outside, the flip is always legal. Since the flip removes an edge of $D$, it is no longer a separating triangle. Now suppose that we flipped $(a,b)$ to a new edge $(x,y)$ and introduced a new separating triangle $D'$. Then $D'$ must be $xyc$.
 But since $n \geq 6$ and our construction so far uses only 5 vertices, one of the faces $ayc$, $byc$, $axc$, or $bcx$ must be a separating triangle as well. This means that either $(a,c)$ or $(b,c)$ is an edge that belongs to multiple separating triangles, while $(a,b)$ only belongs to $D$, which contradicts the choice of $(a,b)$.
\end{proof}

\begin{lemma}[Mori \etal \cite{mori2003diagonal}, Lemma 11]
 \label{lem:fh-mori_4connected}
 Any triangulation on $n$ vertices can be made 4-connected by at most $n - 4$ flips.
\end{lemma}
\begin{proof}
 We will show that a triangulation can have at most $n - 4$ separating triangles, the result follows by Lemma~\ref{lem:fh-mori_flip}. The proof is by induction on $n$. For the base case, let $n = 4$. Then our graph must be $K_4$, which has no separating triangles as required. For the induction we can assume that our graph $G$ has a separating triangle $T$ which partitions $G$ into two components $G_1$ and $G_2$. By induction, $G_1$ and $G_2$ have at most $n_1 - 4$ and $n_2 - 4$ separating triangles, where $n_1$ and $n_2$ are the number of vertices in $G_1$ and $G_2$, respectively, including the vertices of $T$. Therefore $G$ can have at most $n_1 - 4 + n_2 - 4 + 1 = (n_1 + n_2 - 3) - 4 = n - 4$ separating triangles.
\end{proof}

Now we can prove the main result.

\begin{theorem}[Mori \etal \cite{mori2003diagonal}, Theorem 4]
 Any two triangulations on $n$ vertices can be transformed into each other by at most $6n - 30$ flips.
\end{theorem}
\begin{proof}
 The connection between Lemma~\ref{lem:fh-mori_4connected} and Theorem~\ref{thm:fh-mori_hamiltonian} is an old proof by Whitney~\cite{whitney1931theorem} that any 4-connected triangulation is Hamiltonian. Therefore we can transform any triangulation into the canonical form by at most $n - 4 + 2n - 10 = 3n - 14$ flips. By looking carefully at the proof of Theorem~\ref{thm:fh-mori_hamiltonian}, we see that if the graph is 4-connected, the first vertex (vertex $a$) is guaranteed to have degree at least 4, which brings the bound down to $3n - 15$ flips to the canonical triangulation and $6n - 30$ flips between any pair of triangulations.
\end{proof}

Note that, although finding a Hamiltonian cycle is NP-hard in general~\cite{karp1972reducibility}, there exists a linear-time algorithm by Asano~\etal for finding a Hamiltonian cycle in any 4-connected triangulation~\cite{asano1984linear}. Thus, the assumption in the proof that the Hamiltonian cycle is given is not a practical concern when implementing the resulting algorithm.


\section{Lower bounds}
\label{sec:fh-lower}

In addition to the upper bound described in Section~\ref{sec:fh-komuro}, Komuro~\cite{komuro1997diagonal} also gave a lower bound  on the diameter of the flip graph, based on the maximum degree of the vertices in the graph.

\begin{theorem}[Komuro~\cite{komuro1997diagonal}, Theorem 5]
 \label{thm:fh-lb}
 Let $G$ be a triangulation on $n$ vertices. Then at least $2n - 2 \Delta(G) - 3$ flips are needed to transform $G$ into the canonical triangulation, where $\Delta(G)$ denotes the maximum degree of $G$.
\end{theorem}
\begin{proof}
 Let $a$ and $b$ be the two vertices of degree $n - 1$ in the canonical triangulation. Each flip increases the degree in $G$ of either $a$ or $b$ by at most one. The only possible exception is the flip that creates the edge $(a,b)$, which increases the degree of both vertices by one. Since the initial degree of $a$ and $b$ is at most $\Delta(G)$, we need at least $2(n - 1 - \Delta(G)) - 1 = 2n - 2\Delta(G) - 3$ flips.
\end{proof}

Since there are triangulations that have maximum degree 6, this gives a lower bound of $2n - 15$ flips. It is interesting that one of the triangulations in the lower bound is the canonical form. This implies that either the lower bound is very far off, or the canonical triangulation is a bad choice of intermediate triangulation. It also means that as long as we use this canonical form, the best we can hope for is an upper bound of $4n - 30$ flips. Komuro also gave a lower bound on the number of flips required to transform between any pair of triangulations, again based on the degrees of the vertices.

\begin{theorem}[Komuro~\cite{komuro1997diagonal}, Theorem 4]
 Let $G$ and $G'$ be triangulations on $n$ vertices. Let $v_1, \dots, v_n$ and $v_1', \dots, v_n'$ be the vertices of $G$ and $G'$, respectively, ordered by increasing degree. Then at least $\frac{1}{4}D(G, G')$ flips are needed to transform $G$ into $G'$, where $D(G, G') = \sum_{i=1}^n |deg(v_i) - deg(v_i')|$.
\end{theorem}
\begin{proof}
 Let $\sigma$ be a mapping between the vertices of $G$ and $G'$ and suppose we transform $G$ into $G'$ using flips, such that $v_i \in G$ becomes $v_{\sigma(i)}' \in G'$. Since every flip changes the degree of a vertex by one, we need at least $|deg(v_i) - deg(v_{\sigma(i)}')|$ flips to obtain the correct degree for $v_{\sigma(i)}'$. However, each flip affects the degrees of 4 vertices, giving a bound of $\frac{1}{4}\sum_{i=1}^n |deg(v_i) - deg(v_{\sigma(i)}')|$ flips. Our actual lower bound is the minimum of this bound over all mappings $\sigma$. Mapping every vertex to a vertex with the same rank when ordered by degree (i.e. $\sigma(i) = i$) achieves this minimum.
\end{proof}

This was the best known lower bound for almost twenty years, but in a recent pre-print, Frati~\cite{frati2015lower} presented an improved lower bound, based on the notion of common edges.

\begin{theorem}[Frati~\cite{frati2015lower}, Lemma 1]
 \label{thm:fh-frati}
 Let $G$ and $G'$ be triangulations on $n$ vertices. Let $\sigma$ be the bijection between vertices of $G$ and $G'$ that maximizes the number of common edges, and let $c(\sigma)$ be that number. Then any flip sequence that transforms $G$ into $G'$ has length at least $3n - 6 - c(\sigma)$.
\end{theorem}
\begin{proof}
 At the end of the flip sequence, the two graphs will be isomorphic, so they will have all $3n - 6$ edges in common. Since each flip can introduce at most one new common edge, the lower bound follows.
\end{proof}

Frati then constructs a graph $G_{\textsc{lb}}$ that shares at most $\frac{2n}{3}$ edges with the canonical form, regardless of the bijection used. The graph consists of an arbitrary triangulation on $\frac{n}{3} + 2$ vertices with maximum degree six, with a degree-three vertex inserted in each face. The vertices of the original triangulation are colored blue, while the inserted vertices are colored red. Note that the red vertices form an independent set.

\begin{theorem}[Frati~\cite{frati2015lower}, Theorem 1]
 The diameter of the flip graph is at least $\frac{7n}{3} - 34$.
\end{theorem}
\begin{proof}
 Consider any bijection between the vertices of $G_{\textsc{lb}}$ and $\triangle_n$. Since the blue vertices had degree at most six before the red vertices were inserted, the maximum degree of $G_{\textsc{lb}}$ is twelve. Thus, at most 24 of the edges incident to the dominant vertices of $\triangle_n$ can be common. Now consider the chain of edges not incident to the dominant vertices. Since no two red vertices in $G_{\textsc{lb}}$ are adjacent, every edge on this chain must have one endpoint mapped to a blue vertex. But since there are only $\frac{n}{3} + 2$ blue vertices in $G_{\textsc{lb}}$, no more than $\frac{2n}{3} + 4$ edges on the chain can be common. Thus, the maximum number of common edges between $G_{\textsc{lb}}$ and $\triangle_n$ is $\frac{2n}{3} + 28$, which by Theorem~\ref{thm:fh-frati} gives a lower bound of $3n - 6 - (\frac{2n}{3} + 28) = \frac{7n}{3} - 34$ flips.
\end{proof}

\bibliographystyle{plain}
\bibliography{../thesis}
\chapter{Making triangulations 4-connected using flips}
\label{ch:f4c}

In this chapter, we show that any combinatorial triangulation on $n$ vertices can be transformed into a 4-connected one using at most $\lfloor(3n - 9)/5\rfloor$ edge flips. We also give an example of an infinite family of triangulations that requires this many flips to be made 4-connected, showing that our bound is tight. In addition, for $n \geq 19$, we improve the upper bound on the number of flips required to transform any 4-connected triangulation into the canonical triangulation (the triangulation with two dominant vertices), matching the known lower bound of $2n - 15$. Our results imply a new upper bound on the diameter of the flip graph of $5.2 n - 33.6$, improving on the previous best known bound of $6n - 30$.

This chapter was first published in the proceedings of the 23rd Canadian Conference on Computational Geometry (CCCG 2011)~\cite{bose2011making}, and was subsequently invited and accepted to a special issue of Computational Geometry: Theory and Applications~\cite{bose2012making}. It contains joint work with Prosenjit Bose, Dana Jansens, Andr\'e van Renssen and Maria Saumell.

\section{Introduction}
\label{sec:f4c-introduction}

As reviewed in Chapter~\ref{ch:fh}, a lot of research has gone into the following question: ``Given two combinatorial triangulations, how can we transform one into the other using edge flips?" The best known algorithm, developed independently by Sleator~\etal~\cite{sleator1992short} and Mori~\etal~\cite{mori2003diagonal}, consists of two steps. In the first step, the given triangulation is transformed into a 4-connected one, using at most $n - 4$ flips. Since a 4-connected triangulation is always Hamiltonian (an old result by Whitney~\cite{whitney1931theorem}; in fact, the cycle can even be found quickly~\cite{asano1984linear}), the resulting Hamiltonian triangulation is then transformed into the canonical one by at most $2n - 11$ flips, using a decomposition into two outerplanar graphs that share a Hamiltonian cycle as their respective outer faces. Thus $6n - 30$ flips are sufficient to transform any triangulation into any other. The algorithm and analysis are described in more detail in Section~\ref{sec:fh-mori}.

The upper bound on the number of flips used to make the triangulation 4-connected arises from the fact that any separating triangle can be removed by flipping one of its edges, and that a triangulation can have at most $n - 4$ separating triangles. However, this analysis does not take advantage of the fact that if two separating triangles share an edge, a single flip can remove both of them. In this chapter, we combine this observation with an edge charging scheme to show that any triangulation can be made 4-connected using at most $\lfloor(3n - 9)/5\rfloor$ flips.

The problem of making triangulations 4-connected has also been studied in the setting where many edges may be flipped simultaneously, provided none of them are part of the same triangle. Bose~\etal~\cite{bose2007simultaneous} showed that any triangulation can be made 4-connected by one such simultaneous flip and that $O(\log n)$ simultaneous flips are sufficient and sometimes necessary to transform between two given triangulations.

The remainder of this chapter is organized as follows. In Section~\ref{sec:f4c-ub}, we prove the new upper bound on the number of flips to make a triangulation 4-connected, thereby improving the first step of the construction by Mori~\etal For $n \geq 19$, we also improve the bound on the second step of their algorithm to match the lower bound by Komuro~\cite{komuro1997diagonal}. This results in a new upper bound on the diameter of the flip graph of $5.2 n - 33.6$. We then show in Section~\ref{sec:f4c-lb} that, when $n$ is a multiple of 5, there are triangulations that require $(3n - 10)/5 = \lfloor(3n - 9)/5\rfloor$ flips to be made 4-connected, showing that our bound is tight. Section~\ref{sec:f4c-lemmas} contains proofs for various technical lemmas that are used in the proof of the upper bound.

After completion of this chapter, Cardinal~\etal~\cite{cardinal2015arc} further improved the upper bound on the diameter of the flip graph to $5n - 23$ by proving that a triangulation can be directly transformed into a Hamiltonian one using at most $n / 2$ flips. This allowed them to construct an arc drawing (a plane drawing with all vertices on a line and edges represented by a connected sequence of semi-circles centred on the line) for any planar graph, in which all edges are drawn as a single semi-circle, except for $n / 2$ edges that are drawn as a sequence of two semicircles. In addition, they showed that there always exists a single simultaneous flip of fewer than $2n / 3$ edges that makes a triangulation 4-connected, and that this bound is tight up to an additive constant.

\section{Upper bound}
\label{sec:f4c-ub}

In this section we prove an upper bound on the number of flips to make any given triangulation 4-connected. Specifically, we show that \mbox{$\lfloor(3n - 9)/5\rfloor$} flips always suffice. The proof references several technical lemmas whose proofs can be found in Section~\ref{sec:f4c-lemmas}. We also prove that any 4-connected triangulation can be transformed into the canonical form using a worst-case optimal number of $2n - 15$ flips. We start by providing more precise definitions of relevant concepts.

\paragraph{Definitions}

Our input consists of a triangulation $T$, along with a combinatorial embedding specifying the clockwise order of edges around each vertex of $T$. In addition, one of the faces of $T$ is marked as the \emph{outer face}. If an edge of the outer face is flipped, one of the two new faces is designated as the new outer face. A \emph{separating triangle} $D$ is a cycle in $T$ of length three whose removal splits $T$ into two (non-empty) connected components. We call the component that contains vertices of the outer face the \emph{exterior} of $D$, and the other component the \emph{interior} of $D$. A vertex in the interior of $D$ is said to be \emph{inside} $D$ and likewise, a vertex in the exterior of $D$ is said to be \emph{outside} $D$. An edge is inside a separating triangle if one or both endpoints are inside.

A separating triangle $A$ \emph{contains} another separating triangle $B$ if and only if the interior of $B$ is a subgraph of the interior of $A$ with a strictly smaller vertex set. If $A$ contains $B$, $A$ is the \emph{containing} triangle. A separating triangle that is contained by the largest number of separating triangles in $T$ is called \emph{deepest}. Since containment is transitive, a deepest separating triangle cannot contain any separating triangles, as these would have a higher number of containing triangles.

\paragraph{Algorithm}

We use the same general strategy as the earlier algorithms - flip an edge of a separating triangle until there are none left. This strategy is guaranteed to terminate by the following Lemma (see Section~\ref{sec:fh-mori}, Lemma~\ref{lem:fh-mori_flip} for a proof).

\begin{lemma}[Mori~\etal~\cite{mori2003diagonal}, Lemma 11]
 \label{lem:f4c-flip}
 In a triangulation on $n \geq 6$ vertices, flipping any edge of a separating triangle $D$ will remove that separating triangle. This never introduces a new separating triangle, provided that the selected edge belongs to multiple separating triangles or none of the edges of $D$ belong to multiple separating triangles.
\end{lemma}

Since a triangulation is 4-connected if and only if it does not have any separating triangles (see Lemma~\ref{lem:f4c-4connected}), this strategy transforms any triangulation into one that is 4-connected. With this in mind, our algorithm works as follows. The reasoning behind some of the choices will become clear during the analysis.

\begin{algorithm}[Make 4-connected]
\ \\[-2\baselineskip] 
\begin{itemize}
 \renewcommand{\labelitemii}{$\circ$} 
 \item Find a deepest separating triangle $D$, preferring ones that do not use an edge of the outer face.
 \begin{itemize}
 \item If $D$ does not share any edge with other separating triangles, flip an edge of $D$ that is not on the outer face.
 \item If $D$ shares exactly one edge with another separating triangle, flip this edge.
 \item If $D$ shares multiple edges with other separating triangles, flip one of the shared edges that is not shared with a containing triangle (such an edge always exists in this case).
 \end{itemize}
 \item Repeat until $T$ is 4-connected.
\end{itemize}
\end{algorithm}

\paragraph{Analysis}

Fundamentally, our analysis relies on counting edges. We separate the edges into two categories: edges that are part of some separating triangle, and edges that are not. We call the latter \emph{free edges}. If there were no free edges, a triangulation would have sufficiently many edges for each of the maximum $n - 4$ separating triangles (from Lemma~\ref{lem:fh-mori_4connected}) to be edge-disjoint. And since we need to flip at least one edge of every separating triangle in order to make the triangulation 4-connected, we would require $n - 4$ flips. Thus, to get a better upper bound, we need to show that this situation is impossible. The following lemma does so, by showing that the presence of a separating triangle forces some other edges to be free.

 \begin{figure}[htb]
  \centering
  \includegraphics{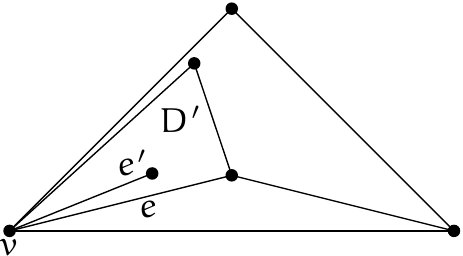}
  \caption{Every vertex of a separating triangle is incident to a free edge inside the triangle.}
  \label{fig:f4c-freeedgeproof}
 \end{figure}

\begin{lemma}
 \label{lem:f4c-freeedge}
 In a triangulation, every vertex $v$ of a separating triangle $D$ is incident to at least one free edge inside $D$.
\end{lemma}
\begin{proof}
 Consider one of the edges of $D$ that is incident to $v$. Since $D$ is separating, its interior cannot be empty and since $D$ is part of a triangulation, there is a triangular face inside $D$ that uses this edge. Let $e$ be the other edge of this face that is incident to $v$ (see Figure~\ref{fig:f4c-freeedgeproof}).

 The remainder of the proof is by induction on the number of separating triangles contained in $D$. For the base case, assume that $D$ does not contain any other separating triangles. Then $e$ must be a free edge and we are done.

 For the induction step, there are two further cases. If $e$ does not belong to a separating triangle, we are again done, so assume that $e$ belongs to a separating triangle $D'$. Since $D'$ is itself a separating triangle contained in $D$ and containment is transitive, the number of separating triangles contained in $D'$ must be strictly smaller than the number contained in $D$. Since $v$ is also a vertex of $D'$, our induction hypothesis tells us that there is a free edge incident to $v$ inside $D'$. Since $D'$ is contained in $D$, this edge is also inside $D$.
\end{proof}

This immediately gives us a better bound on the maximum number of edge-disjoint separating triangles.

 \begin{figure}[htb]
  \centering
  \includegraphics{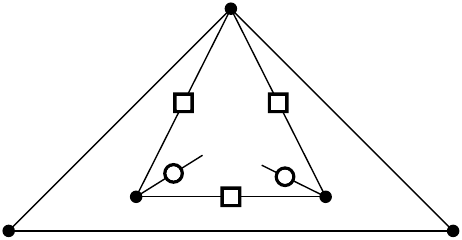}
  \caption{Each edge-disjoint separating triangle can be assigned five edges: the three edges of the triangle (squares) and two free edges incident to unshared vertices (circles).}
  \label{fig:f4c-edgeassignment}
 \end{figure}

\begin{corollary}
    \label{cor:f4c-max-edge-disjoint}
    If all separating triangles are edge-disjoint, a triangulation on $n$ vertices can contain at most $(3n - 10)/5$ separating triangles.
\end{corollary}
\begin{proof}
    We prove this by assigning five edges to each separating triangle without assigning any edge twice. First, each edge of a separating triangle is assigned to that triangle. Since they are all edge-disjoint, this does not assign any edge twice. Next, we turn to the free edges identified by Lemma~\ref{lem:f4c-freeedge}. We start with the topmost separating triangles, i.e. those who are not contained in any other separating triangle, and assign all three free edges (one per vertex) to them. Note that this assigns six edges to each of these triangles, instead of five -- a fact we use later to tighten the bound.
    
    Now consider a separating triangle $D$ that is contained in some other separating triangles. We have to be careful to avoid free edges that have already been assigned to triangles that contain it. But since all separating triangles are edge-disjoint, $D$ can share at most one vertex with a separating triangle that contains it (see~Lemma~\ref{lem:f4c-onecontaining}). Thus, we can safely assign the two free edges that are incident to the unshared vertices to the triangle (see~Figure~\ref{fig:f4c-edgeassignment}).
    
    Since we assigned five edges to each separating triangle, and a triangulation has exactly $3n - 6$ edges, there can be at most $(3n - 6) / 5$ edge-disjoint separating triangles. However, recall that each topmost separating triangle was actually assigned six edges. And any edge that is not contained in a separating triangle has not been assigned at all. To prove a bound of $(3n - 10)/5$ separating triangles, we need to find at least four edges between these two categories.
    
    Consider the edges of the outer face. These edges are either free, or part of a topmost separating triangle. If all three edges of the outer face are free, then either there are no separating triangles, or there is at least one topmost separating triangle whose edge we can also use. In either case, we are done. 
    
    If only two edges of the outer face are free, there is a topmost separating triangle that uses the other edge, giving us three free edges already. But this triangle cannot use the vertex shared by the two free edges. Since this vertex has degree at least three, it is incident to either a free edge, or another topmost separating triangle, both of which give us four free edges.
    
    Finally, if one or no edges of the outer face are free, we have three free edges between the edges of the outer face and the topmost separating triangles, so we just need to find one more. Consider a vertex $v$ shared by two non-free edges of the outer face. Let $D_1$ and $D_2$ be the topmost separating triangles that use these edges. Since $D_1$ and $D_2$ cannot share an edge, there is at least one face adjacent to $v$ that lies between $D_1$ and $D_2$. Consider the edge of that face opposite from $v$. If it is free, we are done. If it is not free, it must be used by another topmost separating triangle that does not use any edge of the outer face, also giving us a fourth free edge. Therefore a triangulation can contain no more than $(3n - 10)/5$ separating triangles.
\end{proof}

Thus, if all separating triangles are edge-disjoint, we only need $(3n - 10)/5$ flips to make a triangulation 4-connected. But what if some of the separating triangles do share edges? As it turns out, we can show a similar upper bound on the number of flips needed by the algorithm we presented earlier.

\begin{theorem}
 \label{thm:f4c-4-connected}
 A triangulation on $n \geq 6$ vertices can be made 4-connected using at most $\lfloor(3n - 9)/5\rfloor$ flips.
\end{theorem}
\begin{proof}
 We prove this using a charging scheme. We begin by placing a coin on every edge of the triangulation. Then we flip the edges indicated by the algorithm until no separating triangles remain, while paying five coins for every flip. The exact charging scheme will be described later. During this process, we maintain two invariants:
 \begin{itemize}
  \item Every edge of a separating triangle has a coin.
  \item Every vertex of a separating triangle has an incident free edge that is inside the triangle and has a coin.
 \end{itemize}

 These invariants have several nice properties. First, an edge can either be a free edge or belong to a separating triangle, but not both. So at any given time, only one invariant applies to an edge. Second, an edge only needs one coin to satisfy the invariants, even if it is on multiple separating triangles or is a free edge for multiple separating triangles. These two properties imply that the invariants hold initially, since by Lemma~\ref{lem:f4c-freeedge}, every vertex of a separating triangle has an incident free edge.
 
 We now show that these invariants are sufficient to guarantee that we can pay five coins for every flip. Consider the situation after we flip an edge that belongs to a deepest separating triangle $D$ and satisfies the criteria of Lemma~\ref{lem:f4c-flip}, but before we remove any coins. Since flipping the edge has removed $D$ and no new separating triangles are introduced, both invariants still hold. We proceed by identifying four types of edges whose coins we can now remove to pay for this flip without upsetting the invariants.
 
  \type{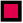} The flipped edge $e$. By Lemma~\ref{lem:f4c-flip}, $e$ cannot belong to any separating triangle after the flip, so the first invariant still holds if we remove $e$'s coin. Before the flip, $e$ was not a free edge, so the second invariant was satisfied even without $e$'s coin. Since the flip did not introduce any new separating triangles, this is still the case.

  \type{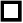} A non-flipped edge $e$ of $D$ that is not shared with any other separating triangle. By Lemma~\ref{lem:f4c-flip}, the flip removed $D$ and did not introduce any new separating triangles. Therefore $e$ cannot belong to any separating triangle, so the first invariant still holds if we remove $e$'s coin. By the same argument as for the previous type, $e$ is also not required to have a coin to satisfy the second invariant.

  \type{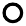} A free edge $e$ of a vertex of $D$ that is not shared with any containing separating triangle. Since $e$ did not belong to any separating triangle and the flip did not introduce any new ones, $e$ is not required to have a coin to satisfy the first invariant. Further, since the flip removed $D$ and $D$ was deepest, $e$ is not incident to a vertex of another separating triangle that contains it. Therefore it is no longer required to have a coin to satisfy the second invariant.

 \begin{figure}[htb]
  \centering
  \includegraphics{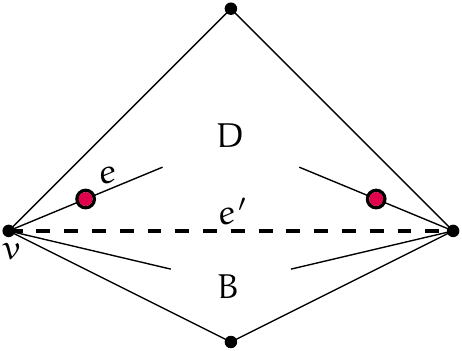}
  \caption{Two type 4 edges.}
  \label{fig:f4c-superfluous}
 \end{figure}

  \type{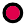} A free edge $e$ incident to a vertex $v$ of $D$, where $v$ is an endpoint of an edge $e'$ of $D$ that is shared with a non-containing separating triangle $B$, provided that we flip $e'$ (illustrated in Figure~\ref{fig:f4c-superfluous}). Any separating triangle that contains $D$ but not $B$ must share $e'$ (Lemma~\ref{lem:f4c-containingshared}) and is therefore removed by the flip.\\
So every separating triangle after the flip that contains $D$ also contains $B$. In particular, this also holds for containing triangles that share $v$. Since the second invariant requires only one free edge with a coin for each vertex of a separating triangle, we can safely charge the one inside $D$, as long as we do not charge the free edge in $B$.
 
 \medskip
 To decide which edges we charge for each flip, we distinguish five cases, based on the number of edges $D$ shares with other separating triangles and whether any of these triangles contain $D$. These cases are illustrated in Figures~\ref{fig:f4c-caseAC}, \ref{fig:f4c-caseB-ab}, and~\ref{fig:f4c-caseB-c}.

\begin{figure}[htb]
 \centering
 \begin{subfigure}[b]{0.48\textwidth}
  \centering
  \includegraphics{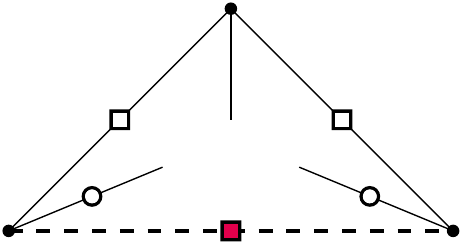}
  \caption{}
  \label{fig:f4c-caseA}
 \end{subfigure}
 \begin{subfigure}[b]{0.48\textwidth}
  \centering
  \includegraphics{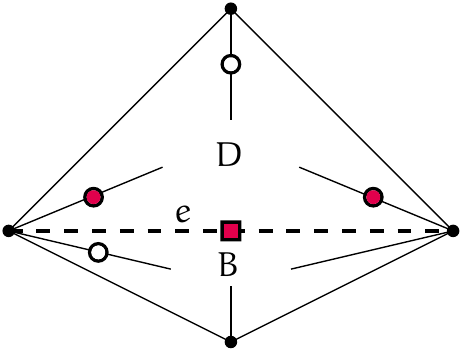}
  \caption{}
  \label{fig:f4c-caseC}
 \end{subfigure}
 \caption{The edges that are charged if (a) the deepest separating triangle does not share any edges with other separating triangles, or (b) the deepest separating triangle only shares edges with non-containing separating triangles. The flipped edge is dashed and the charged edges are marked with filled boxes (Type~1), empty boxes (Type~2), empty disks (Type~3) or filled disks (Type~4).}
 \label{fig:f4c-caseAC}
\end{figure}
 
 \case{1} $D$ does not share any edges with other separating triangles (Figure~\ref{fig:f4c-caseA}). In this case, we flip any of $D$'s edges. By the first invariant, each edge of $D$ has a coin. These edges all fall into Types 1 and 2, so we use their coins to pay for the flip. Further, $D$ can share at most one vertex with a containing triangle (Lemma~\ref{lem:f4c-onecontaining}), so we charge two free edges, each incident to one of the other two vertices (Type~3). 

  \case{2} $D$ does not share any edge with a containing triangle, but shares one or more edges with non-containing separating triangles (Figure~\ref{fig:f4c-caseC}). In this case, we flip one of the shared edges $e$. We charge $e$ (Type~1) and two free edges inside $D$ that are incident to the vertices of $e$ (Type~4). This leaves us with two more coins that we need to charge.

  Let $B$ be the non-containing separating triangle that shares $e$ with $D$. We first show that $B$ must have the same depth as $D$. There can be no separating triangles that contain $D$ but not $B$, as any such triangle would have to share $e$ (Lemma~\ref{lem:f4c-containingshared}) and $D$ does not share any edge with a containing triangle. Therefore any triangle that contains $D$ must contain $B$ as well. Since $D$ is contained in the maximal number of separating triangles, this holds for $B$ as well. This means that $B$ cannot contain any separating triangles and to satisfy the second invariant we only need to concern ourselves with triangles that contain both $B$ and $D$.

  Now consider the number of vertices of the quadrilateral formed by $B$ and $D$ that can be shared with containing triangles. Since $D$ does not share an edge with a containing triangle, it can share at most one vertex with a containing triangle (Lemma~\ref{lem:f4c-onecontaining}). Now suppose that $B$ shares an edge with a containing triangle. Then one of the vertices of this edge is part of $D$ as well. Since the other two vertices of the quadrilateral are both part of $D$, they cannot be shared with containing triangles. On the other hand, if $B$ does not share an edge with a containing triangle, it too can share at most one vertex with containing triangles. Thus, in both cases, at most two vertices of the quadrilateral can be shared with containing triangles, which means that there are at least two vertices that are not shared. For each of these vertices, if it is the vertex of $D$ that is not shared with $B$, we charge the free edge in $D$, otherwise we charge the free edge in $B$ (both Type~3).

\begin{figure}[htb]
 \centering
 \begin{subfigure}[b]{0.48\textwidth}
  \centering
  \includegraphics{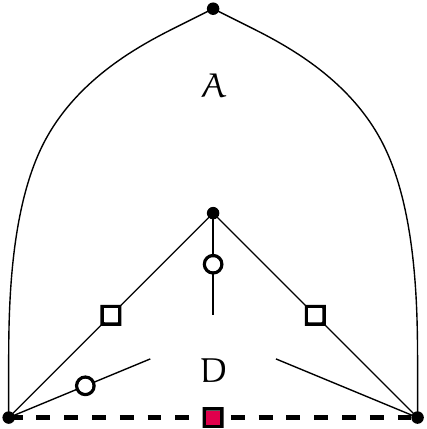}
  \caption{}
  \label{fig:f4c-caseB-a}
 \end{subfigure}
 \begin{subfigure}[b]{0.48\textwidth}
  \centering
  \includegraphics{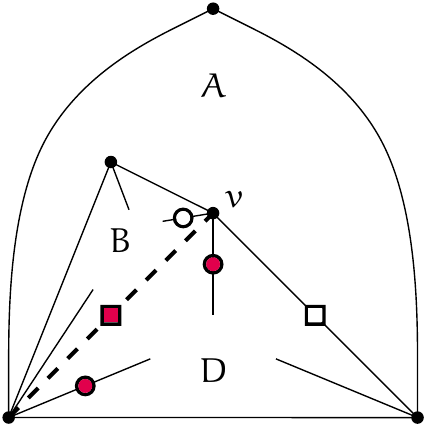}
  \caption{}
  \label{fig:f4c-caseB-b}
 \end{subfigure}
 \caption{The edges that are charged if the deepest separating triangle shares an edge with a containing triangle, and zero (a) or one (b) edges with non-containing separating triangles.}
 \label{fig:f4c-caseB-ab}
\end{figure}

  \case{3} $D$ shares an edge with a containing triangle $A$ and does not share the other edges with any separating triangle (Figure~\ref{fig:f4c-caseB-a}). In this case, we flip the shared edge and charge all of $D$'s edges, since one is the flipped edge (Type~1) and the others are not shared (Type~2). The vertex of $D$ that is not shared with $A$ cannot be shared with any containing triangle (Lemma~\ref{lem:f4c-unsharedvertex}), so we charge a free edge incident to this vertex (Type~3).

  Further, if $A$ shares an edge with a containing triangle, it either shares the flipped edge, which means that the containing triangle is removed by the flip, or it shares another edge, in which case the vertex that is not an endpoint of this edge cannot be shared with any containing triangle. If $A$ does not share an edge with a containing triangle, it can share at most one vertex with a containing triangle (Lemma~\ref{lem:f4c-onecontaining}). In both cases, one of the vertices of the flipped edge is not shared with any containing triangle (Type~3), so we charge a free edge incident to it.

  \case{4} $D$ shares an edge with a containing triangle $A$ and exactly one other edge with a non-containing separating triangle $B$ (Figure~\ref{fig:f4c-caseB-b}). In this case, we flip the edge that is shared with $B$. Let $v$ be the vertex of $D$ that is not shared with $A$. We charge the flipped edge (Type~1), the unshared edge of $D$ (Type~2) and two free edges inside $D$ that are incident to the vertices of the flipped edge (Type~4). We charge the last coin from a free edge in $B$ that is incident to $v$. We can charge it, since $v$ cannot be shared with a triangle that contains $D$ (Lemma~\ref{lem:f4c-unsharedvertex}) and every separating triangle that contains $B$ but not $D$ must share the flipped edge as well (Lemma~\ref{lem:f4c-containingshared}) and is therefore removed by the flip.

  All that is left is to argue that there can be no separating triangle contained in $B$ that requires the coin on this free edge to satisfy the second invariant. Every separating triangle that contains $D$ but not $B$ must share the flipped edge (Lemma~\ref{lem:f4c-containingshared}). Since $D$ already shares another edge with a containing triangle and it cannot share two edges with containing triangles (Lemma~\ref{lem:f4c-onecontainingtriangle}), all separating triangles that contain $D$ must also contain $B$. Since $D$ is deepest, $B$ must be deepest as well and therefore cannot contain any separating triangles.

 \begin{figure}[htb]
  \centering
  \includegraphics{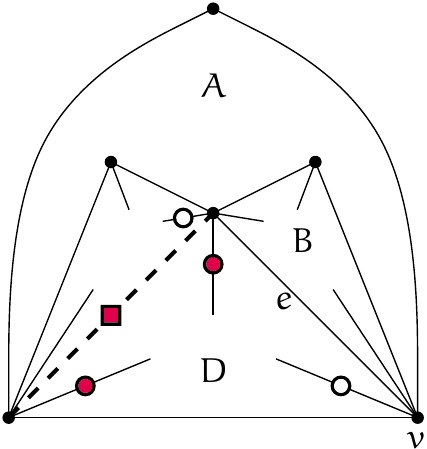}
  \caption{The edges that are charged if the deepest separating triangle shares an edge with a containing triangle and both other edges with non-containing separating triangles.}
  \label{fig:f4c-caseB-c}
 \end{figure}

  \case{5} $D$ shares one edge with a containing triangle $A$ and the other two with non-containing separating triangles (Figure~\ref{fig:f4c-caseB-c}). In this case we also flip the edge shared with one of the non-containing triangles. The charged edges are identical to the previous case, except that there is no unshared edge any more. Instead, we charge the last free edge in $D$.

  Before we argue why we are allowed to charge it, we need to give some names. Let $e$ be the edge of $D$ that is not shared with $A$ and is not flipped. Let $B$ be a non-containing triangle that shares $e$ with $D$ and let $v$ be the vertex that is shared by $A$, $B$ and $D$. Now, any separating triangle that shares $v$ and contains $D$ must contain $B$ as well. If it did not, it would have to share $e$ with $D$, but $D$ already shares an edge with a containing triangle and cannot share more than one (Lemma~\ref{lem:f4c-onecontainingtriangle}). Since the second invariant requires only a single free edge with a coin for each vertex of a separating triangle, it is enough that $v$ still has an incident free edge with a coin in $B$.
 
 \medskip
 This shows that we can charge 5 coins for every flip while maintaining the invariants, but we still need to show that after performing these flips we have indeed removed all separating triangles. So suppose that our graph contains separating triangles. Since each separating triangle is contained in a certain number of other separating triangles (which can be zero), there is at least one deepest separating triangle $D$. Since $D$ shares at most one edge with containing separating triangles (Lemma~\ref{lem:f4c-onecontainingtriangle}), one of the cases above must apply. This gives us an edge of $D$ to flip and five edges to charge, each of which is guaranteed by the invariants to have a coin. Therefore the process stops only after all separating triangles have been removed.
 
 Finally, since we pay 5 coins per flip and there are $3n - 6$ edges, by initially placing a coin on each edge, we flip at most $\lfloor(3n - 6)/5\rfloor$ edges.
 Now consider the edges of the outer face. We show that these still have a coin at the end of the algorithm. By definition, these edges are not inside any separating triangle and since we only charge free edges inside separating triangles, they can only ever be charged as Type~1 or 2. Thus, if an edge of the outer face gets charged, it was part of the deepest separating triangle $D$ that was removed by the flip. Since an edge of the outer face cannot be shared with a non-containing separating triangle and it cannot be contained by any separating triangle, it can only be charged in Case~1 or 3. In Case~1, we charge only two of the free edges inside $D$, since there could be a containing separating triangle that shares just a vertex. However, this is not possible if $D$ uses an edge of the outer face (Lemma~\ref{lem:f4c-outertriangle}), so we can charge this free edge instead of the edge of the outer face. Since we flip one of the edges that is not on the outer face, after the flip, all edges of the outer face still have their coins. In Case~3 we can charge this remaining free edge for the same reasons. However, since in this case we actually flip the edge of the outer face, we are not done yet. The outer face after the flip consists of the flipped edge, one edge of the current outer face and a current interior edge. Charging the extra free edge guarantees that the flipped edge can retain its coin, but we need to ensure that the current interior edge has a coin as well. Let $A$ be the deepest of the separating triangles that contain $D$. Since it, too, uses an edge of the outer face, $A$ can only be contained in triangles that share this edge (Lemma~\ref{lem:f4c-outertriangle}). It also cannot contain any separating triangles other than $D$, as these would be deepest as well and we prefer to remove separating triangles that do not use an edge of the outer face. Therefore there can be no other separating triangle that uses the free edge incident to the vertex of $A$ that is not on the outer face and we can move this coin to the new edge of the outer face. Since this is the only case in which an edge of the outer face is flipped, this shows that the edges of the outer face retain their coins during the entire process. Therefore we actually only need $3n - 9$ coins, resulting in a maximum of $\lfloor(3n - 9)/5\rfloor$ flips.
\end{proof}

 \paragraph{Transforming Hamiltonian triangulations} Now that we improved the bound on the number of flips needed during the first step of the algorithm by Mori~\etal~\cite{mori2003diagonal}, we can turn our attention to the second step. This step consists of transforming the obtained 4-connected triangulation into the canonical form. Mori~\etal showed that this can be done using at most $2n - 11$ flips. We improve this slightly to $2n - 15$ flips, matching the lower bound by Komuro~\cite{komuro1997diagonal} (see~Theorem~\ref{thm:fh-lb}). We first need to prove a few more lemmas.
 
\begin{figure}[htb]
 \centering
 \includegraphics{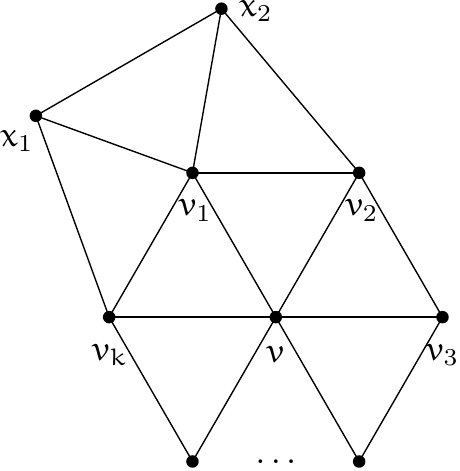}
 \caption{The neighbourhood of a vertex with degree at least 6 in a 4-connected triangulation.}
 \label{fig:f4c-degree}
\end{figure}

\begin{lemma}
 \label{lem:f4c-degree}
 In a 4-connected triangulation on $n \geq 13$ vertices, every vertex of degree at least 6 either has a neighbour of degree at least 6, or it can be connected to a vertex of degree at least 5 by a single flip.
\end{lemma}
\begin{proof}
 Let $v$ be a vertex of degree at least 6. Komuro~\cite{komuro1997diagonal} showed that either the graph consists of a cycle of length $n - 2$ with $v$ and one other vertex connected to every vertex on the cycle, or $v$ has a neighbour with degree at least 5. In the first case, there is a vertex of high degree that can be connected to $v$ by a single flip, so assume that this is not the case. Let $v_1$ be a neighbour of $v$ with degree at least 5 and let $v_2, \dots, v_k$ be the other neighbours of $v$, in clockwise order from $v_1$. Suppose that none of these neighbours have degree at least 6. Since the graph is 4-connected, this means that each has degree 4 or 5 and $v_1$ has degree exactly 5. Furthermore, no edge can connect two non-consecutive neighbours of $v$, as this would create a separating triangle. Let $x_1$ and $x_2$ be the neighbours of $v_1$ that are not adjacent to $v$, in clockwise order (see Figure~\ref{fig:f4c-degree}). We distinguish two cases, based on the degree of $v_2$:
 
 If $v_2$ has degree 4, $x_2$ must be connected to $v_3$. Both $x_1$ and $x_2$ can be connected to $v$ with a single flip, so if either has degree at least 5, we are done. The only way to keep their degree at 4 is to connect both $x_1$ and $v_k$ to $v_3$. But this would give $v_3$ degree at least 6, which is a contradiction. Therefore either $x_1$ or $x_2$ must have degree at least 5.
 
 If $v_2$ has degree 5, let $x_3$ be its new neighbour. Again, if one of $x_1$, $x_2$ or $x_3$ has degree at least 5, we are done. Since $x_2$ already has degree 4, the only way to keep its degree below 5 is to connect $x_1$ and $x_3$ by an edge. But then both $x_1$ and $x_3$ have degree 4 and the only way to keep one at degree 4 is to create an edge to the other. Therefore at least one of $x_1$, $x_2$ or $x_3$ must have degree at least 5.
\end{proof}

In 1931, Whitney \cite{whitney1931theorem} showed that any 4-connected triangulation has a Hamiltonian cycle. The main ingredient of his proof is the following lemma:

\begin{lemma}[Whitney \cite{whitney1931theorem}]
 \label{lem:f4c-cyclepath}
 Consider a cycle $C$ in a 4-connected triangulation, along with two distinct vertices $a$ and $b$ on $C$. These vertices split $C$ into two paths $C_1$ and $C_2$ with $a$ and $b$ as endpoints. Consider all edges on one side of the cycle, say the inside. If no vertex on $C_1$ (resp. $C_2$) is connected to another vertex on $C_1$ (resp. $C_2$) by an edge inside $C$, we can find a path from $a$ to $b$ that passes through each vertex on and inside $C$ exactly once and uses only edges of $C$ and inside $C$.
\end{lemma}

\begin{figure}[htb]
 \centering
 \includegraphics{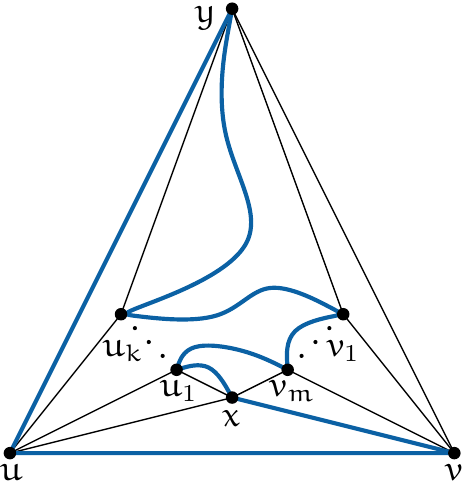}
 \caption{A possible Hamiltonian cycle that uses $(u, v)$ and has all non-cycle edges incident to $u$ on one side of the cycle and all non-cycle edges incident to $v$ on the other.}
 \label{fig:f4c-hamiltoniancycle}
\end{figure}

We use this to prove the following lemma:

\begin{lemma}
 \label{lem:f4c-goodcycle}
 For every edge $(u, v)$ in a 4-connected triangulation, there is a Hamiltonian cycle that uses $(u, v)$ such that all non-cycle edges incident to $u$ are on one side of the cycle and all non-cycle edges incident to $v$ are on the other side.
\end{lemma}
\begin{proof}
 Let $x$ and $y$ be the other vertices of the faces that have $(u, v)$ as an edge. Let $v, x, u_1, \dots , u_k, y$ be the neighbours of $u$ in counter-clockwise order and let $y, v_1, \dots , v_m, x, u$ be the neighbours of $v$ (see Figure~\ref{fig:f4c-hamiltoniancycle}). Note that all the $u_i$ and $v_i$ are distinct vertices, as a vertex other than $x$ or $y$ that is adjacent to both $u$ and $v$ would form a separating triangle. This means that $x, u_1, \dots , u_k, y, v_1, \dots , v_m, x$ forms a cycle. Moreover, no two non-consecutive neighbours of $u$ can be connected by an edge, since this would create a separating triangle as well. Since this holds for the neighbours of $v$ as well, $x$ and $y$ split the cycle into two parts that satisfy the conditions of Lemma~\ref{lem:f4c-cyclepath}. If we call the side of the cycle that does not contain $(u, v)$ the inside, this means that we can find a path from $x$ to $y$ that passes through each vertex on and inside the cycle exactly once and uses only edges of and inside the cycle. This path can be completed to a Hamiltonian cycle that satisfies the conditions by adding the edges $(y, u)$, $(u, v)$ and $(v, x)$.
\end{proof}

\begin{theorem}
\label{thm:f4c-canonical}
Any 4-connected triangulation $T$ on $n \geq 13$ vertices can be transformed into the canonical triangulation using at most $2n - \Delta(T) - 8$ flips, where $\Delta(T)$ is the maximum degree among vertices of $T$.
\end{theorem}
\begin{proof}
We use the same approach as used by Mori~\etal~\cite{mori2003diagonal} in the proof of Theorem~\ref{thm:fh-mori_hamiltonian}, but instead of taking an arbitrary Hamiltonian cycle, we use the preceding lemmas to carefully construct a good cycle.

Let $x$ be a vertex of maximal degree in $T$ and suppose for now that $x$ has a neighbour $y$ with degree at least 6. We use the cycle given by Lemma~\ref{lem:f4c-goodcycle} to decompose $T$ into two outerplanar graphs $T_1$ and $T_2$, each sharing the cycle and having all edges on the inside and outside, respectively. Note that $x$ is an ear in one of these, say $T_2$, while $y$ is an ear in the other. Mori~\etal showed that we can make any vertex $v$ of an outerplanar graph dominant using at most $n - d_v - 1$ flips, where $d_v$ is the degree of $v$. Therefore we can make $x$ dominant in $T_1$ using at most $n - \Delta(T) - 1$ flips. These flips are allowed because $x$ does not have any incident edges in $T_2$. Then we can make $y$ dominant in $T_2$ using at most $n - d_y - 1 \leq n - 7$ flips. Thus we can transform $T$ into the canonical triangulation using at most $2n - \Delta(T) - 8$ flips.

Since any triangulation on $n \geq 13$ vertices has a vertex of degree at least 6, if $x$ does not have a neighbour with degree at least 6, Lemma~\ref{lem:f4c-degree} tells us that there is a vertex $v$ with degree at least 5 that can be connected to $x$ by a single flip. We perform this flip and use $v$ in the place of $y$. Since $x$ now has degree $\Delta(T) + 1$, we can make it dominant using at most $n - \Delta(T) - 2$ flips. Similarly, $v$ has degree at least 6 after the flip, so we can make it dominant using at most $n - 7$ flips. Including the initial flip, we again obtain the canonical triangulation using at most $2n - \Delta(T) - 8$ flips.
\end{proof}

Combining this result with Theorem~\ref{thm:f4c-4-connected} gives the following bound on the maximum flip distance between two triangulations.

\begin{corollary}
 Any two triangulations $T_1$ and $T_2$ can be transformed into each other using at most $5.2 n - 19.6 - \Delta(T_1) - \Delta(T_2)$ flips, where $\Delta(T)$ is the maximum degree among vertices of $T$.
\end{corollary}

Theorem~\ref{thm:f4c-canonical} matches the worst-case lower bound of $2n - 15$ flips if the maximum degree is at least 7, but we need a stronger result if the maximum degree is 6.

\begin{lemma}
 \label{lem:f4c-6-6flip}
 In a 4-connected triangulation on $n \geq 19$ vertices with maximum degree 6, there is always a pair of vertices of degree 6 that can be connected by a flip.
\end{lemma}
\begin{proof}
 Suppose that such a pair does not exist and consider the neighbourhood of a vertex $v$ of degree 6. Each edge incident to $v$ can be flipped, otherwise there would be an edge connecting two non-consecutive neighbours of $v$, forming a separating triangle. Thus there are 6 pairs of vertices that can be connected by a flip and one vertex of each pair needs to have degree at most 5. To realize this, $v$ needs to have at least 4 neighbours of degree at most 5. Similarly, a vertex of degree 5 needs at least 3 such neighbours and a vertex of degree 4 needs at least 2. Therefore each vertex of degree at most 5 can have at most 2 neighbours of degree 6.

 Let $n_d$ be the number of vertices of degree $d$ and let $k$ be the number of edges between vertices of degree 6 and vertices of degree at most 5. Every vertex of degree 6 needs at least 4 neighbours of degree at most 5, so $k \geq 4 n_6$. But every vertex of degree at most 5 can have at most 2 neighbours of degree 6, so $k \leq 2 (n_4 + n_5)$. Combining these inequalities, we get that $n_6 \leq (n_4 + n_5)/2$. Since a triangulation with maximum degree 6 can have at most 12 vertices of degree less than 6, it follows that $n = n_4 + n_5 + n_6 \leq 18$. Thus for $n \geq 19$, there is always a pair of vertices of degree 6 that can be connected by a flip.
\end{proof}

\begin{theorem}
\label{thm:f4c-canonicalmaxdeg6}
Any 4-connected triangulation on $n \geq 19$ vertices with maximum degree 6 can be transformed into the canonical triangulation using at most $2n - 15$ flips.
\end{theorem}
\begin{proof}
By Lemma~\ref{lem:f4c-6-6flip}, there is always a pair of vertices $x$ and $y$ of degree 6 that can be connected by a flip. We first perform the flip that connects \mbox{$x$ and $y$}, giving both vertices degree 7. We then proceed similarly to the proof of Theorem~\ref{thm:f4c-canonical}. We make $x$ dominant in one of the outerplanar graphs using $n - 8$ flips and we make $y$ dominant in the other, also using $n - 8$ flips. Counting the initial flip, we obtain the canonical triangulation using at most $2n - 15$ flips.
\end{proof}

By combining this with Theorem~\ref{thm:f4c-canonical}, we get the following bound.

\begin{corollary}
 \label{cor:f4c-4-contocanon}
 Any 4-connected triangulation $T$ on $n \geq 19$ vertices can be transformed into the canonical triangulation using at most $\min\{2n - 15, 2n - \Delta(T) - 8\}$ flips, where $\Delta(T)$ is the maximum degree among vertices of $T$.
\end{corollary}
\begin{proof}
  This follows from Theorems~\ref{thm:f4c-canonical} and \ref{thm:f4c-canonicalmaxdeg6}, along with the observation that $2n - 15 < 2n - \Delta(T) - 8$ if $\Delta(T)$ is 6 and $2n - \Delta(T) - 8 \leq 2n - 15$ if $\Delta(T) \geq 7$.
\end{proof}

And finally, using our bound from Theorem~\ref{thm:f4c-4-connected} on the number of flips it takes to make triangulations 4-connected, we obtain an improved bound on the diameter of the flip graph.

\begin{corollary}
 The diameter of the flip graph of all triangulations on $n \geq 19$ vertices is at most $5.2 n - 33.6$.
\end{corollary}
\begin{proof}
  By Theorem~\ref{thm:f4c-4-connected}, any triangulation can be made 4-connected using at most $\lfloor (3n - 9)/5 \rfloor$ flips. By Corollary~\ref{cor:f4c-4-contocanon}, we can transform the resulting graph into the canonical triangulation using at most $2n - 15$ flips. Hence, we can transform any triangulation into any other using at most $2 \cdot \big( (3n - 9)/5 + 2n - 15 \big) = 5.2 n - 33.6$ flips.
\end{proof}

\section{Lower bound}
\label{sec:f4c-lb}

In this section we present a lower bound on the number of flips required to remove all separating triangles from a triangulation. Specifically, we present a triangulation that has $(3n - 10)/5$ edge-disjoint separating triangles. This matches our upper bound on the number of edge-disjoint separating triangles in a triangulation and shows that there indeed exist triangulations that require this many flips to make them 4-connected.

\begin{figure}[htb]
 \centering
 \includegraphics{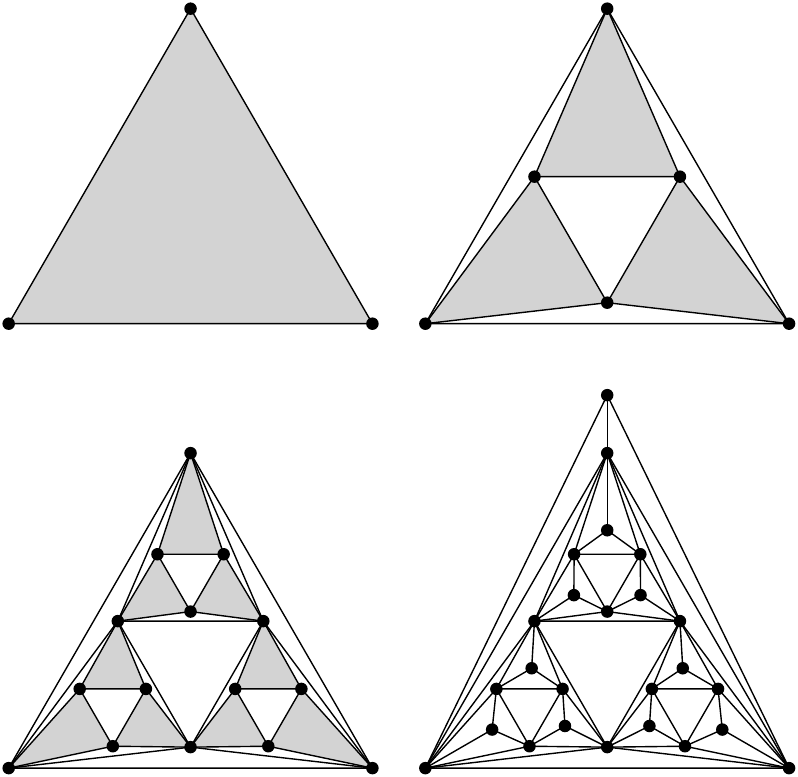}
 \caption{The stepwise construction of $\mathcal{T}_2$. The triangles used in the next recursive step are shaded.}
 \label{fig:f4c-sierpinsky}
\end{figure}

The triangulation that gives rise to the lower bound is constructed recursively and resembles the Sierpi\'nski triangle~\cite{sierpinski1915courbe}. The construction starts with an empty triangle. The recursive step consists of adding an inverted triangle in the interior and connecting each vertex of the new triangle to the two vertices of the opposing edge of the original triangle. This is recursively applied to the three new triangles that share an edge with the inserted triangle, but not to the inserted triangle itself (see~Figure~\ref{fig:f4c-sierpinsky}). After $k$ iterations, instead of applying the recursive step again, we add a single vertex in the interior of each triangle we are recursing on and connect this vertex to each vertex of the triangle. We also add a single vertex in the exterior face so that the original triangle becomes separating. The resulting triangulation is called $\mathcal{T}_k$.

\begin{theorem}
 There are triangulations on $n$ vertices that require $(3n - 10)/5$ flips to make them 4-connected, where $n$ is a multiple of 5.
\end{theorem}
\begin{proof}
 In the construction scheme presented above, each of the triangles we recurse on becomes a separating triangle that does not share any edges with the original triangle or the other triangles that we recurse on. Thus all these separating triangles are edge-disjoint. But how many of these triangles do we get? Let $L_i$ be the number of triangles that we recurse on after $i$ iterations of the construction, so $L_0 = 1$, $L_1 = 3$, etc. Now let $V_i$ be the number of vertices of $\mathcal{T}_i$. We can see that $V_1 = 10$ and if we transform $\mathcal{T}_1$ into $\mathcal{T}_2$, we have to remove each of the interior vertices added in the final step and replace them with a configuration of 6 vertices. So to get $\mathcal{T}_2$, we add 5 vertices in each of the $L_1$ triangles. This is true in general, giving
 \begin{equation}
  V_i~~=~~V_{i-1} + 5 L_{i-1}~~=~~10 + 5 \sum_{j=2}^{i} L_{j-1}. \label{eq:f4c-v}
 \end{equation}
 Let $S_i$ be the number of separating triangles of $\mathcal{T}_i$. We can see that $S_1 = 4$ and each recursive refinement of a separating triangle leaves it intact, while adding 3 new ones. Therefore
 \begin{equation}
  S_i~~=~~S_{i-1} + 3 L_{i-1}~~=~~4 + 3 \sum_{j=2}^{i} L_{j-1}. \label{eq:f4c-s}
 \end{equation}
 From Equation~\eqref{eq:f4c-v}, we get that 
 \[ 
  \sum_{j=2}^{i} L_{j-1}~~=~~\frac{V_i - 10}{5}.
 \]
 Substituting this into Equation~\eqref{eq:f4c-s} gives 
 \[
  S_i~~=~~4 + 3 \cdot \frac{V_i - 10}{5}~~=~~\frac{3 V_i - 10}{5}.
 \]
 Since each flip removes only the separating triangle that the edge belongs to, we need $(3n - 10)/5$ flips to make this triangulation 4-connected. Constructions for multiples of 5 between $V_i$ and $V_{i+1}$ can be obtained by recursing on a subset of the triangles in the final recursion step.
\end{proof}

\begin{figure}[htb]
 \centering
 \includegraphics{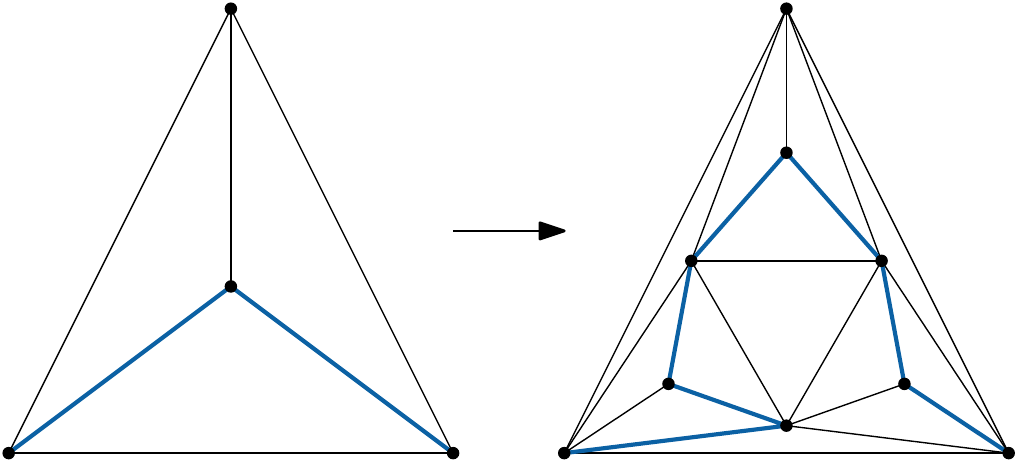}
 \caption{Updating the Hamilton cycle (bold) after a recursive step. The top vertex is visited at some other point on the initial cycle.}
 \label{fig:f4c-hamiltonian_update}
\end{figure}

Note that this triangulation achieves the upper bound on the number of edge-disjoint separating triangles from Corollary~\ref{cor:f4c-max-edge-disjoint}. It is natural to wonder whether this construction also leads to a better lower bound for the diameter of the flip graph in general. This is unfortunately not the case, as the resulting triangulation is Hamiltonian (see Figure~\ref{fig:f4c-hamiltonian_update}). Thus, even though it is not 4-connected, we know that it can be transformed into the canonical triangulation by at most $2n - 11$ flips from the proof by Mori~\etal~\cite{mori2003diagonal}.

\section{Conclusions and open problems}
\label{sec:f4c-conclusion}

We showed that any triangulation on $n$ vertices can be made 4-con\-nec\-ted using at most $\lfloor (3n - 9)/5 \rfloor$ flips, while there are triangulations that require $(3n - 10)/5 = \lfloor (3n - 9)/5 \rfloor$ flips when $n$ is a multiple of 5. This shows that our bound is tight for an infinite family of values for $n$, although a slight improvement to $\lfloor (3n - 10)/5 \rfloor$ is still possible. We believe that this is the true bound. We also showed that any 4-connected triangulation on $n \geq 19$ vertices can be transformed into the canonical form using at most $2n - 15$ flips. This matches the lower bound by Komuro~\cite{komuro1997diagonal} in the worst case where the graph has maximum degree 6 and results in a new upper bound of $5.2 n - 33.6$ on the diameter of the flip graph. It also means that both steps of the algorithm, when considered individually, are now tight in the worst case. Therefore, any further improvement must either merge the two steps in some fashion or employ a different technique.

Since 4-connectivity is not a necessary condition for Hamiltonicity, one possible approach is to show that a triangulation can be made Hamiltonian with fewer than $\lfloor (3n - 9)/5 \rfloor$ flips. Indeed, Cardinal~\etal~\cite{cardinal2015arc} successfully used this approach to lower the bound further to $5n - 23$, by showing that $n / 2$ flips suffice to transform any triangulation into a Hamiltonian one. However, this still leaves a gap with the current best lower bound of $(n - 8)/3$ flips, due to Aichholzer~\etal~\cite{aichholzer2008triangulations}.

Furthermore, all of the current algorithms use the same, single, canonical form. Surprisingly, the best known \emph{lower bound} on the diameter of the flip graph (which was recently improved to $\frac{7n}{3} - 34$ by Frati~\cite{frati2015lower}) actually goes to the canonical form as well. This suggests that at least one of the two bounds still has significant room for improvement. So is there another canonical form that gives a better upper bound? Or can we get a better bound by using multiple canonical forms and picking the closest?

Another interesting problem is to minimize the number of flips to make a triangulation 4-connected. We showed that our technique is worst-case optimal, but there are cases where fewer flips would suffice. There is a natural formulation of the problem as an instance of 3-hitting set, where the subsets correspond to the edges of separating triangles and we need to pick a minimal set of edges such that we include at least one edge from every separating triangle. This gives a simple 3-approximation algorithm that picks an arbitrary separating triangle and flips all shared edges or an arbitrary edge if there are no shared edges. However, it is not clear whether the problem is even NP-hard, as not all instances of 3-hitting set can be encoded as separating triangles in a triangulation. Therefore it might be possible to compute the optimal sequence in polynomial time.

\section{Lemmas and proofs}
\label{sec:f4c-lemmas}

This section contains proofs for the technical lemmas used in the proof of Theorem~\ref{thm:f4c-4-connected}.

\begin{lemma}
 \label{lem:f4c-4connected}
 A triangulation is 4-connected if and only if it contains no separating triangles.
\end{lemma}
\begin{proof}
 The first direction is easy. If a triangulation has a separating triangle, by definition, removing three vertices is sufficient to disconnect the graph, which implies that it is not 4-connected. For the other direction, assume that we have a triangulation $T$ that is not 4-connected. Since any triangulation is 3-connected, $T$ must have a cut set of size three that separates the graph into components $T_1$ and $T_2$. Consider a vertex $v$ in this cut set. This vertex must have neighbours in both $T_1$ and $T_2$. If not, the other two vertices would form a cut set of size two, which cannot exist as $T$ is 3-connected.

 Now look at the clockwise order of the neighbours of $v$, excluding the other vertices in the cut set. At some point, $v$ has a neighbour $v_1$ in $T_1$, followed by a neighbour $v_2$ in $T_2$. If there were no other edges separating these, the edge $(v_1, v_2)$ would be part of our triangulation, contradicting the fact that $v$ is part of a cut set. Therefore these edges must be separated by an edge to a neighbour neither in $T_1$, nor $T_2$: a vertex of the cut set. The same argument holds for the transition from $T_2$ to $T_1$. Thus $v$ is connected to both other vertices in the cut set. And since our choice of $v$ was arbitrary, this same argument applies to them. Therefore this cut set must be a separating triangle.
\end{proof}

\begin{lemma}
 \label{lem:f4c-interior}
 If a separating triangle $A$ contains a separating triangle $B$, then there is a vertex of $B$ inside $A$ and no vertex of $B$ can lie outside $A$.
\end{lemma}
\begin{proof}
 Let $z$ be a vertex in the interior of $B$ and let $y$ be a vertex of $A$ that is not shared with $B$. Since the interior of $B$ is a subgraph of the interior of $A$ and $y$ is not inside $A$, $y$ must be outside $B$. Since every triangulation is 3-connected, there is a path from $z$ to $y$ that stays inside $A$. This path connects the interior of $B$ to the exterior, so there must be a vertex of $B$ on the path and hence inside $A$.
 
 Now suppose that there is another vertex of $B$ outside $A$. Since all vertices of a triangle are connected by an edge, there is an edge between this vertex and the vertex of $B$ inside $A$. This contradicts the fact that $A$ is a separating triangle, so no such vertex can exist.
\end{proof}

\begin{lemma}
 \label{lem:f4c-interiorvertex}
 If a vertex $x$ of a separating triangle $B$ is inside a separating triangle $A$, then $A$ contains $B$.
\end{lemma}
\begin{proof}
 Let $y$ be a vertex of $A$ that is not shared with $B$. There is a path from $y$ to the outer face that stays in the exterior of $A$. There can be no vertex of $B$ on this path, since this would create an edge between the interior and exterior of $A$. Therefore $y$ is outside $B$.
 
 Now suppose that $A$ does not contain $B$. Then there is a vertex $z$ inside $B$ that is not inside $A$. There must be a path from $z$ to $x$ that stays inside $B$. Since $x$ is inside $A$, there must be a vertex of $A$ on this path. But since $y$ is outside $B$, this would create an edge between the interior and exterior of $B$. Therefore $A$ must contain $B$.
\end{proof}

\begin{lemma}
 \label{lem:f4c-onecontainingtriangle}
 A separating triangle can share at most one edge with containing triangles.
\end{lemma}
\begin{proof}
 Suppose we have a separating triangle $D$ that shares two of its edges with separating triangles that contain it. First of all, these triangles cannot be the same, since then they would be forced to share the third edge as well, which means that they are $D$. Since a triangle does not contain itself, this is a contradiction. So call one of these triangles $A$ and call one of the triangles that shares the other edge $B$. Let $x$, $y$ and $z$ be the vertices of $D$, such that $x$ is shared with $A$ and $B$, $y$ is shared only with $A$ and $z$ is shared only with $B$.

 By Lemma~\ref{lem:f4c-interior}, $z$ must be inside $A$, while $y$ must be inside $B$, since in both cases the other two vertices of $D$ are shared and therefore not in the interior. But then by Lemma~\ref{lem:f4c-interiorvertex}, $A$ contains $B$ and $B$ contains $A$. This is a contradiction, since by transitivity it would imply that the interior of $A$ is a subgraph of itself with a strictly smaller vertex set.
\end{proof}

\begin{lemma}
 \label{lem:f4c-onecontaining}
 A separating triangle $D$ that shares no edge with containing triangles can share at most one vertex with containing triangles.
\end{lemma}
\begin{proof}
 Suppose that $D$ shares two of its vertices with containing triangles. First, both vertices cannot be shared with the same containing triangle, since then the edge between these two vertices would also be shared. Now let $A$ be one of the containing triangles and let $B$ be one of the containing triangles sharing the other vertex. By Lemma~\ref{lem:f4c-interior}, there must be a vertex of $D$ inside $A$. So then both vertices of $D$ that are not shared with $A$ must be inside $A$, otherwise there would be an edge between the interior and the exterior of $A$. In particular, the vertex shared by $B$ and $D$ lies inside $A$, which by Lemma~\ref{lem:f4c-interiorvertex} means that $A$ contains $B$. But the reverse is also true, so $B$ contains $A$ as well, which is a contradiction.
\end{proof}

\begin{lemma}
 \label{lem:f4c-unsharedvertex}
 A separating triangle that shares an edge with a containing triangle cannot share the unshared vertex with another containing triangle.
\end{lemma}
\begin{proof}
 Suppose we have a separating triangle $D = (x, y, z)$ that shares an edge $(x, y)$ with a containing triangle $A$ and the other vertex $z$ with another containing triangle $B$. By Lemma~\ref{lem:f4c-interior}, at least one of $x$ and $y$ has to be inside $B$. Since these are vertices of $A$, by Lemma~\ref{lem:f4c-interiorvertex}, $B$ contains $A$. Similarly, $z$ has to be inside $A$ and since it is a vertex of $B$, $A$ contains $B$. This is a contradiction.
\end{proof}

\begin{lemma}
 \label{lem:f4c-containingshared}
 Given two separating triangles $A$ and $B$ that share an edge $e$, any separating triangle that contains $A$ but not $B$ must use $e$.
\end{lemma}
\begin{proof}
 Suppose that we have a separating triangle $D$ that contains $A$, but not $B$ and that does not use one of the vertices $v$ of $e$. By Lemma~\ref{lem:f4c-interior}, $v$ must be inside $D$. But then $D$ would also contain $B$ by Lemma~\ref{lem:f4c-interiorvertex}, as $v$ is a vertex of $B$ as well. Therefore $D$ must share both vertices of $e$ and hence $e$ itself.
\end{proof}

\begin{lemma}
 \label{lem:f4c-outertriangle}
 A separating triangle $D$ that uses an edge $e$ of the outer face cannot be contained in a separating triangle that does not share $e$.
\end{lemma}
\begin{proof}
 Suppose $D$ is contained in a separating triangle $A$. If $A$ does not share $e$, by Lemma~\ref{lem:f4c-interior}, at least one of the vertices of $e$ must be inside $A$. But since $e$ is part of the outer face, this is a contradiction.
\end{proof}

\bibliographystyle{plain}
\bibliography{../thesis}
\chapter{edge-la\-belled flips}
\label{ch:el}

The number of edge flips required to transform one triangulation into another has been studied extensively for unlabelled and vertex-labelled triangulations. In this chapter, we study this question for \emph{edge-la\-belled triangulations}, in which every edge has a unique label that is carried over when the edge is flipped. Specifically, we prove that $O(n \log n)$ flips or $O(\log^2 n)$ simultaneous flips suffice to transform any combinatorial triangulation or triangulation of a convex $n$-gon into any other, and that $\Omega(n \log n)$ flips are sometimes required. For edge-la\-belled pseu\-do-tri\-an\-gu\-la\-tions, we also obtain a $\Theta(n \log n)$ bound, although the upper bound increases to $O(n^2)$ when we restrict ourselves to pointed pseu\-do-tri\-an\-gu\-la\-tions and exchanging flips.

\extraStretch{1em}{The results on pseu\-do-tri\-an\-gu\-la\-tions have been accepted to the 27th Canadian Conference on Computational Geometry (CCCG 2015)~\cite{bose2015flips}.
This chapter is based on joint work with Prosenjit Bose, Anna Lubiw, and Vinayak Pathak.}

\section{Introduction}

Flips have been studied in many different settings. While Wagner~\cite{wagner1936bemerkungen} originally studied them in the context of combinatorial triangulations (as surveyed in Chapter~\ref{ch:fh}), interest in flips increased after Sleator, Tarjan and Thurston~\cite{sleator1988rotation} showed a simple bijection between binary trees and triangulations of a convex polygon (subdivisions of the polygon into triangles using only diagonals), such that a flip in the triangulation corresponds to a rotation in the binary tree. This observation helped them in deriving a precise bound of $2n - 10$ flips on the diameter of the flip graph of an $n$-vertex convex polygon, and thereby on the maximal rotation distance between two binary trees on $n - 2$ nodes.

The same authors also studied flips in combinatorial triangulations with labelled vertices. Whereas in the unlabelled setting two triangulations are considered the same if they are isomorphic, here the isomorphism additionally needs to be consistent with the vertex labels. They proved an $O(n \log n)$ bound on the diameter of the flip graph in this setting~\cite{sleator1992short}. Additionally, they presented a general framework for bounding the number of graphs reachable from an initial graph using simple transformations -- including flips. Applying this framework to the vertex-labelled setting results in a matching $\Omega(n \log n)$ lower bound.

\begin{figure}[htb]
 \centering
 \includegraphics{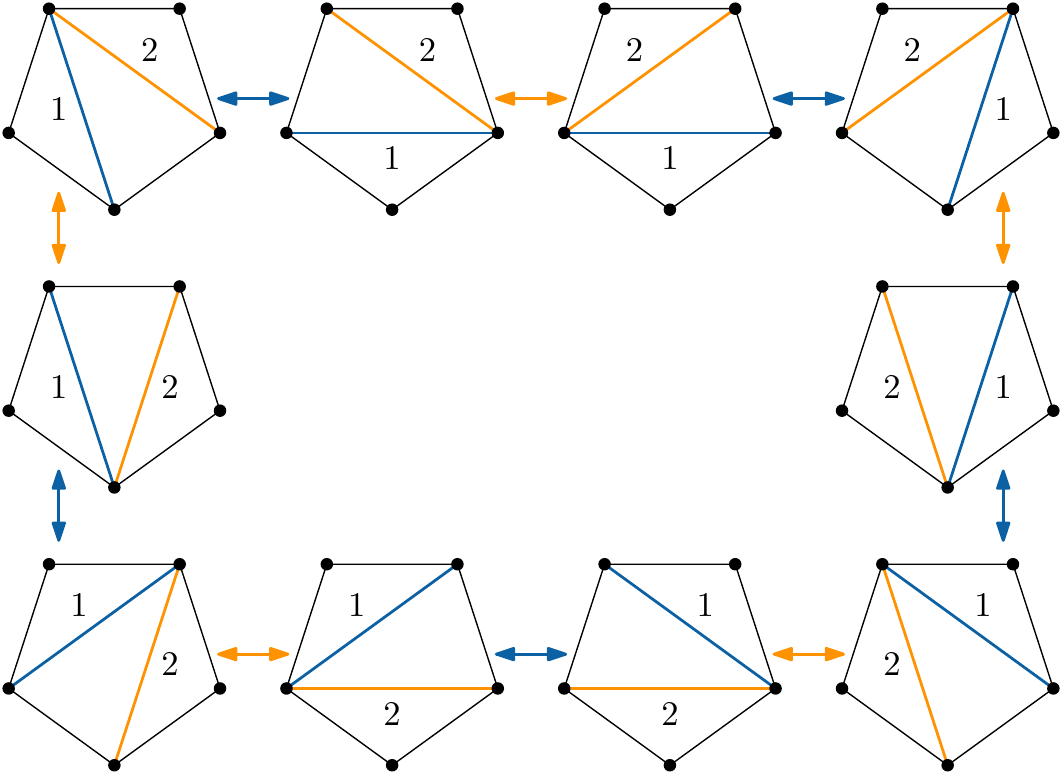}
 \caption{The flip graph of edge-la\-belled triangulations of a convex pentagon. The colour of the arrows corresponds to the colour of the flipped edge.}
 \label{fig:el-pentagon-labelled-flip-graph}
\end{figure}

In this chapter, we consider a natural extension of the work by Sleator~\etal: what happens when, instead of the vertices, the edges are labelled? A flip then reassigns the label of the flipped edge to the new edge, so that the set of labels does not change.

Edge-la\-belled flips in triangulations of a convex polygon have been studied independently by Araujo-Pardo~\etal~\cite{araujo2015colorful}. Their interest lies in the flip graph itself, which they call the \emph{colorful associahedron} (see Figure~\ref{fig:el-pentagon-labelled-flip-graph} for an example). Although they establish that it is connected, their proof only gives a quadratic bound on the diameter. They then proceed to prove various structural properties that relate it to the flip graph in the unlabelled setting.

Edge labels have also been considered in different settings. Hernando~\etal~\cite{hernando2003grafos} investigated edge-la\-belled spanning trees of graphs. They showed that one can transform between any two edge-la\-belled spanning trees of a 2-connected graph by iteratively removing an edge and replacing it elsewhere with the same label while maintaining connectivity. The setting considered by Cano~\etal~\cite{cano2013edge} is different still; they transform between non-maximal plane graphs by `rotating' edges around one of their endpoints. They prove that the corresponding edge rotation graph is connected, both in the labelled and unlabelled setting.

We start with the simplest setting, edge-la\-belled triangulations of an $n$-vertex convex polygon (Section~\ref{sec:el-convex}), and show that this flip graph has a diameter of $\Theta(n \log n)$. We reuse Sleator~\etal's framework for the lower bound, but the proof for the upper bound is new. We then use this result to prove that the same bounds hold for edge-la\-belled combinatorial triangulations (Section~\ref{sec:el-combinatorial}). As an aside, we consider what changes when we allow multiple edges to be flipped simultaneously, as long as they are not incident to the same triangle. In this setting, the upper bound reduces to $O(\log^2 n)$ both for convex polygons and combinatorial triangulations, but we no longer have a matching lower bound.

Finally, we consider edge-la\-belled pseu\-do-tri\-an\-gu\-la\-tions of point sets in the plane (Section~\ref{sec:el-pts}). A \emph{pseu\-do-tri\-an\-gu\-la\-tion} is a subdivision of the convex hull into \emph{pseudo-triangles}: simple polygons with three convex interior angles. We first restrict ourselves to edge-la\-belled \emph{pointed} pseu\-do-tri\-an\-gu\-la\-tions, which have the minimum number of edges. Here, we show that $O(n^2)$ \emph{exchanging flips} (flips that replace one edge with another) suffice to transform between any two edge-la\-belled pointed pseu\-do-tri\-an\-gu\-la\-tions. If we additionally allow flips that only insert or remove an edge, we can transform any edge-la\-belled pseu\-do-tri\-an\-gu\-la\-tion into any other with $O(n \log c + h \log h)$ flips, where $c$ is the number of convex layers of the point set and $h$ is the number of points on the convex hull.

\section{Convex polygons}
\label{sec:el-convex}

An \emph{edge-la\-belled triangulation} of an $n$-vertex convex polygon is a triangulation of the polygon where each diagonal has a unique label in $\{1, \dots, n - 3\}$. In this section, we prove a tight $\Theta(n \log n)$ bound on the diameter of the flip graph of edge-la\-belled triangulations of a convex $n$-gon. We also show that the upper bound decreases to $O(\log^2 n)$ if we allow multiple edges, no two of which are incident on the same triangle, to be flipped simultaneously.

\subsection{Upper bound}

\begin{figure}[htb]
 \centering
 \includegraphics{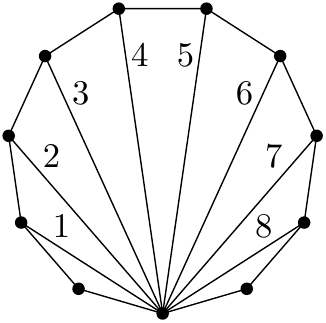}
 \caption{The canonical edge-la\-belled triangulation of a convex polygon with eleven vertices.}
 \label{fig:el-convex-canonical}
\end{figure}

For the upper bound on the diameter of the flip graph, we show how to transform any edge-la\-belled triangulation into a canonical one in $O(n \log n)$ flips. Given two edge-la\-belled triangulations $G_1$ and $G_2$, the result then follows by composing this sequence for $G_1$ with the inverse sequence for $G_2$. First, consider the triangulation where all edges are incident to the vertex with the lowest $y$-coordinate. We call this configuration a \emph{fan}. The canonical triangulation we use is a fan triangulation where the interior edges are labelled $1, \dots, n-3$ in clockwise order around the bottom vertex (see~Figure~\ref{fig:el-convex-canonical}).

\begin{figure}[htb]
 \centering
 \includegraphics{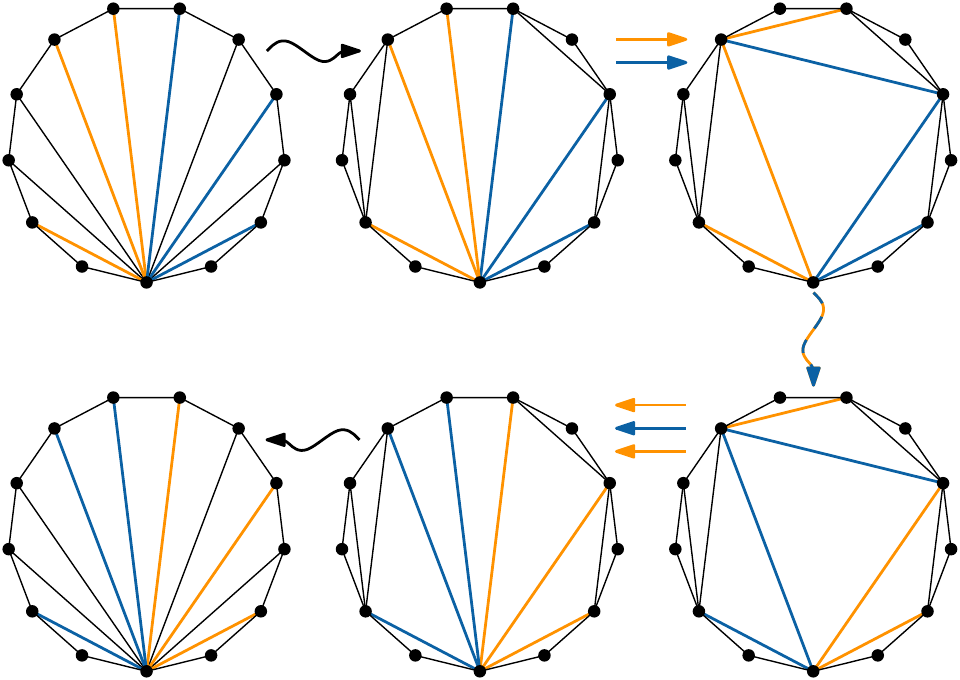}
 \caption{The flip sequence used in Lemma~\ref{lem:el-convex-partition} to swap the labels of the high (lighter) and low (darker) edges.}
 \label{fig:el-convex-partition-sequence}
\end{figure}

As we can transform any triangulation into a fan with $O(n)$ flips~\cite{culik1982note}, the problem essentially reduces to sorting the labels of a fan. In this light, it is not surprising that our solution mimics a well-known sorting algorithm -- quicksort -- with a slight modification: instead of choosing a pivot at random, we always use the median. This guarantees that even in the worst case, we only use $O(n \log n)$ flips. We first show that we can use $2.5 n$ flips to perform the `partition' step of quicksort (ensuring that the first half of the edges of the fan are labelled with the first half of the labels and vice-versa). The flip sequence for this step is illustrated in Figure~\ref{fig:el-convex-partition-sequence}.

\begin{lemma}
 \label{lem:el-convex-partition}
 Let the diagonals of a fan triangulation of an $n$-vertex convex polygon be partitioned into three groups: low, neutral, and high, such that $|\text{low}| = |\text{high}|$ and all high edges occur to the left of any low edge. Then we can exchange the labels on the high and low edges with $2.5 n$ flips, while leaving those on the neutral edges in place.
\end{lemma}
\begin{proof}
 We prove this by induction on $n$. In the base case ($n = 3$), there are no diagonals and we are done. So assume that $n > 3$ and that the lemma holds for any convex polygon with fewer than $n$ vertices.
 
 First, suppose that there is a neutral edge $e$. Flipping $e$ makes it an ear of the current triangulation. Now the remaining diagonals are part of a fan triangulation of a convex polygon with $n - 1$ vertices that has $e$ on the boundary. By induction, we can exchange the labels on the high and low edges in this polygon with $2.5 (n - 1)$ flips. Afterwards, we simply flip $e$ back into place, giving a sequence of $2.5 (n - 1) + 2 \leq 2.5 n$ flips that successfully completes the swap.
 
 Now suppose that there there are no neutral edges. Then the first half of the edges is high and the second half is low. Let $e_1$ and $e_2$ be the two edges in the middle, so that $e_1$ is high and $e_2$ is low. Then we can swap them with five flips, as shown in Figure~\ref{fig:el-pentagon-labelled-flip-graph}. But note that two flips into this sequence, the two edges are out of the way -- the remaining edges form a fan triangulation of a convex polygon with $n - 2$ vertices. As we removed one low and one high edge, the two groups still have the same size in the smaller polygon. Thus, we can swap the labels of all other high and low edges with $2.5 (n - 2)$ flips by induction. Finally, we complete the swap of $e_1$ and $e_2$ with three more flips. This exchanges the labels of all high and low edges with a total of $2.5 (n - 2) + 5 = 2.5 n$ flips and concludes the proof.
\end{proof}

With this in place, we can sort all labels with $O(n \log n)$ flips by recursing on each half.

\begin{lemma}
 \label{lem:el-convex-sort}
 Given an edge-la\-belled fan triangulation of a convex polygon with $n$ vertices, we can sort the labels in ascending order around the bottom vertex with $O(n \log n)$ flips.
\end{lemma}
\begin{proof}
 The proof is by induction on $n$. In the base case $n = 3$ or $n = 4$, so the diagonals are sorted by default. Therefore assume that $n > 4$ and the lemma holds for all convex polygons with fewer than $n$ vertices.

 We identify two groups of edges. Let $m = \lfloor \frac{n - 3}{2} \rfloor$ be the middle label. Then \emph{high} edges have a label in $\{m + 1, \ldots, n - 3\}$, but are among the $m$ leftmost diagonals in the current fan. Conversely, \emph{low} edges have a label in $\{1, \ldots, m\}$, but are not among the $m$ leftmost diagonals. To see that $|\text{high}| = |\text{low}|$, consider the $m$ leftmost diagonals. By definition, these contain $m - |\text{low}|$ edges with a label in $\{1, \ldots, m\}$. Thus, there must be $m - (m - |\text{low}|) = |\text{low}|$ edges among them with a label in $\{m + 1, \ldots, n - 3\}$.
 
 Therefore all conditions of Lemma~\ref{lem:el-convex-partition} are satisfied, and we can swap the labels of the low and high edges with $2.5 n$ flips. This ensures that the first $m$ diagonals contain all labels from $1$ through $m$, while the rightmost $m$ or $m + 1$ diagonals (depending on whether $n$ is even or odd) contain all labels from $m + 1$ through $n - 3$. By induction, we can sort these two halves recursively, thereby sorting all labels. The total number of flips satisfies the recursion $T(n) = T(\lfloor n/2 \rfloor) + T(\lceil n/2 \rceil) + 2.5 n$, which solves to $O(n \log n)$.
\end{proof}

Since $O(n)$ flips suffice to transform any edge-la\-belled triangulation into a fan (by simply ignoring the labels), the upper bound on the diameter of the flip graph follows.

\begin{theorem}
 \label{thm:el-convex-ub}
 Any edge-la\-belled triangulation of a convex polygon with $n$ vertices can be transformed into any other by $O(n \log n)$ flips.
\end{theorem}

\subsection{Lower bound}
\label{sec:el-convex-lb}

The lower bound uses a slightly modified version of the $\Omega(n \log n)$ lower bound for the vertex-labelled setting by Sleator, Tarjan, and Thurston~\cite{sleator1992short}. We first give an overview of their technique, before applying it to edge-la\-belled triangulations of a convex polygon.

\begin{figure}[htb]
 \centering
 \begin{subfigure}[b]{0.4\textwidth}
  \centering
  \includegraphics{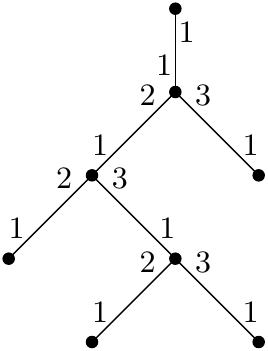}
  \caption{}
  \label{fig:el-half-edge-graph}
 \end{subfigure}
 \begin{subfigure}[b]{0.58\textwidth}
  \centering
  \includegraphics{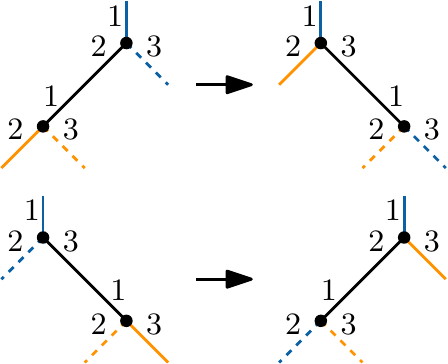}
  \caption{}
  \label{fig:el-rotation-grammar}
 \end{subfigure}
 \caption{(a) A half-edge graph representation of a rooted binary tree. (b) A graph grammar for rotations in binary trees. Correspondence between half-edges is indicated by a combination of colour and line style.}
\end{figure}

Let a \emph{tagged half-edge graph} be an undirected graph with maximum degree $\Delta$, whose vertices have labels called \emph{tags}, and whose edges are split into two \emph{half-edges}. Each half-edge is incident to one endpoint, and labelled with an \emph{edge-end label} in $\{1, \ldots, \Delta\}$, such that all edge-end labels incident on a vertex are distinct (see Figure~\ref{fig:el-half-edge-graph} for an example). A \emph{half-edge part} is a half-edge graph in which some half-edges do not have a twin. Note that tags are not restricted to integers: they could be tuples, or even arbitrary strings.

A \emph{graph grammar} $\Gamma$ is a sequence of production rules $\Gamma_i = (L, \rightarrow, \mathcal{T}, R)$, where $L$ and $R$ are half-edge parts with the same number of half-edges, $\rightarrow$ is a correspondence between the half-edges of $L$ and $R$, and $\mathcal{T}$ is a function that computes the tags of vertices in $R$ from those in $L$. A possible graph grammar for rotations in (unlabelled) binary trees is depicted in Figure~\ref{fig:el-rotation-grammar}.

Sleator, Tarjan, and Thurston prove the following theorem.

\begin{theorem}[Sleator, Tarjan, and Thurston~\cite{sleator1992short}]
 \label{thm:el-stt-reachable}
 Let $G$ be a tagged half-edge graph of $n$ vertices, $\Gamma$ be a graph grammar, $c$ be the number of vertices in left sides of $\Gamma$, and $r$ be the maximum number of vertices in any right side of a production of $\Gamma$. Then $|R(G, \Gamma, m)| \leq (c + 1)^{n + r \cdot m}$, where $R(G, \Gamma, m)$ is the set of graphs obtainable from $G$ by derivations in $\Gamma$ of length at most $m$.
\end{theorem}

We cannot apply this theorem directly to triangulations of a convex polygon, as these do not have bounded degree. Instead, we turn to the dual graph. The \emph{augmented dual graph} of a triangulation of a convex polygon is a tagged half-edge graph $G$ with two sets of vertices: triangle-vertices $T$ corresponding to the triangles of the triangulation, and edge-vertices $E_{\textsc{CH}}$ corresponding to the boundary edges. One edge-vertex is designated as the \emph{root}.

\begin{figure}[htb]
 \centering
 \includegraphics{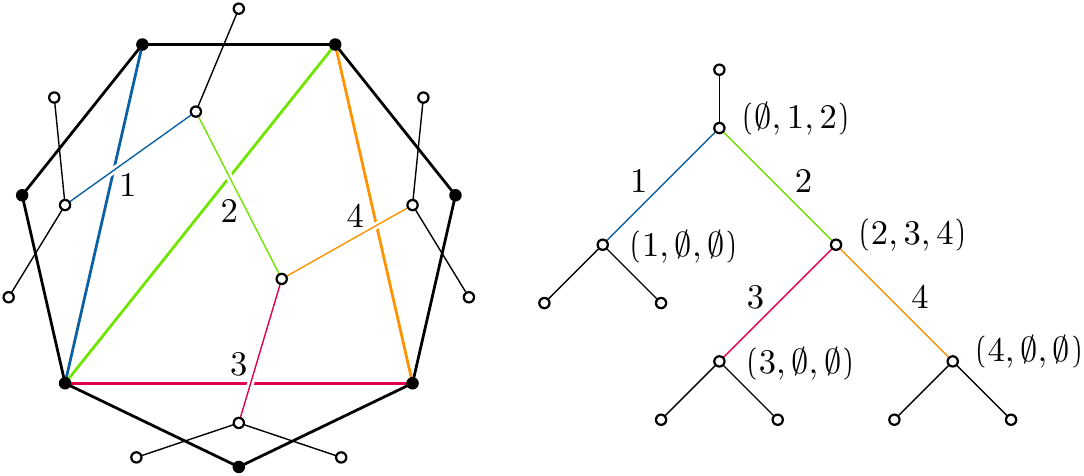}
 \caption{An edge-la\-belled triangulation of a convex polygon with its augmented dual graph. The edges-labels on the dual graph are shown to more clearly indicate the correspondence -- they are actually labelled with edge-end labels as in Figure~\ref{fig:el-half-edge-graph}.}
 \label{fig:el-triangulation-to-tree}
\end{figure}

Two triangle-vertices are connected by an edge if their triangles are adjacent. All edge-vertices are leaves, each connected to the triangle-vertex whose triangle is incident to their corresponding edge (see Figure~\ref{fig:el-triangulation-to-tree}). As every triangle has three edges, the maximum degree of $G$ is three. The edge towards the root receives edge-end label $1$. For a triangle-vertex, the other edge-end labels are assigned in counter-clockwise order, as in Figure~\ref{fig:el-half-edge-graph}.

This is where we deviate slightly from the original paper. Since Sleator, Tarjan, and Thurston were working in the vertex-labelled setting, they used the tags in the augmented dual graph to encode the labels of the vertices around the corresponding triangles. Instead, we use these tags to encode the edge-labels. Specifically, we tag each triangle-vertex with a triple containing the edge-label of each edge of its triangle, starting from the edge closest to the root, and proceeding in counter-clockwise order. Edges of the convex hull are assumed to have label $\emptyset$. Edge-vertices will not be involved in any of the production rules, so they do not need tags.

\begin{figure}[htb]
 \centering
 \includegraphics{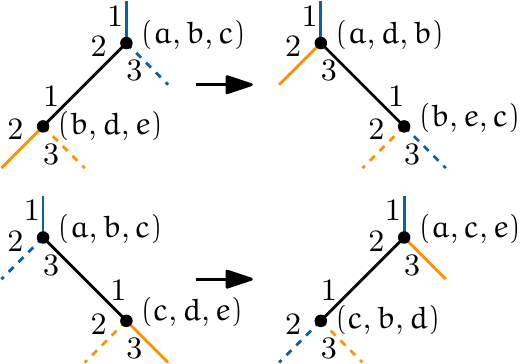}
 \caption{A graph grammar for rotations in augmented dual graphs, which correspond to flips in edge-la\-belled triangulations of a convex polygon.}
 \label{fig:el-labelled-rotation-grammar}
\end{figure}

As flips in the triangulation correspond to rotations in the augmented dual graph~\cite{sleator1988rotation}, the graph grammar is identical to the graph grammar presented before. The only addition is the computation of new tags for the vertices on the right-hand side (see Figure~\ref{fig:el-labelled-rotation-grammar}). This grammar has four vertices in left sides, and a maximum of two vertices in any right side. Since a triangulation of an $n$-vertex convex polygon has $n - 2$ triangles and $n$ convex hull edges, the augmented dual graph has $2n - 2$ vertices. Thus, Theorem~\ref{thm:el-stt-reachable} gives us the following.

\begin{lemma}
 \label{lem:el-convex-reachable}
 Given an edge-la\-belled triangulation $G$ of an $n$-vertex convex polygon, the number of distinct edge-la\-belled triangulations reachable from $G$ in $m$ flips is at most $5^{2n - 2 + 2m}$.
\end{lemma}

This bound can be further refined to $3^{n - 1 + 2m}$, using the leader-follower and zero-elimination techniques from Sleator, Tarjan, and Thurton's paper~\cite{sleator1992short}. However, the cruder bound already suffices to derive the correct asymptotic lower bound.

\begin{theorem}
 \label{thm:el-convex-lb}
 There are pairs of edge-la\-belled triangulations of a convex polygon with $n$ vertices such that transforming one into the other requires $\Omega(n \log n)$ flips.
\end{theorem}
\begin{proof}
 We first estimate the number of edge-la\-belled triangulations. An $n$-vertex convex polygon has $n - 3$ diagonals, and in a fan triangulation, each sequence of labellings results in a new triangulation. Thus, there are at least $(n - 3)!$ edge-la\-belled triangulations.

 Let $d$ be the diameter of the flip graph. Then, for every graph $G$, $d$ flips suffice to reach all edge-la\-belled triangulations. But from Lemma~\ref{lem:el-convex-reachable}, we know that a sequence of $m$ flips can generate at most $5^{2n - 2 + 2m}$ unique edge-la\-belled triangulations. This gives us the following bound.
 \begin{align*}
  5^{2n - 2 + 2d}~~&\geq~~(n - 3)! \\
  \log_5 5^{2n - 2 + 2d}~~&\geq~~\log_5 (n - 3)! \\
  2n - 2 + 2d~~&\geq~~\log_5 (n! / n^3) \\
  2d~~&\geq~~\log_5 n! - \log_5 n^3 - 2n + 2\\
  2d~~&\geq~~\Omega(n \log n) - O(n) \\
  d~~&\geq~~\Omega(n \log n) \qedhere
 \end{align*}
\end{proof}

Combining the upper bound from Theorem~\ref{thm:el-convex-ub} with the lower bound from Theorem~\ref{thm:el-convex-lb} gives us an asymptotically tight bound on the worst-case number of flips required to transform one edge-la\-belled triangulation of a convex polygon into another.

\begin{corollary}
 The flip graph of edge-la\-belled triangulations of a convex polygon with $n$ vertices has diameter $\Theta(n \log n)$.
\end{corollary}

\subsection{Simultaneous flips}

A \emph{simultaneous flip} is a transformation that consists of one or more regular flips that are executed at the same time. For this definition to make sense, it is important that two of these flips do not interfere with each other. Therefore we add the restriction that in a single simultaneous flip, at most one edge of each triangle can be flipped. In other words, no two edges in the same simultaneous flip can be incident to the same triangle. Galtier~\etal~\cite{galtier2003simultaneous} showed that $O(\log n)$ simultaneous flips suffice to transform any triangulation of an $n$-vertex convex polygon into any other.

The bounds on simultaneous flips are clearly connected to those on regular flips, as each simultaneous flip can group only $O(n)$ regular flips. This means that the $\Omega(n \log n)$ lower bound from Theorem~\ref{thm:el-convex-lb} also implies an $\Omega(\log n)$ lower bound for simultaneous flips. This does not apply to the upper bound, however. For example, it is not possible, in general, to simply perform each $c \cdot n$ flips simultaneously (for some constant $c$), as two such flips can easily be incident on the same triangle. The upper bounds do translate in the other direction: proving that $O(k)$ simultaneous flips suffice immediately implies that $O(kn)$ regular flips suffice, simply by performing each set of flips in sequence.

In this section, we show that any edge-la\-belled triangulation of a convex polygon with $n$ vertices can be transformed into any other with $O(\log^2 n)$ simultaneous flips. We show that even the partition step of quicksort already requires $\Omega(\log n)$ simultaneous flips. We start by proving an analogue to Lemma~\ref{lem:el-convex-partition}, showing that this bound on the partition step is tight.

\begin{figure}[htb]
 \centering
 \includegraphics{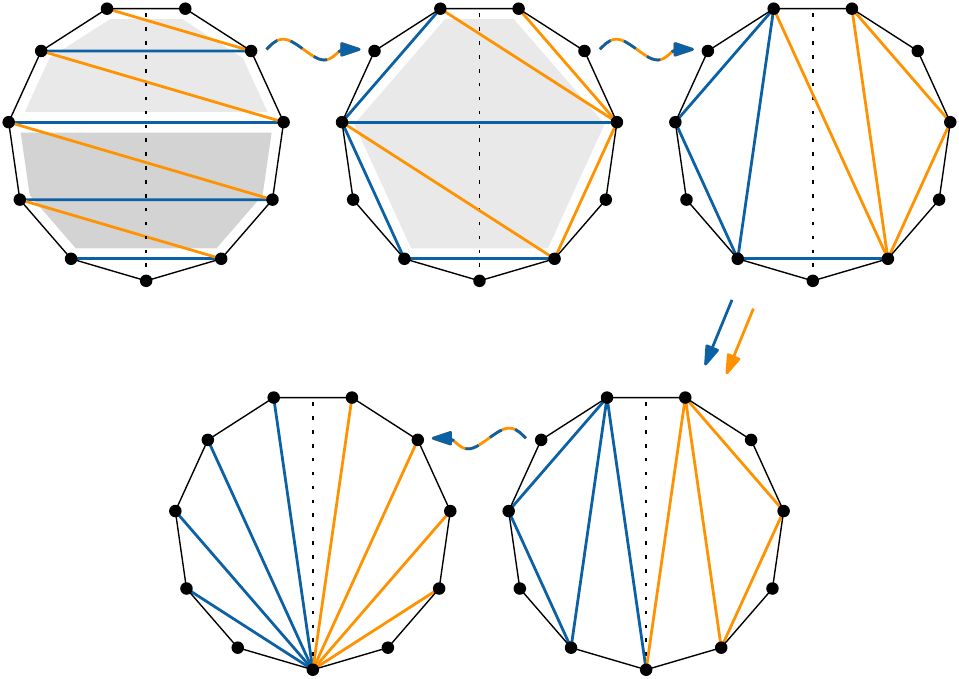}
 \caption{A sequence of simultaneous flips that transforms an alternating zig-zag into a fan partition. We can choose which group of edges ends up to the left of the ray. The hexagons involved in the first two steps are lightly shaded.}
 \label{fig:el-convex-simultaneous-partition-sequence}
\end{figure}

\begin{lemma}
 \label{lem:el-convex-partition-simultaneous}
 Let the diagonals of a fan triangulation of an $n$-vertex convex polygon be partitioned into three groups: low, neutral, and high, such that $|\text{low}| = |\text{high}|$ and all high edges occur to the left of any low edge. Then we can exchange the labels on the high and low edges with $O(\log n)$ simultaneous flips, while leaving those on the neutral edges in place.
\end{lemma}
\begin{proof}
 We first flip each neutral diagonal, so that they are no longer incident to the bottom vertex. Since we can flip every second edge in a single simultaneous flip, $O(\log n)$ simultaneous flips suffice to flip all neutral diagonals. This leaves only the high and low diagonals, inside a smaller convex polygon formed by the original polygon and the neutral edges.

 This reduces our problem to transforming a convex polygon where the first half of the diagonals is high and the second is low to one where these sets are reversed. We do this by transforming a third configuration, called an \emph{alternating zig-zag}, into both, with $O(\log n)$ simultaneous flips. The result then follows by the reversibility of flips.

 Let $r$ be a ray from the bottom vertex that has $\lfloor n/2 \rfloor$ of the remaining vertices to its left. Let $a_i$ be the vertex to the left of $r$ at distance $i$ from the bottom vertex (along the boundary of the polygon), and let $b_i$ be the analogous vertex to the right of $r$. The alternating zig-zag contains the edges $a_1b_1$, $b_1a_2$, $a_2b_2$, etc. Each edge $a_ib_i$ is low, and each edge $b_ia_{i+1}$ is high (see~Figure~\ref{fig:el-convex-simultaneous-partition-sequence}).

 To transform the alternating zig-zag into the fan triangulation with all low edges on the left, we first partition it into hexagons. Each hexagon is formed by the edges $a_ib_i$ and $a_{i+2}b_{i+2}$, along with the boundary edges between them. In this way, each hexagon contains three diagonals: two high diagonals $b_ia_{i+1}$ and $b_{i+1}a_{i+2}$, and one low diagonal $a_{i+1}b_{i+1}$. We now use a constant number of simultaneous flips to transform the triangulation within each hexagon to have the low diagonal at $a_ia_{i+2}$, and the high diagonals at $b_ib_{i+2}$ and $b_ia_{i+2}$. Since the hexagons are separated by low diagonals, none of these simultaneous flips interfere with each other.

 Now consider the edges that intersect $r$ in the resulting triangulation. These are the low edges separating the hexagons, and one high diagonal inside each hexagon. These edges form another alternating zig-zag, half the size of the first one. Therefore we can repeat this procedure until less than three edges intersect $r$, halving the size each time. This requires $O(\log n)$ simultaneous flips. Once few edges intersect $r$, a constant number of simultaneous flips suffice to properly partition these, giving a triangulation where all low edges are to the left of $r$. Similarly, by consistently moving the low edge in each hexagon to the right of $r$, we obtain a sequence of $O(\log n)$ simultaneous flips that constructs a triangulation where all low edges are to the right of $r$.

 Now that we have a triangulation where all low edges are on one side of $r$ and all high edges are on the other, all that is left is to change this into a fan triangulation. Since each group forms a triangulation of a smaller convex polygon, we can use the result by Galtier~\etal~\cite{galtier2003simultaneous} to transform these two parts into a fan triangulation with $O(\log n)$ simultaneous flips. This completes the proof.
\end{proof}

As in the non-simultaneous upper bound, this allows us to simulate a deterministic version of quicksort (that always picks the median as pivot) to sort the labels of a fan, giving the following upper bound.

\begin{theorem}
 \label{thm:el-convex-simultaneous-ub}
 Any edge-la\-belled triangulation of a convex polygon with $n$ vertices can be transformed into any other by $O(\log^2 n)$ simultaneous flips.
\end{theorem}
\begin{proof}
 We first ignore the edge labels and transform both triangulations into a fan triangulation with $O(\log n)$ simultaneous flips~\cite{galtier2003simultaneous}. Next, we show how to sort the labels of a fan. The result follows by the reversibility of flips.

 To sort, we partition the edges into high, neutral, and low edges as in the proof of Lemma~\ref{lem:el-convex-sort}, and use Lemma~\ref{lem:el-convex-partition-simultaneous} to exchange the edges with low labels with those with high labels with $O(\log n)$ flips. Now we can sort each half recursively, combining the simultaneous flips in each part into one larger simultaneous flip. Thus, the total number of simultaneous flips is given by the recurrence $T(n) = T(\lfloor n/2 \rfloor) + O(\log n)$, which comes out to $O(\log^2 n)$ flips.

 Of course, for the combined simultaneous flip to be valid, we must ensure that no two flips use an edge of the same triangle. We do this by finding the edge with the median label and placing it on the right edge. This is a special case of Lemma~\ref{lem:el-convex-partition-simultaneous}, with one high and low edge, and all others neutral, so we can do it with $O(\log n)$ simultaneous flips. By recursing to the left and right of this fixed edge, we can guarantee that the simultaneous flips on each side do not interfere, proving the theorem.
\end{proof}

Unfortunately, we do not have a matching lower bound in this case. In fact, the best asymptotic lower bound we have is that $\Omega(\log n)$ simultaneous flips are sometimes required -- identical to the unlabelled setting! This bound is easily derived from Theorem~\ref{thm:el-convex-lb}, or directly, by observing that there are triangulations with constant maximum degree, and every simultaneous flip can at most double the degree of a vertex. What we \emph{can} prove is that this lower bound holds already for the partition step. This means that at least the result of Lemma~\ref{lem:el-convex-partition-simultaneous} is best possible.

\begin{theorem}
 \label{thm:el-convex-simultaneous-lb}
 Let the first half of the diagonals of a fan triangulation of an $n$-vertex convex polygon be \emph{high}, and the rest \emph{low}. Then any sequence of simultaneous flips that exchanges the labels on the high and low edges must have length $\Omega(\log n)$.
\end{theorem}
\begin{proof}
 Let $r$ be a ray from the bottom vertex that separates the high edges from the low ones, and let $\mathcal{F}$ be a sequence of simultaneous flips that exchanges the high and low edges. Consider an arbitrary label $\ell$. Before executing $\mathcal{F}$, $\ell$ is on one side of $r$ and afterwards it is on the other side. But since an edge that lies completely on one side of $r$ cannot intersect an edge that lies completely on the other side, and a flip always transforms an edge into another that intersects it, we cannot flip directly from the first edge to the second. In particular, at some point during the flip sequence, the edge with label $\ell$ must intersect $r$. This holds for all labels.

 Now let $X_i$ be the number of distinct labels that have intersected $r$ after $i$ simultaneous flips. Clearly $X_0 = 0$, and by the argument above $X_{|\mathcal{F}|} = n - 3$ (the total number of labels). Consider a flip that creates a new edge intersecting $r$. This flip takes place in a quadrilateral that itself intersects $r$. Therefore two of its boundary edges must intersect $r$ as well. Since a boundary edge can be shared by at most two quadrilaterals, and each quadrilateral has two boundary edges that intersect $r$, a simultaneous flip that creates $k$ new crossing must already have had at least $k$ crossings. In other words, $X_{i+1} \leq 2 X_i$. Since $X_{|\mathcal{F}|} = n - 3$, this implies that $|\mathcal{F}| \geq \log_2 (n - 3) = \Omega(\log n)$, proving the lemma.
\end{proof}

\section{Combinatorial triangulations}
\label{sec:el-combinatorial}

In this section, we show that any edge-labelled combinatorial triangulation can be transformed into any other with $O(n \log n)$ flips, and that this bound is tight. Note that we consider two edge-la\-belled triangulations to be equivalent if they have an isomorphism that preserves the edge labels.

\subsection{Upper bound}

\begin{figure}[htb]
  \centering
  \includegraphics{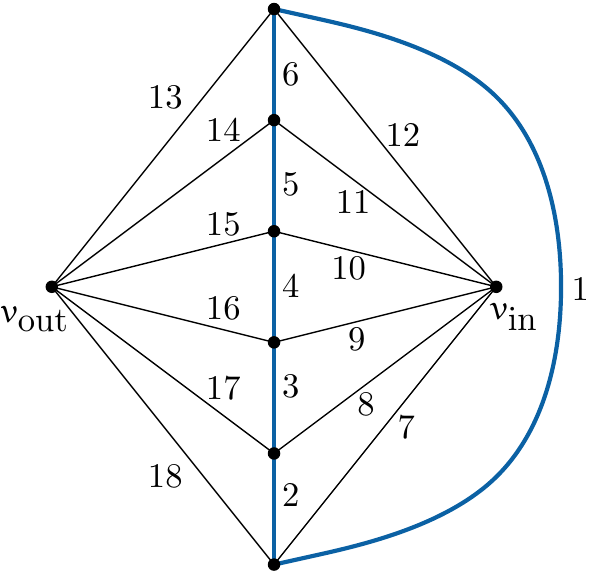}
  \caption{The canonical edge-la\-belled combinatorial triangulation on 8 vertices. The spine is indicated in bold.}
  \label{fig:el-comb-canonical}
 \end{figure}

For the upper bound, we use a canonical triangulation much like the one used by Sleator, Tarjan, and Thurston~\cite{sleator1992short} for the vertex-labelled variant. It is a double wheel: a cycle of length $n - 2$ (called the \emph{spine}), plus a vertex $\vI$ inside the cycle and a vertex $\vO$ outside the cycle, each connected to every vertex on the cycle (see Figure~\ref{fig:el-comb-canonical}). For our canonical labelling, we separate the labels into three groups. We call labels $1, \ldots, n - 2$ group \gS and we place them on the spine edges, starting with the edge on the outer face and continuing in clockwise order around $\vI$. The next $n - 2$ labels make up group \gI and are placed on the edges incident to $\vI$ in clockwise order, starting with the edge incident to the vertex shared by the edges with labels 1 and 2. Finally, group \gO consists of the last $n - 2$ labels, which we place on the edges incident to $\vO$ in clockwise order, starting with the edge that shares a vertex with the edge labelled $2n - 4$.

\begin{figure}[p]
   \centering
   \includegraphics{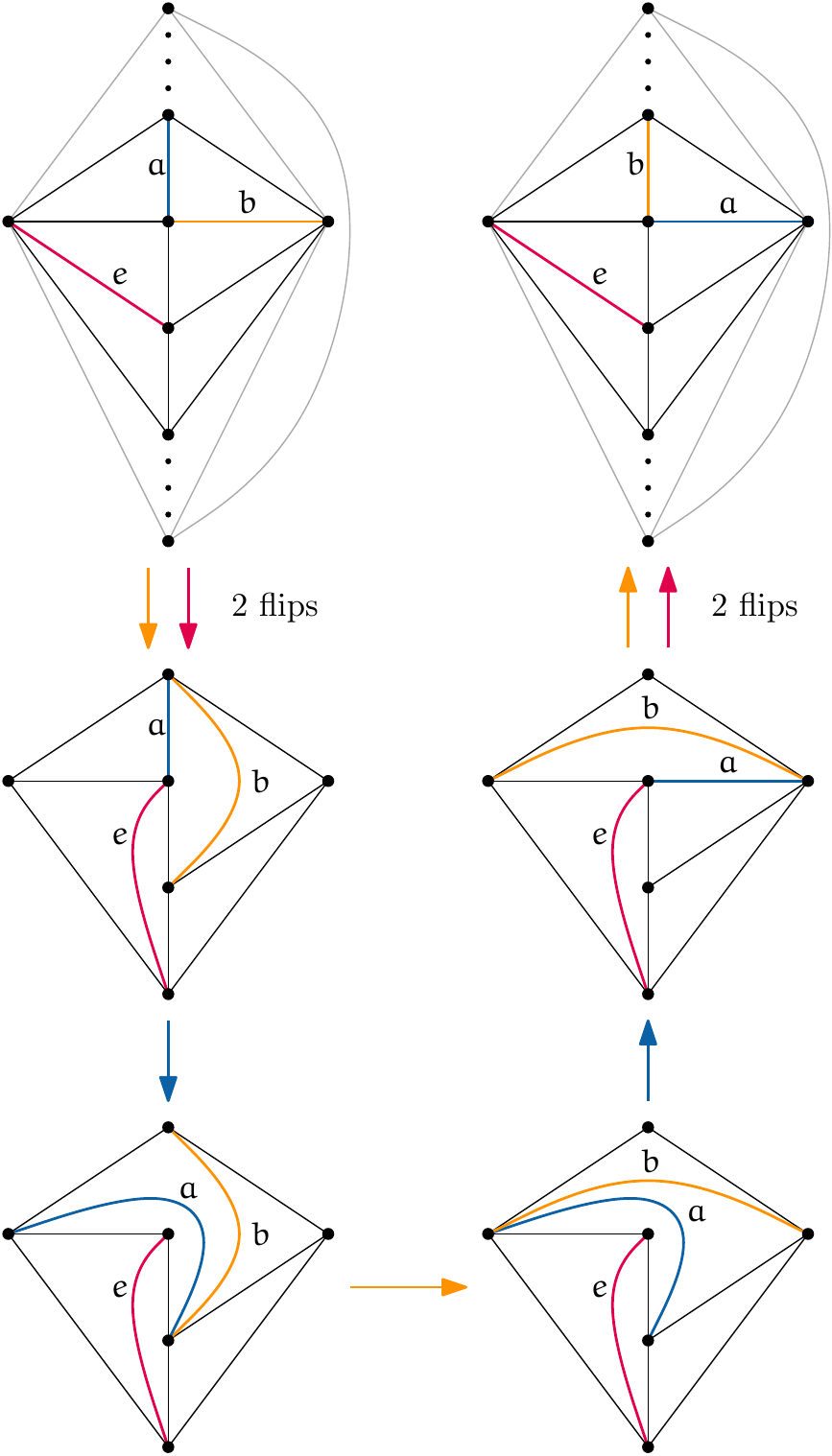}
   \caption{A sequence of seven flips that swaps two edges $a$ and $b$ that are consecutive around a vertex on the spine. Although edge $e$ ends up at the same place as at the start of the sequence, it essentially acts as a catalyst here. If we did not flip it, we would not be able to flip edge $a$ after edge $b$, as that would create a duplicate edge.}
   \label{fig:el-spineswap}
 \end{figure}

\begin{theorem}
 \label{thm:el-combinatorial-ub}
 Any edge-la\-belled combinatorial triangulation with $n$ vertices can be transformed into any other by $O(n \log n)$ flips.
\end{theorem}
\begin{proof}
 We show that we can transform any edge-la\-belled combinatorial triangulation into the canonical one using $O(n\log n)$ flips. As flips are reversible, we can also go from the canonical triangulation to any other, proving the theorem.
 
 As for convex polygons, our algorithm first ignores the labels and transforms the given triangulation into the unlabelled canonical triangulation. This requires $O(n)$ flips~\cite{sleator1992short} and results in the correct graph, although the labels may be in arbitrary positions. To fix the labels, we first get the groups to have the correct set of labels, that is, all labels in group $\gS$ are on the spine, etc., before we rearrange the labels within each group.

 We use two main tools for this. The first is a \emph{swap} that interchanges one spine edge with an incident non-spine edge in seven flips, using the flip sequence depicted in Figure~\ref{fig:el-spineswap}. Our second tool is a \emph{scramble} algorithm that reorders all labels incident to $\vI$ or $\vO$ using $O(n \log n)$ flips. To do this, we first flip the spine edge that is part of the exterior face (labelled 1 in Figure~\ref{fig:el-comb-canonical}) and then apply the algorithm from Theorem~\ref{thm:el-convex-ub} to the outerplanar graph induced by the spine plus $\vI$ (or $\vO$), observing that no flip will create a duplicate edge since the omitted edges are all incident to $\vO$ (resp. $\vI$). Note that this method cannot alter the labels on the two non-spine edges that lie on the exterior face of the outerplanar graph (labelled 7 and 12 in Figure~\ref{fig:el-comb-canonical}), but since there are only two of these, we can move them to their correct places by swapping them along the spine, using $O(n)$ flips total.

 To get the labels of group $\gS$ on the spine, we partner every edge incident to $\vI$ that has a label in $\gS$ with an edge on the spine that has a label in $\gI$ or $\gO$. A scramble at $\vI$ makes each such edge incident to its partner, and then swaps exchange partners. By doing the same at $\vO$, all labels of $\gS$ are placed on the spine. Next we partner every edge incident to $\vI$ that has a label in $\gO$ with an edge incident to $\vO$ that has a label in $\gI$. A scramble at $\vI$ makes partners incident, and three swaps per pair then exchange partners.

 This ensures that each edge's label is in the correct group, but the order of the labels within each group may still be incorrect. Rearranging the labels in $\gI$ and $\gO$ is straightforward, as we can simply scramble at $\vI$ and $\vO$, leaving only the labels on the spine out of order. We then use swaps to exchange the labels on the spine with those incident to $\vI$ in $O(n)$ flips and scramble at $\vI$ to order them correctly. Since this scramble does not affect the order of labels on the spine, we can simply exchange the edges once more to obtain the canonical triangulation.
\end{proof}

\subsection{Lower bound}

The proof for the lower bound for combinatorial triangulations is very similar to the lower bound for triangulations of a convex polygon, described in Section~\ref{sec:el-convex-lb}. We again construct a graph grammar, which describes transformations on the dual graph that correspond to flips.

As our primary graph is a combinatorial triangulation, each vertex of the dual graph corresponds to a triangle and has degree three. As such, there is no distinction between internal nodes and leaves, and no root. This means that we need to adapt our definitions slightly. Without a root, the placement of the edge-end labels is less constrained. We only require that they occur in counter-clockwise order around each vertex. The order of labels in each tag can now follow the placement of the edge-end labels: the first label belongs to the primary edge corresponding to the dual edge with edge-end label 1, and so on.

\begin{figure}[htbp]
 \centering
 \includegraphics{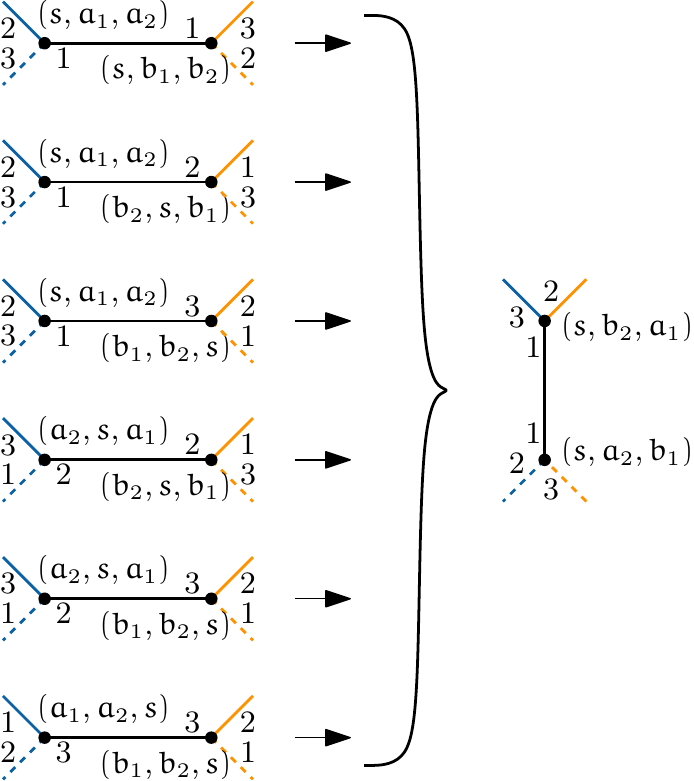}
 \caption{A graph grammar that corresponds to flips in edge-la\-belled combinatorial triangulations. The right-hand side of all productions is the same.}
 \label{fig:el-combinatorial-graph-grammar}
\end{figure}

Finally, we need a few more production rule to deal with all possible rotations of the edge-end labels around the two triangle-vertices involved in the flip. The full collection of rules is shown in Figure~\ref{fig:el-combinatorial-graph-grammar}.

As the dual graph of an $n$-vertex combinatorial triangulation has $2n - 4$ vertices, Theorem~\ref{thm:el-stt-reachable} gives us the following bound.

\begin{lemma}
 \label{lem:el-combinatorial-reachable}
 Given an $n$-vertex edge-la\-belled combinatorial triangulation $G$, the number of distinct edge-la\-belled triangulations reachable from $G$ in $m$ flips is at most $13^{2n - 4 + 2m}$.
\end{lemma}

Again, Sleator, Tarjan, and Thurston~\cite{sleator1992short} show that this bound can be significantly reduced (to $3^{2n - 4}8^m$), but the simple bound suffices for our purposes.

\begin{theorem}
 \label{thm:el-combinatorial-lb}
 There are pairs of edge-la\-belled combinatorial triangulations with $n$ vertices such that transforming one into the other requires $\Omega(n \log n)$ flips.
\end{theorem}
\begin{proof}
 If we fix the labelling of the spine edges in the canonical triangulation from the proof of Theorem~\ref{thm:el-combinatorial-ub}, any relabelling of the remaining edges is unique. Thus, there are at least $(2n - 6)!$ distinct edge-la\-belled combinatorial triangulations. Combined with Lemma~\ref{lem:el-combinatorial-reachable}, this implies that $13^{2n - 4 + 2d} \geq (2n - 6)!$, where $d$ is the diameter of the flip graph. We derive the following.
 \begin{align*}
  13^{2n - 4 + 2d}~~&\geq~~(2n - 6)! \\
  13^{2n - 4 + 2d}~~&\geq~~n!\text{\hspace{8em}(for $n \geq 5$)} \\
  \log_{13} 13^{2n - 4 + 2d}~~&\geq~~\log_{13} n! \\
  2n - 4 + 2d~~&\geq~~\log_{13} n! \\
  2d~~&\geq~~\log_{13} n! - 2n + 4 \\
  2d~~&\geq~~\Omega(n \log n) - O(n) \\
  d~~&\geq~~\Omega(n \log n) \qedhere
 \end{align*}
\end{proof}

Combining the upper and lower bound from Theorem~\ref{thm:el-combinatorial-ub} and~\ref{thm:el-combinatorial-lb} yields the following.

\begin{corollary}
 The flip graph of edge-la\-belled combinatorial triangulations with $n$ vertices has diameter $\Theta(n \log n)$.
\end{corollary}

\subsection{Simultaneous flips}

Recall that, in a triangulation of a convex polygon, a simultaneous flip is a set of flips that are executed in parallel, such that no two flipped edges share a triangle. In a combinatorial triangulation, we have the additional requirement that the resulting graph may not contain duplicate edges.

Simultaneous flips in combinatorial triangulations were first studied by Bose~\etal~\cite{bose2007simultaneous}. They showed a tight $\Theta(\log n)$ bound on the diameter of the flip graph. As part of their proof, they showed that every combinatorial triangulation can be made 4-connected with a single simultaneous flip. Recently, Cardinal~\etal~\cite{cardinal2015arc} proved that it is possible to find such a simultaneous flip that consists of fewer than $2n/3$ individual flips. They used this result to obtain arc drawings of planar graphs in which only $2n/3$ edges are represented by multiple arcs.

In this section, we show that, just as in the non-simultaneous setting, we obtain the same bounds for edge-la\-belled convex polygons and edge-la\-belled combinatorial triangulations. That is, we can transform any edge-la\-belled combinatorial triangulation into any other with $O(\log^2 n)$ simultaneous flips, and $\Omega(\log n)$ simultaneous flips are sometimes necessary. The lower bound holds already in the unlabelled setting, if one vertex has linear degree in the first triangulation, while every vertex has constant degree in the second. We now prove the upper bound.

\begin{figure}[htb]
 \centering
 \includegraphics{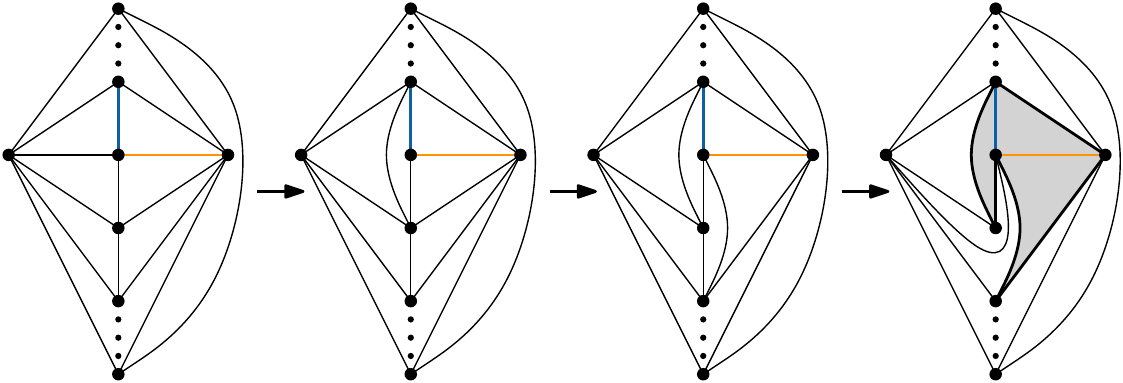}
 \caption{A sequence of three flips that creates a pentagon (shaded) in which the two highlighted edges can be swapped. All new edges and all diagonals of the pentagon are incident to one of the four spine vertices shown.}
 \label{fig:el-spineswap-simultaneous}
\end{figure}

\begin{theorem}
 Any edge-la\-belled combinatorial triangulation with $n$ vertices can be transformed into any other by $O(\log^2 n)$ simultaneous flips.
\end{theorem}
\begin{proof}
 We closely follow the strategy of the proof of Theorem~\ref{thm:el-combinatorial-ub}. We first transform the given triangulations into the canonical one with $O(\log n)$ simultaneous flips, using the result of Bose~\etal~\cite{bose2007simultaneous}. This reduces the problem to sorting the edge labels on the canonical triangulation. In the non-simultaneous setting, we did this by reordering the labels on the edges incident to $\vI$ or $\vO$ (called \emph{scrambling}), and swapping a subset of spine edges with incident non-spine edges. Thus, the theorem follows if we can show how to perform these operations with $O(\log^2 n)$ simultaneous flips.

 Since the sequence of flips from Figure~\ref{fig:el-spineswap}, which swaps a single spine edge with an incident non-spine edge, only involves a constant number of triangles, it is tempting to think we can simply perform many of these swaps simultaneously. Unfortunately, this is not the case, since the sequence creates the edge $(\vO, \vI)$. This means that trying to perform this sequence simultaneously in different locations would create a duplicate edge. Therefore we use a slightly longer sequence that creates a pentagon containing the edges to be swapped (illustrated in Figure~\ref{fig:el-spineswap-simultaneous}), performs the swap inside this pentagon, and restores the canonical triangulation, using a total of eleven flips. The crucial property of this sequence is that it only creates edges incident to four spine vertices near the edge to be swapped. Thus, we can perform any number of swaps simultaneously without creating duplicate edges, as long as each swap is at distance four or more from the others. This means that, given a set of spine edges to swap, we can divide them into four rounds such that the edges to be swapped in each round are at distance four or more, and perform the swaps in each round simultaneously. Thus, we can swap any subset of spine edges with $O(1)$ simultaneous flips.

 To scramble the edges incident on $\vO$, we first flip to create $(\vO, \vI)$ and then apply the algorithm from Theorem~\ref{thm:el-convex-simultaneous-ub} to the outerplanar graph induced by the edges incident to $\vO$. This uses $O(\log^2 n)$ simultaneous flips to rearrange all labels, except for those on the two outermost edges that are part of the boundary. In the non-simultaneous setting, we fixed this by swapping these labels along the spine, but this would take too many flips here. Instead, if the labels that need to be on the outermost edges are in the interior, we use Theorem~\ref{thm:el-convex-simultaneous-ub} to place these labels on the interior edges closest to the outermost edges. Then, we can exchange them with the labels on the outermost edges with only three swaps. This ensures that the outermost edges have the correct labels, so a second application of Theorem~\ref{thm:el-convex-simultaneous-ub} can place the remaining labels in the right order. If the label for one of the outermost edges is not in the interior and not already in place, it must be on the other outermost edge. In this case, we can first exchange it with the label on a nearby interior edge with a constant number of swaps. The entire sequence requires $O(\log^2 n)$ simultaneous flips.

 Since these operations use $O(1)$ and $O(\log^2 n)$ simultaneous flips, and we can sort the labels with a constant number of applications, the theorem follows.
\end{proof}

\section{Pseu\-do-tri\-an\-gu\-la\-tions}
\label{sec:el-pts}

\begin{figure}[htb]
 \centering
 \begin{subfigure}[b]{0.48\textwidth}
  \centering
  \includegraphics{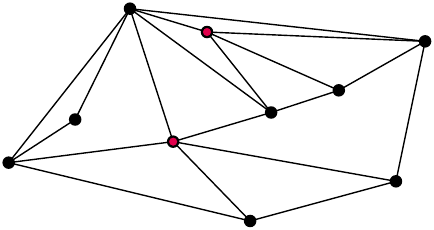}
  \caption{}
  \label{fig:el-pt}
 \end{subfigure}
 \begin{subfigure}[b]{0.48\textwidth}
  \centering
  \includegraphics{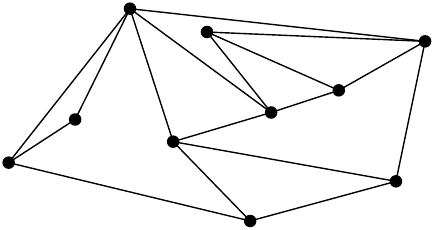}
  \caption{}
  \label{fig:el-pointed-pt}
 \end{subfigure}
 \caption{(a) A pseu\-do-tri\-an\-gu\-la\-tion with two non-pointed vertices. (b) A pointed pseu\-do-tri\-an\-gu\-la\-tion.}
\end{figure}

A \emph{pseudo-triangle} is a simple polygon with three convex interior angles, called \emph{corners}, that are connected by reflex chains. Given a set $P$ of $n$ points in the plane, a \emph{pseu\-do-tri\-an\-gu\-la\-tion} of $P$ is a subdivision of its convex hull into pseudo-triangles, using all points of $P$ as vertices (see~Figure~\ref{fig:el-pt}). A pseu\-do-tri\-an\-gu\-la\-tion is \emph{pointed} if all vertices are incident to a reflex angle in some face (including the outer face; see Figure~\ref{fig:el-pointed-pt} for an example). Pseu\-do-tri\-an\-gu\-la\-tions find applications in areas such as kinetic data structures~\cite{kirkpatrick2002kinetic} and rigidity theory~\cite{streinu2005pseudo}. More information on pseu\-do-tri\-an\-gu\-la\-tions can be found in a survey by Rote, Santos, and Streinu~\cite{rote2007pseudotriangulations}.

Since a regular triangle is also a pseudo-triangle, pseu\-do-tri\-an\-gu\-la\-tions generalize triangulations (subdivisions of the convex hull into triangles). In a triangulation, a flip is a local transformation that removes one edge, leaving an empty quadrilateral, and inserts the other diagonal of that quadrilateral. Note that this is only possible if the quadrilateral is convex. Lawson~\cite{lawson1972transforming} showed that any triangulation with $n$ vertices can be transformed into any other with $O(n^2)$ flips, and Hurtado, Noy, and Urrutia~\cite{hurtado1999flipping} gave a matching $\Omega(n^2)$ lower bound.

\begin{figure}[htb]
 \centering
 \begin{subfigure}[b]{0.48\textwidth}
  \centering
  \includegraphics{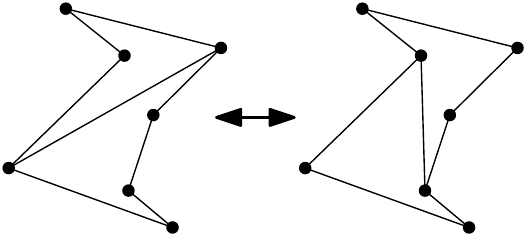}
  \caption{}
  \label{fig:el-flip}
 \end{subfigure}
 \begin{subfigure}[b]{0.48\textwidth}
  \centering
  \includegraphics{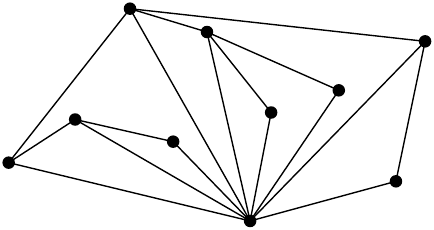}
  \caption{}
  \label{fig:el-left-shelling}
 \end{subfigure}
 \caption{(a) A flip in a pseudo-quadrilateral. (b) A left-shelling pseu\-do-tri\-an\-gu\-la\-tion.}
\end{figure}

Pointed pseu\-do-tri\-an\-gu\-la\-tions support a similar type of flip, but before we can introduce this, we need to generalize the concept of pseudo-triangles to \emph{pseudo-$k$-gons}: weakly simple polygons with $k$ convex interior angles. A diagonal of a pseudo-$k$-gon is called a \emph{bitangent} if the pseudo-$k$-gon remains pointed after insertion of the diagonal. In a pointed pseu\-do-tri\-an\-gu\-la\-tion, \emph{flipping} an edge removes the edge, leaving a pseudo-quadrilateral, and inserts the unique other bitangent of the pseudo-quadrilateral (see Figure~\ref{fig:el-flip}). In contrast with triangulations, all internal edges of a pointed pseu\-do-tri\-an\-gu\-la\-tion are flippable. Bereg~\cite{bereg2004transforming} showed that $O(n \log n)$ flips suffice to transform any pseu\-do-tri\-an\-gu\-la\-tion into any other.

Aichholzer~\etal~\cite{aichholzer2003pseudotriangulations} showed that the same result holds for all pseu\-do-tri\-an\-gu\-la\-tions (including triangulations) if we allow two more types of flips: \emph{insertion} and \emph{deletion} flips. As the name implies, these either insert or delete one edge, provided that the result is still a pseu\-do-tri\-an\-gu\-la\-tion. To disambiguate, they call the other flips \emph{exchanging} flips. In a later paper, this bound was refined to $O(n \log c)$~\cite{aichholzer2006transforming}, where $c$ is the number of convex layers of the point set.

In this section, we investigate flips in \emph{edge-la\-belled pseu\-do-tri\-an\-gu\-la\-tions}: pseu\-do-tri\-an\-gu\-la\-tions where each internal edge has a unique label in $\{1, \ldots, 3n - 3 - 2h\}$, where $h$ is the number of vertices on the convex hull ($3n - 3 - 2h$ is the number of internal edges in a triangulation). In the case of an exchanging flip, the new edge receives the label of the old edge. For a deletion flip, the edge and its label are simply removed, and for an insertion flip, the new edge receives an unused label from the set of all possible labels.

Our results are the following: using only exchanging flips, we show that $O(n^2)$ flips suffice to transform any edge-la\-belled pointed pseu\-do-tri\-an\-gu\-la\-tion into any other with the same set of labels. By using insertion, deletion and exchanging flips, we can transform any edge-la\-belled pseu\-do-tri\-an\-gu\-la\-tion into any other with $O(n \log c + h \log h)$ flips.

Before we can start the proof, we need a few more definitions. Given a set of points in the plane, let $v_0$ be the point with the lowest $y$-coordinate, and let $v_1, \ldots, v_n$ be the other points in clockwise order around $v_0$. The \emph{left-shelling} pseu\-do-tri\-an\-gu\-la\-tion is the union of the convex hulls of $v_0, \ldots, v_i$, for all $2 \leq i \leq n$ (see Figure~\ref{fig:el-left-shelling}). Thus, every vertex after $v_1$ is associated with two edges: a \emph{bottom} edge connecting it to $v_0$ and a \emph{top} edge that is tangent to the convex hull of the earlier vertices. The \emph{right-shelling} pseu\-do-tri\-an\-gu\-la\-tion is similar, with the vertices added in counter-clockwise order instead.

\subsection{Pointed pseu\-do-tri\-an\-gu\-la\-tions}
\label{sec:el-pointed}

In this section, we show that every edge-la\-belled pointed pseu\-do-tri\-an\-gu\-la\-tion can be transformed into any other with the same set of labels by $O(n^2)$ exchanging flips. We do this by showing how to transform a given edge-la\-belled pointed pseu\-do-tri\-an\-gu\-la\-tion into a \emph{canonical} one. The result then follows by the reversibility of flips. As canonical pseu\-do-tri\-an\-gu\-la\-tion, we use the left-shelling pseu\-do-tri\-an\-gu\-la\-tion, with the bottom edges labelled in clockwise order around $v_0$, followed by the internal top edges in the same order (based on their associated vertex).

Since we can transform any pointed pseu\-do-tri\-an\-gu\-la\-tion into the left-shelling pseu\-do-tri\-an\-gu\-la\-tion with $O(n \log n)$ flips~\cite{bereg2004transforming}, the main part of the proof lies in reordering the labels of a left-shelling pseu\-do-tri\-an\-gu\-la\-tion. We use two tools for this, called a \emph{sweep} and a \emph{shuffle}, that are implemented by a sequence of flips. A sweep interchanges the labels of some internal top edges with their respective bottom edges, while a shuffle permutes the labels on all bottom edges.

\begin{lemma}
\label{lem:el-sort-left-shelling}
We can transform any left-shelling pseu\-do-tri\-an\-gu\-la\-tion into the canonical one with $O(1)$ shuffle and sweep operations.
\end{lemma}
\begin{proof}
In the canonical pseu\-do-tri\-an\-gu\-la\-tion, we call the labels assigned to bottom edges \emph{low}, and the labels assigned to top edges \emph{high}. In the first step, we use a shuffle to line up every bottom edge with a high label with a top edge with a low label. Then we exchange these pairs of labels with a sweep. Now all bottom edges have low labels and all top edges have high labels, so all that is left is to sort the labels. We can sort the low labels with a second shuffle. To sort the high labels, we sweep them to the bottom edges, shuffle to sort them there, then sweep them back.
\end{proof}

The remainder of this section describes how to perform a sweep and a shuffle with flips.

\begin{figure}[htb]
 \centering
 \includegraphics{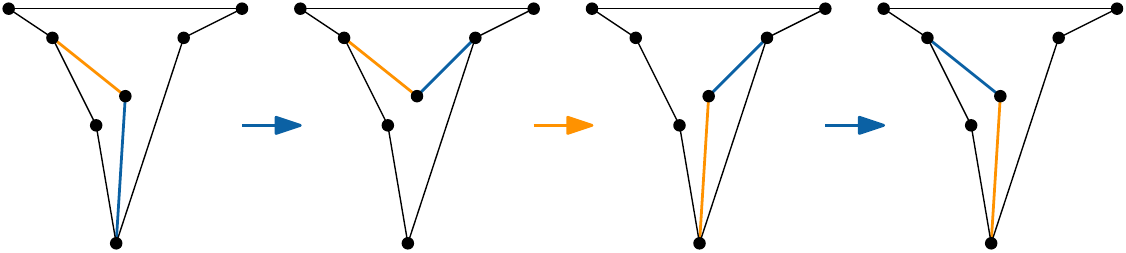}
 \caption{Interchanging the labels of the edges incident to a vertex of degree two.}
 \label{fig:el-degree-2-swap}
\end{figure}

\begin{lemma}
\label{lem:el-degree-2-swap}
We can interchange the labels of the edges incident to an internal vertex $v$ of degree two with three exchanging flips.
\end{lemma}
\begin{proof}
Consider what happens when we remove $v$. Deleting one of its edges leaves a pseudo-quadrilateral. Removing the second edge then either merges two corners into one, or removes one corner, leaving a pseudo-triangle $T$. There are three bitangents that connect $v$ to $T$, each corresponding to the geodesic between $v$ and a corner of $T$. Any choice of two of these bitangents results in a pointed pseu\-do-tri\-an\-gu\-la\-tion. When one of them is flipped, the only new edge that can be inserted so that the result is still a pointed pseu\-do-tri\-an\-gu\-la\-tion is the bitangent that was not there before the flip. Thus, we can interchange the labels with three flips (see Figure~\ref{fig:el-degree-2-swap}).
\end{proof}

\begin{lemma}[Sweep]
\label{lem:el-sweep}
In the left-shelling pseu\-do-tri\-an\-gu\-la\-tion, we can interchange the labels of any number of internal top edges and their corresponding bottom edges with $O(n)$ exchanging flips.
\end{lemma}
\begin{proof}
Let $S$ be the set of vertices whose internal top edge should have its label swapped with the corresponding bottom edge. Consider a ray $L$ from $v_0$ that starts at the positive $x$-axis and sweeps through the point set to the negative $x$-axis. We will maintain the following invariant: the graph induced by the vertices to the left of $L$ is their left-shelling pseu\-do-tri\-an\-gu\-la\-tion and the graph induced by the vertices to the right of $L$ is their right-shelling pseu\-do-tri\-an\-gu\-la\-tion (both groups include $v_0$). Furthermore, the labels of the top edges of the vertices in $S$ to the right of $L$ have been interchanged with their respective bottom edges. This invariant is satisfied at the start.

Suppose that $L$ is about to pass a vertex $v_k$. If $v_k$ is on the convex hull, its top edge is not internal and no action is required for the invariant to hold after passing $v_k$. So assume that $v_k$ is not on the convex hull and consider its incident edges. It is currently part of the left-shelling pseu\-do-tri\-an\-gu\-la\-tion of points to the left of $L$, where it is the last vertex. Thus, $v_k$ is connected to $v_0$ and to one vertex to its left. It is not connected to any vertex to its right, since there are $2n - 3$ edges in total, and the left- and right-shelling pseu\-do-tri\-an\-gu\-la\-tions to each side of $L$ contribute $2(k + 1) - 3 + 2(n - k) - 3 = 2n - 4$ edges. So the only edge that crosses $L$ is an edge of the convex hull. Therefore $v_k$ has degree two, which means that we can use Lemma~\ref{lem:el-degree-2-swap} to swap the labels of its top and bottom edge with three flips if $v_k \in S$.

Furthermore, the sides of the pseudo-triangle that remains if we were to remove $v_k$, form part of the convex hull of the points to either side of $L$. Thus, flipping the top edge of $v_k$ results in the tangent from $v_k$ to the convex hull of the points to the right of $L$ -- exactly the edge needed to add $v_k$ to their right-shelling pseu\-do-tri\-an\-gu\-la\-tion. Therefore we only need $O(1)$ flips to maintain the invariant when passing $v_k$.

At the end, we have constructed the right-shelling pseu\-do-tri\-an\-gu\-la\-tion and swapped the desired edges. An analogous transformation without any swapping can transform the graph back into the left-shelling pseu\-do-tri\-an\-gu\-la\-tion with $O(n)$ flips in total.
\end{proof}

\begin{figure}[htb]
 \centering
 \includegraphics{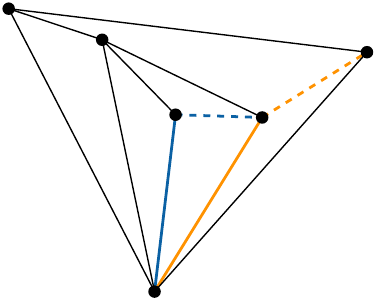}
 \caption{A pseudo-pentagon with four bitangents. It is impossible to swap the two diagonals without flipping an edge of the pseudo-pentagon, as they just flip back and forth between the solid bitangents and the dotted ones, regardless of the position of the other diagonal.}
 \label{fig:el-four-bitangents}
\end{figure}

\begin{figure}[htb]
 \centering
 \includegraphics{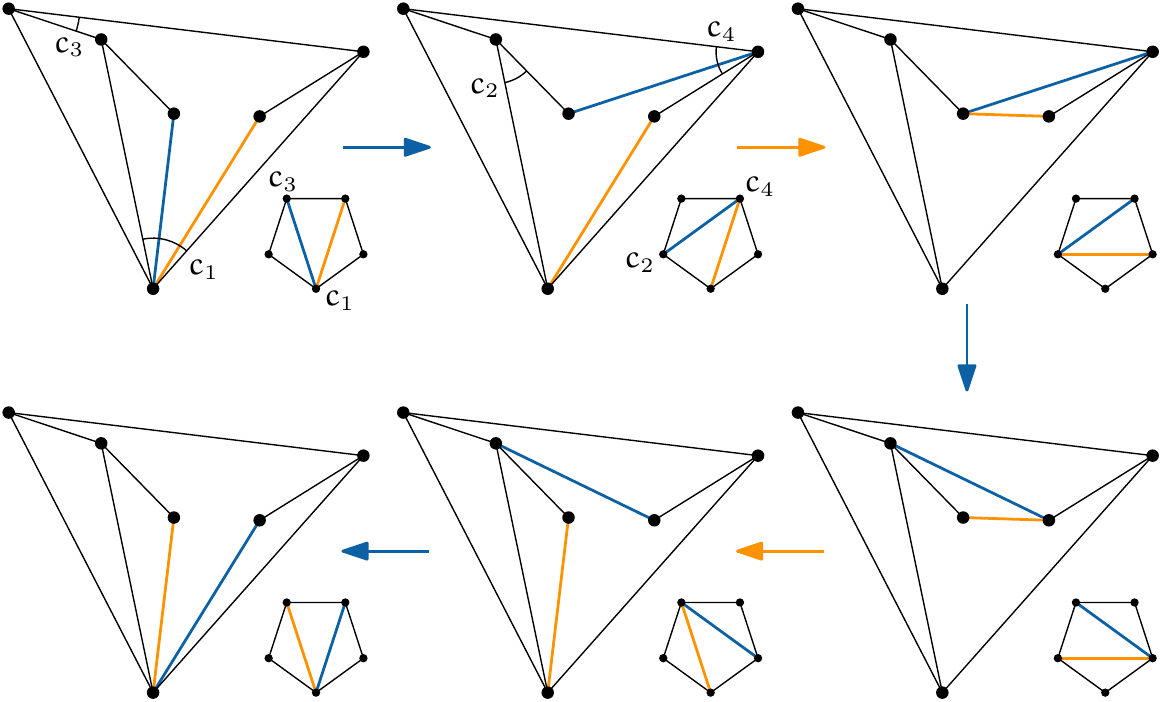}
 \caption{Interchanging the labels of two bitangents of a pseudo-pen\-tagon with five bitangents. An edge in the pentagon corresponds to a geodesic between two corners of the pseudo-pentagon.}
 \label{fig:el-pentagon-swap}
\end{figure}

\begin{lemma}
\label{lem:el-consecutive-swap}
In the left-shelling pseu\-do-tri\-an\-gu\-la\-tion, we can interchange the labels of two consecutive bottom edges with $O(1)$ exchanging flips.
\end{lemma}
\begin{proof}
When we remove the two consecutive bottom edges (say $a$ and $b$), we are left with a pseudo-pentagon $X$. A pseudo-pentagon can have up to five bitangents, as each bitangent corresponds to a geodesic between two corners. If $X$ has exactly five bitangents, this correspondence is a bijection. This implies that the bitangents of $X$ can be swapped just like diagonals of a convex pentagon (see Figure~\ref{fig:el-pentagon-swap}). On the other hand, if $X$ has only four bitangents, it is impossible to swap $a$ and $b$ without flipping an edge of $X$ (see~Figure~\ref{fig:el-four-bitangents}).

Fortunately, we can always transform $X$ into a pseudo-pentagon with five bitangents. If the pseudo-triangle to the right of $b$ is a triangle, $X$ already has five bitangents (see~Lemma~\ref{lem:el-five-bitangents-a} in Section~\ref{sec:el-deferred}). Otherwise, the top endpoint of $b$ is an internal vertex of degree two and we can flip its top edge to obtain a new pseudo-pentagon that does have five bitangents (see~Lemma~\ref{lem:el-five-bitangents-b} in Section~\ref{sec:el-deferred}). After swapping the labels of $a$ and $b$, we can flip this top edge back. Thus, in either case we can interchange the labels of $a$ and $b$ with $O(1)$ flips.
\end{proof}

We can use Lemma~\ref{lem:el-consecutive-swap} to reorder the labels of the bottom edges with insertion or bubble sort, as these algorithms only swap adjacent values.

\begin{corollary}[Shuffle]
\label{cor:el-shuffle}
In the left-shelling pseu\-do-tri\-an\-gu\-la\-tion, we can reorder the labels of all bottom edges with $O(n^2)$ exchanging flips.
\end{corollary}

Combining this with Lemmas~\ref{lem:el-sort-left-shelling} and \ref{lem:el-sweep}, and the fact that we can transform any pointed pseu\-do-tri\-an\-gu\-la\-tion into the left-shelling one with $O(n \log n)$ flips~\cite{bereg2004transforming}, gives the main result.

\begin{theorem}
\label{thm:el-upper-bound}
We can transform any edge-la\-belled pointed pseu\-do-tri\-an\-gu\-la\-tion with $n$ vertices into any other with $O(n^2)$ exchanging flips.
\end{theorem}

The following lower bound follows from the $\Omega(n \log n)$ lower bound on the flip distance between edge-la\-belled triangulations of a convex polygon (Theorem~\ref{thm:el-convex-lb}).

\begin{theorem}
\label{thm:el-lower-bound}
There are pairs of edge-la\-belled pointed pseu\-do-tri\-an\-gu\-la\-tions with $n$ vertices that require $\Omega(n \log n)$ exchanging flips to transform one into the other.
\end{theorem}

\subsubsection{Deferred proofs}
\label{sec:el-deferred}

This section contains a few technical lemmas that were omitted from the previous section.

\begin{figure}[htb]
 \centering
 \begin{subfigure}[b]{0.48\textwidth}
  \centering
  \includegraphics{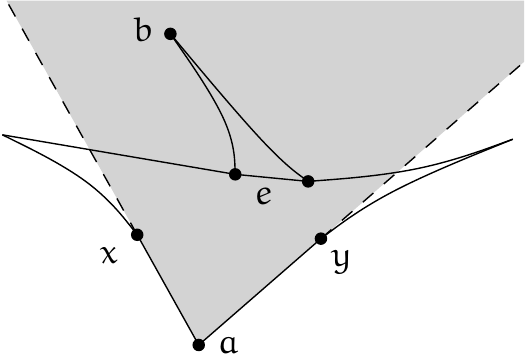}
  \caption{}
  \label{fig:el-opposite-flip}
 \end{subfigure}
 \begin{subfigure}[b]{0.48\textwidth}
  \centering
  \includegraphics{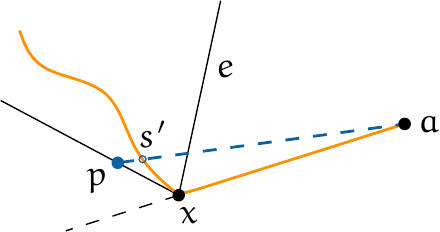}
  \caption{}
  \label{fig:el-shorter-path}
 \end{subfigure}
 \caption{(a) A corner of a pseudo-triangle and an edge such that the entire pseudo-triangle on the other side of the edge lies inside the corner's wedge. (b) If $a$ can see a point past $x$, then the geodesic does not contain $x$.}
\end{figure}

\begin{lemma}
\label{lem:el-opposite-flip}
Let $a$ be a corner of a pseudo-triangle with neighbours $x$ and $y$, and let $e$ be an edge on the chain opposite $a$. If all vertices of the other pseudo-triangle containing $e$ lie in the wedge formed by extending the edges $ax$ and $ay$ into half-lines (see Figure~\ref{fig:el-opposite-flip}), then flipping $e$ will result in an edge incident on $a$.
\end{lemma}
\begin{proof}
Let $T$ be the pseudo-triangle on the other side of $e$, and let $b$ be the corner of $T$ opposite $e$. Then flipping $e$ inserts the geodesic between $a$ and $b$. This geodesic must intersect $e$ in a point $s$ and then follow the shortest path from $s$ to $a$. If $s$ lies strictly inside the wedge, nothing can block $as$, thus the new edge will contain $as$ and be incident on $a$.

Now, if all of $e$ lies strictly inside the wedge, our result follows. But suppose that $e$ has $x$ as an endpoint and the geodesic between $a$ and $b$ intersects $e$ in $x$. As $a$ can see $x$ and all of $T$ lies inside the wedge, there is an $\varepsilon > 0$ such that $a$ can see the point $X$ on the boundary of $T$ at distance $\varepsilon$ from $x$ (see Figure~\ref{fig:el-shorter-path}). The line segment $ap$ intersects the geodesic at a point $s'$. By the triangle inequality, $s'a$ is shorter than following the geodesic from $s'$ via $x$ to $a$. But then this would give a shorter path between $a$ and $b$, by following the geodesic to $s'$ and then cutting directly to $a$. As the geodesic is the shortest path by definition, this is impossible. Thus, the geodesic cannot intersect $e$ at $x$ and the new edge must be incident to $a$.
\end{proof}

\begin{lemma}
\label{lem:el-five-bitangents-a}
Let $a$ and $b$ be two consecutive internal bottom edges in the left-shelling pseu\-do-tri\-an\-gu\-la\-tion, such that the pseudo-triangle to the right of $b$ is a triangle. Then the pseudo-pentagon $X$ formed by removing $a$ and $b$ has five bitangents.
\end{lemma}
\begin{proof}
Let $c_0, \ldots, c_4$ be the corners of $X$ in counter-clockwise order around the boundary. By Lemma~\ref{lem:el-opposite-flip}, flipping $b$ results in an edge $b'$ that intersects $b$ and is incident on $c_1$. This edge is part of the geodesic between $c_1$ and $c_3$, and as such it is tangent to the convex chain $v_0, v_a, \ldots, c_3$, where $v_a$ is the top endpoint of $a$ ($v_a$ could be $c_3$). Therefore it is also the tangent from $c_1$ to the convex hull of $\{v_0, \ldots, v_a\}$. This means that the newly created pseudo-triangle with $c_1$ as corner and $a$ on the opposite pseudo-edge also meets the conditions of Lemma~\ref{lem:el-opposite-flip}. Thus, flipping $a$ results in another edge, $a'$, also incident on $c_1$. As $b$ separates $c_1$ from all vertices in $\{v_0, \ldots, v_a\}$, $a'$ must also intersect $b$. This gives us four bitangents, of which two are incident on $v_0$ ($a$ and $b$), and two on $c_1$ ($a'$ and $b'$). Finally, flipping $a$ before flipping $b$ results in a bitangent that is not incident on $v_0$ (as $v_0$ is a corner and cannot be on the new geodesic), nor on $c_1$ (as $b$ separates $a$ from $c_1$). Thus, $X$ has five bitangents.
\end{proof}

\begin{lemma}
\label{lem:el-five-bitangents-b}
Let $a$ and $b$ be two consecutive internal bottom edges in the left-shelling pseu\-do-tri\-an\-gu\-la\-tion, such that the pseudo-triangle to the right of $b$ is not a triangle. Then the pseudo-pentagon $X$ formed by flipping the corresponding top edge of $b$ and removing $a$ and $b$ has five bitangents.
\end{lemma}
\begin{proof}
Let $v_a$ and $v_b$ be the top endpoints of $a$ and $b$. By Lemma~\ref{lem:el-opposite-flip} and since $b$ had degree two, flipping the top edge of $b$ results in the edge $v_bc_1$. We get three bitangents for free: $a$, $b$, and $b'$ -- the old top edge of $b$ and the result of flipping $b$.

$X$ consists of a reflex chain $C$ that is part of the convex hull of the points to the left of $a$, followed by three successive tangents to $C$, $v_a$, or $v_b$. Since $C$ lies completely to the left of $a$, it cannot significantly alter any of the geodesics or bitangents inside the polygon, so we can reduce it to a single edge. Now, $X$ consists either of a triangle with two internal vertices, or a convex quadrilateral with one internal vertex.

If $X$ is a triangle with two internal vertices, the internal vertices are $v_a$ and $v_b$. Let its exterior vertices be $v_0$, $x$, and $y$. Then there are seven possible bitangents: $a = v_0v_a, b = v_0v_b, xv_a, xv_b, yv_a, yv_b$, and $v_av_b$. We know that $xv_a$ and $yv_b$ are edges, so there are five possible bitangents left. As all vertices involved are either corners or have degree one in $X$, the only condition for an edge to be a bitangent is that it does not cross the boundary of $X$. Since the exterior boundary is a triangle, this reduces to it not crossing $xv_a$ and $yv_b$. Two line segments incident to the same vertex cannot cross. Thus, $xv_b$, $yv_a$, and $v_av_b$ cannot cross $xv_a$ and $yv_b$, and $X$ has five bitangents.

If $X$'s convex hull has four vertices, the internal vertex is $v_b$ (otherwise the pseudo-triangle to the right of $b$ would be a triangle). Let its exterior vertices be $v_0$, $x$, $v_a$, and $y$. Then there are six possible bitangents: $a = v_0v_a, b = v_0v_b, xy, xv_b, yv_b$, and $v_av_b$, of which one ($yv_b$) is an edge of $X$. Since $a$ and $b$ are guaranteed to be bitangents, and $xy$, $xv_b$, and $v_av_b$ all share an endpoint with $yv_b$, the arguments from the previous case apply and we again have five bitangents.
\end{proof}

\subsection{General pseu\-do-tri\-an\-gu\-la\-tions}
\label{sec:el-general-pts}

In this section, we extend our results for edge-la\-belled pointed pseu\-do-tri\-an\-gu\-la\-tions to all edge-la\-belled pseu\-do-tri\-an\-gu\-la\-tions. Since not all pseu\-do-tri\-an\-gu\-la\-tions have the same number of edges, we need to allow flips that change the number of edges. In particular, we allow a single edge to be deleted or inserted, provided that the result is still a pseu\-do-tri\-an\-gu\-la\-tion.

Since we are dealing with edge-la\-belled pseu\-do-tri\-an\-gu\-la\-tions, we need to determine what happens to the edge labels. It is useful to first review the properties we would like these flips to have. First, a flip should be a local operation -- it should affect only one edge. Second, a labelled edge should be flippable if and only if the edge is flippable in the unlabelled setting. This allows us to re-use the existing results on flips in pseu\-do-tri\-an\-gu\-la\-tions. Third, flips should be reversible. Like most proofs about flips, our proof in the previous section crucially relies on the reversibility of flips.

With these properties in mind, the edge-deletion flip is rather straightforward -- the labelled edge is removed, and other edges are not affected. Since the edge-insertion flip needs to be the inverse of this, it should insert the edge and assign it a \emph{free label} -- an unused label in $\{1, \ldots, 3n - 3 - 2h\}$, where $h$ is the number of vertices on the convex hull ($3n - 3 - 2h$ is the number of internal edges in a triangulation).

With the definitions out of the way, we can turn our attention to the number of flips required to transform any edge-la\-belled pseu\-do-tri\-an\-gu\-la\-tions into any other. In this section, we show that by using insertion and deletion flips, we can shuffle (permute the labels on bottom edges) with $O(n + h \log h)$ flips. Combined with the unlabelled bound of $O(n \log c)$ flips by Aichholzer~\etal~\cite{aichholzer2006transforming}, this brings the total number of flips down to $O(n \log c + h \log h)$. Note that, by Theorem~\ref{thm:el-convex-ub}, this holds for a set of points in convex position ($h = n$). In the remainder of this section we assume that $h < n$. As before, we first build a collection of simple tools that help prove the main result.

\begin{figure}[htb]
 \centering
 \includegraphics{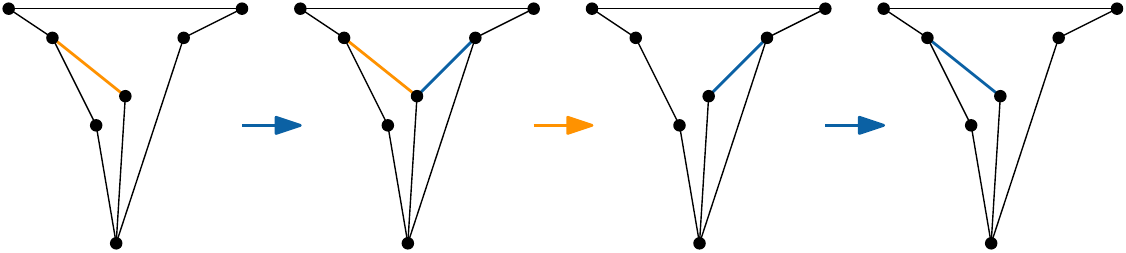}
 \caption{Interchanging the label of an edge incident to a vertex of degree two with a free label.}
 \label{fig:el-degree-2-free}
\end{figure}

\begin{lemma}
 \label{lem:el-degree-2-free}
 With $O(1)$ flips, we can interchange the label of an edge incident to an internal vertex of degree two with a free label.
\end{lemma}
\begin{proof}
 Let $v$ be a vertex of degree two and let $e$ be an edge incident to $v$. Since $v$ has degree two, its removal leaves an empty pseudo-triangle $T$. There are three bitangents that connect $v$ to $T$, one for each corner. Thus, we can insert the third bitangent $f$ with the desired free label, making $v$ non-pointed (see~Figure~\ref{fig:el-degree-2-free}). Flipping $e$ now removes it and frees its label. Finally, flipping $f$ moves it into $e$'s starting position, completing the exchange.
\end{proof}

This implies that, using an arbitrary free label as placeholder, we can swap any two edges incident to internal degree-two vertices -- no matter where they are in the pseu\-do-tri\-an\-gu\-la\-tion.

\begin{corollary}
 \label{lem:el-two-degree-2-swap}
 We can interchange the labels of two edges, each incident to some internal vertex of degree two, with $O(1)$ flips.
\end{corollary}

Recall that during a sweep (Lemma~\ref{lem:el-sweep}), each internal vertex has degree two at some point. Since the number of free labels for a pointed pseu\-do-tri\-an\-gu\-la\-tion is equal to the number of internal vertices, this means that we can use Lemma~\ref{lem:el-degree-2-free} to swap every label on a bottom edge incident to an internal vertex with a free label by performing a single sweep. Afterwards, a second sweep can replace these labels on the bottom edges in any desired order. Thus, permuting the labels on bottom edges incident to internal vertices can be done with $O(n)$ flips. Therefore, the difficulty in permuting the labels on all bottom edges lies in bottom edges that are not incident to an internal vertex, that is, chords of the convex hull. If there are few such chords, a similar strategy (free them all and replace them in the desired order) might work. Unfortunately, the number of free labels can be far less than the number of chords.

We now consider operations on maximal groups of consecutive chords, which we call \emph{fans}. As the vertices of a fan are in convex position, fans behave in many ways like triangulations of a convex polygon, which can be rearranged with $O(n \log n)$ flips (Theorem~\ref{thm:el-convex-ub}). The problem now becomes getting the right set of labels on the edges of a fan.

\begin{figure}[htb]
 \centering
 \begin{subfigure}[b]{0.38\textwidth}
  \centering
  \includegraphics{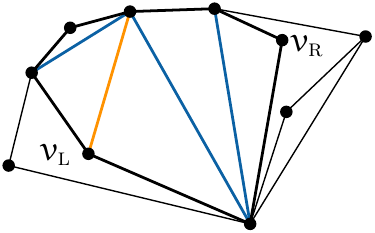}
  \caption{}
  \label{fig:el-indexed-fan}
 \end{subfigure}
 \begin{subfigure}[b]{0.58\textwidth}
  \centering
  \includegraphics{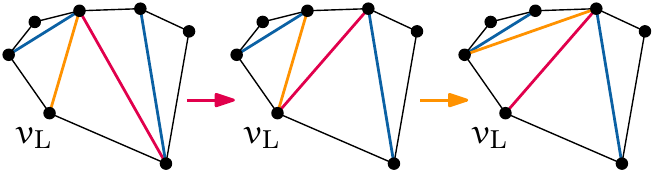}
  \caption{}
  \label{fig:el-increase-index}
 \end{subfigure}
 \caption{(a) An indexed fan. (b) Shifting the index ($\vl$) from the yellow edge to the red edge.}
\end{figure}

Consider the internal vertices directly to the left ($\vl$) and right ($\vr$) of a fan $F$, supposing both exist. Vertex $\vl$ has degree two and forms part of the reflex chain of the first pseudo-triangle to the left of $F$. Thus, flipping $\vl$'s top edge connects it to the leftmost vertex of $F$ (excluding $v_0$). Vertex $\vr$ is already connected to the rightmost vertex of $F$, so we just ensure that it has degree two. To do this, we flip all incident edges from vertices further to the right, from the bottom to the top. Now the diagonals of $F$ form a triangulation of a convex polygon whose boundary consists of $v_0$, $\vl$, the top endpoints of the chords, and $\vr$ (see~Figure~\ref{fig:el-indexed-fan}). It is possible that there is no internal vertex to one side of $F$. In that case, there is only one vertex on that side of $F$, which is part of the convex hull, and we can simply use that vertex in place of $\vl$ or $\vr$ without flipping any of its edges. Since there is at least one internal vertex by assumption, either $\vl$ or $\vr$ is an internal vertex. This vertex is called the \emph{index} of $F$. If a vertex is the index of two fans, it is called a \emph{shared index}.

A triangulated fan is called an \emph{indexed fan} if there is one edge incident to the index, the \emph{indexed edge}, and the remaining edges are incident to one of the neighbours of the index on the boundary. Initially, all diagonals of $F$ are incident to $v_0$, so we transform it into an indexed fan by flipping the diagonal of $F$ closest to the index. Next, we investigate several operations on indexed fans that help us move labels between fans.
  
\begin{lemma}[Shift]
 In an indexed fan, we can shift the indexed edge to the next diagonal with $O(1)$ flips.
\end{lemma}
\begin{proof}
 Suppose that $\vl$ is the index (the proof for $\vr$ is analogous). Let $e$ be the current indexed edge, and $f$ be the leftmost diagonal incident to $v_0$. Then flipping $f$ followed by $e$ makes $f$ the only edge incident to the index and $e$ incident to the neighbour of the index (see~Figure~\ref{fig:el-increase-index}). Since flips are reversible, we can shift the index the other way too.
\end{proof}

\begin{figure}[htb]
 \centering
 \includegraphics{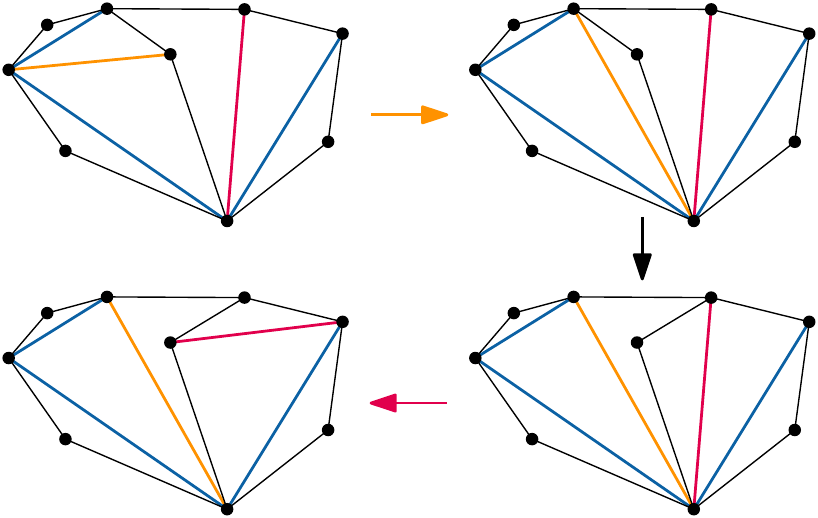}
 \caption{Changing which side a shared index indexes.}
 \label{fig:el-swap-shared-index}
\end{figure}

\begin{lemma}
 We can switch which fan a shared index currently indexes with $O(1)$ flips.
\end{lemma}
\begin{proof}
 Flipping the current indexed edge ``parks" it by connecting it to the two neighbours of the index, and reduces the degree of the index to two (see~Figure~\ref{fig:el-swap-shared-index}). Now, flipping the top edge of the index connects it to the other fan, where we parked the previously indexed edge. Flipping that edge connects it to the index again.
\end{proof}

\begin{lemma}
 \label{lem:el-deg-3-to-2}
 In a pointed pseu\-do-tri\-an\-gu\-la\-tion, we can always decrease the degree of a vertex $v$ of degree three by flipping one of the edges incident to its reflex angle.
\end{lemma}
\begin{proof}
 Consider the geodesic from $v$ to the opposite corner $c$ of the pseudo-triangle $v$ is pointed in. The line supporting the part of the geodesic when it reaches $v$ splits the edges incident to $v$ into two groups. As there are three edges, one of these groups must contain multiple edges. Flipping the edge incident to its reflex angle in the group with multiple edges results in a geodesic to $c$. If this geodesic passed through $v$, it would insert the missing edges along the geodesic from $v$ to $c$ (otherwise we could find a shorter path). But inserting this geodesic would make $v$ non-pointed. Thus, $v$ cannot be on this geodesic. Therefore the new edge is not incident to $v$ and the flip reduces the degree of $v$.
\end{proof}

Since the index always has degree three, this allows us to extend the results from Lemmas~\ref{lem:el-degree-2-free} and \ref{lem:el-degree-2-swap} regarding vertices of degree two to indexed edges.

\begin{corollary}
 \label{cor:el-index-free}
 In an indexed fan, we can interchange the label of the indexed edge with a free label in $O(1)$ flips.
\end{corollary}

\begin{corollary}
 \label{cor:el-index-swap}
 Given two indexed fans, we can interchange the labels of the two indexed edges with $O(1)$ flips.
\end{corollary}

Now we have enough tools to shuffle the bottom edges.
 
\begin{lemma}[Shuffle]
\label{lem:el-shuffle-non-pointed}
 In the left-shelling pseu\-do-tri\-an\-gu\-la\-tion, we can reorder the labels of all bottom edges with $O(n + h \log h)$ flips, where $h$ is the number of vertices on the convex hull.
\end{lemma}
\begin{proof}
 In the initial pseu\-do-tri\-an\-gu\-la\-tion, let $B$ and $\free$ be the sets of labels on bottom edges and free labels, respectively. Let $F_i$ be the set of labels on the $i$-th fan (in some fixed order), and let $\nf$ be the set of labels on non-fan bottom edges. Let $F_i'$ and $\nf'$ be these same sets in the target pseu\-do-tri\-an\-gu\-la\-tion. As we are only rearranging the bottom labels, we have that $B = F_1 \cup \ldots \cup F_k \cup \nf = F_1' \cup \ldots \cup F_k' \cup \nf'$, where $k$ is the number of fans.

 We say that a label $\ell$ \emph{belongs to} fan $i$ if $\ell \in F_i'$. At a high level, the reordering proceeds in four stages. In stage one, we free all labels in $\nf$. In stage two, we place each label from $B \setminus \nf'$ in the fan it belongs to, leaving the labels in $\nf'$ free. Then, in stage three, we correct the order of the labels within each fan. Finally, we place the labels in $\nf'$ correctly.

 Since each internal vertex contributes exactly one top edge, one bottom edge, and one free label, we have that $|\nf| = |\free|$. To free all labels in $\nf$, we perform a sweep (see~Lemma~\ref{lem:el-sweep}). As every internal vertex has degree two at some point during the sweep, we can exchange the label on its bottom edge with a free label at that point, using Lemma~\ref{lem:el-degree-2-free}. This requires $O(n)$ flips. The labels in $\free$ remain on the bottom edges incident to internal vertices throughout stage two and three, as placeholders.
 
 To begin stage two, we index all fans with $O(n)$ flips and shift these indices to the first `foreign' edge: the first edge whose label does not belong to the current fan. If no such edge exists, we can ignore this fan for the remainder of stage two, as it already has the right set of labels. Now suppose that there is a fan $F_i$ whose indexed edge $e$ is foreign: $\ell_e \notin F_i'$. Then either $\ell_e \in F_j'$ for some $j \neq i$, or $\ell_e \in \nf'$. In the first case, we exchange $\ell_e$ with the label on the indexed edge of $F_j$, and shift the index of $F_j$ to the next foreign edge. In the second case, we exchange $\ell_e$ with a free label in $B \setminus \nf'$. If this label belongs to $F_i$, we shift its index to the next foreign edge. In either case, we increased the number of correctly placed labels by at least one. Thus $n - 1$ repetitions suffice to place all labels in the fan they belong to, wrapping up stage two. Since we perform a linear number of swaps and shifts, and each takes a constant number of flips, the total number of flips required for stage two is $O(n)$.

 For stage three, we note that each indexed fan corresponds to a triangulation of a convex polygon. As such, we can rearrange the labelled diagonals of a fan $F_i$ into their desired final position with $O(|F_i| \log |F_i|)$ flips (Theorem~\ref{thm:el-convex-ub}). Thus, if we let $h$ be the number of vertices on the convex hull, the total number of flips for this step is bounded by
 \[ 
  \sum_i O(|F_i| \log |F_i|)~~\leq~~\sum_i O(|F_i| \log h)~~=~~O(h \log h).
 \]
 For stage four, we first return to a left-shelling pseu\-do-tri\-an\-gu\-la\-tion by un-indexing each fan, using $O(n)$ flips. After stage two, the labels in $\nf'$ are all free, so all that is left is to place these on the correct bottom edges, which we can do with a final sweep. Thus, we can reorder all bottom labels with $O(n + h \log h)$.
\end{proof}

This leads to the following bound.

\begin{theorem}
 We can transform any edge-la\-belled pseu\-do-tri\-an\-gu\-la\-tion with $n$ vertices into any other with $O(n \log c + h \log h)$ flips, where $c$ is the number of convex layers and $h$ is the number of vertices on the convex hull.
\end{theorem}
\begin{proof}
 Using the technique by Aichholzer~\etal~\cite{aichholzer2006transforming}, we first transform the pseu\-do-tri\-an\-gu\-la\-tion into the left-shelling pseu\-do-tri\-an\-gu\-la\-tion $T$ with $O(n \log c)$ flips. Our canonical pseu\-do-tri\-an\-gu\-la\-tion contains the labels $\{1, \ldots, 2n - h - 3\}$, but it is possible for $T$ to contain a different set of labels. Since all labels are drawn from $\{1, \ldots, 3n - 2h - 3\}$, at most $n - h$ labels differ. This is exactly the number of internal vertices. Thus, we can use $O(n + h \log h)$ flips to shuffle (Lemma~\ref{lem:el-shuffle-non-pointed}) all non-canonical labels on fan edges to bottom edges incident to an internal vertex. Once there, we use a sweep (Lemma~\ref{lem:el-sweep}) to ensure that every internal vertex has degree two at some point, at which time we replace its incident non-canonical labels with canonical ones with a constant number of flips (Lemma~\ref{lem:el-degree-2-free}). Once our left-shelling pseu\-do-tri\-an\-gu\-la\-tion has the correct set of labels, we use a constant number of shuffles and sweeps to sort the labels (Lemma~\ref{lem:el-sort-left-shelling}). Since we can shuffle and sweep with $O(n + h \log h)$ and $O(n)$ flips, respectively, the total number of flips reduces to $O(n \log c + n + h \log h) = O(n \log c + h \log h)$.
\end{proof}

The correspondence between triangulations of a convex polygon and pseu\-do-tri\-an\-gu\-la\-tions gives us the following lower bound.

\begin{theorem}
\label{thm:el-non-pointed-lower-bound}
There are pairs of edge-la\-belled pseu\-do-tri\-an\-gu\-la\-tions with $n$ vertices such that any sequence of flips that transforms one into the other has length $\Omega(n \log n)$.
\end{theorem}

\section{Conclusions and open problems}

We initiated the study of the diameter of the flip graph of edge-labelled triangulations in various settings. For edge-labelled triangulations of a convex polygon, we presented matching upper and lower bounds of $\Theta(n \log n)$. Allowing simultaneous flips brings the upper bound down to $O(\log^2 n)$, while our best lower bound in that setting is $\Omega(\log n)$. For edge-labelled combinatorial triangulations, we obtained the same tight $\Theta(n \log n)$ bound on the diameter of the flip graph, as well as the $O(\log^2 n)$ and $\Omega(\log n)$ upper- and lower bounds for simultaneous flips.

We also studied the diameter of the flip graph of edge-labelled pseudo-triangulations. Here, we showed that $O(n^2)$ exchanging flips suffice to transform any edge-labelled pointed pseudo-triangulation into any other, while $\Omega(n \log n)$ flips are sometimes necessary. By allowing insertion and deletion flips in addition to exchanging flips, we obtained a tight $\Theta(n \log n)$ bound on the shortest flip sequence between any two edge-labelled pseudo-triangulations, even non-pointed ones.

There is still a lot of room for future work. The most obvious set of open problems is closing the gaps between the $O(\log^2 n)$ and $\Omega(\log n)$ bounds in the simultaneous setting, and the $O(n^2)$ and $\Omega(n \log n)$ bounds for pointed pseudo-triangulations. The most likely solution for the latter is to prove a result similar to Lemma~\ref{lem:el-convex-partition} in this setting. This is easy when the vertices in the subsequence form a convex or concave chain, but handling alternations will be trickier.

The next set of open problems is to study what happens to the edge-labelled flip graph in non-convex polygons or point sets. Since it is possible for edges to be unflippable in these settings, we can never change the label of such edges with flips alone, resulting in a disconnected flip graph. In fact, each edge has a fixed set of other edges that it can be transformed into via flips. We call this set the \emph{orbit} of the edge. This gives rise to the following conjecture.

\begin{conjecture}[Orbit Conjecture]
 Given two edge-labelled triangulations, $T$ and $T'$, we can transform $T$ into $T'$ if and only if for every label $\ell$, the edge with label $\ell$ in $T'$ is in the orbit of the edge with label $\ell$ in $T$.
\end{conjecture}

For convex polygons, the Orbit Conjecture is implied by Theorem~\ref{thm:el-convex-ub}, since the orbit of each edge contains all other edges. In addition to convex polygons, the Orbit Conjecture also holds for polygons with a single reflex chain~\cite{pathak2014reconfiguring}. For general polygons, it is fairly easy to show that the condition is necessary, but we have not been able to show that it is also sufficient. Since triangulations of a set of points in the plane face the same difficulties, we believe that the Orbit Conjecture holds for that setting as well.

The final set of open problems relates to the computational hardness of finding the shortest sequence of flips that transforms one edge-labelled triangulation into the other. For unlabelled convex polygons, this question has been open for over 30 years~\cite{culik1982note}. Recently, the problem was shown to be NP-hard for triangulations of point sets~\cite{lubiw2012flip} and simple polygons~\cite{aichholzer2013flip}, but the case of convex polygons remains open.

We see two ways in which edge-labelled triangulations can help here. First, if we lift the restriction that all edge-labels have to be unique, the problem directly generalizes the unlabelled setting. Thus, an NP-hardness proof in this setting might generate techniques that can be reused for the unlabelled setting.

Second, note that every flip sequence in the unlabelled setting defines a bijection between edges of the initial and final triangulations. We can view the flip distance problem in the unlabelled setting as consisting of two steps: find the best bijection, and find a minimum flip sequence that realizes this bijection. The edge-labelled version of the problem is just the second step in this chain. Thus, settling the complexity of the edge-labelled version could give insight into which part of the unlabelled problem generates the complexity.

\bibliographystyle{plain}
\bibliography{../thesis}

\begin{thebibliography}{10}

\bibitem{avis1996reverse}
David Avis and Komei Fukuda.
\newblock Reverse search for enumeration.
\newblock {\em Discrete Applied Mathematics}, 65(1--3):21--46, 1996.

\bibitem{bern1992mesh}
Marshall Bern and David Eppstein.
\newblock Mesh generation and optimal triangulation.
\newblock In {\em Computing in {E}uclidean geometry}, volume~1 of {\em Lecture
  Notes Series on Computing}, pages 23--90. 1992.

\bibitem{bose2015optimal}
Prosenjit Bose, Rolf Fagerberg, Andr\'e van Renssen, and Sander Verdonschot.
\newblock Optimal local routing on {D}elaunay triangulations defined by empty
  equilateral triangles.
\newblock {\em SIAM Journal on Computing}.
\newblock Forthcoming.

\bibitem{bose2012competitive}
Prosenjit Bose, Rolf Fagerberg, Andr\'e van Renssen, and Sander Verdonschot.
\newblock Competitive routing in the half-{$\theta_6$}-graph.
\newblock In {\em Proceedings of the 23rd ACM-SIAM Symposium on Discrete
  Algorithms (SODA 2012)}, pages 1319--1328, 2012.

\bibitem{bose2012competitive2}
Prosenjit Bose, Rolf Fagerberg, Andr\'e van Renssen, and Sander Verdonschot.
\newblock Competitive routing on a bounded-degree plane spanner.
\newblock In {\em Proceedings of the 24th Canadian Conference on Computational
  Geometry (CCCG 2012)}, pages 299--304, 2012.

\bibitem{bose2011making}
Prosenjit Bose, Dana Jansens, Andr\'e van Renssen, Maria Saumell, and Sander
  Verdonschot.
\newblock Making triangulations 4-connected using flips.
\newblock In {\em Proceedings of the 23rd Canadian Conference on Computational
  Geometry (CCCG 2011)}, pages 241--247, 2011.

\bibitem{bose2012making}
Prosenjit Bose, Dana Jansens, Andr\'e van Renssen, Maria Saumell, and Sander
  Verdonschot.
\newblock Making triangulations 4-connected using flips.
\newblock {\em Computational Geometry: Theory and Applications},
  47(2A):187--197, 2014.
\newblock Special issue for CCCG 2011.

\bibitem{bose2013theta5}
Prosenjit Bose, Pat Morin, Andr\'e van Renssen, and Sander Verdonschot.
\newblock The {$\theta_5$}-graph is a spanner.
\newblock In {\em Proceedings of the 39th International Workshop on
  Graph-Theoretic Concepts in Computer Science (WG 2013)}, pages 100--114,
  2013.

\bibitem{bose2013theta5journal}
Prosenjit Bose, Pat Morin, Andr\'e van Renssen, and Sander Verdonschot.
\newblock The {$\theta_5$}-graph is a spanner.
\newblock {\em Computational Geometry: Theory and Applications},
  48(2):108--119, 2015.

\bibitem{bose2012history}
Prosenjit Bose and Sander Verdonschot.
\newblock A history of flips in combinatorial triangulations.
\newblock In {\em Proceedings of the XIV Spanish Meeting on Computational
  Geometry (EGC 2011)}, volume 7579 of {\em Lecture Notes in Computer Science},
  pages 29--44. 2012.

\bibitem{bose2015flips}
Prosenjit Bose and Sander Verdonschot.
\newblock Flips in edge-labelled pseudo-triangulations.
\newblock In {\em Proceedings of the 27th Canadian Conference on Computational
  Geometry (CCCG 2015)}, pages 63--69, 2015.

\bibitem{cooper2009flip}
Colin Cooper, Martin Dyer, and Andrew~J Handley.
\newblock The flip {M}arkov chain and a randomising {P2P} protocol.
\newblock In {\em Proceedings of the 28th ACM Symposium on Principles of
  Distributed Computing}, pages 141--150, 2009.

\bibitem{dekok2007generating}
Thierry de~Kok, Marc van Kreveld, and Maarten L{\"o}ffler.
\newblock Generating realistic terrains with higher-order {D}elaunay
  triangulations.
\newblock {\em Computational Geometry: Theory and Applications}, 36(1):52--65,
  2007.

\bibitem{lin1965computer}
Shen Lin.
\newblock Computer solutions of the traveling salesman problem.
\newblock {\em The Bell System Technical Journal}, 44(10):2245--2269, 1965.

\end{thebibliography}


\begin{thebibliography}{10}

\bibitem{althofer1993sparse}
Ingo Alth{\"o}fer, Gautam Das, David Dobkin, Deborah Joseph, and Jos{\'e}
  Soares.
\newblock On sparse spanners of weighted graphs.
\newblock {\em Discrete \& Computational Geometry}, 9(1):81--100, 1993.

\bibitem{barba2013new}
Luis Barba, Prosenjit Bose, Mirela Damian, Rolf Fagerberg, Wah~Loon Keng,
  Joseph O'Rourke, Andr\'e van Renssen, Perouz Taslakian, Sander Verdonschot,
  and Ge~Xia.
\newblock New and improved spanning ratios for {Y}ao graphs.
\newblock {\em Journal of Computational Geometry}, 6(2):19--53, 2015.
\newblock Special issue for SoCG 2014.

\bibitem{bose2012piArxiv}
Prosenjit Bose, Mirela Damian, Karim Dou{\"{\i}}eb, Joseph O'Rourke, Ben
  Seamone, Michiel Smid, and Stefanie Wuhrer.
\newblock {$\pi/2$}-angle {Y}ao graphs are spanners.
\newblock {\em ArXiv e-prints}, 2010.
\newblock \href{http://arxiv.org/abs/1001.2913}{arXiv:1001.2913} [cs.CG].

\bibitem{bose2012optimal}
Prosenjit Bose, Jean-Lou De~Carufel, Pat Morin, Andr\'e van Renssen, and Sander
  Verdonschot.
\newblock Optimal bounds on {T}heta-graphs: More is not always better.
\newblock In {\em Proceedings of the 24th Canadian Conference on Computational
  Geometry (CCCG 2012)}, pages 305--310, 2012.

\bibitem{bose2014towards}
Prosenjit Bose, Jean-Lou De~Carufel, Pat Morin, Andr\'e van Renssen, and Sander
  Verdonschot.
\newblock Towards tight bounds on {T}heta-graphs.
\newblock {\em Theoretical Computer Science}, 2014.
\newblock Forthcoming.

\bibitem{bose2004approximating}
Prosenjit Bose, Anil Maheshwari, Giri Narasimhan, Michiel Smid, and Norbert
  Zeh.
\newblock Approximating geometric bottleneck shortest paths.
\newblock {\em Computational Geometry: Theory and Applications},
  29(3):233--249, 2004.

\bibitem{chew1986there}
L.~Paul Chew.
\newblock There is a planar graph almost as good as the complete graph.
\newblock In {\em Proceedings of the 2nd Annual ACM symposium on Computational
  Geometry (SoCG 1986)}, pages 169--177, 1986.

\bibitem{chew1989there}
L.~Paul Chew.
\newblock There are planar graphs almost as good as the complete graph.
\newblock {\em Journal of Computer and System Sciences}, 39(2):205--219, 1989.

\bibitem{clarkson1987approximation}
Kenneth~L. Clarkson.
\newblock Approximation algorithms for shortest path motion planning.
\newblock In {\em Proceedings of the 19th ACM Symposium on the Theory of
  Computing (STOC 1987)}, pages 56--65, 1987.

\bibitem{dobkin1987delaunay}
David~P. Dobkin, Steven~J. Friedman, and Kenneth~J. Supowit.
\newblock Delaunay graphs are almost as good as complete graphs.
\newblock In {\em Proceedings of the 28th Annual Symposium on Foundations of
  Computer Science (FOCS 1987)}, pages 20--26, 1987.

\bibitem{eppstein1999spanning}
David Eppstein.
\newblock Spanning trees and spanners.
\newblock In J.R. Sack and J.~Urrutia, editors, {\em Handbook of Computational
  Geometry}, pages 425--461. Elsevier Science, 1999.

\bibitem{flinchbaugh1981strong}
B.~E. Flinchbaugh and L.~K. Jones.
\newblock Strong connectivity in directional nearest-neighbor graphs.
\newblock {\em SIAM Journal on Algebraic and Discrete Methods}, 2(4):461--463,
  1981.

\bibitem{keil1988approximating}
J.~Mark Keil.
\newblock Approximating the complete {E}uclidean graph.
\newblock In {\em Proceedings of the 1st Scandinavian Workshop on Algorithm
  Theory (SWAT 88)}, pages 208--213, 1988.

\bibitem{keil1989delaunay}
J.~Mark Keil and Carl~A. Gutwin.
\newblock The {D}elaunay triangulation closely approximates the complete
  {E}uclidean graph.
\newblock In {\em Proceedings of the 1st Workshop on Algorithms and Data
  Structures (WADS 1989)}, pages 47--56, 1989.

\bibitem{keil1992classes}
J.~Mark Keil and Carl~A. Gutwin.
\newblock Classes of graphs which approximate the complete {E}uclidean graph.
\newblock {\em Discrete \& Computational Geometry}, 7(1):13--28, 1992.

\bibitem{keng2013yao}
Wah~Loon Keng and Ge~Xia.
\newblock The {Y}ao graph {$Y_5$} is a spanner.
\newblock {\em ArXiv e-prints}, 2013.
\newblock \href{http://arxiv.org/abs/1307.5030}{arXiv:1307.5030} [cs.CG].

\bibitem{lukovski1999new}
Tam{\'a}s Lukovski.
\newblock {\em New results on geometric spanners and their applications}.
\newblock PhD thesis, Universit{\"a}t Paderborn, 1999.

\bibitem{narasimhan2007geometric}
Giri Narasimhan and Michiel Smid.
\newblock {\em Geometric spanner networks}.
\newblock Cambridge University Press, 2007.

\bibitem{ruppert1991approximating}
Jim Ruppert and Raimund Seidel.
\newblock Approximating the {$d$}-dimensional complete {E}uclidean graph.
\newblock In {\em Proceedings of the 3rd Canadian Conference on Computational
  Geometry (CCCG 1991)}, pages 207--210, 1991.

\bibitem{R2014ConstrainedSpanners}
Andr\'e van Renssen.
\newblock {\em Theta-graphs and other constrained spanners}.
\newblock {PhD} thesis, Carleton University, 2014.

\bibitem{xia2011improved}
Ge~Xia.
\newblock Improved upper bound on the stretch factor of {D}elaunay
  triangulations.
\newblock In {\em Proceedings of the 27th Annual Symposium on Computational
  Geometry (SoCG 2011)}, pages 264--273, 2011.

\bibitem{yao1982constructing}
Andrew Chi-Chih Yao.
\newblock On constructing minimum spanning trees in {$k$}-dimensional spaces
  and related problems.
\newblock {\em SIAM Journal on Computing}, 11(4):721--736, 1982.

\end{thebibliography}


\begin{thebibliography}{10}

\bibitem{barba2013new}
Luis Barba, Prosenjit Bose, Mirela Damian, Rolf Fagerberg, Wah~Loon Keng,
  Joseph O'Rourke, Andr\'e van Renssen, Perouz Taslakian, Sander Verdonschot,
  and Ge~Xia.
\newblock New and improved spanning ratios for {Y}ao graphs.
\newblock {\em Journal of Computational Geometry}, 6(2):19--53, 2015.
\newblock Special issue for SoCG 2014.

\bibitem{barba2013stretch}
Luis Barba, Prosenjit Bose, Jean-Lou De~Carufel, Andr\'e van Renssen, and
  Sander Verdonschot.
\newblock On the stretch factor of the {T}heta-4 graph.
\newblock In {\em Proceedings of the 13th Algorithms and Data Structures
  Symposium (WADS 2013)}, pages 109--120, 2013.

\bibitem{bonichon2010connections}
Nicolas Bonichon, Cyril Gavoille, Nicolas Hanusse, and David Ilcinkas.
\newblock Connections between theta-graphs, {D}elaunay triangulations, and
  orthogonal surfaces.
\newblock In {\em Proceedings of the 36th International Workshop on
  Graph-Theoretic Concepts in Computer Science (WG 2010)}, pages 266--278,
  2010.

\bibitem{bose2012pi}
Prosenjit Bose, Mirela Damian, Karim Dou{\"{\i}}eb, Joseph O'Rourke, Ben
  Seamone, Michiel Smid, and Stefanie Wuhrer.
\newblock {$\pi/2$}-angle {Y}ao graphs are spanners.
\newblock {\em International Journal of Computational Geometry \&
  Applications}, 22(1):61--82, 2012.

\bibitem{bose2013theta5}
Prosenjit Bose, Pat Morin, Andr\'e van Renssen, and Sander Verdonschot.
\newblock The {$\theta_5$}-graph is a spanner.
\newblock In {\em Proceedings of the 39th International Workshop on
  Graph-Theoretic Concepts in Computer Science (WG 2013)}, pages 100--114,
  2013.

\bibitem{bose2013theta5journal}
Prosenjit Bose, Pat Morin, Andr\'e van Renssen, and Sander Verdonschot.
\newblock The {$\theta_5$}-graph is a spanner.
\newblock {\em Computational Geometry: Theory and Applications},
  48(2):108--119, 2015.

\bibitem{chew1989there}
L.~Paul Chew.
\newblock There are planar graphs almost as good as the complete graph.
\newblock {\em Journal of Computer and System Sciences}, 39(2):205--219, 1989.

\bibitem{damian2012yao}
Mirela Damian and Kristin Raudonis.
\newblock Yao graphs span theta graphs.
\newblock {\em Discrete Mathematics, Algorithms and Applications},
  4(2):1250024, 16, 2012.

\bibitem{el2009yao}
Nawar~M. El~Molla.
\newblock {\em Yao spanners for wireless ad hoc networks}.
\newblock PhD thesis, Villanova University, 2009.

\bibitem{kanj2013geometric}
Iyad Kanj.
\newblock Geometric spanners: Recent results and open directions.
\newblock In {\em Proceedings of the 3rd International Conference on
  Communications and Information Technology (ICCIT 2013)}, pages 78--82, 2013.

\bibitem{morin2014average}
Pat Morin and Sander Verdonschot.
\newblock On the average number of edges in {T}heta graphs.
\newblock In {\em Proceedings of the 11th Meeting on Analytic Algorithmics and
  Combinatorics (ANALCO14)}, 2014.

\bibitem{dist2}
SATEL Oy.
\newblock What is a radio modem?
\newblock http://www.satel.com/products/what-is-a-radio-modem.
\newblock Accessed 13 Dec 2013.

\bibitem{ruppert1991approximating}
Jim Ruppert and Raimund Seidel.
\newblock Approximating the {$d$}-dimensional complete {E}uclidean graph.
\newblock In {\em Proceedings of the 3rd Canadian Conference on Computational
  Geometry (CCCG 1991)}, pages 207--210, 1991.

\bibitem{dist1}
Peter Rysavy.
\newblock Wireless broadband and other fixed-wireless systems.
\newblock http://www.networkcomputing.com/netdesign/bb1.html.
\newblock Accessed 13 Dec 2013.

\end{thebibliography}


\begin{thebibliography}{10}

\bibitem{angelini2010algorithm}
Patrizio Angelini, Fabrizio Frati, and Luca Grilli.
\newblock An algorithm to construct greedy drawings of triangulations.
\newblock {\em Journal of Graph Algorithms and Applications}, 14(1):19--51,
  2010.

\bibitem{baezayates1993searching}
Ricardo~A. Baeza-Yates, Joseph~C. Culberson, and Gregory J.~E. Rawlins.
\newblock Searching in the plane.
\newblock {\em Information and Computation}, 106(2):234--252, 1993.

\bibitem{barba2013new}
Luis Barba, Prosenjit Bose, Mirela Damian, Rolf Fagerberg, Wah~Loon Keng,
  Joseph O'Rourke, Andr\'e van Renssen, Perouz Taslakian, Sander Verdonschot,
  and Ge~Xia.
\newblock New and improved spanning ratios for {Y}ao graphs.
\newblock {\em Journal of Computational Geometry}, 6(2):19--53, 2015.
\newblock Special issue for SoCG 2014.

\bibitem{bonichon2010connections}
Nicolas Bonichon, Cyril Gavoille, Nicolas Hanusse, and David Ilcinkas.
\newblock Connections between theta-graphs, {D}elaunay triangulations, and
  orthogonal surfaces.
\newblock In {\em Proceedings of the 36th International Workshop on
  Graph-Theoretic Concepts in Computer Science (WG 2010)}, pages 266--278,
  2010.

\bibitem{bonichon2010plane}
Nicolas Bonichon, Cyril Gavoille, Nicolas Hanusse, and Ljubomir Perkovi{\'c}.
\newblock Plane spanners of maximum degree six.
\newblock In {\em Proceedings of the 37th International Colloquium on Automata,
  Languages and Programming (ICALP 2010 (1))}, pages 19--30, 2010.

\bibitem{bose2002online}
Prosenjit Bose, Andrej Brodnik, Svante Carlsson, Erik~D. Demaine, Rudolf
  Fleischer, Alejandro L{\'o}pez-Ortiz, Pat Morin, and J.~Ian Munro.
\newblock Online routing in convex subdivisions.
\newblock {\em International Journal of Computational Geometry {\&}
  Applications}, 12(4):283--295, 2002.

\bibitem{bose2012optimal}
Prosenjit Bose, Jean-Lou De~Carufel, Pat Morin, Andr\'e van Renssen, and Sander
  Verdonschot.
\newblock Optimal bounds on {T}heta-graphs: More is not always better.
\newblock In {\em Proceedings of the 24th Canadian Conference on Computational
  Geometry (CCCG 2012)}, pages 305--310, 2012.

\bibitem{bose2015optimal}
Prosenjit Bose, Rolf Fagerberg, Andr\'e van Renssen, and Sander Verdonschot.
\newblock Optimal local routing on {D}elaunay triangulations defined by empty
  equilateral triangles.
\newblock {\em SIAM Journal on Computing}.
\newblock Forthcoming.

\bibitem{bose2012competitive}
Prosenjit Bose, Rolf Fagerberg, Andr\'e van Renssen, and Sander Verdonschot.
\newblock Competitive routing in the half-{$\theta_6$}-graph.
\newblock In {\em Proceedings of the 23rd ACM-SIAM Symposium on Discrete
  Algorithms (SODA 2012)}, pages 1319--1328, 2012.

\bibitem{bose2012competitive2}
Prosenjit Bose, Rolf Fagerberg, Andr\'e van Renssen, and Sander Verdonschot.
\newblock Competitive routing on a bounded-degree plane spanner.
\newblock In {\em Proceedings of the 24th Canadian Conference on Computational
  Geometry (CCCG 2012)}, pages 299--304, 2012.

\bibitem{bose2004online}
Prosenjit Bose and Pat Morin.
\newblock Online routing in triangulations.
\newblock {\em SIAM Journal on Computing}, 33(4):937--951, 2004.

\bibitem{bose2013spanning}
Prosenjit Bose, Andr\'e van Renssen, and Sander Verdonschot.
\newblock On the spanning ratio of {T}heta-graphs.
\newblock In {\em Proceedings of the 13th Algorithms and Data Structures
  Symposium (WADS 2013)}, pages 182--194, 2013.

\bibitem{chew1989there}
L.~Paul Chew.
\newblock There are planar graphs almost as good as the complete graph.
\newblock {\em Journal of Computer and System Sciences}, 39(2):205--219, 1989.

\bibitem{dhandapani2010greedy}
Raghavan Dhandapani.
\newblock Greedy drawings of triangulations.
\newblock {\em Discrete {\&} Computational Geometry}, 43(2):375--392, 2010.

\bibitem{dillencourt1990realizability}
Michael~B. Dillencourt.
\newblock Realizability of {D}elaunay triangulations.
\newblock {\em Information Processing Letters}, 33(6):283--287, 1990.

\bibitem{goodrich2009succinct}
Michael~T. Goodrich and Darren Strash.
\newblock Succinct greedy geometric routing in the {E}uclidean plane.
\newblock In {\em Proceedings of the 20th International Symposium on Algorithms
  and Computation (ISAAC 2009)}, pages 781--791, 2009.

\bibitem{he2010schnyder}
Xin He and Huaming Zhang.
\newblock Schnyder greedy routing algorithm.
\newblock In {\em Proceedings of the 7th annual conference on Theory and
  Applications of Models of Computation (TAMC 2010)}, pages 271--283, 2010.

\bibitem{he2011succinct}
Xin He and Huaming Zhang.
\newblock On succinct convex greedy drawing of 3-connected plane graphs.
\newblock In {\em Proceedings of the 22nd annual ACM-SIAM Symposium on Discrete
  Algorithms (SODA 2011)}, pages 1477--1486, 2011.

\bibitem{keil1992classes}
J.~Mark Keil and Carl~A. Gutwin.
\newblock Classes of graphs which approximate the complete {E}uclidean graph.
\newblock {\em Discrete \& Computational Geometry}, 7(1):13--28, 1992.

\bibitem{leighton2010some}
Tom Leighton and Ankur Moitra.
\newblock Some results on greedy embeddings in metric spaces.
\newblock {\em Discrete {\&} Computational Geometry}, 44(3):686--705, 2010.

\bibitem{misra2009guide}
Sudip Misra, Isaac Woungang, and Subhas~Chandra Misra.
\newblock {\em Guide to wireless sensor networks}.
\newblock Springer, 2009.

\bibitem{papadimitriou2005conjecture}
Christos~H. Papadimitriou and David Ratajczak.
\newblock On a conjecture related to geometric routing.
\newblock {\em Theoretical Computer Science}, 344(1):3--14, 2005.

\bibitem{racke2009survey}
Harald R{\"a}cke.
\newblock Survey on oblivious routing strategies.
\newblock In {\em Proceedings of the 5th Conference on Computability in Europe
  (CiE 2009)}, pages 419--429, 2009.

\bibitem{schnyder1990embedding}
Walter Schnyder.
\newblock Embedding planar graphs on the grid.
\newblock In {\em Proceedings of the 1st annual ACM-SIAM Symposium on Discrete
  Algorithms (SODA 1990)}, pages 138--148, 1990.

\end{thebibliography}


\begin{thebibliography}{10}

\bibitem{asano1984linear}
Takao Asano, Shunji Kikuchi, and Nobuji Saito.
\newblock A linear algorithm for finding {H}amiltonian cycles in 4-connected
  maximal planar graphs.
\newblock {\em Discrete Applied Mathematics}, 7(1):1--15, 1984.

\bibitem{bose2009flips}
Prosenjit Bose and Ferran Hurtado.
\newblock Flips in planar graphs.
\newblock {\em Computational Geometry: Theory and Applications}, 42(1):60--80,
  2009.

\bibitem{bose2012history}
Prosenjit Bose and Sander Verdonschot.
\newblock A history of flips in combinatorial triangulations.
\newblock In {\em Proceedings of the XIV Spanish Meeting on Computational
  Geometry (EGC 2011)}, volume 7579 of {\em Lecture Notes in Computer Science},
  pages 29--44. 2012.

\bibitem{frati2015lower}
Fabrizio Frati.
\newblock A lower bound on the diameter of the flip graph.
\newblock {\em ArXiv e-prints}, 2015.
\newblock \href{http://arxiv.org/abs/1508.03473}{arXiv:1508.03473} [cs.CG].

\bibitem{gao2001diagonal}
Zhicheng Gao, Jorge Urrutia, and Jianyu Wang.
\newblock Diagonal flips in labelled planar triangulations.
\newblock {\em Graphs and Combinatorics}, 17(4):647--657, 2001.

\bibitem{karp1972reducibility}
Richard~M. Karp.
\newblock Reducibility among combinatorial problems.
\newblock In Raymond~E. Miller and James~W. Thatcher, editors, {\em Complexity
  of Computer Computations}, pages 85--103. 1972.

\bibitem{komuro1997diagonal}
Hideo Komuro.
\newblock The diagonal flips of triangulations on the sphere.
\newblock {\em Yokohama Mathematical Journal}, 44(2):115--122, 1997.

\bibitem{mori2003diagonal}
Ryuichi Mori, Atsuhiro Nakamoto, and Katsuhiro Ota.
\newblock Diagonal flips in {H}amiltonian triangulations on the sphere.
\newblock {\em Graphs and Combinatorics}, 19(3):413--418, 2003.

\bibitem{negami1993diagonal}
Seiya Negami and Atsuhiro Nakamoto.
\newblock Diagonal transformations of graphs on closed surfaces.
\newblock {\em Science Reports of the Yokohama National University. Section I.
  Mathematics, Physics, Chemistry}, (40):71--97, 1993.

\bibitem{sleator1992short}
Daniel~D. Sleator, Robert~E. Tarjan, and William~P. Thurston.
\newblock Short encodings of evolving structures.
\newblock {\em SIAM Journal on Discrete Mathematics}, 5(3):428--450, 1992.

\bibitem{wagner1936bemerkungen}
Klaus Wagner.
\newblock Bemerkungen zum vierfarbenproblem.
\newblock {\em Jahresbericht der Deutschen Mathematiker-Vereinigung},
  46:26--32, 1936.

\bibitem{whitney1931theorem}
Hassler Whitney.
\newblock A theorem on graphs.
\newblock {\em Annals of Mathematics, Second Series}, 32(2):378--390, 1931.

\end{thebibliography}


\begin{thebibliography}{10}

\bibitem{aichholzer2008triangulations}
Oswin Aichholzer, Clemens Huemer, and Hannes Krasser.
\newblock Triangulations without pointed spanning trees.
\newblock {\em Computational Geometry: Theory and Applications}, 40(1):79--83,
  2008.

\bibitem{asano1984linear}
Takao Asano, Shunji Kikuchi, and Nobuji Saito.
\newblock A linear algorithm for finding {H}amiltonian cycles in 4-connected
  maximal planar graphs.
\newblock {\em Discrete Applied Mathematics}, 7(1):1--15, 1984.

\bibitem{bose2007simultaneous}
Prosenjit Bose, Jurek Czyzowicz, Zhicheng Gao, Pat Morin, and David~R. Wood.
\newblock Simultaneous diagonal flips in plane triangulations.
\newblock {\em Journal of Graph Theory}, 54(4):307--330, 2007.

\bibitem{bose2011making}
Prosenjit Bose, Dana Jansens, Andr\'e van Renssen, Maria Saumell, and Sander
  Verdonschot.
\newblock Making triangulations 4-connected using flips.
\newblock In {\em Proceedings of the 23rd Canadian Conference on Computational
  Geometry (CCCG 2011)}, pages 241--247, 2011.

\bibitem{bose2012making}
Prosenjit Bose, Dana Jansens, Andr\'e van Renssen, Maria Saumell, and Sander
  Verdonschot.
\newblock Making triangulations 4-connected using flips.
\newblock {\em Computational Geometry: Theory and Applications},
  47(2A):187--197, 2014.
\newblock Special issue for CCCG 2011.

\bibitem{cardinal2015arc}
Jean Cardinal, Michael Hoffmann, Vincent Kusters, Csaba~D. T{\'o}th, and Manuel
  Wettstein.
\newblock Arc diagrams, flip distances, and {H}amiltonian triangulations.
\newblock In {\em Proceedings of the 32nd International Symposium on
  Theoretical Aspects of Computer Science (STACS 2015)}, pages 197--210, 2015.

\bibitem{frati2015lower}
Fabrizio Frati.
\newblock A lower bound on the diameter of the flip graph.
\newblock {\em ArXiv e-prints}, 2015.
\newblock \href{http://arxiv.org/abs/1508.03473}{arXiv:1508.03473} [cs.CG].

\bibitem{komuro1997diagonal}
Hideo Komuro.
\newblock The diagonal flips of triangulations on the sphere.
\newblock {\em Yokohama Mathematical Journal}, 44(2):115--122, 1997.

\bibitem{mori2003diagonal}
Ryuichi Mori, Atsuhiro Nakamoto, and Katsuhiro Ota.
\newblock Diagonal flips in {H}amiltonian triangulations on the sphere.
\newblock {\em Graphs and Combinatorics}, 19(3):413--418, 2003.

\bibitem{sierpinski1915courbe}
Wac{\l}aw Sierpi{\'n}ski.
\newblock Sur une courbe dont tout point est un point de ramification.
\newblock {\em Compte Rendus hebdomadaires des s{\'e}ance de l'Acad{\'e}mie des
  Science de Paris}, 160:302--305, 1915.

\bibitem{sleator1992short}
Daniel~D. Sleator, Robert~E. Tarjan, and William~P. Thurston.
\newblock Short encodings of evolving structures.
\newblock {\em SIAM Journal on Discrete Mathematics}, 5(3):428--450, 1992.

\bibitem{whitney1931theorem}
Hassler Whitney.
\newblock A theorem on graphs.
\newblock {\em Annals of Mathematics, Second Series}, 32(2):378--390, 1931.

\end{thebibliography}


\begin{thebibliography}{10}

\bibitem{aichholzer2006transforming}
Oswin Aichholzer, Franz Aurenhammer, Clemens Huemer, and Hannes Krasser.
\newblock Transforming spanning trees and pseudo-triangulations.
\newblock {\em Information Processing Letters}, 97(1):19--22, 2006.

\bibitem{aichholzer2003pseudotriangulations}
Oswin Aichholzer, Franz Aurenhammer, Hannes Krasser, and Peter Brass.
\newblock Pseudotriangulations from surfaces and a novel type of edge flip.
\newblock {\em SIAM Journal on Computing}, 32(6):1621--1653, 2003.

\bibitem{aichholzer2013flip}
Oswin Aichholzer, Wolfgang Mulzer, and Alexander Pilz.
\newblock Flip distance between triangulations of a simple polygon is
  {NP}-complete.
\newblock {\em Discrete {\&} Computational Geometry}, 54(2):368--389, 2015.

\bibitem{araujo2015colorful}
Gabriela Araujo-Pardo, Isabel Hubard, Deborah Oliveros, and Egon Schulte.
\newblock Colorful associahedra and cyclohedra.
\newblock {\em Journal of Combinatorial Theory, Series A}, 129:122--141, 2015.

\bibitem{bereg2004transforming}
Sergey Bereg.
\newblock Transforming pseudo-triangulations.
\newblock {\em Information Processing Letters}, 90(3):141--145, 2004.

\bibitem{bose2007simultaneous}
Prosenjit Bose, Jurek Czyzowicz, Zhicheng Gao, Pat Morin, and David~R. Wood.
\newblock Simultaneous diagonal flips in plane triangulations.
\newblock {\em Journal of Graph Theory}, 54(4):307--330, 2007.

\bibitem{bose2015flips}
Prosenjit Bose and Sander Verdonschot.
\newblock Flips in edge-labelled pseudo-triangulations.
\newblock In {\em Proceedings of the 27th Canadian Conference on Computational
  Geometry (CCCG 2015)}, pages 63--69, 2015.

\bibitem{cano2013edge}
Javier Cano, Jos{\'e}-Miguel D{\'\i}az-B{\'a}{\~n}ez, Clemens Huemer, and Jorge
  Urrutia.
\newblock The edge rotation graph.
\newblock {\em Graphs and Combinatorics}, 29(5):1207--1219, 2013.

\bibitem{cardinal2015arc}
Jean Cardinal, Michael Hoffmann, Vincent Kusters, Csaba~D. T{\'o}th, and Manuel
  Wettstein.
\newblock Arc diagrams, flip distances, and {H}amiltonian triangulations.
\newblock In {\em Proceedings of the 32nd International Symposium on
  Theoretical Aspects of Computer Science (STACS 2015)}, pages 197--210, 2015.

\bibitem{culik1982note}
Karel Culik, II and Derick Wood.
\newblock A note on some tree similarity measures.
\newblock {\em Information Processing Letters}, 15(1):39--42, 1982.

\bibitem{galtier2003simultaneous}
Jer{\^o}me Galtier, Ferran Hurtado, Marc Noy, St{\'e}phane P{\'e}rennes, and
  Jorge Urrutia.
\newblock Simultaneous edge flipping in triangulations.
\newblock {\em International Journal of Computational Geometry \&
  Applications}, 13(2):113--133, 2003.

\bibitem{hernando2003grafos}
Carmen Hernando, Ferran Hurtado, Merc{\`e} Mora, and Eduardo Rivera-Campo.
\newblock Grafos de {\'a}rboles etiquetados y grafos de {\'a}rboles
  geom{\'e}tricos etiquetados.
\newblock In {\em Proceedings of the X Spanish Meeting on Computational
  Geometry (Encuentros de Geometra Computacional)}, pages 13--19, 2003.

\bibitem{hurtado1999flipping}
Ferran Hurtado, Marc Noy, and Jorge Urrutia.
\newblock Flipping edges in triangulations.
\newblock {\em Discrete \& Computational Geometry}, 22(3):333--346, 1999.

\bibitem{kirkpatrick2002kinetic}
David Kirkpatrick, Jack Snoeyink, and Bettina Speckmann.
\newblock Kinetic collision detection for simple polygons.
\newblock {\em International Journal of Computational Geometry \&
  Applications}, 12(1-2):3--27, 2002.

\bibitem{lawson1972transforming}
Charles~L. Lawson.
\newblock Transforming triangulations.
\newblock {\em Discrete Mathematics}, 3(4):365--372, 1972.

\bibitem{lubiw2012flip}
Anna Lubiw and Vinayak Pathak.
\newblock Flip-distance between two triangulations of a point-set is
  {NP}-complete.
\newblock In {\em Proceedings of the 23rd Canadian Conference on Computational
  Geometry (CCCG 2012)}, pages 119--124, 2012.

\bibitem{pathak2014reconfiguring}
Vinayak Pathak.
\newblock {\em Reconfiguring triangulations}.
\newblock PhD thesis, University of Waterloo, 2014.

\bibitem{rote2007pseudotriangulations}
G{\"u}nter Rote, Francisco Santos, and Ileana Streinu.
\newblock Pseudo-triangulations---a survey.
\newblock In {\em Surveys on discrete and computational geometry}, volume 453
  of {\em Contemporary Mathematics}, pages 343--410. 2008.

\bibitem{sleator1988rotation}
Daniel~D. Sleator, Robert~E. Tarjan, and William~P. Thurston.
\newblock Rotation distance, triangulations, and hyperbolic geometry.
\newblock {\em Journal of the American Mathematical Society}, 1(3):647--681,
  1988.

\bibitem{sleator1992short}
Daniel~D. Sleator, Robert~E. Tarjan, and William~P. Thurston.
\newblock Short encodings of evolving structures.
\newblock {\em SIAM Journal on Discrete Mathematics}, 5(3):428--450, 1992.

\bibitem{streinu2005pseudo}
Ileana Streinu.
\newblock Pseudo-triangulations, rigidity and motion planning.
\newblock {\em Discrete \& Computational Geometry}, 34(4):587--635, 2005.

\bibitem{wagner1936bemerkungen}
Klaus Wagner.
\newblock Bemerkungen zum vierfarbenproblem.
\newblock {\em Jahresbericht der Deutschen Mathematiker-Vereinigung},
  46:26--32, 1936.

\end{thebibliography}


\begin{thebibliography}{}

\end{thebibliography}

\cleardoublepage\pagestyle{empty}

\hfill

\vfill

\pdfbookmark[0]{Colophon}{colophon}
\section*{Colophon}
This document was typeset using the typographical look-and-feel \texttt{classicthesis} developed by Andr\'e Miede. 
The style was inspired by Robert Bringhurst's seminal book on typography ``\emph{The Elements of Typographic Style}''. 
\texttt{classicthesis} is available for both \LaTeX\ and \mLyX: 
\begin{center}
\url{http://code.google.com/p/classicthesis/}
\end{center}
Happy users of \texttt{classicthesis} usually send a real postcard to the author, a collection of postcards received so far is featured here: 
\begin{center}
\url{http://postcards.miede.de/}
\end{center}
 
\bigskip

\noindent\finalVersionString

\end{document}